\documentclass[11pt]{article}
\usepackage[utf8]{inputenc}
\usepackage[T1]{fontenc}
\usepackage{lmodern}
\usepackage{amsmath, amsthm, amssymb, mathtools}
\usepackage{float}
\usepackage{hyperref}
\usepackage[letterpaper,margin=1in]{geometry}
\usepackage{enumitem}
\usepackage[oldenum,olditem]{paralist}
\usepackage{xcolor}
\usepackage{hyperref}
\usepackage{cancel}
\usepackage[suppress]{color-edits}
\usepackage{bbm}
\usepackage[nameinlink,noabbrev,capitalise]{cleveref}
\usepackage{physics}
\usepackage{graphicx}
\usepackage{pgfplots}
\pgfplotsset{compat=1.18}
\usepgfplotslibrary{fillbetween}
\usetikzlibrary{arrows.meta,decorations.pathreplacing}
\usepackage[outline]{contour}
\usepackage{transparent}
\usepackage{thmtools}
\usepackage{thm-restate}
\usepackage{tablefootnote}
\usepackage[normalem]{ulem}

\usepackage{import}
\usepackage{fancyhdr}

\hypersetup{
  colorlinks   = true,
  linkcolor    = blue,
  citecolor    = black,
  filecolor    = blue,
  urlcolor     = blue
}

\newtheorem{theorem}{Theorem}[section]

\newtheorem{lemma}[theorem]{Lemma}
\newtheorem{observation}[theorem]{Observation}
\newtheorem{proposition}[theorem]{Proposition}
\newtheorem{corollary}[theorem]{Corollary}
\newtheorem{definition}[theorem]{Definition}

\crefname{observation}{Observation}{Observations}
\Crefname{observation}{Observation}{Observations}
\crefname{claim}{Claim}{Claims}
\Crefname{claim}{Claim}{Claims}
\crefname{conjecture}{Conjecture}{Conjectures}
\Crefname{conjecture}{Conjecture}{Conjectures}

\DeclarePairedDelimiter{\paren}{\lparen}{\rparen}

\DeclarePairedDelimiter{\bracket}{[}{]}
\DeclarePairedDelimiter{\oc}{(}{]}
\DeclarePairedDelimiter{\co}{[}{)}
\DeclarePairedDelimiter{\set}{\lbrace}{\rbrace}

\def\aug{\mathsf{Aug}}

\newcommand{\residualValue}[2]{\ensuremath{#1\vert_{#2}}}
\newcommand{\effectiveValue}[2]{\ensuremath{#1\vert_{#2}}}

\newcommand{\utility}{\mathsf{util}}
\newcommand{\utilityMechanism}[2]{\ensuremath{\utility_{#1}\!\left(#2\right)}}
\newcommand{\allocationUtilityMechanism}[3]{\ensuremath{\utility_{#1}\!\left(#2, #3\right)}}
\newcommand{\utilityAB}[2]{\ensuremath{\utility_{\AB}\!}\left(#1,#2\right)}
\newcommand{\utilityBA}[2]{\ensuremath{\utility_{\BA}\!}\left(#1,#2\right)}

\newcommand{\utilityDuo}[2]{\ensuremath{\utility\!\left(#1,#2\right)}}
\newcommand{\utilityTri}[3]{\ensuremath{\utility\!\left(#1,#2,#3\right)}}

\newcommand{\allocDuo}[3]{\ensuremath{\mathsf{x}_{\mathsf{#1}}\!\left(#2,#3\right)}}
\newcommand{\allocTri}[4]{\ensuremath{\mathsf{x}_{\mathsf{#1}}\!\left(#2,#3,#4\right)}}
\newcommand{\allocTriCond}[5]{\ensuremath{\mathsf{x}^{\mathsf{#1}}_{\mathsf{#2}}\!\left(#3,#4,#5\right)}}
\newcommand{\zmarg}[1]{\ensuremath{\mathtt{z}_{#1}}}
\newcommand{\pmarg}[1]{\ensuremath{\mathtt{pay}_{#1}}}
\newcommand{\Pmarg}[1]{\ensuremath{P_{#1}}}
\newcommand{\AliceMechanism}{\ensuremath{\mathcal{A}}}
\newcommand{\BobMechanism}{\ensuremath{\mathcal{B}}}

\newcommand{\rev}{\mathsf{rev}}
\newcommand{\revDuo}[3]{\ensuremath{\rev_{\mathsf{#1}}\!\left(#2, #3\right)}}
\newcommand{\revDuoOrder}[4]{\ensuremath{\rev^{#1}_{\mathsf{#2}}\!\left(#3, #4\right)}}
\newcommand{\revTri}[4]{\ensuremath{\rev_{\mathsf{#1}}\!\left(#2, #3, #4\right)}}

\newcommand{\opt}[2]{\ensuremath{\mathsf{opt}_{\mathsf{A}}\!\left(#1;#2\right)}}
\newcommand{\optOrder}[3]{\ensuremath{\mathsf{opt}^{#3}_{\mathsf{A}}\!\left(#1;#2\right)}}

\newcommand{\SL}{\mathsf{SL}}
\newcommand{\DL}{\mathsf{DL}}
\newcommand{\p}{\mathtt{p}}
\newcommand{\vthreshold}{\mathtt{v}}
\newcommand{\x}{\mathtt{x}}

\newcommand{\del}{\mathrm{d}}

\newcommand{\FP}[1]{\ensuremath{\mathsf{FP}(#1)}}

\newcommand{\Alice}{\ensuremath{A}}
\newcommand{\Bob}{\ensuremath{B}}

\newcommand{\AllMechanisms}{\ensuremath{\mathcal{C}}}
\newcommand{\AllStackelberg}{\ensuremath{\mathcal{S}}}
\newcommand{\ActionPlans}{\ensuremath{\Pi}}

\newcommand{\AB}{\ensuremath{\mathsf{AB}}}
\newcommand{\BA}{\ensuremath{\mathsf{BA}}}

\addauthor[Moshe]{mb}{red}          %

\newcommand{\mbc}[1]{\mbcomment{}}
\newcommand{\mbd}[1]{\mbdelete{}}
\addauthor[AI]{ai}{brown}         %
\addauthor[Yuval]{yg}{teal}         %
\addauthor[Brendan]{bl}{purple}      %
\addauthor[Raghuvansh]{rrs}{orange}  %
\newcommand{\rrsnote}[1]{\rrscomment{}}

\title{On Competing Sellers with Substitutable Goods}
\author{Moshe Babaioff\thanks{{Hebrew University of Jerusalem.}} \and Yuval Grofman\footnotemark[1]
\and Brendan Lucier\thanks{{Microsoft Research New England.}} \and
Raghuvansh R. Saxena\thanks{{Tata Institute of Fundamental Research, Mumbai.}}}

\date{November 4, 2026}

\begin{document}

\hypersetup{pageanchor=false}
\pagenumbering{gobble}

\begingroup
\makeatletter
\let\@fnsymbol\@arabic
\makeatother
\maketitle
\endgroup

\begin{abstract}
We study mechanism design by an incumbent seller facing competition from an entrant seller of a substitute good.
The incumbent commits to a potentially randomized mechanism before her competitor responds. A unit-demand buyer with fixed, publicly known values for the two goods observes both mechanisms and may interact with them in either order. How should a revenue-maximizing incumbent design her mechanism when her product is superior or inferior to her competitor's?

We determine the incumbent's exact optimal revenue and characterize all mechanisms that obtain it. Commitment and randomization allow her to earn strictly positive revenue even against an arbitrarily superior product, despite earning nothing under posted-price (i.e., Bertrand) competition.
Optimal mechanisms can 
require a continuum of lottery options, even though the buyer selects only one in equilibrium: {unselected} %
options shape the competitor's incentives rather than screen private buyer information.
We also show that the buyer's ability to choose the interaction order can be strictly beneficial to the incumbent, even though the incumbent is always visited first at equilibrium. Finally, we show that without commitment, every pure Nash equilibrium yields Bertrand revenues even with randomized mechanisms.

\end{abstract}

\clearpage
\tableofcontents
\newpage

\clearpage
\pagenumbering{arabic}
\hypersetup{pageanchor=true}
\newpage
\section{Introduction}
\label{sec:introduction}

Sellers nearly always face competitive pressure from other {sellers who offer similar goods that might be} 
better or worse. A buyer choosing between two software products, for example, may prefer the features of one but ultimately purchase the other if its price is sufficiently attractive. In such environments, a seller's ability to generate revenue depends both on the appeal of her own product and on the offers available from her competitors. 
We therefore model the sellers %
as competitors in a game where each is an unrestricted mechanism designer:
rather than restricting competition to posted prices, we allow each seller to choose an arbitrary sales mechanism, possibly involving multiple lotteries. How should a seller design her mechanism in anticipation of an imperfectly competitive environment, where the buyer could also purchase a competing product {that is an imperfect substitute}
from a rival seller who can design his mechanism as well?

Imagine an incumbent seller, Alice, who can commit to a mechanism for selling her good but faces competition from another seller, Bob, {who sells an imperfect substitute for Alice's good}.
Bob observes Alice's mechanism and then chooses his own mechanism in response, giving rise to a Stackelberg game between the sellers. A single potential buyer observes both mechanisms and chooses which seller to approach first. After interacting with that seller, possibly obtaining a good and paying a transfer, the buyer may approach the other seller as well. For example, the buyer might purchase a lottery from Alice and, conditional on walking away empty-handed, turn to Bob. The goods are {imperfect
substitutes, so} %
the buyer may prefer either seller's product. Our focus is on how Alice should use her commitment power when her product is superior or inferior to Bob's from the buyer's perspective. Importantly, Alice's ability to commit first \emph{does not} require the buyer to visit her first. %

To make these questions concrete, suppose the buyer's values for Alice's and Bob's goods are fixed and publicly known, and write them as $a>0$ and $b>0$, respectively. 
{The buyer is unit-demand, so if the buyer obtains both goods, his value is $\max\{a,b\}$.}
The familiar posted-price benchmark is Bertrand competition: Alice earns revenue $a-b$ if her good is more valuable, and zero
otherwise. This quantity $(a-b)_+=\max\{a-b,0\}$ is her \emph{Bertrand revenue}.
Restricting the sellers to posted prices gives Alice this same Bertrand revenue at {every} equilibrium; {this is true even when she can commit to her price before Bob responds, where the equilibrium concept becomes subgame perfect equilibrium.}
\emph{Does allowing arbitrary, possibly randomized mechanisms change the {equilibrium} outcome?  
Can Alice generate positive revenue when her product is inferior, or more than the difference in values when it is superior?  More generally, what is the maximum revenue she can obtain, and what must her mechanism look like to attain it?}

One might think that fixed and publicly known buyer values make this design problem trivial. After all, a monopolist facing a buyer with a known value could simply post a price equal to that value, bypassing the mechanism design problem entirely. Under competition, however, the buyer's willingness to pay for Alice's good depends on what Bob offers. A more attractive fallback from Bob can reduce what the buyer is willing to pay Alice. With unequal values, the interaction is richer still: after obtaining the less valuable good, the buyer may remain willing to purchase the more valuable one. The buyer's choice of interaction order and subsequent purchases must therefore be considered together. Even in this public-information setting, willingness to pay is endogenous to the competing mechanisms.

This dependence makes Alice's design problem very different from selling to a single, known-value buyer in isolation. Different mechanisms offered by Bob can lead the buyer to select different options from Alice's menu. Alice can therefore strategically offer lotteries that are not chosen on the equilibrium path but nevertheless affect the profitability of possible deviations by Bob. An optimal mechanism must account not only for the offer the buyer eventually selects, but also for how the entire menu shapes the competitor's incentives. 
As we will show, attaining the optimum can require a continuum of menu options,
despite the buyer's publicly known values. These options do not screen privately known preferences; instead, they {influence how the buyer would behave if Bob were to deviate to other potential mechanisms, which in turn guides Bob's choice of mechanism at equilibrium.}
This creates a connection to screening problems with multiple buyer types   
that we exploit in our analysis, even though there is only one buyer type in our model.

Lucier and Saxena~\cite{lucier2025approximately} study competition between mechanisms for homogeneous goods, corresponding to the case $a=b$. They show that commitment and randomization can together sustain substantial revenue, obtaining approximation guarantees relative to a seller's revenue as a monopolist. Their work leaves open the exact optimal mechanism. We study {a more general model, allowing also for differentiated goods, and we} determine Alice's \emph{exact optimal revenue}, characterizing all mechanisms that attain it. Our aim is thus not only to understand whether mechanism design can improve on Bertrand competition, but to identify how the optimal mechanism uses commitment to shape competition, and how its structure depends on the relative values of the sellers' goods.

\subsection{Our Contribution.}
\label{sec:introduction-contribution}

We first consider the model described above, which we call our \emph{main model}, in which Alice first commits to her (possibly randomized) mechanism, then Bob chooses his mechanism in response, and then finally the buyer is free to choose which seller to visit first, {and he sequentially interacts with the two mechanisms of the sellers in the chosen order}.
We determine Alice's maximum expected revenue over all subgame-perfect equilibria and characterize all mechanisms attaining it.
For the statements below, we incorporate the buyer's responses into
the sellers' payoffs and consider the resulting Stackelberg game between Alice and Bob.
We assume the buyer breaks ties by first maximizing Bob's
expected revenue and then Alice's.\footnote{
  Remaining tie-breaking rules are specified in the model section.
}
The sellers' mechanisms
then uniquely determine the buyer's
expected utility, each seller's expected revenue, the marginal
allocation probability from each seller, and the interaction order.
We call a mechanism \emph{optimal for Alice}
if it occurs in a Stackelberg equilibrium that 
maximizes her expected
revenue among all such equilibria.

\begin{figure}[H]
    \centering
    \begingroup%
\definecolor{MainBertrandBlue}{HTML}{3F51A6}%
\definecolor{MainBertrandCurve}{HTML}{5F79E6}%
\definecolor{MainBertrandOrange}{HTML}{C58A20}%
\definecolor{MainBertrandGrid}{HTML}{D5D5D5}%
\definecolor{MainBertrandGuide}{HTML}{AEBBC6}%
\def\MainBertrandWidth{10cm}%
\def\MainBertrandHeight{6cm}%
\def\MainBertrandXAxisTitleY{-0.12}%
\def\MainBertrandYAxisTitleX{-0.085}%
\def\MainBertrandLegendX{0.02}%
\def\MainBertrandLegendY{0.91}%
\def\MainBertrandLegendRowGap{0.095}%
\def\MainBertrandOptLabelX{0.175}%
\def\MainBertrandOptLabelY{0.317}%
\def\MainBertrandOptLabelSize{10}%
\def\MainBertrandOptLineWidth{2.2pt}%
\def\MainBertrandBaseLineWidth{1.2pt}%
\def\MainBertrandCutoffGuideWidth{0.65pt}%
\def\MainBertrandCoincidenceLineLength{10pt}%
\def\MainBertrandCoincidenceLineWidth{1.3pt}%
\def\MainBertrandEqualityX{0.717}%
\def\MainBertrandEqualityY{0.294}%
\def\MainBertrandEqualityArrowEndX{\MainBertrandC}%
\def\MainBertrandEqualityArrowEndY{\MainBertrandC-1}%
\def\MainBertrandEqualityArrowGap{7pt}%
\def\MainBertrandHaloScale{2.2}%
\def\MainBertrandHaloStrength{3.5}%
\pgfmathsetlengthmacro{\MainBertrandHaloOuter}{max(0,\MainBertrandHaloScale)*0.65pt}%
\pgfmathsetlengthmacro{\MainBertrandHaloMiddle}{max(0,\MainBertrandHaloScale)*0.40pt}%
\pgfmathsetlengthmacro{\MainBertrandHaloCore}{max(0,\MainBertrandHaloScale)*0.18pt}%
\pgfmathsetmacro{\MainBertrandHaloOpacityOuter}{min(1,max(0,0.15*\MainBertrandHaloStrength))}%
\pgfmathsetmacro{\MainBertrandHaloOpacityMiddle}{min(1,max(0,0.30*\MainBertrandHaloStrength))}%
\pgfmathsetmacro{\MainBertrandHaloOpacityCore}{min(1,max(0,0.70*\MainBertrandHaloStrength))}%
\newcommand*\MainBertrandHalo[1]{%
  \begingroup
  \ifdim\MainBertrandHaloOuter>0pt\relax
    \rlap{\begingroup\transparent{\MainBertrandHaloOpacityOuter}\color{white}%
      \contourlength{\MainBertrandHaloOuter}\contour{white}{#1}\endgroup}%
    \rlap{\begingroup\transparent{\MainBertrandHaloOpacityMiddle}\color{white}%
      \contourlength{\MainBertrandHaloMiddle}\contour{white}{#1}\endgroup}%
    \rlap{\begingroup\transparent{\MainBertrandHaloOpacityCore}\color{white}%
      \contourlength{\MainBertrandHaloCore}\contour{white}{#1}\endgroup}%
  \fi
  #1\endgroup
}%
\def\MainBertrandC{3.1461932206205825}%
\def\MainBertrandAtEquality{0.351693340639234}%
\begin{tikzpicture}[
  font=\fontsize{8}{10}\selectfont,
  baseline=(MainBertrandAxis.south west),
  main bertrand text/.style={align=center,inner sep=1.5pt,outer sep=0pt},
  main bertrand arrow/.style={draw=black!80,line width=0.95pt,
    -{Stealth[length=5pt,width=3.5pt]},shorten >=3.5pt},
]
\begin{axis}[
  name=MainBertrandAxis,
  width=\MainBertrandWidth,height=\MainBertrandHeight,scale only axis,
  xmin=0,xmax=4.05,ymin=0,ymax=3.15,
  axis x line*=bottom,axis y line*=left,
  axis line style={black,line width=0.7pt},
  tick align=outside,tick style={black,line width=0.6pt},
  major tick length=3pt,
  xtick={0,1,2,\MainBertrandC,4},
  xticklabels={$0$,$1$,$2$,{$C\approx3.146$},$4$},
  ytick={0,1,2,3},yticklabels={$0$,$b$,$2b$,$3b$},
  tick label style={font=\fontsize{8}{9}\selectfont},
  scaled ticks=false,
  xlabel={Value ratio $a/b$},ylabel={Alice's revenue},
  label style={font=\fontsize{9}{11}\selectfont},
  xlabel style={at={(axis description cs:0.5,\MainBertrandXAxisTitleY)},
    anchor=north,inner sep=0pt},
  ylabel style={at={(axis description cs:\MainBertrandYAxisTitleX,0.5)},
    anchor=south,inner sep=0pt},
  grid=major,grid style={MainBertrandGrid,line width=0.35pt,opacity=0.7},
  set layers=standard,axis on top=false,clip=false,
  every axis plot/.append style={forget plot},
]
\begin{pgfonlayer}{axis background}
  \path[fill=MainBertrandBlue!5,draw=none]
    (axis cs:0,0) rectangle (axis cs:1,3.15);
\end{pgfonlayer}
\addplot[name path=MainBertrandOptPath,draw=none]
  table[x=ratio,y=optimal] {observations/introduction_tikz_plots/main_vs_bertrand_revenue_data.dat};
\path[name path=MainBertrandBasePath]
  (axis cs:0,0)--(axis cs:1,0)--(axis cs:4,3);
\addplot[fill=MainBertrandBlue,fill opacity=0.17,draw=none]
  fill between[of=MainBertrandOptPath and MainBertrandBasePath,
    soft clip={domain=0:\MainBertrandC}];
\draw[MainBertrandGuide,line width=0.65pt,dash pattern=on 2pt off 3pt]
  (axis cs:1,0)--(axis cs:1,3.15);
\draw[MainBertrandGuide,line width=\MainBertrandCutoffGuideWidth,
  dash pattern=on 2pt off 3pt]
  (axis cs:\MainBertrandC,0)--(axis cs:\MainBertrandC,3.15);
\addplot[MainBertrandCurve,line width=\MainBertrandOptLineWidth,no marks]
  table[x=ratio,y=optimal] {observations/introduction_tikz_plots/main_vs_bertrand_revenue_data.dat};
\addplot[MainBertrandOrange,line width=\MainBertrandBaseLineWidth,
  dash pattern=on 4pt off 3pt,no marks]
  coordinates {(0,0) (1,0) (4,3)};
\addplot[only marks,mark=*,mark size=1.8pt,MainBertrandCurve]
  coordinates {(1,\MainBertrandAtEquality)};
\draw[black,line width=\MainBertrandCoincidenceLineWidth]
  ([yshift={-0.5*\MainBertrandCoincidenceLineLength}]axis cs:\MainBertrandC,{\MainBertrandC-1})--
  ([yshift={0.5*\MainBertrandCoincidenceLineLength}]axis cs:\MainBertrandC,{\MainBertrandC-1});

\draw[MainBertrandCurve,line width=\MainBertrandOptLineWidth]
  (axis description cs:\MainBertrandLegendX,\MainBertrandLegendY)--
  (axis description cs:{\MainBertrandLegendX+0.075},\MainBertrandLegendY);
\node[main bertrand text,anchor=west,text=MainBertrandBlue,
  font=\fontsize{9.5}{11}\selectfont]
  at (axis description cs:{\MainBertrandLegendX+0.095},\MainBertrandLegendY)
  {\MainBertrandHalo{Optimal revenue $\opt{a}{b}$}};
\draw[MainBertrandOrange,line width=\MainBertrandBaseLineWidth,
  dash pattern=on 4pt off 3pt]
  (axis description cs:\MainBertrandLegendX,
    {\MainBertrandLegendY-\MainBertrandLegendRowGap})--
  (axis description cs:{\MainBertrandLegendX+0.075},
    {\MainBertrandLegendY-\MainBertrandLegendRowGap});
\node[main bertrand text,anchor=west,text=MainBertrandOrange,
  font=\fontsize{9.5}{11}\selectfont]
  at (axis description cs:{\MainBertrandLegendX+0.095},
    {\MainBertrandLegendY-\MainBertrandLegendRowGap})
  {\MainBertrandHalo{Bertrand revenue $(a-b)_+$}};

\node[main bertrand text,text=MainBertrandBlue,
  font=\fontsize{\MainBertrandOptLabelSize}{12}\selectfont]
  (MainBertrandOptLabel)
  at (axis description cs:\MainBertrandOptLabelX,\MainBertrandOptLabelY)
  {\MainBertrandHalo{$\opt{a}{b}\approx0.352b$}};
\draw[main bertrand arrow,MainBertrandBlue!85!black]
  (MainBertrandOptLabel.south) to[out=-90,in=130]
  (axis cs:1,\MainBertrandAtEquality);

\node[main bertrand text,text=MainBertrandCurve,font=\fontsize{8.5}{10}\selectfont]
  (MainBertrandEquality)
  at (axis description cs:\MainBertrandEqualityX,\MainBertrandEqualityY)
  {\MainBertrandHalo{Revenues coincide}\\\MainBertrandHalo{thereafter}};
\draw[main bertrand arrow,shorten >=\MainBertrandEqualityArrowGap]
  (MainBertrandEquality.north) to[out=70,in=-100]
  (axis cs:\MainBertrandEqualityArrowEndX,{\MainBertrandEqualityArrowEndY});
\end{axis}
\end{tikzpicture}%
\endgroup%
    \caption{\small{Alice's {\textcolor[HTML]{5F79E6}{optimal revenue $\mathsf{opt}(a;b)$}} in the main model (solid indigo) and her {\textcolor[HTML]{C58A20}{Bertrand revenue $(a-b)_+$}} (dashed amber), for fixed $b>0$. Vertical ticks are in units of $b$.}}
    \label{fig:main-vs-bertrand-revenue}
\end{figure}
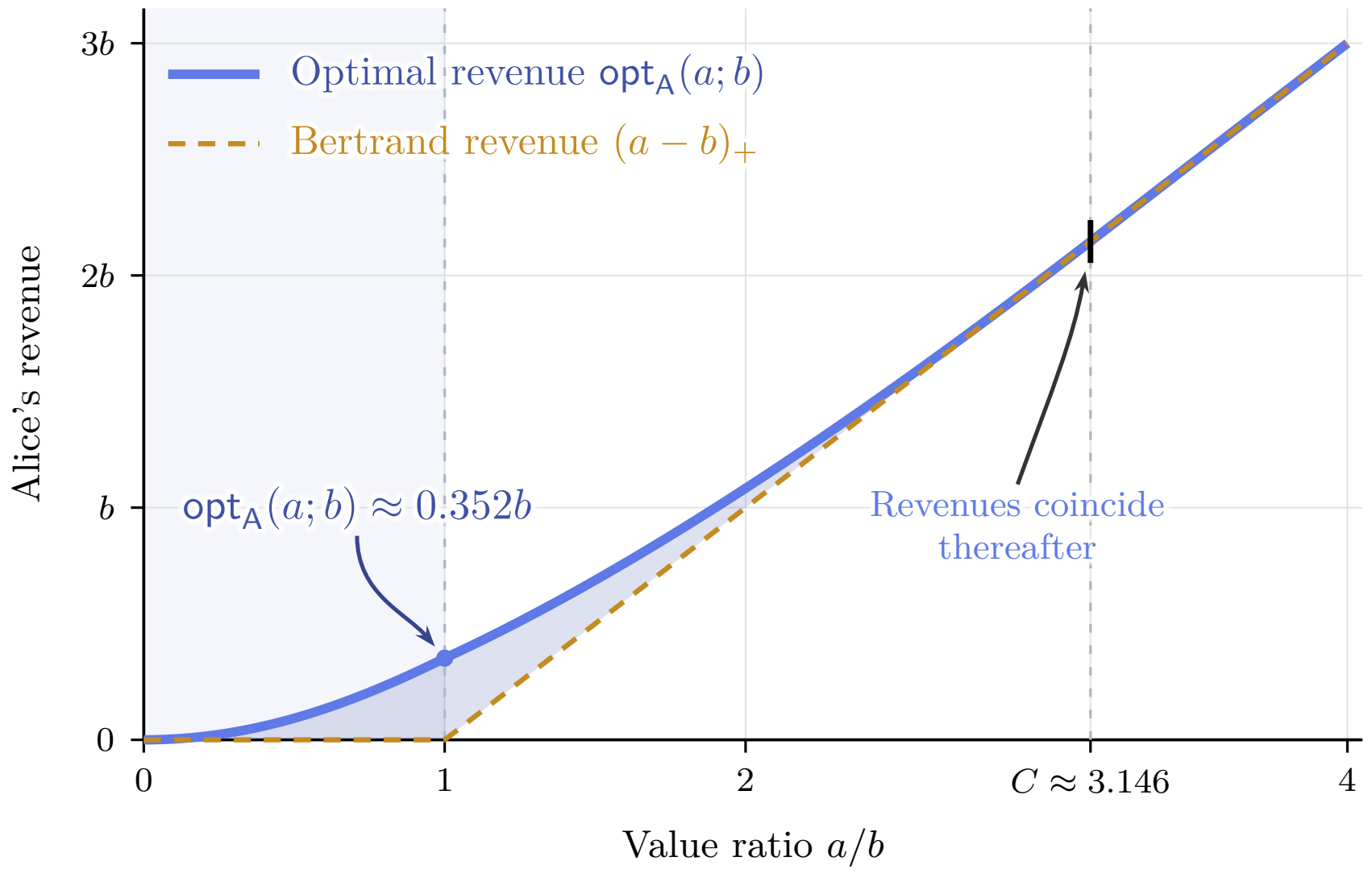

We find Alice's optimal revenue (see \cref{fig:main-vs-bertrand-revenue}) 
and characterize all
optimal mechanisms of Alice.
Surprisingly,
Alice exceeds her Bertrand revenue even when
her good's value is
arbitrarily small relative to Bob's.
For fixed $b$, this difference decreases as $a$ rises from $b$,
until it vanishes.

\begin{restatable}[Main Model]{theorem}{introductionmainmodel}    \label{thm:introduction-variable-order-rephrased}
    Consider an instance of the duopoly game in the main model
    with fixed buyer values $a,b>0$.
    {{There exists
    a complete and explicit characterization of all mechanisms that are Alice-optimal in a Stackelberg equilibrium 
    {(\cref{prop:main-optimal-mechanisms})},
    and it defines the optimal revenue for Alice.}
    The equilibrium utilities of all three agents (Alice, Bob and the buyer) are independent of the equilibrium selection.

    For each Alice-optimal mechanism, posting a price of 
    $b$ is a best response
    for Bob.
    Under this response, it is optimal for
    the buyer to visit Alice first, select a specific lottery from her (point $E$ in \cref{fig:introduction-alice-pricing-menu}), 
    and buy from Bob if and only if
    {he got nothing from Alice's lottery.}
    
    Alice's optimal revenue strictly exceeds her Bertrand revenue $(a-b)_+$ if $a<C \cdot b$, 
    and equals it otherwise, where
    $C\in(3.146,3.147)$ satisfies $C-\ln C=2$.
    When $a<C\cdot b$, all these mechanisms \emph{share}  an infinite continuum of menu
    options
    {%
    ending at 
    the lottery selected 
      by the buyer (point $E$).}%
      \footnote{%
      Different optimal menus can offer menu entries with a price 
      greater than that of $E$,  
      which are irrelevant to the payoffs of all three agents.
    }
    }

\end{restatable}
We conclude that commitment power
combined with randomization
within the mechanism 
allows Alice to exceed
her Bertrand revenue if and only if
$a<C\cdot b$,
that is, when
her good is not too much more
valuable than Bob's.
This reveals an asymmetry between the sellers, 
{due to Alice's position as the leader in the Stackelberg game. Alice}
{manages} to earn positive revenue even when
her good's value
is arbitrarily small compared to
the value of Bob's good.
When
Bob's good is sufficiently less valuable to the buyer than
Alice's ($a\geq C\cdot b$), however,
she optimally posts the price $a-b$
and sells with certainty, leaving Bob with zero revenue.
For fixed $b$, Alice's gain over Bertrand decreases as $a$ rises
from $b$, until it vanishes.

To provide some intuition into Theorem~\ref{thm:introduction-variable-order-rephrased}, it turns out that an optimal strategy for Alice is to arrange for Bob's best response to be the same mechanism he would use if Alice weren't present: posting a fixed price of $b$. Such a mechanism from Bob generates $0$ utility for the buyer, which maximizes the buyer's willingness to purchase from Alice. While this is convenient for Alice, inducing this behavior from Bob limits the set of menus that Alice can pick from. Our characterization shows that this strategy is nevertheless optimal over all possible mechanisms that Alice could choose.

When $a \geq C \cdot b$,
Alice's unique\footnote{
  {Throughout the paper, uniqueness claims and statements specifying
  a seller's mechanism (e.g. ``Bob posts a price of $b$'') are understood
  up to \emph{{essential equivalence}}.}
  {Informally, this is \emph{outcome equivalence} in the
  Stackelberg game between Alice and Bob: 
  two mechanisms are essentially equivalent
  if they induce the
  same expected payoffs, visiting order, and allocation behavior against
  every mechanism of the other seller under the stated tie-breaking rules.}
  See \cref{def:strategic-equivalence} for the formal definition.
} 
optimal mechanism
is a posted price of $a-b$.
For $a < C \cdot b$, 
we find that every optimal mechanism of Alice 
offers an infinite continuum of lottery options,
even though the buyer selects only one
in equilibrium: the off-path options are necessary, 
even though they are never chosen by the buyer, to deter profitable deviations by Bob.

More concretely,
when visiting Alice first,
the buyer's willingness to pay for
her good
depends on his value $a$ and the continuation utility available
from Bob. We call this adjusted willingness to pay his
\emph{effective buyer type}
when interacting with Alice.
The effective buyer type
lies between zero and $a$ and decreases as the utility
he forgoes from Bob upon receiving
Alice's good increases.
These effective types vary across Bob's possible responses;
they are not private information about the buyer.

To induce Bob to post price $b$, Alice must defend against \bledit{potential deviating responses. This involves striking a balance on how much revenue to leave to Bob at equilibrium:\footnote{Note that in any equilibrium where the buyer visits Alice first and Bob sets a posted price of $b$, Bob's revenue depends only on the probability of allocation in the lottery the buyer chooses from Alice's menu.} higher revenue to Bob leaves less for Alice, but also makes it easier to prevent Bob from obtaining yet higher revenue from deviations (including those that lead the buyer to visit Alice first as well as responses that attract the buyer to Bob first).} For example, Bob might sell at a fixed price less than $b$, offering positive utility to entice the buyer away from Alice's mechanism.
Alice's lower-priced lotteries would be selected under such a deviation. Her menu controls how the allocation bought from her changes as Bob cuts his price, so that his increased probability of sale does not make the price cut profitable.

\begin{figure}[ht]
    \centering
    \resizebox{0.96\textwidth}{!}{%
        \begingroup%
\definecolor{csrevblue}{HTML}{246298}%
\definecolor{csrevorange}{HTML}{B95B17}%
\definecolor{csrevgrid}{HTML}{D5D5D5}%
\def\RevenueXAxisTitleY{-0.17}%
\def\RevenueD{0.66625979861295}%
\def\RevenueC{3.1461932206205825}%
\def\RevenueCGap{-0.0136722582928751}%
\begin{tikzpicture}[font=\fontsize{8}{10}\selectfont,
  baseline=(current axis.south west),
  csrevarrow/.style={draw=black!85,line width=0.95pt,
    -{Stealth[length=2mm,width=1.4mm]},shorten <=2pt,shorten >=3.5pt}]
\begin{axis}[
  width=5.6cm, height=4.8cm, scale only axis,
  xmin=0, xmax=8, ymin=-0.029, ymax=0.004,
  axis x line*=bottom, axis y line*=left,
  axis line style={black, line width=0.6pt},
  tick align=outside, tick style={black, line width=0.5pt},
  major tick length=2.5pt,
  xtick={0,2,4,6,8},
  ytick={-0.02,-0.01,0},
  yticklabels={$-2\%$,$-1\%$,$0\%$},
  scaled ticks=false,
  set layers=standard,
  grid=major, grid style={csrevgrid, line width=0.3pt},
  xlabel={Value ratio $a/b$},
  ylabel={$\Delta_{\mathsf{opt}}(a;b)/a$},
  label style={font=\fontsize{9}{11}\selectfont},
  xlabel style={at={(axis description cs:0.5,\RevenueXAxisTitleY)}, anchor=north,
    inner sep=0pt},
  ylabel style={at={(axis description cs:-0.16,0.5)}, anchor=south,
    inner sep=0pt},
  clip=false,
]
  \begin{pgfonlayer}{axis background}
    \path[fill=csrevblue!6, draw=none]
      (axis cs:0,-0.029) rectangle (axis cs:1,0.004);
  \end{pgfonlayer}
  \addplot[name path=csrevcurve, draw=none, forget plot]
    table[x=ratio,y=gap] {observations/introduction_tikz_plots/revenue_difference_data.dat};
  \path[name path=csrevzero] (axis cs:0,0) -- (axis cs:8,0);
  \addplot[fill=csrevblue!40, draw=none, forget plot]
    fill between[of=csrevcurve and csrevzero,
      soft clip={domain=0:\RevenueD}];
  \addplot[fill=csrevorange!24, draw=none, forget plot]
    fill between[of=csrevcurve and csrevzero,
      soft clip={domain=\RevenueD:8}];
  \draw[black!60,line width=0.45pt]
    (axis cs:0,0) -- (axis cs:8,0);
  \draw[black!45,densely dashed,line width=0.4pt]
    (axis cs:1,-0.029) -- (axis cs:1,0.004);
  \draw[black!45,densely dashed,line width=0.4pt]
    (axis cs:\RevenueC,-0.029) -- (axis cs:\RevenueC,0.004);
  \addplot[csrevblue,line width=1.25pt,no marks]
    table[x=ratio,y=gap] {observations/introduction_tikz_plots/revenue_difference_data.dat};
  \addplot[only marks,mark=*,mark size=1.5pt,black!75]
    coordinates {(\RevenueD,0) (\RevenueC,\RevenueCGap)};

  \node[csrevblue,anchor=east,inner sep=0pt]
    at (axis cs:7.85,0.0022)
    {$\mathsf{opt}_{\mathsf A}(a;b)>
       \mathsf{opt}^{\mathsf{AB}}_{\mathsf A}(a;b)$};
  \node[csrevorange,anchor=east,inner sep=0pt]
    at (axis cs:7.85,-0.0105)
    {$\mathsf{opt}_{\mathsf A}(a;b)<
       \mathsf{opt}^{\mathsf{AB}}_{\mathsf A}(a;b)$};
  \node[anchor=west,inner sep=1pt]
    (csrevDlabel) at (axis cs:1.30,-0.0061) {$D\approx0.6663$};
  \draw[csrevarrow]
    (csrevDlabel.north west) -- (axis cs:\RevenueD,0);
  \node[anchor=west,inner sep=1pt,black!85]
    (csrevClabel) at (axis cs:4.2,-0.0254) {$C\approx3.1462$};
  \draw[csrevarrow]
    (csrevClabel.north west) -- (axis cs:\RevenueC,\RevenueCGap);
\end{axis}
\end{tikzpicture}%
\endgroup%
%
        \hspace{0.70cm}%
        \begingroup%
\providecommand{\aiedit}[1]{#1}%
\definecolor{AliceMenuBlue}{HTML}{246298}%
\definecolor{AliceMenuGreen}{HTML}{08745A}%
\definecolor{AliceMenuOrange}{HTML}{B95B17}%
\definecolor{AliceMenuGuide}{HTML}{C9D1D9}%
\definecolor{AliceMenuGrid}{HTML}{D5D5D5}%
\def\MenuXAxisTitleY{-0.17}%
\def\MenuSegmentLabelY{-0.05}%
\def\MenuSegmentLabelXShift{-2pt}%
\def\MenuStraightLabelXShift{2pt}%
\def\MenuStraightLabelYShift{-1.5pt}%
\def\MenuCurvedLabelXShift{1pt}%
\def\MenuCurvedLabelYShift{0pt}%
\def\MenuExtensionLabelXShift{1.5pt}%
\def\MenuExtensionLabelYShift{0pt}%
\def\MenuShowXOne{0}%
\def\MenuSharedX{0.428}%
\def\MenuSharedY{0.687}%
\def\MenuDiscountX{0.175}%
\def\MenuDiscountY{0.36}%
\def\MenuLegendY{0.94}%
\def\MenuParametersX{0.035}%
\def\MenuParametersY{0.88}%
\def\MenuParametersGap{0.06em}%
\def\MenuParameterAXShift{0pt}%
\def\MenuTextHaloScale{2.5}%
\def\MenuTextHaloStrength{3}%
\def\MenuTextHaloRadius{0.65pt}%
\def\MenuTextHaloMiddle{0.40pt}%
\def\MenuTextHaloCore{0.18pt}%
\pgfmathsetlengthmacro{\MenuHaloOuterWidth}{max(0,\MenuTextHaloScale)*\MenuTextHaloRadius}%
\pgfmathsetlengthmacro{\MenuHaloMiddleWidth}{max(0,\MenuTextHaloScale)*\MenuTextHaloMiddle}%
\pgfmathsetlengthmacro{\MenuHaloInnerWidth}{max(0,\MenuTextHaloScale)*\MenuTextHaloCore}%
\pgfmathsetmacro{\MenuHaloOuterOpacity}{min(1,max(0,0.15*\MenuTextHaloStrength))}%
\pgfmathsetmacro{\MenuHaloMiddleOpacity}{min(1,max(0,0.30*\MenuTextHaloStrength))}%
\pgfmathsetmacro{\MenuHaloInnerOpacity}{min(1,max(0,0.70*\MenuTextHaloStrength))}%
\newcommand*\MenuTextHalo[1]{%
  \begingroup
  \ifdim\MenuHaloOuterWidth>0pt\relax
  \rlap{\begingroup\transparent{\MenuHaloOuterOpacity}\color{white}%
    \contourlength{\MenuHaloOuterWidth}\contour{white}{#1}\endgroup}%
  \rlap{\begingroup\transparent{\MenuHaloMiddleOpacity}\color{white}%
    \contourlength{\MenuHaloMiddleWidth}\contour{white}{#1}\endgroup}%
  \rlap{\begingroup\transparent{\MenuHaloInnerOpacity}\color{white}%
    \contourlength{\MenuHaloInnerWidth}\contour{white}{#1}\endgroup}%
  \fi
  #1%
  \endgroup
}%
\def\MenuBLabelXShift{-6pt}%
\def\MenuBLabelYShift{0.5pt}%
\def\MenuELabelXShift{-6pt}%
\def\MenuELabelYShift{0.5pt}%
\def\AliceMenuRevenue{0.1587351932054160}%
\def\AliceMenuUtility{0.5991031710146508}%
\def\AliceMenuBase{0.5991031710146508}%
\def\AliceMenuBasePayment{0.6942019286156388}%
\def\AliceMenuEquilibrium{0.8412648067945840}%
\def\AliceMenuEquilibriumPayment{1.0834264425745173}%
\def\AliceMenuCorner{0.2701731743288348}%
\def\AliceMenuTerminalPayment{1.4008968289853492}%
\begin{tikzpicture}[
  baseline=(AliceMenuMain.south west),
  menu text halo/.style={inner sep=1.5pt,outer sep=0pt},
  declare function={
    menuoriginal(\x)=(1+\AliceMenuRevenue)*\x;
    menulower(\x)=\x+max(0,\AliceMenuRevenue*
      ln((1-\AliceMenuCorner)/(1-\x)));
    menucore(\x)=\AliceMenuBasePayment+(\x-\AliceMenuBase)
      +\AliceMenuRevenue*ln((1-\AliceMenuBase)/(1-\x));
  },
]
\begin{axis}[
  name=AliceMenuMain,
  width=7.2cm, height=4.8cm, scale only axis,
  xmin=0, xmax=1.025, ymin=0, ymax=1.52,
  axis lines=left, axis line style={line width=0.8pt,-},
  tick align=outside, tick style={black,line width=0.8pt},
  major tick length=3.5pt,
  xtick={0,0.25,0.5,0.75,1},
  xticklabels={$0$,{},$1/2$,{},{$\ifnum\MenuShowXOne=1\relax 1\fi$}},
  ytick={0,0.5,1,1.5},
  yticklabels={$0$,$a/4$,$a/2$,$3\cdot a/4$},
  tick label style={font=\fontsize{8}{9}\selectfont},
  xticklabel style={yshift=-1pt},
  yticklabel style={xshift=-1pt},
  xlabel={Allocation probability, $x$},
  xlabel style={at={(axis description cs:0.5,\MenuXAxisTitleY)},anchor=north,inner sep=0pt,
    font=\fontsize{9}{10}\selectfont},
  ylabel={Expected payment, $\tau$},
  ylabel style={at={(axis description cs:-0.11,0.5)},anchor=south,
    font=\fontsize{9}{10}\selectfont},
  grid=major, grid style={AliceMenuGrid,line width=0.45pt,opacity=0.65},
  axis on top=false, clip=false,
  every axis plot/.append style={forget plot},
]
\addplot[name path=AliceMenuUpper,draw=none,domain=0:\AliceMenuBase,
  samples=2] {menuoriginal(x)};
\addplot[name path=AliceMenuLower,draw=none,
  samples at={0,\AliceMenuCorner,0.28,0.29,...,0.59,\AliceMenuBase}]
  {menulower(x)};
\addplot[fill=AliceMenuGreen,fill opacity=0.32,draw=none]
  fill between[of=AliceMenuUpper and AliceMenuLower];

\draw[AliceMenuGuide,line width=0.65pt,dash pattern=on 2pt off 4pt]
  (axis cs:\AliceMenuBase,0) --
  (axis cs:\AliceMenuBase,\AliceMenuBasePayment);
\draw[AliceMenuGuide,line width=0.65pt,dash pattern=on 2pt off 4pt]
  (axis cs:\AliceMenuEquilibrium,0) --
  (axis cs:\AliceMenuEquilibrium,\AliceMenuEquilibriumPayment);
\draw[AliceMenuGuide,line width=0.65pt,dash pattern=on 2pt off 4pt]
  (axis cs:1,0) -- (axis cs:1,\AliceMenuTerminalPayment);

\addplot[AliceMenuGreen,line width=0.65pt,opacity=0.70,
  samples at={0,\AliceMenuCorner,0.28,0.29,...,0.59,\AliceMenuBase}]
  {0.5*menuoriginal(x)+0.5*menulower(x)};
\addplot[AliceMenuGreen,line width=0.65pt,dash pattern=on 3pt off 1.7pt,
  samples at={0,\AliceMenuCorner,0.28,0.29,...,0.59,\AliceMenuBase}]
  {menulower(x)};
\addplot[AliceMenuBlue,line width=1.6pt,domain=0:\AliceMenuBase,
  samples=2] {menuoriginal(x)};
\addplot[AliceMenuBlue,line width=1.6pt,
  domain=\AliceMenuBase:\AliceMenuEquilibrium,samples=101]
  {menucore(x)};
\addplot[AliceMenuOrange,line width=1.4pt,dash pattern=on 5pt off 3pt,
  domain=\AliceMenuEquilibrium:1,samples=2]
  {2*x-\AliceMenuUtility};
\addplot[only marks,mark=*,mark size=1.5pt,AliceMenuBlue]
  coordinates {(0,0) (\AliceMenuBase,\AliceMenuBasePayment)
    (\AliceMenuEquilibrium,\AliceMenuEquilibriumPayment)};
\addplot[only marks,mark=o,mark size=1.7pt,AliceMenuOrange,
  mark options={solid,fill=white,line width=0.8pt}]
  coordinates {(1,\AliceMenuTerminalPayment)};

\node[menu text halo,font=\fontsize{9}{10}\selectfont,text=AliceMenuBlue,
  anchor=south,xshift=\MenuBLabelXShift,yshift=\MenuBLabelYShift]
  at (axis cs:\AliceMenuBase,\AliceMenuBasePayment) {\MenuTextHalo{$B$}};
\node[menu text halo,font=\fontsize{9}{10}\selectfont,text=AliceMenuBlue,
  anchor=south,xshift=\MenuELabelXShift,yshift=\MenuELabelYShift]
  at (axis cs:\AliceMenuEquilibrium,\AliceMenuEquilibriumPayment) {\MenuTextHalo{$E$}};

\node[menu text halo,align=center,font=\fontsize{8}{9}\selectfont\bfseries,
  text=AliceMenuBlue] (AliceMenuShared)
  at (axis description cs:\MenuSharedX,\MenuSharedY)
  {\MenuTextHalo{Shared by all}\\\MenuTextHalo{Optimal Mechanisms}};
\draw[AliceMenuBlue!85!black,line width=0.95pt,
  -{Stealth[length=5pt,width=3.5pt]},
  shorten >=3pt]
  (AliceMenuShared.south) to[out=-90,in=155]
  (axis cs:{(\AliceMenuBase+\AliceMenuEquilibrium)/2},
    {menucore((\AliceMenuBase+\AliceMenuEquilibrium)/2)});
\node[menu text halo,align=center,font=\fontsize{8}{9}\selectfont,
  text=AliceMenuGreen] (AliceMenuDiscount)
  at (axis description cs:\MenuDiscountX,\MenuDiscountY)
  {\MenuTextHalo{Optional discount}\\\MenuTextHalo{by Alice}};
\coordinate (AliceMenuDiscountTarget)
  at (axis cs:0.26,{0.5*menuoriginal(0.26)+0.5*menulower(0.26)});
\path (AliceMenuDiscountTarget) ++(117:7pt)
  coordinate (AliceMenuDiscountJoin);
\draw[AliceMenuGreen!85!black,line width=0.95pt,
  line cap=round,line join=round,
  -{Stealth[length=5pt,width=3.5pt]},shorten >=1pt]
  (AliceMenuDiscount.south)
  .. controls +(0,-3pt) and +(117:3pt) ..
  (AliceMenuDiscountJoin) -- (AliceMenuDiscountTarget);

\draw[AliceMenuGreen,line width=2.1pt]
  (axis cs:0,0) -- (axis cs:\AliceMenuBase,0);
\draw[AliceMenuBlue,line width=2.1pt]
  (axis cs:\AliceMenuBase,0) -- (axis cs:\AliceMenuEquilibrium,0);
\draw[AliceMenuOrange,line width=2.1pt]
  (axis cs:\AliceMenuEquilibrium,0) -- (axis cs:1,0);
\node[anchor=north,align=center,inner sep=0pt,font=\fontsize{7}{8}\selectfont,
  text=AliceMenuGreen,
  xshift={\MenuSegmentLabelXShift+\MenuStraightLabelXShift},
  yshift=\MenuStraightLabelYShift]
  at (axis cs:{\AliceMenuBase/2},\MenuSegmentLabelY)
  {Straight segment};
\node[anchor=north,align=center,inner sep=0pt,font=\fontsize{7}{8}\selectfont,
  text=AliceMenuBlue,
  xshift={\MenuSegmentLabelXShift+\MenuCurvedLabelXShift},
  yshift=\MenuCurvedLabelYShift]
  at (axis cs:{(\AliceMenuBase+\AliceMenuEquilibrium)/2},\MenuSegmentLabelY)
  {Curved\\segment};
\node[anchor=north,align=center,inner sep=0pt,font=\fontsize{7}{8}\selectfont,
  text=AliceMenuOrange,
  xshift={\MenuSegmentLabelXShift+\MenuExtensionLabelXShift},
  yshift=\MenuExtensionLabelYShift]
  at (axis cs:{(\AliceMenuEquilibrium+1)/2},\MenuSegmentLabelY)
  {Optional\\extension};

\draw[AliceMenuBlue,line width=1.6pt]
  (axis description cs:0.035,\MenuLegendY) --
  (axis description cs:0.115,\MenuLegendY);
\node[menu text halo,anchor=west,font=\fontsize{8}{9}\selectfont,xshift=3pt]
  at (axis description cs:0.115,\MenuLegendY) {{\MenuTextHalo{Canonical menu}}};
\node[menu text halo,anchor=north west,font=\fontsize{8}{9}\selectfont]
  at (axis description cs:\MenuParametersX,\MenuParametersY)
  {\MenuTextHalo{$\rlap{\hspace*{\MenuParameterAXShift}$a=2$}%
    \phantom{a=2},\hspace{\MenuParametersGap} b=1$}};
\end{axis}
\end{tikzpicture}%
\endgroup%
%
    }
    \caption{\small{{\textbf{Left.} }
    Difference in Alice's optimal revenue as a percentage of $a$.
    The curve shows
    $\Delta_{\mathsf{opt}}(a;b)/a$, expressed as a percentage, against $a/b$.
    The main model yields greater revenue for $a/b<D$, and the aligned model
    for $a/b>D$, with equality at $a/b=D\approx0.6663$.}
    {\textbf{Right.} Alice's optimal pricing menu in the main model
    when $0<a<C \cdot b$.
    {The solid \textcolor{blue}{blue line} depicts the canonical menu.
    Every optimal menu shares its curved portion from $B$ to $E$;
    its initial straight segment from $(0,0)$ to $B$ may be replaced
    by permitted lower curves, illustrated in \textcolor[HTML]{08745A}{green}.
    Optional} %
    extensions follow the \textcolor{orange}{orange tangent.}}}
    \label{fig:introduction-aligned-variable-revenue-difference}
    \label{fig:introduction-alice-pricing-menu}
\end{figure}
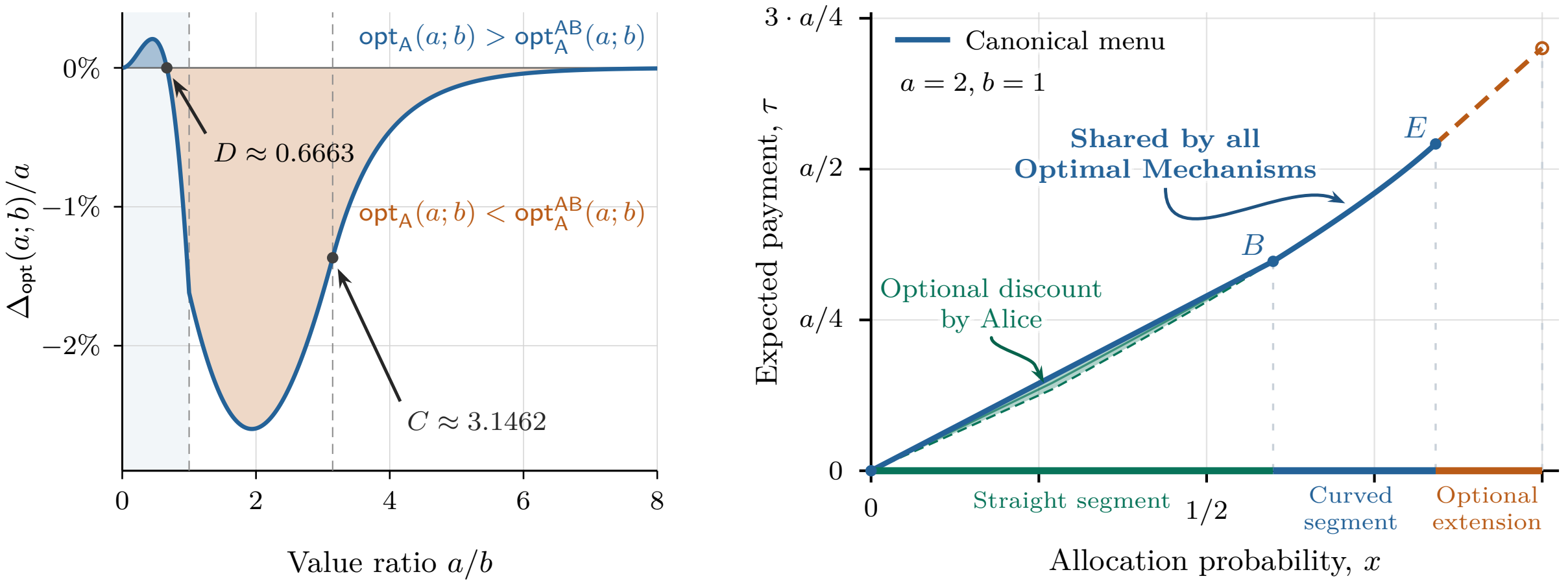

We first focus on one specific optimal mechanism for Alice, which we call her \emph{canonical menu}; we later use it to characterize \emph{all} optimal mechanisms.
Alice's canonical menu
consists of a straight segment followed
by a strictly convex curve
(\cref{fig:introduction-alice-pricing-menu}, right).
A lottery $(x,\tau)$ has allocation probability $x$
and expected payment $\tau$.
The menu starts
at the opt-out option $(0,0)$ and follows a
straight line
to a \emph{base lottery}
$B = (x_B,\tau_B)$.
The curve then leads to the
\emph{equilibrium lottery}
$E=(x_E,\tau_E)$:
equal increases in allocation probability require progressively
larger increases in expected payment.

Bob might instead offer a lottery aimed at attracting the buyer to visit his mechanism first, then move to Alice's mechanism second. Defending against this strategy imposes a 
constraint on the {minimal} utility
Alice's menu offers the buyer. The initial portion of her menu helps meet this constraint, but also affects the utility from options on the curved portion; it is not an independent extra lottery. If $a<b$, Bob can also offer lotteries to discriminate between buyers who did and did not obtain
Alice's good, whose values for
his good are $b-a$ and $b$, respectively.
Alice must deter these responses together with those inducing the Bob-first order, rather than defend only against posted prices. The resulting constraints jointly shape her optimal menu. Its curved portion makes Bob indifferent across a range of
competing responses, an \emph{equalizer} property we use in the analysis.

We 
\ygedit{now extend our description}
to all optimal menus of Alice.
Every optimal
mechanism of Alice
\ygedit{offers the equilibrium lottery $E$, and}
contains \emph{exactly the same curved portion}
connecting the base lottery $B$ to the equilibrium lottery $E$.
The initial straight segment may be replaced by certain pricing
curves with the same endpoints.
These curves lie weakly below the canonical straight segment,
so lotteries with smaller allocation probabilities can be offered at lower prices.
Finally, after discarding irrelevant menu entries,\footnote{
  Specifically, entries that are never utility-maximizing 
  for any effective buyer type
  between $0$ and $a$
  (removing these entries preserves
  essential equivalence).
}
the only possible extension beyond the equilibrium lottery is a
straight segment, charging $a$ per additional unit of allocation
probability. This extension may end anywhere up to allocation
probability one.

These variations preserve
the equilibrium lottery
and the on-path behavior of Bob and the buyer, 
as well as each seller's
expected revenue and the buyer's expected utility.
Thus, for fixed $a,b>0$ with $a< C \cdot b$, 
every Alice-optimal equilibrium has Bob post price $b$
and the buyer select the common lottery $E$ from Alice first,
buying from Bob exactly when Alice's lottery fails.
We conclude that when $a < C \cdot b$, to 
attain her optimal revenue,
Alice must offer a mechanism with infinitely many
menu options\footnote{Here and throughout, requiring infinitely many
menu options means that no finite menu attains 
Alice's exact optimal revenue.}
and arrange for Bob to post a price of $b$.

Next, we examine whether Alice could have exceeded her Bertrand revenue 
had she lacked commitment power.
We consider the \emph{simultaneous model},
in which Alice and Bob
choose their \emph{randomized} mechanisms
simultaneously.
We establish the following result.

\begin{restatable}[Simultaneous Model]{theorem}{introductionsimultaneous}
    \label{thm:introduction-simultaneous}
    Consider an instance of the duopoly game in the 
    \textit{simultaneous model}, in which Alice and
    Bob select their {randomized} mechanisms simultaneously. 
    {Then, every pure Nash equilibrium yields the Bertrand revenues 
    {for both sellers.}
    }
\end{restatable}

We conclude that commitment power is crucial 
in Alice's quest to beat her Bertrand revenue.

We move on to the \emph{aligned model},
in which Alice commits first and the buyer must
visit her first.
In the main model, Alice must induce the buyer to visit her first
to attain her optimum.
Since the aligned model guarantees this order, one might expect it
to raise her optimal revenue; Alice no longer needs to arrange it.
We find Alice's optimal revenue and characterize all
optimal mechanisms of Alice. 
Maybe surprisingly, we find that this revenue can be 
lower than her optimal revenue in the main model 
{(in which the buyer is free to pick the order of interaction)}.
Let $e$ denote the base of the natural logarithm
{and let $\optOrder{a}{b}{\AB}$ 
denote Alice's optimal
expected revenue in the aligned model. }

\begin{restatable}[Aligned Model]{theorem}{introductionaligned}    \label{thm:introduction-aligned-rephrased}
    Consider an instance of the duopoly game in the aligned model
    with fixed buyer values $a,b>0$.
    {There exists a complete and explicit characterization}
    of all mechanisms that are Alice-optimal in a Stackelberg equilibrium
    \ygedit{(\cref{prop:aligned-optimal-mechanisms})},
    and {it defines} 
    the optimal revenue for Alice:
    \begin{align*}
        \ygedit{\optOrder{a}{b}{\AB} = }
        \begin{cases}
            a-b+b \cdot e^{-a/b} & \textnormal{for}\ a\geq b,\\[3pt]
            \dfrac{a^2}{e \cdot b} & \textnormal{for}\ a<b,
        \end{cases}
    \end{align*}
    which strictly exceeds her Bertrand revenue of $(a-b)_+$, for all $a,b>0$. 
    
    {The equilibrium utilities of all three agents (Alice, Bob and the buyer) are independent of the equilibrium selection.}
    \ygedit{All} the mechanisms Alice offers in these equilibria
    \emph{share} an infinite continuum of menu options;
    posting a price of
    $b$ is a best response
    for Bob and under this response,
    it is optimal for the buyer
    to select the lottery %
    at the top of this shared continuum
    ({point $E'$ 
    in \cref{fig:aligned-optimal-pricing-menus}})\footnote{
      {Different optimal menus can offer menu entries with a price 
      greater than that of $E'$,  
      which are irrelevant to the payoffs of all three agents.}
    }
    and buy from Bob if and only if he got nothing from Alice's lottery.
\end{restatable}

Thus, when the buyer is forced to 
visit Alice first, commitment power and randomization 
within the mechanism always provide Alice extra revenue.
Unlike in the main model, 
both agents manage to achieve
positive revenue for all values $a,b > 0$.

The optimal menus of Alice in
the aligned model have a similar
structure to their counterpart in the main
model for $a<C\cdot b$, except that there is
no initial straight segment.
The opt-out option $(0,0)$ plays the role of 
the base lottery,
and the menus follow a strictly convex curve
from this point to the equilibrium lottery.
All optimal menus of Alice share this exact curve.
Similarly to before, after disregarding irrelevant entries.\footnote{
  {Those that are never utility-maximizing 
  for any effective buyer type
  between $0$ and $a$ 
  (again, removing these options
  preserves essential equivalence).}
}
Alice may extend the pricing rule
beyond the equilibrium
lottery only along a straight segment,
charging $a$ per additional unit
of allocation probability.
The extension may end anywhere up 
to allocation probability one.
This structure applies for every $a,b>0$.
We conclude that in order to achieve her optimal
revenue, Alice must offer a mechanism with infinitely 
many menu options
and arrange for a posted price of $b$ to be a 
best response of Bob.

Finally, we compare Alice's optimal revenues in the two models.
Perhaps surprisingly (as we discuss below), there exist values\footnote{For example, $a = 1$ and $b = 2$.} of $a,b$ such that Alice achieves a strictly
greater optimal revenue in the main model.
{Moreover, we observe that there exists a} threshold {$D\approx 0.6663$}
that determines which model gives Alice higher optimal revenue:
{$D$ is set to be such that at $a = D \cdot b$ the two revenues coincide.}
For $a<D\cdot b$, Alice's optimal revenue
is strictly greater in the main model,
whereas for $a>D\cdot b$, it is 
strictly greater in the aligned model.
Let $\opt{a}{b}$ 
denote Alice's optimal
expected revenue in the main model. 
We write $\Delta_{\mathsf{opt}}(a;b):=\opt{a}{b}-\optOrder{a}{b}{\AB}$
for the difference between these optimal revenues.
For every fixed $b>0$, this difference approaches zero from 
below as $a\to\infty$.
We summarize this comparison in \cref{fig:introduction-aligned-variable-revenue-difference},
with the difference normalized by $a$ and expressed as a percentage.

It might at first seem surprising that forcing the buyer to visit Alice first can reduce Alice's revenue, given that the buyer chooses to visit Alice first on the equilibrium path in the main model.  The reason is that this restriction changes Bob's off-path incentives. Although the buyer visits Alice first on the equilibrium path in the main model, a deviation by Bob can change that order. Fixing the order opens up new strategic options for Bob: he can offer lotteries that discriminate between buyers who did and did not obtain
Alice's good without worrying about inducing
the buyer to visit him first. These additional responses can impose stronger constraints on
Alice's menu. This does not require Bob to use a multi-lottery mechanism on path: posting price $b$ remains a best response in both models.

Finally, we study the \emph{misaligned model}, in which
Alice commits first but the buyer must visit Bob first.
Here, Alice cannot exceed her Bertrand revenue in equilibrium. 
{We conclude that commitment by itself is not enough for Alice 
to exceed her Bertrand revenue: when the buyer 
is forced to visit Bob first, commitment 
is essentially working against Alice, as Bob gets to 
undercut Alice after she commits. Thus, the ability of 
Alice to extract any extra revenue (on top of Bertrand) 
with commitment crucially depends on the fact she
has an opportunity to be approached first by the buyer.}

\begin{restatable}[Misaligned Model]{theorem}{introductionmisaligned}
   \label{thm:introduction-misaligned-rephrased}
    Consider an instance of the duopoly game in the 
    misaligned model with fixed buyer values $a,b>0$.
    In every Stackelberg equilibrium, Alice
    earns her Bertrand revenue $(a-b)_+$.
    When $a\leq b$, the revenue of Bob and 
    the utility of the buyer are \emph{not} uniquely determined.
    The set of equilibrium payoff triples,
    Alice's expected revenue, Bob's expected revenue, and the buyer's
    expected utility, is exactly
    \begin{align*}
      \begin{cases}
        \{(a-b,0,b)\},&a>b,\\[3pt]
        {\{(0,b-u,u):u\in[0,a]\}},&a\le b
      \end{cases}.
    \end{align*}
    {Here, $u$ denotes the buyer's expected utility.}
\end{restatable}

Notably, when $a \leq b$, the payoffs of Bob and
the buyer are not uniquely determined, unlike in all other
models considered in the paper. 
Every admissible Alice mechanism is optimal
(as they all earn a revenue of zero),
yet her choice affects Bob's expected 
revenue and the buyer's expected utility. 
In order to gain revenue,
Bob must match the utility Alice offers.
The on-path pair
$(\FP{a - u},\FP{b-u})$, where $u\in[0,a]$
represents the utility offered to the buyer, 
yields the payoff triple
$(0,b - u,u)$ for Alice, Bob, and the buyer, respectively.
As $u$ ranges over $[0,a]$, 
these triples exactly cover all possible equilibrium payoffs.
We conclude that when $a \leq b$, 
Alice's arbitrary choice of mechanism determines how Bob 
and the buyer divide the surplus without affecting her own revenue.
When $a>b$, Alice's
unique optimal mechanism is a posted price of $a - b$, 
while Bob is unrestricted on path (as he always receives a revenue of zero).

To summarize, we determine Alice's optimal revenue in 
the simultaneous, main, aligned, and misaligned models.
For the latter three, we also characterize all of Alice's
optimal mechanisms and the associated equilibrium outcomes.
The following table summarizes when Alice can exceed her
Bertrand revenue.
\par\vspace{7pt}
\noindent
{\begin{minipage}{\linewidth}
\footnotesize\centering
\setlength{\tabcolsep}{5pt}
\renewcommand{\arraystretch}{1}
\begin{tabular}{@{}lllll@{}}
\hline
\textbf{Model} & \textbf{Commitment order} & \textbf{Buyer order} & \textbf{Equilibria} &
\textbf{Exceeds Bertrand} \\
\hline
Simultaneous & Simultaneous & Buyer chooses & All pure Nash & Never \\
Main & Alice, then Bob & Buyer chooses & Alice-optimal Stackelberg & $a<C \cdot b$ \\
Aligned & Alice, then Bob & Alice first & Alice-optimal Stackelberg & All $a,b>0$ \\
Misaligned & Alice, then Bob & Bob first & All Stackelberg & Never \\
\hline
\end{tabular}
\end{minipage}}
\par\vspace{6pt}
{Note that without randomization, Alice cannot exceed her Bertrand revenue,
with or without commitment.}

\par

\subsection{Overview of our Proofs}
\label{sec:introduction-proof-overview}

In this section, we present an overview of the main 
ideas used in our proofs. Further details and refinements can be found 
in subsequent sections. We begin with the main model
(\cref{thm:introduction-variable-order-rephrased}), then extend
its analysis to the aligned model
(\cref{thm:introduction-aligned-rephrased}).
Next, we analyze the misaligned model
(\cref{thm:introduction-misaligned-rephrased}).
Finally, we turn to the simultaneous model
(\cref{thm:introduction-simultaneous}), highlighting the new ideas
used to extend the result of Lucier and Saxena~\cite{lucier2025approximately}
to unequal values.

\subsubsection{Analyzing the Main Model}
\label{sec:introduction-proof-overview:main}
Recall that our setting has three parties, Alice, Bob, and the buyer.
The buyer's values for Alice's and Bob's goods are $a$ and $b$, respectively. Throughout, we use $\residualValue{a}{b}$ and $\residualValue{b}{a}$
as a shorthand for $\max\paren*{ 0, a - b }$ and $\max\paren*{ 0, b - a }$ respectively. In our setting, first, Alice commits to a mechanism $\AliceMechanism$, then based on $\AliceMechanism$, Bob determines the mechanism $\BobMechanism$ that he wants to use, and finally, the buyer who knows both mechanisms decides on his action plan. This involves deciding
which seller to go to first, picking an option from their mechanism, and then, based on whether or not he receives
the good from this seller, deciding which option to pick from the mechanism of the other seller.

\paragraph{Buyer's behavior in a duopolist setting.} A key insight in our proof is that the mechanism used by the other seller affects what option the buyer picks from this seller. For example, suppose that Alice is using the mechanism $\AliceMechanism$ and Bob decides not to sell
his good (equivalently, uses the mechanism that charges a price of $\infty$). This effectively removes Bob from the market and the buyer buys from Alice with the same probability as when she was a monopolist. In contrast, suppose that Alice is again using the mechanism $\AliceMechanism$ but Bob is giving
his good for free (that is, using the mechanism that charges no price). Now, if $a < b$, then the buyer can simply go to Bob and get
his good for free, which would mean that he no longer needs to interact with Alice and get
her good. Even if $a > b$, the fact that he can get
Bob's good for free means that his effective value for
Alice's good is only $a - b$, and therefore he may pick an option from her menu that is different from the option that he would have picked if Alice were a monopolist.

In \cref{sec:stackelberg:buyer}, we describe exactly how the mechanism used by the other seller affects the original seller. Specifically, let $\zmarg{ \mathcal{M} }^{+}\paren*{ v }$ denote the probability that the buyer with value $v$ buys the good when faced with a monopolist using the mechanism $\mathcal{M}$. We show that the buyer behaves {as if he is approaching a mechanism picked by a monopolist seller} with the seller he interacts with second. That is, if the buyer interacts with Alice first and does not get
the good, then he will buy
the good from Bob with probability $\zmarg{ \BobMechanism }^{+}\paren*{ b }$, and if he gets
Alice's good, will buy
Bob's good with probability $\zmarg{ \BobMechanism }^{+}\paren*{ \residualValue{b}{a} }$, as his value for
Bob's good given that he has
Alice's good is just $\residualValue{b}{a}$.

With the first seller, the buyer's behavior might be different from that in the monopolist setting. To understand how it may be different, suppose for concreteness that the
buyer first visits Alice.
\cref{lemma:buyer} shows that the relevant property from Bob's mechanism is the utility difference it gives to a buyer with value $b$ and a buyer with value $\residualValue{b}{a}$ in a monopolist setting. We denote this utility difference by $\aug_{ \BobMechanism }\paren*{ a, b }$ and formally define it in \cref{eq:aug}. For the purposes of this sketch, it suffices to note that this difference is always non-negative and is determined by Bob's mechanism. Roughly speaking, \cref{lemma:buyer} shows that when the buyer interacts with Alice, his effective value is $\effectiveValue{a}{\BobMechanism} = a - \aug_{ \BobMechanism }\paren*{ a, b }$ instead of being $a$. Thus, the presence of Bob means that the buyer's effective value for
Alice's good is now lower, with the amount of decrease equal to the augmentation-induced utility gap of the buyer between receiving and not receiving
Alice's good.

\paragraph{Bob's best response to Alice's mechanism.} The upshot of the foregoing discussion is that when faced with a mechanism $\AliceMechanism$ by Alice, Bob knows that he can either design his mechanism so that the buyer {visits} him first or design it so that the buyer {visits} Alice first. In the former case, Bob will face a buyer with an effective value of $\effectiveValue{b}{\AliceMechanism}$ which, depending on $\AliceMechanism$, may be much lower than $b$ and thus give him much lower revenue compared to a monopolist. In the second case, Bob will face a buyer with value $b$ if he did not receive
Alice's good, and a buyer with value $\residualValue{b}{a}$ if he received
Alice's good. It can be shown that $\residualValue{b}{a} \leq \effectiveValue{b}{\AliceMechanism} \leq b$ and so, in this case, Bob is counting on the probabilistic event that the buyer did not receive
the good from Alice so he can face a buyer with a larger value.

It is {a priori} unclear whether Bob should design for the buyer to approach him first or second, and to get bounds on the highest revenue Bob can attain, we consider these two cases separately. \cref{sec:stackelberg:bob:bob} deals with the case where the buyer approaches Bob first, while \cref{sec:stackelberg:bob:alice} deals with the case where the buyer approaches Alice first.
\ygedit{For the case where Bob is approached first, he} is designing a mechanism for buyers with a single effective value $\effectiveValue{b}{\AliceMechanism}$, so his mechanism only needs to have a single option. Similarly, in the latter case, Bob is designing his mechanism for only two values, $\residualValue{b}{a}$ and $b$, and thus, his mechanism only needs to have two options.

\paragraph{Designing for the order Bob \textrightarrow\ Alice.} As mentioned above, in this case, Bob's mechanism only needs to have a single option, meaning that all Bob needs to do is to decide on an allocation probability $x \in \bracket*{ 0, 1 }$ and a {per-unit-}price\footnote{The per-unit-price is defined as the ratio of the buyer's payment to the probability that he is allocated
the good.} $p \geq 0$ {(so the expected payment is $x \cdot p$)} such that \begin{inparaenum}[(1)] \item When he offers this option, the buyer approaches him first and picks this lottery. \item The values $x$ and $p$ are large so that he generates a lot of revenue. \end{inparaenum} We call such mechanisms single lottery mechanisms and denote them by $\SL_{ x, p }$. For these mechanisms, we show in \cref{lemma:sl} that, given any $x \in \bracket*{ 0, 1 }$,
{there is a trade-off between these goals.}
Concretely, for any given $x \in \bracket*{ 0, 1 }$, increasing $p$ makes it so that Bob gets more revenue, but increasing it beyond a certain threshold that is determined by Alice's mechanism and denoted by $\p_{ \AliceMechanism }(x)$ would mean that the buyer would buy from Alice first. Importantly, this threshold may be smaller than $\effectiveValue{b}{\AliceMechanism}$, the highest per-unit-price the buyer is willing to pay for
Bob's good, and may depend on $x$. Thus, for any $x \in \bracket*{ 0, 1 }$, the maximum revenue Bob can achieve using single lottery mechanisms is $x \cdot \p_{ \AliceMechanism }(x)$.

\paragraph{Designing for the order Alice \textrightarrow\ Bob.} Unlike the previous case, here the buyer may or may not have received
Alice's good before interacting with Bob, and his effective value is $\residualValue{b}{a}$ if he did and $b$ if he did not receive
Alice's good. In turn, we get that Bob needs to design his mechanism for two values, which means that his mechanism needs to have at most two options. This might seem more complicated than the previous case at first, but there is a crucial difference.

In the previous case, Bob was designing his mechanism with only one value $\effectiveValue{b}{\AliceMechanism}$, but he could not increase his per-unit-price all the way to $\effectiveValue{b}{\AliceMechanism}$, as doing so may lead the buyer to go to Alice first. In this case, we can show that for the option for the value $\residualValue{b}{a}$, Bob can in fact raise the per-unit-price corresponding to the value $\residualValue{b}{a}$ all the way up to $\residualValue{b}{a}$. Showing this formally requires a proof, but the intuition is that doing so keeps $\effectiveValue{a}{\BobMechanism}$ unchanged, which means that the buyer does not have an incentive to switch the order in which he approaches the sellers. Using similar reasoning, we can also show that the allocation probability of the larger value $b$ can be increased all the way to $1$.

Thus, we are again left with only two arguments, the allocation probability of the smaller value, which we denote by $x$, and the price\footnote{As the allocation probability of the larger value is $1$, the price is the same as the per-unit-price.} for the larger probability, which we denote by $q$. These two parameters determine our double-lottery mechanism, which we denote using $\DL_{ x, q }$ and, similarly to the previous case,  Bob's goal is to determine the pair $x$ and $q$ such that \begin{inparaenum}[(1)] \item When he offers the mechanism $\DL_{ x, q }$, the buyer buys from Alice first before interacting with Bob. \item The values $x$ and $q$ are large so that Bob gets more revenue. \end{inparaenum}

In \cref{lemma:dl}, we show that, 
once again, {there is a trade-off between these goals.}
Specifically, we show that for any given $q$, there is a threshold $x$ given by $\x_{ \AliceMechanism }(q)$ such that increasing $x$ beyond this threshold would make the buyer buy from Bob first.
We also show that fixing $q$ \ygcomment{}
fixes the probability the buyer will buy from Alice, and thus the maximum revenue $\mathtt{R}_{ \AliceMechanism }(q)$ possible with a given value of $q$ is determined by $q$ and $\x_{ \AliceMechanism }(q)$.

\paragraph{Alice's optimal mechanism.} With this understanding of the behavior of Bob and the buyer given a mechanism $\AliceMechanism$ by Alice, we can proceed to understand the nature of Alice's optimal mechanism. To this end, let $\AliceMechanism$ be a mechanism for Alice that gives her the optimum revenue and let $\BobMechanism$ be Bob's best response to $\AliceMechanism$ that gives him a revenue of $R$. This means that no matter whether Bob designs for the buyer to approach him first or not, his revenue is upper bounded by $R$. In turn, we can use our discussion about Bob's best responses above to constrain Alice's mechanism. 
\ygedit{Against Bob's strategy aimed at attracting the buyer to visit him first, Alice must}
ensure that for all $x \in \bracket*{ 0, 1 }$, the largest revenue possible with this value of $x$ must be at most $R$. That is, we must have $x \cdot \p_{ \AliceMechanism }(x) \leq R$. 
Similarly, \ygedit{to handle Bob designing for the buyer to visit Alice first, she} 
must \ygedit{ensure} 
that 
$\mathtt{R}_{ \AliceMechanism }(q) \leq R$ \ygedit{for all $q$}.

Interestingly, we show that both these constraints can be equivalently cast as constraints on the (monopolist) allocation rule and the (monopolist) utility function of the mechanism $\AliceMechanism$: 
\ygedit{To defend against Bob's strategy aimed at the order Bob \textrightarrow\ Alice, she must offer a minimal utility to the buyer.}
\ygedit{In particular,} the constraint $x \cdot \p_{ \AliceMechanism }(x) \leq R$ for all $x \in \bracket*{ 0, 1 }$ 
turns out to be the same as saying that Alice's 
\ygedit{monopolist} utility function \ygedit{must}
lie above a certain line that is determined by $R$ and the maximum utility her mechanism gives to the buyer. 
\ygedit{On the other hand, Alice's approach against Bob's second strategy} 
changes depending on
which good the buyer values more. If the buyer values
Alice's good more,
\ygedit{Bob has no ability to screen and therefore only considers fixed prices. By 
increasing his utility (and lowering his price), Bob entices the buyer away from Alice's mechanism. Alice's lower-priced lotteries would be selected. To defend against this, Alice controls the change in her allocation rule as Bob cuts his price, ensuring that such a cut is not profitable.}
\ygedit{Thus, the condition $\mathtt{R}_{\AliceMechanism}(q) \leq R$ for all $q$} translates to a pointwise lower bound on the allocation rule of Alice's mechanism. 
We call this case the Large-Alice case. A similar lower bound also holds for the Large-Bob case where the buyer values
Bob's good more, except that the lower bound in this case is a bit more complex and also involves Alice's utility function.

For each case, we then design mechanisms for Alice that match these lower bounds, and show that these give her the optimal revenue. As we match these lower bounds pointwise, our mechanisms ensure that for a wide range of $q$, Bob's revenue when he uses the mechanism $\DL_{ \x_{ \AliceMechanism }(q), q }$ is the exact same,
and equal to $R$ (see \cref{fig:bob-equalizer-main}). Because of this property, we call the mechanisms for Alice `equalizer mechanisms'. In the Large-Alice case, we call them Large-Alice-equalizer mechanisms and in the Large-Bob case, we call them Large-Bob-equalizer mechanisms. These are formally defined in \cref{def:la-equalizer} and \cref{def:lb-equalizer} respectively. An illustration of these mechanisms can be found in \cref{fig:introduction-alice-pricing-menu}. As depicted there, these mechanisms consist of a straight segment and a curved segment. We show that the curved segment must be exactly the same for all mechanisms that are optimal for Alice. In contrast, there is some wiggle room in the straight segment but it does not affect the buyer's payments and allocations to either seller.
\ygedit{Additional wiggle room exists beyond the curved segment as well, but this also has no effect on the buyer's payments to or allocations from the sellers.}

\subsubsection{Analyzing the Aligned Model}
\label{sec:introduction-aligned-proof-overview}

Our proof in the aligned model follows the same outline as in the main model. We observe that since the buyer has to interact with Alice's mechanism before approaching Bob, Bob can never design for the order Bob \textrightarrow\ Alice. This also means that when he designs for the order Alice \textrightarrow\ Bob, he does not need to ensure that the parameters $x$ and $q$ of the mechanism $\DL_{ x, q }$ are such that the buyer will interact {with Alice} first when Alice and Bob use the mechanisms $\AliceMechanism$ and $\DL_{ x, q }$ respectively.

Operationally, this means that Bob can increase $x$ beyond the threshold $\x_{ \AliceMechanism }(q)$ described above to its largest permissible value of $\frac{ b - q }{ b - \residualValue{b}{a} }$ without worrying about it causing the buyer to interact with him first. This changes the value of the revenue $\mathtt{R}_{ \AliceMechanism }(q)$ possible with a given value of $q$ and therefore the constraints on Alice's mechanism are now different. However, we can still design mechanisms that make these constraints tight. We call such mechanisms Aligned-Large-Alice-equalizers and Aligned-Large-Bob-equalizers; see \cref{sec:alice-opt-stalkeberg-AB} for more details.

\subsubsection{Analyzing the Misaligned Model}

{While an approach similar to the one used for the aligned case could potentially work for the misaligned case as well, 
we present a different proof for simplicity.}
{Here,}
the buyer visits Bob first, so Alice's
commitment determines a single effective type $\effectiveValue{b}{\AliceMechanism}\ge0$ at Bob,
which captures the buyer's willingness to
pay for
Bob's good after accounting for the outside options
provided by Alice.
Bob can extract his maximal revenue of $\effectiveValue{b}{\AliceMechanism}$ by posting that price.
If $\effectiveValue{b}{\AliceMechanism} > 0$,
this is Bob's 
unique best response.
The buyer then receives
Bob's good and reaches Alice with residual
value $(a-b)_+$ for
her good.
If $\effectiveValue{b}{\AliceMechanism}=0$, Alice must offer standalone utility at least $b$
(this can only occur when $a\ge b$). 
Thus, Alice's revenue {by any mechanism, even with lotteries,}
is limited to $(a - b)_+$.  
Posting a price of $(a-b)_+$ attains this bound, 
showing that this is her optimal revenue. 

When $a \leq b$, every admissible Alice mechanism is optimal
(as they all earn a revenue of zero%
).
Her choice nevertheless affects Bob's expected 
revenue and the buyer's expected utility. 
Bob must match the utility Alice offers in order to gain revenue.
The on-path pair
$(\FP{a - u},\FP{b-u})$, where $u\in[0,a]$
represents the utility offered to the buyer, 
yields the payoff triple
$(0,b - u,u)$ for Alice, Bob, and the buyer, respectively.
As $u$ ranges over $[0,a]$, 
these triples exactly cover all possible equilibrium payoffs.
When $a>b$, Alice's
unique optimal mechanism is a posted price of $a - b$, 
while Bob is unrestricted on path (as he always receives a revenue of zero).
Thus the unique equilibrium payoff triple is $(a-b,0,b)$.

\subsubsection{Analyzing the Simultaneous Model}

For the simultaneous model, we show that at every pure Nash equilibrium both Alice and Bob obtain their Bertrand revenue.  This extends a result of Lucier and Saxena~\cite{lucier2025approximately}, who show that when $a=b$ every pure Nash equilibrium yields revenue zero for both sellers (even when randomized mechanisms are allowed).
Our result extends their analysis to unequal values and the corresponding Bertrand revenues. Assume without loss of generality that $a\ge b$ so that 
$\residualValue{a}{b}=a-b$ and $\residualValue{b}{a}=0$.

To describe how the asymmetric case differs, we'll first summarize part of the symmetric analysis.
{Roughly speaking, the argument of~\cite{lucier2025approximately} for the case $a=b$ first establishes that if \emph{either} seller obtains strictly positive revenue at equilibrium then \emph{both} obtain strictly positive revenue. This part of the analysis is straightforward: if we suppose Bob obtains any positive revenue at equilibrium, it cannot be that all of the surplus is going to the buyer, so there must be some sufficiently small (but positive) price that Alice could offer to attract the buyer and obtain positive revenue. This establishes that \emph{both} sellers obtain more than their Bertrand revenue in this hypothetical equilibrium. The remainder (and more technical part) of their argument goes on to analyze the structure of the two sellers' lotteries given their positive revenues, ultimately inferring the existence of a deviating response.}

{A key difference in the asymmetric case $a > b$ is in this first step of the argument. It is no longer immediate that if one seller obtains strictly more than their Bertrand revenue then the other must as well. In particular, new ideas are needed to rule out equilibria where Alice obtains revenue $\residualValue{a}{b}$, Bob obtains strictly positive revenue, and the buyer visits Bob first. Unlike in the symmetric case, a posted price slightly higher than $\residualValue{a}{b}$ is not necessarily a profitable deviation for Alice: even if this offer is attractive to a buyer, it can cause Alice to lose revenue from buyers who first obtain Bob's good and would have considered purchasing from Alice afterwards.}

To rule out such equilibria, we recall that the buyer may still pay Alice after winning
Bob's good, unlike in the symmetric setting. This motivates a new double-lottery deviation by Alice that generates revenue by distinguishing buyers that won
Bob's good from those that did not.  Alice offers two menu options: one targets a revenue of (slightly less than) $\residualValue{a}{b}$ if Bob's lottery causes an allocation, and the other targets a revenue of (slightly more than)
\ygedit{$\residualValue{a}{b}$} if it does not.
Careful design of the lotteries, including control of the buyer's response 
to any potential free lotteries offered by Bob, leads to a strict revenue gain for Alice, generating the
desired deviation and contradicting the existence of the proposed equilibrium.

\subsection{Additional Related Work.}
\label{sec:introduction-related-work}

Our work relates to classical models of duopoly,
particularly Bertrand price competition, 
and also relates to Stackelberg games.
For general treatments of classical
market models and oligopoly theory,
see~\cite[Chapter~4]{jehle2001advanced}
and~\cite{vives1999oligopoly}.
\bledit{As discussed above, our work builds directly on Lucier and Saxena~\cite{lucier2025approximately}
who consider sellers with identical goods in a Bayesian model. We focus on a complete information setting with non-identical goods; we leave an extension of our characterization to models of incomplete information as a direction for future work.}
\bldelete{}

\paragraph[Competition through prices.]{Competition through prices.}
A related literature studies posted-price competition, in which
sellers compete by choosing prices for their goods.
Tauman, Urbano, and Watanabe~\cite{tauman1997multiproduct}
extend the classic Bertrand model to heterogeneous goods, 
studying single-product firms facing a representative buyer with values for bundles.
\bldelete{}
They prove that pure equilibria
exist and maximize total surplus.
Arribas and Urbano~\cite{arribas2005nash}
characterize the equilibrium prices and purchased bundles under the same
equilibrium condition.
Babaioff, Nisan, and Paes Leme~\cite{babaioff2014price}
focus on the case of a buyer with submodular valuation \bledit{and further characterize the structure of equilibrium prices.}
{Lev, Oren, Boutilier, and Rosenschein}~\cite{lev2015pricing} extend this setting to sellers owning
multiple items and show that pure equilibria can fail to exist.
Network models examine prices for complementary resources that buyers
combine into routes~\cite{chawla2008bertrand,chawla2009price}.
Babaioff, Lucier, and Nisan~\cite{babaioff2013bertrand}
study price competition among sellers of a homogeneous good but where each seller can sell only to a specified subset of
buyers. 
\bldelete{}
\bledit{Relative to this literature, our work focuses on the particular case of two goods and a unit-demand buyer, but allows sellers to employ lotteries in addition to posted prices.}

\paragraph[Competing mechanisms and buyer participation.]{Competing mechanisms and buyer participation.}
\bledit{Another related literature studies
how sellers design selling mechanisms while competing for a pool of buyers who each choose one seller.}
McAfee~\cite{mcafee1993mechanism}
studies a dynamic market in which sellers \bledit{choose between auction rules and reserve prices, and buyers then choose which seller
to visit.}
\bldelete{}
This analysis was later extended to a limit market 
as the number of buyers and sellers grows large~\cite{peters1997competition}. The case in which sellers simultaneously choose reserve prices in competing second-price auctions has received particular attention~\cite{burguet1999imperfect,virag2010competing}.
\bldelete{}
Damianov~\cite{damianov2012seller} allows anonymous, Bayesian
incentive-compatible mechanisms specifying allocation probabilities and
payments, including allocation lotteries; in this model, buyers choose their seller \emph{before}
learning their values.
\bledit{Finally, while they don't explicitly model a game between sellers,} {Gonczarowski, Immorlica, Li, and Lucier}~\cite{gonczarowski2024revenue} study buyers
with private values and participation costs, which can represent outside options forgone upon participation (such as another seller).
This outside option affects the buyer's willingness to pay,
since participation must be at least as attractive as taking that option.
\bledit{In contrast to these works, the buyer in our model retains access to the second seller
after participating in the first seller's mechanism, so the choice of seller is not exclusive.}
This continuation opportunity still shapes the buyer's willingness to pay at the first seller.

\paragraph[Product differentiation and screening.]{Product differentiation and screening.}
Our model is also connected to the literature on product
differentiation. \bldelete{} 
Models of quality differentiation~\cite{gabszewicz1979price,shaked1982relaxing}
{study price competition between firms selling different products
to heterogeneous buyers.}
{Mussa and Rosen~\cite{mussa1978monopoly} study a single monopolist
serving buyers who differ in their willingness to pay for quality.
The monopolist offers a menu of qualities and prices, letting buyers
select their preferred option. Here, product differentiation occurs
within one firm's product line and serves to screen buyers.
Champsaur and Rochet~\cite{champsaur1989multiproduct} connect these themes
by studying two competing firms that each offer a range of qualities
to heterogeneous buyers.}
{Subsequent work studies competitive quality--price menus when
buyers privately know their preferences or outside options, so rival
offers affect both participation and screening~\cite{stole1995nonlinear,rochet2002nonlinear}.}
Moscarini and Ottaviani~\cite{moscarini2001price} study price competition
for a single buyer with a private signal about relative quality.
\bldelete{}
\bledit{In contrast, our model treats good quality and buyer willingness to pay as public knowledge, so no screening occurs; rather, each seller's choice of mechanism serves to impact inter-seller competition by shaping buyer behavior under different potential choices of the competitor's mechanism.}

\paragraph[Lotteries and menu complexity.]{{Lotteries and menu complexity.}}
The use of lotteries in menus connects our problem to the complexity
of revenue maximization. 
It is a classical result that a monopolist selling a single item
to one risk-neutral buyer with a privately known value can maximize
expected revenue by posting a price~\cite{Mye81}.
With multiple goods, lotteries can be essential
for optimal revenue~\cite{Thanassoulis04,ManelliV07,DaskalakisDT17}.
{In such settings, privately known buyer values can lead to complex
optimal menus, motivating work on menu complexity and simple approximately
optimal mechanisms}~\cite{HartN13,CaiZ17,BabaioffGN17}.
Our paper,
which assumes publicly known values,
identifies a different source of menu complexity:
Alice may need to commit to a continuum of lotteries
to deter profitable deviations by Bob.

\paragraph[Stackelberg games and commitment.]{Stackelberg games and commitment.}
Our model naturally relates to Stackelberg games, in which a leader
commits before a follower observes the commitment and \bledit{best responds~\cite{stackelberg1934marktform}.}
In our case, Alice chooses her menu anticipating Bob's response.
\bledit{A Stackelberg game related to ours was studied by} Martimort and Stole~\cite{martimort2003contractual}, who identify a related
commitment role in common-agency games where a single agent can accept
contracts from several principals. In their model, contracts are offered
simultaneously, \bledit{and (as in our game) unused menu options} can deter rivals' deviations
even under complete information.
The role of commitment in shaping competitors' responses also connects
our work to Cai, Babaioff, and Lucier~\cite{cai2024bundling}, who study
a seller committing to a deterministic menu of bundle prices
before single-item competitors set prices for an additive buyer with
independent private values. Under price-sensitivity and variance assumptions
on the value distributions, a single grand-bundle price earns a constant
fraction of optimal revenue in every equilibrium of the ensuing price
competition.

\paragraph{Scope and Organization.}

{We formalize the models and equilibrium conventions in
\cref{sec:prelim}.
Our analysis of the main model appears in
\cref{sec:alice-opt-stackelberg}, where we prove
\cref{thm:introduction-variable-order-rephrased}.
For the aligned model, \cref{sec:alice-opt-stalkeberg-AB}
proves \cref{thm:introduction-aligned-rephrased}, characterizing
Alice's optimal mechanisms and revenue.
For the misaligned model, \cref{sec:alice-opt-stalkeberg-BA}
proves 
\cref{thm:introduction-misaligned-rephrased}.
We then analyze the simultaneous model in \cref{sec:nash}
and prove \cref{thm:introduction-simultaneous}.
A methodology note describes our use of AI assistance.
Deferred proofs for the preliminaries, main model, and simultaneous
model appear in the appendices.}

\section{Model and Preliminaries}
\label{sec:prelim}

Two sellers, Alice and Bob, 
each sell one indivisible good for which they 
have no value. There is a single unit-demand buyer 
with value $a > 0$ 
for Alice's good, value $b > 0$ for Bob's good, 
and value $\max\left\{ a,b \right\}$ for both.  Values $a$ and $b$ are fixed and known to the buyer and both sellers.
We denote with $\Delta = \left|a - b\right|$ the absolute 
difference between the two values.
For $t\in\mathbb R$, we write $(t)_+=\max\{t,0\}$ for its positive part.

\paragraph{Mechanisms.} 
\label{par:prelim-mechanisms}
Alice and Bob each choose a \emph{mechanism} with which to sell their good. A (possibly randomized) mechanism takes as input a message $\sigma \in \Sigma$ from the buyer, where $\Sigma$ is an arbitrary \ygedit{uncountable standard Borel} message space, and samples
an outcome $(q,t) \in \mathcal{O} = \{0,1\} \times \mathbb{R}_{\geq 0}$ from some distribution over outcomes,
where $q \in \{0,1\}$ indicates whether the buyer receives 
the seller's good and $t$ represents a transfer from the 
buyer to the seller.\footnote{As usual, this formalization also captures multi-round mechanisms by having the message space $\Sigma$ encode contingent responses to the mechanism across rounds.} Formally, a mechanism $\mathcal{M}$ is then a measurable mapping $\mathcal{M} \colon \Sigma \to \Delta(\mathcal{O})$ with compact image under weak convergence of outcome distributions.\footnote{We emphasize that the outcome of a mechanism depends only on the message sent to that mechanism.
This means, for example, that the outcome of Bob's mechanism is not impacted directly by any interactions between the buyer and Alice's mechanism, but only indirectly via changes to the buyer's interaction with Bob.}
{The buyer may decide not to participate in any mechanism. Thus, we} %
will assume for convenience that every mechanism accepts a null message $\emptyset$, equivalent to non-participation, upon which the mechanism returns outcome $(0,0)$.
{We denote with $\AllMechanisms$ the set of 
all mechanisms described above}.

\paragraph{Timing.}\label{par:prelim-timing} Informally, the timing of our
market proceeds as follows.
Alice selects her mechanism first 
and Bob responds by selecting his mechanism.\footnote{We will later consider a variant with simultaneous mechanism selection.
} 
The buyer then observes both mechanisms and 
chooses one of the sellers to 
approach first.  The buyer interacts with the chosen mechanism 
and observes the outcome.  The buyer can then choose to 
interact with the remaining seller's mechanism (or not), 
after which the game ends. 
More formally:  
\begin{enumerate}
    \item Alice selects her mechanism $\mathcal{M}_\Alice: \Sigma \to \Delta(\mathcal{O})$. 
    
    \item Bob observes $\mathcal{M}_\Alice$ and then selects his mechanism $\mathcal{M}_\Bob: \Sigma \to \Delta(\mathcal{O})$.\footnote{\ygedit{We assume that all mechanisms available to either
    seller use the same fixed message space $\Sigma$, containing the
    null message $\emptyset$.}}
    \item The buyer observes $\mathcal{M}_\Alice$ and $\mathcal{M}_\Bob$.    
    \item The buyer selects a seller 
    $i \in \{ \Alice, \Bob\}$ to approach first. 
    Write $-i$ for the remaining seller.
    \item The buyer sends a 
    message $\sigma_i$ 
    to $\mathcal{M}_i$, 
    {and observes the outcome} %
    $(q_i, t_i)$ 
    that is sampled from the distribution
    $\mathcal{M}_i(\sigma_i)$.  
    \item 
    Next, the buyer sends a 
    message $\sigma_{-i}$ to $\mathcal{M}_{-i}$, yielding 
    an outcome $(q_{-i}, t_{-i})$
    that is sampled from the distribution
    $\mathcal{M}_{-i}({\sigma_{-i}} %
    )$.\footnote{Recall 
    that {$\sigma_{-i}$} and/or $\sigma_i$
    can be the null message $\emptyset$, 
    interpreted as non-participation. 
    }
\end{enumerate}
{We refer to this as the \emph{main model}. 
It has \emph{sequential timing}, meaning that the sellers choose their mechanisms sequentially; 
and it has an \emph{unrestricted buyer}, meaning that the buyer can choose the order of interaction with the sellers.}
\ygedit{We denote the buyer's interaction orders by
$\AB$ (Alice followed by Bob) and
$\BA$ (Bob followed by Alice).}
All players are risk neutral with quasi-linear utilities.
Alice obtains utility 
$t_\Alice$, Bob obtains utility $t_\Bob$, and the buyer's utility is 
$\max\{a \cdot q_\Alice, b \cdot q_\Bob\} - t_\Alice - t_\Bob$.

\paragraph{Subgame Perfect Equilibrium.}
Our solution concept is a subgame perfect equilibrium, which we now
describe informally in terms of the three players' strategies.
The buyer's strategy specifies, for every 
pair of mechanisms chosen by Alice and Bob, his behaviour: which seller to approach first and which messages to send, with the second message allowed to depend on the first outcome. Bob’s strategy specifies a mechanism for every possible mechanism chosen by Alice. Alice’s strategy is her choice of mechanism. These strategies form a \textit{subgame perfect equilibrium} 
if (a) for every pair of mechanisms, the 
buyer maximizes his expected utility from 
each decision onward; (b) given the buyer's
 strategy, Bob maximizes his expected revenue
 for every mechanism chosen by Alice; and 
 (c) given the buyer's and Bob's strategies, 
 Alice maximizes her expected revenue. 
 {We seek a subgame perfect equilibrium maximizing Alice's
 expected revenue.\footnote{A priori, a sequence of equilibria could give her
 strictly increasing revenues approaching a supremum that no
 equilibrium attains. Our final result rules out this possibility.}
}

\paragraph{The Buyer's Strategy.} 
{We proceed by formally defining 
the buyer's behaviour and strategy.}
Once Alice and Bob have chosen the pair of mechanisms $(\mathcal{M}_\Alice, \mathcal{M}_\Bob)$, 
all subsequent actions in the game are taken by the buyer, i.e., choosing the order in which to approach
the sellers, and then choosing the messages 
to provide to each seller --- with the message $\sigma_{-i}$ potentially depending on the outcome $(q_i, t_i)$.  We refer to this contingent sequence of choices as the buyer's \emph{action plan}, denoted $\pi(\mathcal{M}_\Alice, \mathcal{M}_\Bob)$.  We will sometimes omit the dependence on $\mathcal{M}_\Alice, \mathcal{M}_\Bob$ when clear from context.

\phantomsection\label{par:prelim-reduced-plans}
{Our first observation is that since the transfer $t_i$ does not 
influence the buyer's relative value of different outcomes 
$(q_{-i}, t_{-i})$ from mechanism $\mathcal{M}_{-i}$, 
the buyer always has a best response action plan in which 
$\sigma_{-i}$ does not depend on $t_{i}$. We
establish this formally in \cref{obs:independent-of-transfer}.}
Such an action plan $\pi$ for the buyer can therefore be described 
as a tuple $(i,\sigma_i, \sigma^W_{-i},\sigma^L_{-i}) \in 
\ActionPlans = \{ \Alice, \Bob\} \times \Sigma^3$, where $i$ is 
the seller approached first, $\sigma_i$ is the message sent to $i$, 
$\sigma^W_{-i}$ is the message sent to $-i$ when 
$q_i = 1$ {({\bf W}ins the good)}, and $\sigma^L_{-i}$ is the
message sent to $-i$ when $q_i = 0$ {({\bf L}oses the good)}.
{We treat the choice of an action plan as a single action, 
as this action plan directly determines the expected utility
of the buyer and the expected revenues of the sellers.\footnote{
\ygedit{We complete optimal action plans with optimal choices at every
unreached buyer decision, including branches of probability zero,
using the stated tie-breaking when imposed. This leaves expected
payoffs unchanged.}}} 

Given the mechanisms $\mathcal{M}_\Alice,\mathcal{M}_\Bob$
and an action plan $\pi \in \ActionPlans$, 
we denote the buyer's expected utility with 
$\utilityTri{\mathcal{M}_\Alice}{\mathcal{M}_\Bob}{\pi}$.
The buyer's strategy 
is a mapping $\hat{\mathbb{M}} : \AllMechanisms \times \AllMechanisms \to \ActionPlans$
from pairs of mechanisms $(\mathcal{M}_\Alice, \mathcal{M}_\Bob)$ 
chosen by Alice and Bob, respectively, to an action plan $\pi$.
At subgame perfect equilibrium, the buyer's choice of strategy 
$\hat{\mathbb{M}}$ 
{selects an action plan which maximizes the 
buyer's expected utility.}

{Consider a tie-breaking rule 
under which the buyer breaks ties first in favor
of maximizing the expected revenue of 
{Bob,}
then 
in favor of maximizing the expected revenue of 
{Alice.}
{If ties still remain, they are broken in favor of maximizing 
the probability of obtaining the good from 
{Bob,}
then in favor of 
maximizing the probability of obtaining the good from 
{Alice.}}
{Any remaining tie is broken in favor of the order $\AB$.}
{In \cref{obs:buyer-tie-breaks-for-alice}, we show that there exists
a buyer strategy selecting, for every pair of mechanisms, a
utility-maximizing action plan satisfying these tie-breaking rules.}
{\Cref{obs:tie-breaking-revenue-comparison} shows that, whenever
the buyer maximizes utility and Bob best-responds, replacing the buyer's 
strategy by one following these tie-breaking rules 
and replacing Bob's strategy by one which 
tie-breaks in Alice's favor weakly increases her expected revenue for every
mechanism she might choose.}}
Thus, 
\ygedit{to find Alice's optimal revenue and characterize 
all optimal mechanisms of Alice,}
it suffices to
maximize her revenue under these tie-breaking rules and show that
the maximum is attained in equilibrium.

\phantomsection\label{par:prelim-unique-outcomes}
{Under these tie-breaking assumptions, the sellers' mechanisms
$(\mathcal{M}_\Alice,\mathcal{M}_\Bob)$ uniquely determine the buyer's
expected utility, each seller's expected revenue, the marginal
allocation probability from each seller, and the interaction order.}
{We establish 
this formally in \cref{obs:uniqe-transfer-and-alloc-rules}}.
We therefore define 
$ \rev_{\mathsf{A}}(\mathcal{M}_\Alice, \mathcal{M}_\Bob)$ 
to denote the expected payment the buyer makes to Alice, given equilibrium play by the buyer,
when his value for Alice's good is $a$, for Bob's good is $b$, and
Alice's and Bob's mechanisms are $\mathcal{M}_\Alice$ 
and $\mathcal{M}_\Bob$ respectively.
Similarly, 
$\rev_{\mathsf{B}} (\mathcal{M}_\Alice, \mathcal{M}_\Bob)$ 
will be the 
{expected 
payment the buyer makes to}
Bob under equilibrium 
play by the buyer.

{Additionally, for the seller $i$ approached first, we write
$\allocDuo{i}{\mathcal{M}_\Alice}{\mathcal{M}_\Bob}$ for the
probability that the buyer obtains that seller's good.}

{From here on out, we leave the buyer's strategy implicit:
whenever we refer to an equilibrium or to the revenues induced by a pair
of mechanisms, the buyer is understood to play
according to a utility-maximizing strategy
with the tie-breaking rules stated above.
This induces a two-agent game between the sellers Alice and Bob.
}{Observe that strategies of Alice and Bob that form a subgame-perfect equilibrium in the three-player game (with the specified buyer behavior)  correspond to strategies that form a Stackelberg Equilibrium in the two-player game (induced by the buyer behavior) between Alice and Bob.}

\paragraph{Induced Stackelberg Equilibrium.} 

Having described the equilibrium behavior of the buyer, 
{
we now fix this behavior and consider the induced two-player game between Alice and Bob.}
We describe the equilibrium conditions that apply to Alice and Bob.
Recall that
Alice selects her 
mechanism first, 
then Bob responds by selecting his mechanism.
Formally, Alice's 
strategy is a mechanism $\mathcal{M}_\Alice$ and
Bob's strategy is a mapping $\mathbb{M} : \AllMechanisms \to \AllMechanisms$ 
from Alice's choice of mechanism to a mechanism for Bob.
We say these strategies of the sellers 
$(\mathcal{M}_\Alice, \mathbb{M})$ 
form a                       
{Stackelberg equilibrium (in the induced two-agent game)}
if 
{(a)}
$\mathbb{M}(\mathcal{M}') \in \arg\sup_{\mathcal{M}}
\rev_{\mathsf{B}}(\mathcal{M}', \mathcal{M})$ 
for all mechanisms 
$\mathcal{M}'$ (Bob best-responds to any possible mechanism $\mathcal{M}'$ 
chosen by Alice) 
and 
{(b)}
$\mathcal{M}_\Alice \in \arg\sup_{\mathcal{M}}
\rev_{\mathsf{A}}(\mathcal{M}, \mathbb{M}(\mathcal{M}))$ 
(Alice is choosing the revenue-maximizing mechanism in 
anticipation of Bob's best response). 
In a slight abuse of notation, we will sometimes say that mechanisms 
$(\mathcal{M}_\Alice, \mathcal{M}_\Bob)$ are in Stackelberg equilibrium if 
$\mathcal{M}_\Bob = \mathbb{M}(\mathcal{M}_\Alice)$; i.e., when 
$\mathcal{M}_\Bob$ is the on-path mechanism chosen by Bob at 
Stackelberg equilibrium.

For notational convenience, we will often denote the mechanism chosen by Alice by 
$\AliceMechanism$ instead of $\mathcal{M}_\Alice$ 
and the mechanism chosen by Bob by $\BobMechanism$ instead of $\mathcal{M}_\Bob$. 

{
\paragraph{Bob's Tie-Breaking.}
At Stackelberg equilibrium, Bob's choice of strategy 
$\mathbb{M}$ 
selects a mechanism which maximizes his expected revenue. 
Under the specified buyer behavior, every Alice mechanism admits a
best response for Bob, and Alice's revenue attains a maximum over
these best responses; see \cref{obs:buyer-tie-breaks-for-alice}.
{\Cref{obs:tie-breaking-revenue-comparison} shows that selecting
these best responses, together with the specified buyer behavior,
weakly increases Alice's revenue at the mechanism she chooses in any
subgame perfect equilibrium.}
{Thus, in seeking Alice's optimal mechanism, we may assume that Bob
selects, after every Alice mechanism, a best response maximizing her
expected revenue among his best responses.}
{Henceforth, Bob is assumed to follow this tie-breaking rule.}
}

\paragraph{Alice's Optimal Revenue.}

We wish to find the optimal revenue which Alice can achieve.
To this end, 
let $\AllStackelberg$ be the set of all Stackelberg equilibria.
We define \textit{Alice's optimal revenue}
as $\opt{a}{b} = 
\sup_{(\AliceMechanism, \mathbb{M}) \in \AllStackelberg} \revDuo{A}{\AliceMechanism}{\mathbb{M}(\AliceMechanism)}$. 
We say that a mechanism $\AliceMechanism$ is \textit{optimal} for Alice
(with respect to $(a,b)$) if there exists a strategy $\mathbb{M}$ for Bob
such that $(\AliceMechanism, \mathbb{M})$ is a Stackelberg equilibrium 
in the induced game, and for any Stackelberg equilibrium   
$(\AliceMechanism', \mathbb{M}') \in \AllStackelberg$ 
it holds that 
$\revDuo{A}{\AliceMechanism}{\mathbb{M}(\AliceMechanism)} \geq 
\revDuo{A}{\AliceMechanism'}{\mathbb{M}'(\AliceMechanism')}$. 
{For such an optimal mechanism,
$\revDuo{A}{\AliceMechanism}{\mathbb{M}(\AliceMechanism)}=\opt{a}{b}$;
that is, the supremum is attained.}
{\cref{cor:optimal-mechanisms-tie-breaking} establishes that 
a mechanism is optimal under the prescribed tie-breaking rules 
if and only if it is optimal over all subgame-perfect equilibria: 
There exists a subgame-perfect equilibrium, in which Alice offers this mechanism, 
which maximizes her expected revenue over all such equilibria. 
Consequently, characterizing all optimal mechanisms in the induced game also characterizes 
all mechanisms optimal for Alice over unrestricted subgame-perfect equilibria.
Moreover, the existence of an optimal mechanism for Alice implies that 
Alice's optimal expected revenue under the prescribed tie-breaking rules is equal to Alice's 
maximal revenue over all subgame-perfect equilibria.}

\paragraph{Pricing Functions and the Taxation Principle.} 
\label{par:prelim-pricing-functions}
We next consider an analogue of the taxation principle: 
any mechanism can equivalently be defined by a pricing function (i.e., a menu).
Formally, a \emph{pricing mechanism} is described by a pricing function 
$P \colon [0,1] \to \mathbb{R}_{\geq 0} \cup \{\infty\}$. 
The buyer can choose any $x\in[0,1]$ with $P(x)<\infty$, make an
expected payment of $P(x)$, and receive the good with probability $x$.
A price of $\infty$ means that the allocation is unavailable.
We say that a pricing function is \emph{proper}
if (a) $P(0) = 0$, (b) $P$ is non-decreasing, 
and (c) $P$ is continuous and weakly convex on the interval
$[0,\bar{x}]$, where $\bar{x} = \sup \{ x \colon P(x) < \infty \}$ 
is the supremum of the allocation probabilities available at finite prices.\footnote{
    \ygedit{Endpoint continuity is understood in the extended-real sense.
    For pricing functions used as sellers' mechanisms, we additionally
    require $P(\bar x)<\infty$, so that the finite-price graph is compact
    and admits an implementation with compact image.
    This additional condition is not part of properness, allowing
    the characterization presented in
    \cref{lem:prelim-MYERSON} to hold.}
}

\begin{definition}[{Essential Equivalence}]
    \label{def:strategic-equivalence}
    Fix
    a seller
    $i\in\{\Alice,\Bob\}$. Two mechanisms $\mathcal{M}_i$ and
    $\mathcal{M}'_i$ offered by seller $i$ are \emph{{essentially equivalent}}
    if, for every mechanism $\mathcal{M}_{-i}$ offered
    by the other seller, they induce the same expected utility for
    the buyer, the same expected revenue for each seller, and the
    same interaction order under the stated buyer tie-breaking rules.
    {They must also induce the same allocation probability from the
    first seller and the same conditional allocation probability from
    the second seller on each branch reached with positive probability,
    according to whether the buyer receives the first seller's good.}
\end{definition}

\ygedit{Informally, this is \emph{outcome equivalence} in the
induced game between Alice and Bob with respect to expected payoffs,
interaction order, and allocation behavior.}
\ygedit{Throughout the paper, mechanism uniqueness is understood up to
{essential equivalence} for the fixed buyer values and in the model under
consideration.}

The following observation shows that each mechanism has a proper
pricing representation that is {essentially equivalent} against
every mechanism of the other seller.

\begin{restatable}{observation}{taxationprinciple}
    \label{obs:taxation}
    \emph{(Taxation Principle)}
    {For a buyer with values $a,b$, every mechanism $\mathcal M$
    offered by either seller has an {essentially equivalent} proper
    pricing mechanism $\Pmarg{\mathcal M}$. Consequently, restricting
    both sellers to such pricing
    mechanisms preserves their best-response and revenue-optimization
    problems.}
\end{restatable}

See \cref{sec:prelim-app} for the proof of \cref{obs:taxation}. Given \cref{obs:taxation}, 
we will sometimes consider the pricing mechanism 
corresponding to the mechanisms chosen by the sellers.  
For an arbitrary mechanism $\mathcal{M}$, 
we will write $\Pmarg{\mathcal{M}}$ to denote the essentially equivalent 
proper pricing mechanism.

\blcomment{}

\ygcomment{}\blcomment{}

\paragraph{Allocation Rules and Implementability.}
\label{sec:prelim-allocation-payment-rules}
Throughout our analysis, it will be helpful to describe a pricing function through allocation and payment 
rules\ygedit{, even though the buyer's values are public and fixed}.
An \emph{allocation rule} is a function
$\zmarg{}:[0,\infty)\to[0,1]$. Its input $\hat{v}$ is the buyer's reported value (bid)
for a single seller's good, and its output $\zmarg{}(\hat{v})$ specifies
the probability that the buyer receives that good when reporting $\hat{v}$.
An allocation rule $\zmarg{}$ \emph{implements} 
a proper pricing function $P_{\mathcal{M}}$ if
\begin{equation*}
    \zmarg{}(\hat{v})
    \in \operatorname*{arg\,max}_{x\in[0,1]}
      \{\hat{v} \cdot x-P_{\mathcal{M}}(x)\}
    \qquad\text{for all }\hat{v}\geq0.
\end{equation*}
Thus, in a single-seller setting, 
a buyer with true value $v$ maximizes expected utility 
by reporting $\hat{v}=v$, obtaining allocation
$\zmarg{}(\hat{v})$ at expected payment $P_{\mathcal{M}}(\zmarg{}(\hat{v}))$.
An allocation rule is \emph{implementable} if it implements some proper pricing function.
A mechanism $\mathcal{M}$ may admit several implementing allocation rules when
the buyer is indifferent between options.
{We define its \emph{upper induced allocation rule}
$\zmarg{\mathcal M}^{+}$ and \emph{lower induced allocation rule}
$\zmarg{\mathcal M}^{-}$ to select, respectively, the largest and
smallest allocation probabilities $x\in[0,1]$ maximizing
$\hat v\cdot x-P_{\mathcal M}(x)$, for each $\hat v\geq0$.
These rules describe the extremal utility-maximizing allocations
in the auxiliary single-seller problem.}
{Both rules are well defined: since $P_{\mathcal M}$ is
proper, it is lower semicontinuous on $[0,1]$, so
$\hat v\cdot x-P_{\mathcal M}(x)$ is upper semicontinuous
on this compact set. Its maximizers therefore form a nonempty
compact set, which has both a largest and a smallest element.}
A \emph{payment rule} is a function
$\pmarg{}:[0,\infty)\to\mathbb{R}_{\geq0}$, where $\pmarg{}(\hat{v})$
specifies the expected payment of a buyer reporting $\hat{v}$.
{The corresponding induced payment rules are
$\pmarg{\mathcal M}^{+}(\hat v)=P_{\mathcal M}(\zmarg{\mathcal M}^{+}(\hat v))$
and
$\pmarg{\mathcal M}^{-}(\hat v)=P_{\mathcal M}(\zmarg{\mathcal M}^{-}(\hat v))$.}
The following version of Myerson's characterization
identifies which allocation rules are implementable and how they
determine prices and payment rules. 

\begin{restatable}[{{Implementable Allocation Rules (Myerson~\cite{Mye81})}}]{lemma}{myersonlemma}
    \label{lem:prelim-MYERSON}
    An allocation rule $\zmarg{}:[0,\infty)\to[0,1]$ is implementable
    if and only if it is non-decreasing.
    In this case, there exists a unique proper pricing
    function $P$ implemented by $\zmarg{}$.
    Moreover,
    \begin{equation*}
        P(\zmarg{}(\hat{v}))
        =\hat{v}\cdot\zmarg{}(\hat{v})-\int_0^{\hat{v}}\zmarg{}(u)\,du,
        \textnormal{ for all } \hat{v}\geq0.
    \end{equation*}
\end{restatable}

\ygedit{For an auxiliary single-seller value $v\ge0$, let
$\utilityMechanism{\mathcal M}{v}
=\max_{x\in[0,1]}\{v \cdot x-P_{\mathcal M}(x)\}
=v \cdot \zmarg{\mathcal M}^{+}(v)-\pmarg{\mathcal M}^{+}(v)$
denote the buyer's maximum utility from $\mathcal M$ alone.
It is well known that this function is non-decreasing, convex, and $1$-Lipschitz,
and \cref{lem:prelim-MYERSON} gives
$\utilityMechanism{\mathcal M}{v}
=\int_0^v\zmarg{\mathcal M}^{+}(u)\,du$.}

\ygedit{Henceforth, we identify each mechanism $\mathcal M$ with its
essentially equivalent proper pricing rule:
$\mathcal M(x) = P_{\mathcal{M}}(x)$ is the expected payment for
allocation probability $x\in[0,1]$, with $\mathcal M(x)=\infty$ for
unavailable allocations.}

\section{The {Main Model}}

\label{sec:stackelberg}
\label{sec:alice-opt-stackelberg}

The goal of this section is to show our main result, which is a complete characterization of optimal mechanisms for Alice in the Stackelberg setting. We restate the result below for convenience. Recall that $a > 0$ and $b > 0$ denote the buyer's values for Alice's and Bob's items respectively and $\Delta = \abs*{ a - b }$ denotes their difference. 

\introductionmainmodel*

We note that our characterization is different in the cases $a < b$ and $a \geq b$, and both of these characterizations coincide when $a = b$. Throughout, we make use of the following notation:
\begin{equation}
\label{eq:primes}
\residualValue{a}{b} = \max\paren*{ 0, a - b } \hspace{1cm}\text{and}\hspace{1cm} \residualValue{b}{a} = \max\paren*{ 0, b - a } .
\end{equation}

As stated in the informal overview in \cref{sec:introduction-proof-overview:main}, the optimal mechanisms in the $a \geq b$ case are the Large-Alice equalizer mechanisms defined in \cref{def:la-equalizer} and the optimal mechanisms in the $a < b$ case are the Large-Bob equalizer mechanisms defined in \cref{def:lb-equalizer}. As a first step in our proof, we show an optimal (Large-Alice/Large-Bob) equalizer mechanism in both cases. This constitutes a major chunk of our proof. Afterwards, we extend the ideas to give a proof of \cref{thm:introduction-variable-order-rephrased} and summarize our characterization in \cref{prop:main-optimal-mechanisms}. This part is presented at the end of \cref{sec:stackelberg:alice:la,sec:stackelberg:alice:lb}. We note that all the missing proofs are supplied in the appendix.

\subsection{Understanding Buyer Behavior}
\label{sec:stackelberg:buyer}

Recall that in our setting, the buyer faces mechanisms $\AliceMechanism$ and $\BobMechanism$ by Alice and Bob respectively and has the freedom to approach either seller first and pick any option from their menu. As this option may be a lottery, the buyer may or may not receive the item after picking this option. Based on whether or not he receives the item, his value for the other seller's item changes and he interacts with their mechanism accordingly. The buyer's choice of which seller to approach first and what options he chooses from the sellers' mechanisms is done by considering his utility function.

The first step in our proof is to connect the buyer's behavior in our duopolistic setting to his behavior in the monopolist setting. Suppose, for concreteness, that the buyer interacts with Alice first before going to Bob. Then, we show that if the buyer does not receive Alice's item, he will interact with Bob as if Bob was a monopolist. This means that he will purchase Bob's item with the monopolist allocation probability of $\zmarg{ \BobMechanism }^{+}\paren*{ b }$ and make the corresponding payment. His interaction with Bob when he does receive Alice's item is similar. After he receives Alice's item, his value for Bob's item is $\residualValue{b}{a}$ and the option he picks from Bob's menu is exactly $\zmarg{ \BobMechanism }^{+}\paren*{ \residualValue{b}{a} }$.

In contrast, the fact that the buyer can interact with Bob afterwards means that the option he picks from Alice may be different from the option he picks from her in the monopolist setting. We show that, in fact, his behavior with Alice is as if Alice was a monopolist selling an item he values at $\effectiveValue{a}{\BobMechanism}$, which is always at most $a$, but may be smaller depending on Bob's mechanism. The difference between $a$ and $\effectiveValue{a}{\BobMechanism}$ is exactly the difference in the (monopolist) utility of the buyer from Bob's mechanism when his value is augmented from $\residualValue{b}{a}$ to $b$. We call this the Augmentation-induced Utility Gap ($\aug$). Formally, for values $v, w \geq 0$ and a monopolist using the mechanism $\mathcal{M}$, this gap is defined as:
\begin{equation}
\label{eq:aug}
\aug_{ \mathcal{M} }\paren*{ v, w } = \utilityMechanism{ \mathcal{M} }{ w } - \utilityMechanism{ \mathcal{M} }{ \max\paren*{ 0, w - v } } .
\end{equation}
Observe from the fact that utilities are monotone and $1$-Lipschitz that $0 \leq \aug_{ \mathcal{M} }\paren*{ v, w } \leq \min\paren*{ v, w }$. Using this definition, we can now write the buyer's behavior in our duopolistic setting in terms of his behavior in the monopolist setting as follows:

\begin{restatable}{lemma}{buyerbehavior}
\label{lemma:buyer}
Let Alice and Bob use mechanisms $\AliceMechanism$ and $\BobMechanism$ respectively. Define the quantities $\effectiveValue{a}{\BobMechanism} = a - \aug_{ \BobMechanism }\paren*{ a, b }$ and $\effectiveValue{b}{\AliceMechanism} = b - \aug_{ \AliceMechanism }\paren*{ b, a }$.
\begin{enumerate}
\item \label{item:buyer:util} The maximum utility the buyer can obtain when interacting with Alice first (denoted $\utilityAB{ \AliceMechanism }{ \BobMechanism }$) and when interacting with Bob first (denoted $\utilityBA{ \AliceMechanism }{ \BobMechanism }$) respectively is: 
\[
\utilityAB{ \AliceMechanism }{ \BobMechanism } = \utilityMechanism{ \AliceMechanism }{ \effectiveValue{a}{\BobMechanism} } + \utilityMechanism{ \BobMechanism }{ b } \qquad \text{and} \qquad \utilityBA{ \AliceMechanism }{ \BobMechanism } = \utilityMechanism{ \BobMechanism }{ \effectiveValue{b}{\AliceMechanism} } + \utilityMechanism{ \AliceMechanism }{ a } .
\]
\item \label{item:buyer:allocation} When the buyer interacts with Alice first, the utility in \cref{item:buyer:util} is attained by first picking the option $x_A$ from Alice, where
\[
x_A = \begin{cases}
\zmarg{ \AliceMechanism }^-\paren*{ \effectiveValue{a}{\BobMechanism} } , &\text{~if~} \zmarg{ \BobMechanism }^{+}\paren*{ b } > \zmarg{ \BobMechanism }^{+}\paren*{ \residualValue{b}{a} } \\
\zmarg{ \AliceMechanism }^{+}\paren*{ \effectiveValue{a}{\BobMechanism} } , &\text{~if~} \zmarg{ \BobMechanism }^{+}\paren*{ b } = \zmarg{ \BobMechanism }^{+}\paren*{ \residualValue{b}{a} }
\end{cases} ,
\]
and then picking the option $\zmarg{ \BobMechanism }^{+}\paren*{ \residualValue{b}{a} }$ from Bob if he receives Alice's item, and the option $\zmarg{ \BobMechanism }^{+}\paren*{ b }$ from Bob otherwise. When the buyer interacts with Bob first, our tie breaking implies that an analogous result holds except that the buyer always picks the option $\zmarg{ \BobMechanism }^{+}\paren*{ \effectiveValue{b}{\AliceMechanism} }$ from Bob.
\item \label{item:buyer:payment} When the buyer interacts with Alice first, his payments to Alice and Bob are (here, $x_A$ is as defined above):
\begin{align*}
\revDuo{ A }{ \AliceMechanism }{ \BobMechanism } &= \AliceMechanism\paren*{ x_A } \\
\revDuo{ B }{ \AliceMechanism }{ \BobMechanism } &= \paren*{ 1 - \zmarg{ \AliceMechanism }^-\paren*{ \effectiveValue{a}{\BobMechanism} } } \cdot \BobMechanism\paren*{ \zmarg{ \BobMechanism }^{+}\paren*{ b } } + \zmarg{ \AliceMechanism }^-\paren*{ \effectiveValue{a}{\BobMechanism} } \cdot \BobMechanism\paren*{ \zmarg{ \BobMechanism }^{+}\paren*{ \residualValue{b}{a} } } .
\end{align*}
The case when the buyer interacts with Bob first is analogous.
\end{enumerate}
\end{restatable}

\subsection{The Structure of Bob's Response}
\label{sec:stackelberg:bob}

\cref{lemma:buyer} gives us a handle on the buyer's behavior when faced with mechanisms $\AliceMechanism$ and $\BobMechanism$ from Alice and Bob respectively. In this section, we use it to get a handle on Bob's best response to a given mechanism $\AliceMechanism$ by Alice. Given such an $\AliceMechanism$, Bob can design his mechanism such that the buyer will approach him first and can also design it so that the buyer approaches him second. We handle these two cases separately.

\subsubsection{When the Buyer Interacts with Bob First}
\label{sec:stackelberg:bob:bob}

Suppose Bob decides to design a mechanism $\BobMechanism$ that will make the buyer visit him first. In this case, we know from \cref{lemma:buyer} that the buyer will interact with this mechanism as if his value for Bob's item is $\effectiveValue{b}{\AliceMechanism}$, where $\effectiveValue{b}{\AliceMechanism}$ depends on Alice's mechanism and is defined in \cref{lemma:buyer}. As Bob knows Alice's mechanism, he also knows $\effectiveValue{b}{\AliceMechanism}$ and one might initially try to simply have Bob sell his item at a fixed price of $\effectiveValue{b}{\AliceMechanism}$, thereby extracting the highest revenue he can possibly extract.

The problem with this approach is that there is no guarantee that selling the item at a fixed price of $\effectiveValue{b}{\AliceMechanism}$ would ensure that the buyer approaches Bob first. We adopt a different approach, using the observation that any mechanism to a buyer with a known type is equivalent to a mechanism with a single lottery option. Thus, it suffices to consider single-lottery mechanisms that are determined by a parameter $x \in \bracket*{ 0, 1 }$ and a per-unit-price $p \geq 0$, where the only option the buyer has (other than the null option of not interacting with the mechanism at all) is to take the item with probability $x$ by paying the amount $xp$. Formally, we define these mechanisms as follows.

\begin{definition}
\label{def:sl}
Let $x \in \bracket*{ 0, 1 }$ and $p \geq 0$. We define the mechanism $\SL_{ x, p }$ such that, for all $z \in \bracket*{ 0, 1 }$, we have:
\[
\SL_{ x, p }\paren*{ z } = \begin{cases}
zp, &\text{~if~} 0 \leq z \leq x \\
\infty , &\text{~if~} x < z \leq 1
\end{cases} .
\]
For all $p \geq 0$, we also define $\FP{p} = \SL_{ 1, p }$ to be the fixed price mechanism with price $p$.
\end{definition}

Note that the $\infty$ in the above definition is to conveniently prohibit the buyer from getting the item with probability more than $x$. We can replace it by any number larger than $b$ and have the same effect. With this definition, we are now faced with the problem of determining for what values of the parameters $x$ and $p$ the buyer would interact with Bob first when he uses the mechanism $\SL_{ x, p }$, and among these values, what values will maximize the revenue Bob gets.

To understand this, we fix the parameter $x \in \bracket*{ 0, 1 }$ arbitrarily and show that there is a well-defined threshold $\p_{ \AliceMechanism }(x)$ determined by $x$ and the mechanism used by Alice, such that any price $p$ larger than $\p_{ \AliceMechanism }(x)$ would make the buyer approach Alice first and (except for some corner cases) any price $p$ smaller than $\p_{ \AliceMechanism }(x)$ would make the buyer interact with Bob first. The threshold $\p_{ \AliceMechanism }(x)$ can also be viewed as the price for which the buyer is indifferent between which seller to approach first and is defined next.

\begin{definition}
\label{def:sl-params}
Let $\AliceMechanism$ be a mechanism for Alice and $x \in \bracket*{ 0, 1 }$. Define $\vthreshold_{ \AliceMechanism }(x)$ to be the largest value $v \in \bracket*{ 0, a }$ satisfying: 
\[
\utilityMechanism{ \AliceMechanism }{ v } \leq \utilityMechanism{ \AliceMechanism }{ a } - x \cdot \aug_{ \AliceMechanism }\paren*{ b, a } .
\]
Note that $\vthreshold_{ \AliceMechanism }(x)$ is well defined as $v = 0$ is one such value, and that the continuity of $\utilityMechanism{ \AliceMechanism }{ v }$ implies that there is a largest value satisfying the above inequality. Using this, define:
\[
\p_{ \AliceMechanism }(x) = \begin{cases}
b - \frac{ a - \vthreshold_{ \AliceMechanism }(x) }{ x }, &\text{~if~} x > 0 \\
b - \aug_{ \AliceMechanism }\paren*{ b, a } , &\text{~if~} x = 0
\end{cases} .
\]
\end{definition}
Note that the mechanism $\SL_{ x, p }$ is independent of $p$ when $x = 0$. Thus, as far as the mechanism $\SL_{ x, \p_{ \AliceMechanism }(x) }$ is concerned, the second case in the above definition is irrelevant and we define it to be $b - \aug_{ \AliceMechanism }\paren*{ b, a }$ only for technical convenience. \cref{lemma:sl-params} establishes some basic properties of the quantities defined above. In particular, it implies that the price $\p_{ \AliceMechanism }(x)$ always lies in the range $ \bracket*{ \residualValue{b}{a}, b - \aug_{ \AliceMechanism }\paren*{ b, a } }$.

In the following lemma, we show that $\p_{ \AliceMechanism }(x)$ indeed satisfies the property we desire, namely that the buyer goes to Alice first if $p > \p_{ \AliceMechanism }(x)$ and barring some special cases, goes to Bob first whenever $p < \p_{ \AliceMechanism }(x)$. It follows that the highest revenue Bob can get from a mechanism of the form $\SL_{ x, p }$ is $x \cdot \p_{ \AliceMechanism }(x)$. Then, we can maximize over $x$ to get the highest revenue possible by a mechanism that makes the buyer approach him first. This is captured in \cref{lemma:bob-first}.

\begin{lemma}
\label{lemma:sl}
Let $\AliceMechanism$ be a mechanism for Alice and $x \in \bracket*{ 0, 1 }$. For all $p \in \bracket*{ \residualValue{b}{a}, b - \aug_{ \AliceMechanism }\paren*{ b, a } }$, it holds that:
\[
p \leq \p_{ \AliceMechanism }(x) \iff \utilityAB{ \AliceMechanism }{ \SL_{ x, p } } \leq \utilityBA{ \AliceMechanism }{ \SL_{ x, p } } .
\]
Moreover, if we have $\utilityMechanism{ \AliceMechanism }{ a } > 0$ and $0 < x < 1$, then, for all $p \in \bracket*{ \residualValue{b}{a}, b - \aug_{ \AliceMechanism }\paren*{ b, a } }$, it holds that:
\[
p < \p_{ \AliceMechanism }(x) \iff \utilityAB{ \AliceMechanism }{ \SL_{ x, p } } < \utilityBA{ \AliceMechanism }{ \SL_{ x, p } } .
\]
\end{lemma}

\begin{lemma}
\label{lemma:bob-first}
Let $\AliceMechanism$ and $\BobMechanism$ be mechanisms for Alice and Bob respectively such that the buyer, when faced with this pair of mechanisms, interacts with Bob first. Then, there exists $x \in \bracket*{ 0, 1 }$ such that $\revDuo{ B }{ \AliceMechanism }{ \BobMechanism } \leq x \cdot \p_{ \AliceMechanism }(x)$.
\end{lemma}

\subsubsection{When the Buyer Interacts with Alice First}
\label{sec:stackelberg:bob:alice}

In this section, we consider the case where, when faced with a mechanism $\AliceMechanism$ from Alice, Bob designs his best response so that the buyer approaches him second. As explained in \cref{sec:stackelberg:buyer}, this means that when the buyer receives Alice's item, he interacts with Bob as if Bob was a monopolist selling an item he values at $\residualValue{b}{a}$, and when the buyer does not receive Alice's item, he interacts with Bob as if Bob was a monopolist selling an item he values at $b$. Of course, when $b \leq a$, we have $\residualValue{b}{a} = 0$, and the buyer need not participate in Bob's mechanism given that he has received the item from Alice.

This situation is similar to the situation in the previous section except that Bob now faces a buyer that can have two values instead of one. This might suggest that we now need four parameters to describe Bob's mechanism, an allocation probability and a per-unit-price for the buyer with value $\residualValue{b}{a}$, and another such pair for the buyer with value $b$. 

A closer look at \cref{lemma:buyer} reveals that we actually do not need all these parameters. Indeed, \cref{lemma:buyer} says that the only utilities the buyer cares about when deciding which seller to go to first are the utilities at values $\residualValue{b}{a}$, $\effectiveValue{b}{\AliceMechanism}$, and $b$. The last two are directly used in the utility computation in \cref{item:buyer:util} of \cref{lemma:buyer} and the first one is used implicitly in the computation of the value $\effectiveValue{a}{\BobMechanism}$. Thus, if we change Bob's mechanism while maintaining the value of these utilities relative to each other, the buyer's maximum utilities in the two orders will remain unchanged.

As $\residualValue{b}{a} \leq \effectiveValue{b}{\AliceMechanism} \leq b$, one such change we can do is to increase the payment uniformly so that no buyer with value smaller than $\residualValue{b}{a}$ participates in Bob's mechanism. This process decreases the utility of all the remaining buyers by the same amount and makes sure that the buyer with value $\residualValue{b}{a}$ pays a per-unit-price of $\residualValue{b}{a}$. Similarly, one can also increase the allocation probability of the buyer with value $b$ all the way to $1$ in a way that does not affect any of the utilities of the values at most $b$. This again will not change the relative utilities and only increase Bob's revenue.

With these transformations, two of the four parameters are already fixed. The per-unit-price of the buyer with value $\residualValue{b}{a}$ is always $\residualValue{b}{a}$, and the allocation probability of the buyer with value $b$ is always $1$. The remaining parameters are the allocation probability of the buyer with value $\residualValue{b}{a}$, which we will denote by $x$, and the per-unit-price (which is the same as the price as the allocation probability is $1$) of the buyer with value $b$, which we will denote by $q$. Myerson incentive compatibility constraints imply that $\residualValue{b}{a} \leq q \leq b - x \cdot \paren*{ b - \residualValue{b}{a} }$. We now define double lottery mechanisms that are parametrized by $x$ and $q$.

\begin{restatable}{definition}{doublelotterymechanism}
\label{def:dl}
Let $x \in \bracket*{ 0, 1 }$ and $q \in \bracket*{ \residualValue{b}{a}, b - x \cdot \paren*{ b - \residualValue{b}{a} } }$. We define the mechanism $\DL_{ x, q }$ such that, for all $z \in \bracket*{ 0, 1 }$, we have:
\[
\DL_{ x, q }\paren*{ z } = \begin{cases}
z\residualValue{b}{a}, &\text{~if~} 0 \leq z \leq x \\
z \cdot \frac{ q - x\residualValue{b}{a} }{ 1 - x } - x \cdot \frac{ q - \residualValue{b}{a} }{ 1 - x } , &\text{~if~} x < z \leq 1
\end{cases} .
\]
\end{restatable}
Observe that $\DL_{ 0, q } = \FP{q}$ for all $q \in \bracket*{ \residualValue{b}{a}, b }$. Intuitively, the mechanism $\DL_{ x, q }$ has exactly the two options for the buyer that we described above: pay $x\residualValue{b}{a}$ and get the item with probability $x$, or pay $q$ and get the item with probability $1$. The remaining values of $z$ are obtained by interpolating between these two options and the null option. 

As in the previous section, the task now is to determine the values of the parameters $x$ and $q$ for which the buyer, when faced with the mechanism $\AliceMechanism$ from Alice, and the mechanism $\DL_{ x, q }$ from Bob, decides to interact with Alice first. We adopt a similar approach, this time parameterizing based on $q$ instead of $x$. Specifically, we show that, for all $q \in \bracket*{ \residualValue{b}{a}, b }$, there is a threshold $\x_{ \AliceMechanism }(q)$ determined by the value $q$ and $\AliceMechanism$ such that whenever $x$ is larger than this threshold, the buyer buys from Bob first. The following definition defines this threshold, while \cref{lemma:dl-util,lemma:dl-params} capture some basic properties of the two definitions we made in this section.

\begin{definition}
\label{def:dl-params}
Let $\AliceMechanism$ be a mechanism for Alice and $q \in \bracket*{ \residualValue{b}{a}, b }$. Define the value:
\[
\x_{ \AliceMechanism }(q) = \begin{cases}
\frac{ b - q - \aug_{ \AliceMechanism }\paren*{ b - q, a } }{ \min\paren*{ a, b } - \aug_{ \AliceMechanism }\paren*{ \min\paren*{ a, b }, a } }, &\text{~if~} \aug_{ \AliceMechanism }\paren*{ b, a } < \min\paren*{ a, b } \\
\frac{ b - q }{ b - \residualValue{b}{a} }, &\text{~if~} \aug_{ \AliceMechanism }\paren*{ b, a } = \min\paren*{ a, b } 
\end{cases} .
\]
\end{definition}
Note that the cases above are exhaustive as $\aug_{ \AliceMechanism }\paren*{ b, a }$ can never be larger. The interesting case in this definition is the first one. It remains to show that $\x_{ \AliceMechanism }(q)$ actually does what it is supposed to do. Namely, that when $x > \x_{ \AliceMechanism }(q)$, then the buyer will buy from Bob first, and when $x < \x_{ \AliceMechanism }(q)$, then the buyer will buy from Alice first. This is done in the next lemma. We show the former claim directly, while the second claim is implicit in the ``moreover'' part of the lemma statement, which calculates Bob's revenue when he uses the mechanism $\DL_{ \x_{ \AliceMechanism }(q), q }$ while treating the cases $a \leq b$ and $a \geq b$ separately. These calculated values are consistent with the values Bob would get if he were to be approached after Alice, and thus capture the essence of the latter claim. Later, in \cref{lemma:alice-first}, we show that maximizing over $q$ captures exactly the maximum revenue that Bob can get by designing a mechanism such that the buyer approaches him second.

\begin{lemma}
\label{lemma:dl}
Let $\AliceMechanism$ be a mechanism for Alice and $q \in \bracket*{ \residualValue{b}{a}, b }$. For all $x \in \bracket*{ 0, \frac{ b - q }{ b - \residualValue{b}{a} } }$, we have:
\[
x \leq \x_{ \AliceMechanism }(q) \iff \utilityBA{ \AliceMechanism }{ \DL_{ x, q } } \leq \utilityAB{ \AliceMechanism }{ \DL_{ x, q } } .
\]
Moreover, the following hold:
\begin{enumerate}
\item \label{item:dl:la} If $a \geq b$, for all $x \in \bracket*{ 0, \x_{ \AliceMechanism }(q) }$, we have $\revDuo{ B }{ \AliceMechanism }{ \DL_{ x, q } } = q \cdot \paren*{ 1 - \zmarg{ \AliceMechanism }^-\paren*{ \Delta + q } }$.
\item \label{item:dl:lb} If $a \leq b$, we have $\revDuo{ B }{ \AliceMechanism }{ \DL_{ \x_{ \AliceMechanism }(q), q } } = q - \zmarg{ \AliceMechanism }^-\paren*{ q - \Delta } \cdot \paren*{ q - \Delta \cdot \x_{ \AliceMechanism }(q) }$.
\end{enumerate}
\end{lemma}

\begin{lemma}
\label{lemma:alice-first}
Let $\AliceMechanism$ and $\BobMechanism$ be mechanisms for Alice and Bob respectively such that the buyer, when faced with this pair of mechanisms, interacts with Alice first. Then, for $q = b - a + \effectiveValue{a}{\BobMechanism} \in \bracket*{ \residualValue{b}{a}, b }$, we have that $\revDuo{ B }{ \AliceMechanism }{ \BobMechanism } \leq \revDuo{ B }{ \AliceMechanism }{ \DL_{ \x_{ \AliceMechanism }(q), q } }$.
\end{lemma}

\subsection{Alice's Optimal Mechanism}
\label{sec:stackelberg:alice}

Now that we know how Bob and the buyer respond given any mechanism by Alice, we are ready to describe her optimal mechanism. For this, we divide the proof into two cases. The case when $a \geq b$ is called the Large-Alice case and is discussed in \cref{sec:stackelberg:alice:la} while the case where $a < b$ is called the Large-Bob case and is discussed in \cref{sec:stackelberg:alice:lb}. Both these cases build on a common theme, which we discuss here.

\ygedit{Suppose that $\AliceMechanism$ is an arbitrary mechansim of Alice and $\BobMechanism$ is a best-response of Bob. Let $R$ be Bob's revenue.} Then, 
no mechanism for Bob can get him revenue larger than $R$.
Applying this property to the single-lottery and double-lottery mechanisms discussed above (\cref{lemma:sl,lemma:dl}) gives us constraints on Alice's mechanism. For example, when $a \geq b$, then the fact that the revenue calculation in \cref{item:dl:la} of \cref{lemma:dl} is at most $R$ gives us a lower bound on the allocation rule of Alice's mechanism for a range of values. We use these constraints to show that Alice's mechanism must give a buyer with value $a$ for her item a large (monopolist) utility. That is, we get that $\utilityMechanism{ \AliceMechanism }{ a }$ has a lower bound in terms of $R$.

Next, we use the fact that there is a mechanism, namely, the mechanism $\BobMechanism$, that gives Bob a revenue of $R$ to get an upper bound on Alice's revenue. This upper bound is in terms of $R$ and $\utilityMechanism{ \AliceMechanism }{ a }$ and the bound gets smaller as $\utilityMechanism{ \AliceMechanism }{ a }$ increases. Then, we can plug in our lower bound for $\utilityMechanism{ \AliceMechanism }{ a }$ in terms of $R$ to get an upper bound on Alice's revenue only in terms of $R$. To finish the proof, we construct mechanisms that match this upper bound for all $R$.
\ygedit{These mechanisms ensure that for a wide range of $q$, Bob's revenue when he uses the mechanism $\DL_{ \x_{ \AliceMechanism }(q), q }$ is the exact same, and equal to $R$.
The same holds for $\SL_{x,\p_{\AliceMechanism}(x)}$ with a range 
of values $x$ (see \cref{fig:bob-equalizer-main}). So,}
we call these mechanisms Large-Alice-Equalizer mechanisms and Large-Bob-Equalizer mechanisms in the two cases. As they match our upper bound for all $R$, an optimization over $R$ gives us the optimum Stackelberg revenue for Alice.

\begin{figure}[H]
  \centering
  \newsavebox{\BobEqualizerNativeBox}
  \newsavebox{\BobEqualizerScaledBox}
  \sbox{\BobEqualizerNativeBox}{%
    \input{observations/introduction_tikz_plots/bob_equalizer_main_a2_b1.tikz.tex}%
  }%
  \sbox{\BobEqualizerScaledBox}{%
    \resizebox{0.96\textwidth}{!}{\usebox{\BobEqualizerNativeBox}}%
  }%
  \pgfmathsetlengthmacro{\BobEqualizerScaledOffset}{%
    (\wd\BobEqualizerScaledBox/\wd\BobEqualizerNativeBox)
    *\BobEqualizerPlotCenterOffset}%
  \noindent\makebox[\linewidth][l]{%
    \hspace*{\dimexpr\linewidth/2-\wd\BobEqualizerScaledBox/2
      -\BobEqualizerScaledOffset\relax}%
    \usebox{\BobEqualizerScaledBox}%
  }%
  \caption{Bob's revenue curves for Alice's canonical optimal menu
  ($a=2$, $b=1$). \textbf{Left:} plots $\revDuo{B}{\AliceMechanism}{\DL_{\x_{\AliceMechanism}(q),q}}$ as $q$ varies.
  As $a\ge b$, the same curve is also Bob's revenue from 
  fixed prices (see \cref{item:dl:la} of \cref{lemma:dl}).
  \textbf{Right:} plots
  $\revDuo{B}{\AliceMechanism}{\SL_{x,\p_{\AliceMechanism}(x)}}$
  as $x$ varies.
  Shading marks intervals of best responses.}
  \label{fig:bob-equalizer-main}
\end{figure}

\subsubsection{When $a \geq b$}
\label{sec:stackelberg:alice:la}

We now instantiate the above procedure for the case of $a \geq b$. In this case, we have from \cref{eq:primes} that $\residualValue{a}{b} = a - b = \Delta$ and $\residualValue{b}{a} = 0$. \ygedit{Consider any mechanism $\AliceMechanism$ and a best response $\BobMechanism$ yielding Bob revenue $R$.} As mentioned above, the first step in the procedure is to use the fact that Bob cannot get revenue larger than $R$ regardless of what mechanism he chooses to get a lower bound on Alice's monopolist utility for a buyer with value $a$.

We do this in two steps. First, for any $x \in \bracket*{ \frac{ R }{ b }, 1 }$, we note that \cref{lemma:bob-first} gives an upper bound on $x \cdot \p_{ \AliceMechanism }(x)$ which by \cref{def:sl-params} is an upper bound on $\vthreshold_{ \AliceMechanism }(x)$ and implies that Alice's utility at a certain point (determined by $x$) is large. From \cref{item:dl:la} of \cref{lemma:dl}, we can lower bound Alice's allocation rule beyond that point and combining these gives the following lower bound on $\utilityMechanism{ \AliceMechanism }{ a }$.

\begin{lemma}
\label{lemma:alice-structure-1:la}
Let $R > 0$ and $\AliceMechanism$ be a mechanism for Alice such that $\utilityMechanism{ \AliceMechanism }{ a } > 0$ and for all mechanisms $\BobMechanism$ for Bob, we have $\revDuo{ B }{ \AliceMechanism }{ \BobMechanism } \leq R$. For all $x \in \bracket*{ \frac{ R }{ b }, 1 }$, we have: 
\[
\utilityMechanism{ \AliceMechanism }{ a } \geq \frac{ 1 }{ x } \cdot \int_{ R + \Delta + b \cdot \paren*{ 1 - x } }^a \paren*{ 1 - \frac{ R }{ v - \Delta } } \del v .
\]
\end{lemma}

Next, we use the fact that Bob can get revenue $R$ by using the mechanism $\BobMechanism$ to get the following upper bound on Alice's revenue.

\begin{lemma}
\label{lemma:alice-structure-2:la}
Let $a \geq b$.
Let $\AliceMechanism$ and $\BobMechanism$ be an arbitrary pair of mechanisms for Alice and Bob respectively and let $R = \revDuo{ B }{ \AliceMechanism }{ \BobMechanism }$. We have:
\[
\revDuo{ A }{ \AliceMechanism }{ \BobMechanism } \leq a \cdot \paren*{ 1 - \frac{ R }{ b } } - \utilityMechanism{ \AliceMechanism }{ a }
\]
\end{lemma}

Combining these two lemmas (and using the appropriate $x$ in \cref{lemma:alice-structure-1:la}) gives us the upper bound on Alice's revenue that we need. To state this upper bound, we define $s = s(R) \in \bracket*{ R + \Delta, a }$ to be the unique value that satisfies 
\begin{equation}
\label{eq:s:la-main}
\frac{ b + R }{ s - \Delta } - \ln \frac{ b }{ s - \Delta } = 2 .
\end{equation}
The proofs of the existence and uniqueness of this value and other details can be found in the appendix. Setting $x$ so that the lower limit of the integral in \cref{lemma:alice-structure-1:la} is $s$, we get the following upper bound on Alice's revenue in terms of $R$.
\begin{equation}
\label{eq:arev-ub-la}
\revDuo{ A }{ \AliceMechanism }{ \BobMechanism } \leq a \cdot \paren*{ 1 - \frac{ R }{ b } } - b \cdot \paren*{ 1 - \frac{ R }{ s - \Delta } } .
\end{equation}
To finish the proof, we define Large-Alice-equalizer mechanisms that take $R$ as a parameter and achieve the bound above. These are defined as follows.
\begin{definition}[Large-Alice-equalizer mechanisms]
\label{def:la-equalizer}
Let $a \geq b$. Given a parameter $R \in \bracket*{ 0, b }$, let $s \in \bracket*{ R + \Delta, a }$ be as in \cref{eq:s:la-main}. We define the mechanism $\AliceMechanism_R^{ \mathsf{LA} }$ to be the mechanism with the following allocation rule:
\[
\zmarg{ \AliceMechanism_R^{ \mathsf{LA} } }^{+}\paren*{ v } = \begin{cases}
0, &\text{~if~} 0 \leq v < R + \Delta \\
1 - \frac{ R }{ s - \Delta }, &\text{~if~} R + \Delta \leq v < s \\
1 - \frac{ R }{ v - \Delta }, &\text{~if~} s \leq v \leq a 
\end{cases} .
\]
After $a$, the allocation retains its value at $a$. The mechanisms $\AliceMechanism_R^{ \mathsf{LA} }$ are called Large-Alice-equalizer mechanisms.
\end{definition}
In words, the allocation of a Large-Alice-equalizer mechanism is $0$ until $R + \Delta$, a constant in the interval $\bracket*{ R + \Delta, s }$, and gradually rises beyond that, before plateauing at $a$. It is easy to see that this allocation rule is monotone, implying by \cref{lem:prelim-MYERSON} that there is a proper pricing rule that implements it. Also, note that the above definition implies:
\begin{equation}
\label{eq:la-equalizer-util}
\utilityMechanism{ \AliceMechanism_R^{ \mathsf{LA} } }{ a } = \frac{ \paren*{ s - R - \Delta }^2 }{ s - \Delta } + a - s - R \cdot \ln \frac{ b }{ s - \Delta } = b \cdot \paren*{ 1 - \frac{ R }{ s - \Delta } } ,
\end{equation}
by \cref{eq:s:la-main}. We next show that when Alice uses the mechanism $\AliceMechanism_R^{ \mathsf{LA} }$, Bob can never get revenue more than $R$. Moreover, using the mechanism $\FP{b}$, Bob can attain this revenue.
\begin{lemma}
\label{lemma:la-equalizer}
Let $a \geq b$ and $R \in \bracket*{ 0, b }$. For all mechanisms $\BobMechanism$ for Bob, it holds that $\revDuo{ B }{ \AliceMechanism_R^{ \mathsf{LA} } }{ \BobMechanism } \leq \revDuo{ B }{ \AliceMechanism_R^{ \mathsf{LA} } }{ \FP{b} } = R$. Moreover, when $\BobMechanism$ is the best-response picked by Bob, we have
\[
\revDuo{ A }{ \AliceMechanism_R^{ \mathsf{LA} } }{ \BobMechanism } = a \cdot \paren*{ 1 - \frac{ R }{ b } } - \utilityMechanism{ \AliceMechanism_R^{ \mathsf{LA} } }{ a } = a \cdot \paren*{ 1 - \frac{ R }{ b } } - b \cdot \paren*{ 1 - \frac{ R }{ s - \Delta } } .
\]
\end{lemma}
As this revenue matches the above upper bound exactly, we get that it is Alice's optimum revenue for any $R$. An optimization over $R$ finishes the proof. The formal proof can be found in \cref{app:stackelberg:alice:la}.

\paragraph{A characterization of all equilibria (Proof of \cref{thm:introduction-variable-order-rephrased}).} \label{par:la-characterize} We now extend these ideas to prove \cref{thm:introduction-variable-order-rephrased} in the case $a \geq b$. Specifically, let $R^* \in \bracket*{ 0, b }$ be such that $\AliceMechanism^* = \AliceMechanism_{ R^* }^{ \mathsf{LA} }$ maximizes Alice's revenue, {\em i.e.}, $\AliceMechanism^*$ and $\FP{b}$ constitute a Stackelberg equilibrium. Define $s^* = s\paren*{ R^* }$ and observe that $s^* < a$. We will show that a mechanism $\AliceMechanism$ for Alice is optimal in the Stackelberg setting if and only if the following conditions hold:
\begin{align*}
\utilityMechanism{ \AliceMechanism }{ v } &= \utilityMechanism{ \AliceMechanism^* }{ v } && \forall v \in \set*{ \Delta } \cup \bracket*{ s^*, a } \\
\zmarg{ \AliceMechanism }^{+}\paren*{ v } &\in \bracket*{ 1 - \frac{ R^* }{ v - \Delta }, \frac{ \utilityMechanism{ \AliceMechanism^* }{ a } }{ b } } && \forall v \in \oc*{ \Delta, s^* } .
\end{align*}
Next, we show that these conditions are necessary and sufficient for $\AliceMechanism$ to be optimal for Alice in the Stackelberg setting. We first show that they are necessary. Let $\AliceMechanism$ be optimal for Alice and let $\BobMechanism$ be Bob's best response to $\AliceMechanism$. Let $R = \revDuo{ B }{ \AliceMechanism }{ \BobMechanism }$ be Bob's revenue at equilibrium. Then, as our proof shows that $R^*$ is the unique maximizer of the right hand side of \cref{eq:arev-ub-la}, we must have $R = R^*$.
{First suppose that $a\geq Cb$, so $R=R^*=0$.
For $q\in\oc*{0,b}$, \cref{item:dl:la} of \cref{lemma:dl} applied to $\FP{q}$ gives
$q\cdot\paren*{1-\zmarg{\AliceMechanism}^-\paren*{\Delta+q}}=0$.
Letting $q\downarrow0$ in the optimality inequalities for allocation one yields
$\AliceMechanism(1)-\AliceMechanism(x)\leq\Delta(1-x)$ for all $x\in[0,1]$.
Taking $x=0$ gives $\AliceMechanism(1)\leq\Delta$, while monotonicity bounds Alice's revenue $\Delta$ by $\AliceMechanism(1)$.
Hence $\AliceMechanism(1)=\Delta$ and $\AliceMechanism(x)\geq\Delta x$.
Convexity and $\AliceMechanism(0)=0$ give the reverse inequality.
Hence $\AliceMechanism=\FP{a-b}$, which satisfies both conditions above.}

{For the remaining case $b\leq a<Cb$, we have $R>0$.}
\cref{item:dl:la} of \cref{lemma:dl} then implies the lower bound on $\zmarg{ \AliceMechanism }^{+}\paren*{ v }$. Next, note that the optimality of Alice's revenue and \cref{lemma:alice-structure-2:la} imply that $\utilityMechanism{ \AliceMechanism }{ a } \leq \utilityMechanism{ \AliceMechanism^* }{ a } = b \cdot \paren*{ 1 - \frac{ R }{ s - \Delta } }$. This means that all the calculations in the proof of \cref{lemma:alice-structure-1:la} must be tight implying the required utility constraints. It follows that $\zmarg{ \AliceMechanism }^{+}\paren*{ s^* } = \frac{ \utilityMechanism{ \AliceMechanism^* }{ a } }{ b }$. Finally, the upper bound on $\zmarg{ \AliceMechanism }^{+}\paren*{ v }$ follows because of the monotonicity of the allocation rule.

To verify the sufficiency of these conditions, consider any mechanism that satisfies the above conditions and observe that the proof of \cref{lemma:la-equalizer} goes through using just these conditions. The only step that does not follow directly is the lower bound on $\utilityMechanism{ \AliceMechanism }{ R + a - bx }$ in the displayed equation in the proof. We can substitute that by using the upper bound on $\zmarg{ \AliceMechanism }^{+}\paren*{ v }$. It follows that posting a fixed price of $b$ is optimal for Bob and that the buyer visits Alice first and picks the option $\zmarg{ \AliceMechanism }^-\paren*{ a }$, finishing the proof.

{For $b \leq a < Cb$, we now show that in every Stackelberg equilibrium with an optimal mechanism of Alice under our tie-breaking rules, Bob offers $\FP{b}$, the buyer visits Alice first, and all three players receive the same payoffs as when Alice offers $\AliceMechanism^*$ and Bob offers $\FP{b}$.
Let $\paren*{ \AliceMechanism, \BobMechanism }$ be such an equilibrium, and let $R=\revDuo{ B }{ \AliceMechanism }{ \BobMechanism }$ denote Bob's revenue. The characterization gives $R=R^*>0$, $\utilityMechanism{ \AliceMechanism }{ \Delta }=0$, $\utilityMechanism{ \AliceMechanism }{ a }>0$, and $x^*=\zmarg{ \AliceMechanism }^-\paren*{ a }=1-R^*/b<1$.
Monotonicity of the allocation rule therefore implies $\utilityMechanism{ \AliceMechanism }{ a }-\utilityMechanism{ \AliceMechanism }{ a-\utilityMechanism{ \BobMechanism }{ b } }\leq x^*\utilityMechanism{ \BobMechanism }{ b }<\utilityMechanism{ \BobMechanism }{ b }$ whenever $\utilityMechanism{ \BobMechanism }{ b }>0$, since $\utilityMechanism{ \BobMechanism }{ b }\leq b\leq a$. Alice's optimal revenue makes the bound in \cref{lemma:alice-structure-2:la} tight.}
{The buyer must visit Alice first.
Suppose otherwise, and let $y=\zmarg{ \BobMechanism }^{+}\paren*{ \effectiveValue{b}{\AliceMechanism} }$ be the allocation he selects from Bob, where $\effectiveValue{b}{\AliceMechanism}=b-\utilityMechanism{ \AliceMechanism }{ a }$.
Tightness of the final Bob-first step in the proof of \cref{lemma:alice-structure-2:la} forces $R=\BobMechanism(y)=y\effectiveValue{b}{\AliceMechanism}$, so $\utilityMechanism{ \BobMechanism }{ \effectiveValue{b}{\AliceMechanism} }=0$.
The same option at value $b$ gives $\utilityMechanism{ \BobMechanism }{ b }\geq by-\BobMechanism(y)=y\utilityMechanism{ \AliceMechanism }{ a }>0$.
By \cref{lemma:buyer}, visiting Bob first therefore gives utility $\utilityMechanism{ \AliceMechanism }{ a }$, whereas visiting Alice first gives $\utilityMechanism{ \AliceMechanism }{ a-\utilityMechanism{ \BobMechanism }{ b } }+\utilityMechanism{ \BobMechanism }{ b }>\utilityMechanism{ \AliceMechanism }{ a }$ by the strict inequality above. This contradicts the buyer's choice to visit Bob first.}
{Equality in the Alice-first bound of \cref{lemma:alice-structure-2:la} also requires $\utilityMechanism{ \BobMechanism }{ b }=0$.
Indeed, if $\utilityMechanism{ \BobMechanism }{ b }>0$, the inequality above gives $\utilityMechanism{ \AliceMechanism }{ a-\utilityMechanism{ \BobMechanism }{ b } }>\utilityMechanism{ \AliceMechanism }{ a }-\utilityMechanism{ \BobMechanism }{ b }$. This makes the $1$-Lipschitz step in the Alice-first part of the lemma's proof strict, contradicting Alice's optimality.}
{Thus, $\BobMechanism(q)\geq bq$ for every $q$, and the buyer chooses allocation zero from Bob after Alice's success. Let $y=\zmarg{ \BobMechanism }^{+}\paren*{ b }>0$, where positivity follows from $R>0$. The buyer pays $by$ after Alice's failure and selects $x^*$ from Alice by \cref{lemma:buyer}. Consequently, $R=(1-x^*)by=R^*=b(1-x^*)$, so $y=1$ and $\BobMechanism(1)=b$. Convexity and $\BobMechanism(0)=0$ give $\BobMechanism(q)\leq bq$, proving $\BobMechanism=\FP{b}$. The revenues of Alice and Bob and the buyer's utility are therefore $ax^*-\utilityMechanism{ \AliceMechanism }{ a }$, $R^*$, and $\utilityMechanism{ \AliceMechanism }{ a }$, respectively,
as when Alice offers $\AliceMechanism^*$ and Bob offers $\FP{b}$. This argument includes $a=b$.}

{For $a\geq Cb$, Alice posts $a-b$ and sells with certainty, so these payoffs are $a-b$, $0$, and $b$, respectively; Bob's zero-revenue mechanism need not be unique.}

\subsubsection{When $a < b$}
\label{sec:stackelberg:alice:lb}

We adopt a similar approach to handle the Large-Bob case, where $a < b$. Note from \cref{eq:primes} that when $a < b$, we have $\residualValue{a}{b} = 0$ and $\residualValue{b}{a} = b - a = \Delta$. \ygedit{As above, fix $\AliceMechanism$ and a best response $\BobMechanism$ yielding Bob revenue $R$; no mechanism can give Bob more than $R$ against $\AliceMechanism$.} We first use this fact and \cref{lemma:sl,lemma:dl} to get constraints on the allocation rule and monopolist utilities of Alice's mechanism. The main difference here is that we keep the constraints from \cref{lemma:sl,lemma:dl} separate instead of combining them into one constraint like above. We get:

\begin{lemma}
\label{lemma:alice-structure-1:lb}
Let $R \geq \Delta$ and $\AliceMechanism$ be a mechanism for Alice such that $0 < \utilityMechanism{ \AliceMechanism }{ a } < a$ and for all mechanisms $\BobMechanism$ for Bob, we have $\revDuo{ B }{ \AliceMechanism }{ \BobMechanism } \leq R$. 
\begin{enumerate}
\item \label{item:alice-structure-1:lb:1} For all $v \in \bracket*{ R - \Delta, a }$, we have: 
\[
\utilityMechanism{ \AliceMechanism }{ v } \geq \frac{ v + \Delta - R }{ b } \cdot \utilityMechanism{ \AliceMechanism }{ a } .
\]
\item \label{item:alice-structure-1:lb:2} For all $v \in \oc*{ 0, a }$, we have: 
\[
\zmarg{ \AliceMechanism }^-\paren*{ v } \geq \frac{ v + \Delta - R }{ v + \Delta - \Delta \cdot \x_{ \AliceMechanism }\paren*{ v + \Delta } } = \frac{ v + \Delta - R }{ v + \Delta \cdot \frac{ v - \utilityMechanism{ \AliceMechanism }{ v } }{ a - \utilityMechanism{ \AliceMechanism }{ a } } } .
\]
\end{enumerate}
\end{lemma}

As above, the next step is to upper bound Alice's revenue in terms of her utility and $R$. Here, we deviate slightly from the previous section. In the previous section, all we needed for this step is that Bob's {revenue} is $R$. Here, we also assume that Bob's \ygedit{selected best-response mechanism} is just a take-it-or-leave-it offer at price $b$. In other words, his mechanism $\BobMechanism$ is the mechanism $\FP{b} = \DL_{ 0, b }$. Under this stronger assumption, we can compute Alice's revenue exactly, as is done in the following lemma.

\begin{lemma}
\label{lemma:alice-structure-2:lb}
Let $\AliceMechanism$ be a mechanism for Alice and let $R = \revDuo{ B }{ \AliceMechanism }{ \DL_{ 0, b } }$. We have:
\[
\revDuo{ A }{ \AliceMechanism }{ \DL_{ 0, b } } = a \cdot \paren*{ 1 - \frac{ R }{ b } } - \utilityMechanism{ \AliceMechanism }{ a } .
\]
\end{lemma}

Because \cref{lemma:alice-structure-2:lb} is weaker than its counterpart in the Large-Alice case and only works when Bob's best response is $\FP{b}$, we have to work slightly harder to finish the proof, which proceeds as follows. First, we show that \ygedit{if $\BobMechanism$ best responds to $\AliceMechanism$ and Alice earns at least $\frac{a^2}{4b}$, then} the buyer must approach Alice first. For this we show that if the buyer approaches Bob first, then Alice's revenue is guaranteed to be smaller than $\frac{ a^2 }{ 4b }$. Later in the proof, when we define Large-Bob-equalizer mechanisms, we will show that Alice's revenue is at least $\frac{ a^2 }{ 4b }$. Thus, any equilibrium must give Alice at least this much revenue and therefore the buyer must approach Alice first at equilibrium.

Having proved that the buyer approaches Alice first, let $x \in \bracket*{ 0, 1 }$ be the option the buyer chooses from Alice. In the next step of our proof, we construct a different mechanism $\tilde{ \AliceMechanism }$ for Alice that is the same as $\AliceMechanism$ except that an allocation probability larger than $x$ is disallowed. We call this mechanism a ``capped'' mechanism as it caps Alice's allocation at $x$. Formally, given $x$ and $\AliceMechanism$, for all $z \in \bracket*{ 0, 1 }$, we define:
\begin{equation}
\label{eq:cap}
\tilde{ \AliceMechanism }\paren*{ z } = \begin{cases}
\AliceMechanism\paren*{ z }, &\text{~if~} 0 \leq z \leq x \\
\infty, &\text{~if~} x < z \leq 1
\end{cases} .
\end{equation}

With this definition of a capped mechanism, we show that if \ygedit{$\BobMechanism$ best responds to $\AliceMechanism$ and} the buyer goes to Alice first and gets the item with probability $x$, then \ygedit{$\DL_{0,b}$ best responds to $\tilde{\AliceMechanism}$ with the same revenue for Alice}. For this, we show the following lemma.
\begin{lemma}
\label{lemma:cap}
Let $\AliceMechanism$, $\BobMechanism$, $R$, and $x$ be as above and assume that $R > \Delta$. For all mechanisms $\tilde{ \BobMechanism }$ for Bob, it holds that $\revDuo{ B }{ \tilde{ \AliceMechanism } }{ \tilde{ \BobMechanism } } \leq \revDuo{ B }{ \tilde{ \AliceMechanism } }{ \DL_{ 0, b } } = b \cdot \paren*{ 1 - x }$. Moreover, we have $\revDuo{ A }{ \tilde{ \AliceMechanism } }{ \DL_{ 0, b } } = ax - \utilityMechanism{ \tilde{ \AliceMechanism } }{ a }$.
\end{lemma}

To see why this lemma \ygedit{preserves Alice's revenue}, note that our capped allocation function and the fact that the buyer went to Alice first and picked option $x$ from Alice's menu imply that Alice's revenue in the original \ygedit{pair} is exactly $ax - \utilityMechanism{ \tilde{ \AliceMechanism } }{ a }$, which is the revenue Alice acheives when offering $\tilde{ \AliceMechanism }$ and Bob best-responding with $\FP{b}$. 
Now that we have \ygedit{moved to a pair where Bob best responds with $\FP{b}$}, we can use \cref{lemma:alice-structure-2:lb} and proceed similarly to the previous case.

\cref{lemma:cap} says that when Bob offers $\FP{b}$, revenue equals $b \cdot \paren*{ 1 - x }$. Applying \cref{lemma:alice-structure-1:lb} with this value of the revenue, it is possible to show that Alice's monopolist utility at $a$ is lower bounded by $bu$, where $u = u(x) \in \paren*{ 0, \frac{ ax }{ b } }$ is the unique value that satisfies\footnote{The formal proof shows that $x < \frac{ a }{ b }$ so the denominator on the left hand side below is positive.}:
\begin{equation}
\label{eq:u:lb-copy}
\ln \frac{ a - \paren*{ 2b - a } u }{ a - bx } = \int_u^x \frac{ 1 + \alpha_u - \alpha_u y }{ \paren*{ 1 - y } \paren*{ 1 - \alpha_u y } } \del y , \qquad \text{where} \qquad \alpha_u = \frac{ b - a }{ a - bu } , 
\end{equation}
Now, \cref{lemma:alice-structure-2:lb} says that Alice's revenue \ygedit{in this pair (where Alice offers $\tilde{A}$ and Bob $\FP{b}$)} is at most $ax - bu$. The fact that $u$ exists and is unique follows from \cref{lemma:zeta:lb} in the appendix. As $u$ is determined by $x$, which in turn is determined by Bob's revenue \ygedit{when offering the pair of mechanisms $\tilde{A}$ and $\FP{b}$}, we are in a similar situation as before where we have an upper bound on Alice's revenue in terms of Bob's revenue. Now, we define Large-Bob-equalizer mechanisms that match this upper bound for all values of $x$.
\begin{definition}[Large-Bob-equalizer mechanisms]
\label{def:lb-equalizer}
Let $a < b$. Given a parameter $R \in \oc*{ \Delta, b }$, let $x = 1 - \frac{ R }{ b } \in \co*{ 0, \frac{ a }{ b } }$ and $u = u(x)$ be as in \cref{eq:u:lb-copy} if $x > 0$ and $0$ otherwise. Let $\alpha_u$ be as above and $s_u = \paren*{ a - bx } \cdot \frac{ a - \paren*{ b - a } u }{ a - \paren*{ 2b - a } u }$. We define the mechanism $\AliceMechanism_R^{ \mathsf{LB} }$ to be the mechanism with the following allocation rule:
\[
\zmarg{ \AliceMechanism_R^{ \mathsf{LB} } }^{+}\paren*{ v } = \begin{cases}
0, &\text{~if~} 0 \leq v < R - \Delta \\
u, &\text{~if~} R - \Delta \leq v < s_u \\
\frac{ v - R + \Delta }{ v + \alpha_u \cdot \paren*{ v - \utilityMechanism{ \AliceMechanism_R^{ \mathsf{LB} } }{ v } } }, &\text{~if~} s_u \leq v \leq a 
\end{cases} .
\]
After $a$, the allocation retains its value at $a$. The mechanisms $\AliceMechanism_R^{ \mathsf{LB} }$ are called Large-Bob-equalizer mechanisms.
\end{definition}
Note that the third case above defines the allocation through a differential equation. We have from \cref{lemma:solve-ode} that this solution is well-defined. Also, note that the above definition and \cref{lemma:solve-ode} imply $\utilityMechanism{ \AliceMechanism_R^{ \mathsf{LB} } }{ a } =  bu$. We next show that when Alice uses the mechanism $\AliceMechanism_R^{ \mathsf{LB} }$, Bob can never get revenue more than $R$, and that he can get $R$ by using the mechanism $\FP{b}$.

\begin{lemma}
\label{lemma:lb-equalizer}
Let $a < b$ and $R \in \oc*{ \Delta, b }$. For all mechanisms $\BobMechanism$ for Bob, we have $\revDuo{ B }{ \AliceMechanism_R^{ \mathsf{LB} } }{ \BobMechanism } \leq \revDuo{ B }{ \AliceMechanism_R^{ \mathsf{LB} } }{ \FP{b} } = R$. Moreover, when $\BobMechanism$ is the best-response picked by Bob, we have $\revDuo{ A }{ \AliceMechanism_R^{ \mathsf{LB} } }{ \BobMechanism } = ax - bu$, where $x = 1 - \frac{ R }{ b }$.
\end{lemma}

As in the previous case, this revenue matches the above upper bound exactly, so it is Alice's optimum revenue for any $x$. An optimization over $x$ finishes the proof. The formal proof can be found in \cref{app:stackelberg:alice:lb}.

\paragraph{A characterization of all equilibria (Proof of \cref{thm:introduction-variable-order-rephrased}).} \label{par:lb-characterize} We now extend these ideas to prove \cref{thm:introduction-variable-order-rephrased} in the case $a < b$. Specifically, let $R^* \in \bracket*{ 0, b }$ be such that $\AliceMechanism^* = \AliceMechanism_{ R^* }^{ \mathsf{LB} }$ maximizes Alice's revenue, {\em i.e.}, $\AliceMechanism^*$ and $\FP{b}$ constitute a Stackelberg equilibrium. Define $x^* = 1 - \frac{ R^* }{ b }$ and $u^* = u\paren*{ x^* }$ and $s^* = s_{ u^* }$. Observe that $s^* < a$. We will show that a mechanism $\AliceMechanism$ for Alice is optimal in the Stackelberg setting if and only if the following conditions hold:
\begin{align*}
\utilityMechanism{ \AliceMechanism }{ v } &= \utilityMechanism{ \AliceMechanism^* }{ v } && \forall v \in \bracket*{ s^*, a } \\
\zmarg{ \AliceMechanism }^{+}\paren*{ v } \cdot \paren*{ v + \alpha_{ u^* } \paren*{ v - \utilityMechanism{ \AliceMechanism }{ v } } } &\geq v - R^* + \Delta && \forall v \in \bracket*{ 0, s^* } .
\end{align*}
Next, we show that these conditions are necessary and sufficient for $\AliceMechanism$ to be optimal for Alice in the Stackelberg setting. We first show that they are necessary. Let $\AliceMechanism$ be optimal for Alice and let $\BobMechanism$ be Bob's best response to $\AliceMechanism$. Let $R = \revDuo{ B }{ \AliceMechanism }{ \BobMechanism }$ be Bob's revenue at equilibrium. We already showed that the buyer must approach first in any Stackelberg equilibrium. Using a simple argument, one can show that $R > \Delta$. Let $x$ be the option picked by the buyer from $\AliceMechanism$. Then, as in the proof in \cref{app:stackelberg:alice:lb}, we have from $R > \Delta$ that $R \leq b \cdot \paren*{ 1 - x }$.

Let $\tilde{ \AliceMechanism }$ be $\AliceMechanism$ capped at $x$. Then, from \cref{lemma:cap}, we have that $\tilde{ \AliceMechanism }$ and $\DL_{ 0, b }$ form a Stackelberg equilibrium where Alice gets revenue $ax - \utilityMechanism{ \tilde{ \AliceMechanism } }{ a }$ and Bob gets revenue $b \cdot \paren*{ 1 - x }$. Now, as the proof of this case uses the standard comparison theorem for ordinary differential equations, which is tight if and only if the functions match pointwise, and as $x^*$ is the unique maximizer in the proof, we must have $x = x^*$ and that the utilities of $\tilde{ \AliceMechanism }$ and $\AliceMechanism^*$ match in the interval $\bracket*{ s^*, a }$.
This is possible only if the capping did not change the utility in the interval $\bracket*{ s^*, a }$. Indeed, the capped utility is affine above the effective value at which $x^*$ is selected. Since the equalizer utility is strictly convex on $\bracket*{ s^*, a }$, this effective value must be $a$.
{Consequently, any option with allocation larger than $x^*$ gives no more utility than $x^*$ at any effective value $v \leq a$, so removing these options leaves utility unchanged. Thus,
$\utilityMechanism{ \AliceMechanism }{ v } 
= \utilityMechanism{ \tilde{ \AliceMechanism } }{ v } 
= \utilityMechanism{ \AliceMechanism^* }{ v }$ 
for all $v \in \bracket*{ s^*, a }$, giving the first condition.}
{Since $\zmarg{ \AliceMechanism }^-\paren*{ a }=x^*$, offering $\FP{b}$ against $\AliceMechanism$ gives Bob revenue $R^*=b\cdot\paren*{1-x^*}$. Bob's revenue-maximization combined with the preceding upper bound implies $R^*\leq R\leq b\cdot\paren*{1-x}=R^*$.}
{The second condition follows from item~\ref{item:alice-structure-1:lb:2} of \cref{lemma:alice-structure-1:lb}, using $R = R^*$, $\utilityMechanism{ \AliceMechanism }{ a } = bu^*$, and $\zmarg{ \AliceMechanism }^{+} \geq \zmarg{ \AliceMechanism }^{-}$; the case $v = 0$ is immediate.}

To verify the sufficiency of these conditions, consider any mechanism that satisfies the above conditions and observe that, just like the Large-Alice case, the proof of \cref{lemma:lb-equalizer} goes through using just these conditions. It follows that posting a fixed price of $b$ is optimal for Bob and that the buyer visits Alice first and picks the option $\zmarg{ \AliceMechanism }^-\paren*{ a }$, finishing the proof.

{We now show that, in every Stackelberg equilibrium with an optimal mechanism of Alice under our tie-breaking rules, Bob offers $\FP{b}$, the buyer visits Alice first, and all three players receive the same payoffs as when Alice offers $\AliceMechanism^*$ and Bob offers $\FP{b}$.
{Let $\paren*{ \AliceMechanism, \BobMechanism }$ be such an equilibrium, and let $R=\revDuo{ B }{ \AliceMechanism }{ \BobMechanism }$ denote Bob's revenue. We have already shown that the buyer visits Alice first, selects $x^*$ at effective value $a$, and leaves Bob revenue $R=R^*=b(1-x^*)$.}
{By \cref{eq:aug}, we have $\utilityMechanism{ \BobMechanism }{ b }=\utilityMechanism{ \BobMechanism }{ \Delta }$. Since $\utilityMechanism{ \BobMechanism }{ 0 }=0$ and $\Delta<b$, convexity and nonnegativity imply that both values are zero.} Hence $\BobMechanism(q)\geq bq$ for every $q$, and the residual type $\Delta<b$ chooses allocation zero from Bob after Alice's success. Let $y=\zmarg{ \BobMechanism }^{+}\paren*{ b }$. The buyer pays $by$ after Alice's failure, so $R^*=(1-x^*)by=b(1-x^*)$. Since $x^*<1$, we obtain $y=1$ and $\BobMechanism(1)=b$. Convexity and $\BobMechanism(0)=0$ now give $\BobMechanism(q)\leq bq$, proving $\BobMechanism=\FP{b}$. Thus, the revenues of Alice and Bob and the buyer's utility are $ax^*-bu^*$, $R^*$, and $bu^*$, respectively,
{as when Alice offers $\AliceMechanism^*$ and Bob offers $\FP{b}$}.}

\subsubsection{\texorpdfstring{{Characterization of Optimal Mechanisms}}{Characterization of Optimal Mechanisms}}
\label{sec:stackelberg:alice:summary}

We now summarize our characterization of Alice's optimal mechanisms by proving \cref{prop:main-optimal-mechanisms}. 
To do this, we first define Alice's canonical mechanisms,
one for the case where $a \geq b$
and another for $a < b$. We characterize all Alice-optimal mechanisms by comparing them to these.

\mbc{}
\ygcomment{}

\begin{definition}[Optimal equalizers]
    \label{def:canonical-optimal-equalizers}
    Fix buyer values $a,b>0$ in the main model, and write $\Delta=|a-b|$.
    \begin{enumerate}
      \item \label{item:canonical-optimal-equalizers:lb}
      {For $a<b$, let $R_{\mathsf{LB}}^\ast\in\oc*{\Delta,b}$
      maximize Alice's revenue within the Large-Bob-equalizer family (\cref{def:lb-equalizer}),
      and define $\AliceMechanism_{\mathsf{LB}}^\ast
      =\AliceMechanism_{R_{\mathsf{LB}}^\ast}^{\mathsf{LB}}$.}

      \item \label{item:canonical-optimal-equalizers:la}
      {For $a\geq b$, let $R_{\mathsf{LA}}^\ast\in\bracket*{0,b}$
      maximize Alice's revenue within the Large-Alice-equalizer family (\cref{def:la-equalizer}),
      and define $\AliceMechanism_{\mathsf{LA}}^\ast
      =\AliceMechanism_{R_{\mathsf{LA}}^\ast}^{\mathsf{LA}}$.}
    \end{enumerate}
\end{definition}

We are now ready for \cref{prop:main-optimal-mechanisms}.
Notably, the characterization below shows that optimal mechanisms need not be essentially equivalent to the canonical mechanisms defined above.

\begin{proposition}
    \label{prop:main-optimal-mechanisms}
    Consider an instance of the duopoly game in the main model
    with fixed buyer values $a,b>0$ and let $\Delta=|a-b|$.
    A mechanism $\AliceMechanism$ is optimal for Alice
    if and only if the following condition holds, depending
    on the relation between $a$ and $b$:

    \begin{enumerate}
      \item When $0<a<b$, its allocation rule is monotone and satisfies:
      \begin{align*}
          &\utilityMechanism{ \AliceMechanism }{ v }
          = \utilityMechanism{ \AliceMechanism_{\mathsf{LB}}^\ast }{ v }
          && \forall v \in \bracket*{s^\ast,a},\\
          &{\zmarg{ \AliceMechanism }^{+}\paren*{v}
          \in \bracket*{\frac{v-R_{\mathsf{LB}}^\ast+\Delta}{v+\alpha^\ast\paren*{v-\utilityMechanism{ \AliceMechanism }{ v }}},
          \frac{\utilityMechanism{ \AliceMechanism_{\mathsf{LB}}^\ast }{ a }}{b}}}
          && \forall v \in \oc*{0,s^\ast}.
      \end{align*}
      where $\AliceMechanism_{\mathsf{LB}}^\ast$ is the optimal
      Large-Bob-Equalizer of \cref{def:canonical-optimal-equalizers},
      $R_{\mathsf{LB}}^\ast$ is the revenue of Bob when best-responding to it,
      $u^\ast=\frac{\utilityMechanism{ \AliceMechanism_{\mathsf{LB}}^\ast }{ a }}{b}$,
      $\alpha^\ast=\frac{\Delta}{a-b \cdot u^\ast}$, and
      $s^\ast=(R_{\mathsf{LB}}^\ast-\Delta)\cdot \frac{a-\Delta \cdot u^\ast}{a-(2b-a) \cdot u^\ast}$.

    \item When $b\leq a<C\cdot b$, where $C\in(3.146,3.147)$ satisfies $C-\ln C=2$, its allocation rule is monotone and satisfies:
    \begin{align*}
        &\utilityMechanism{ \AliceMechanism }{ v }
        = \utilityMechanism{ \AliceMechanism_{\mathsf{LA}}^\ast }{ v }
        && \forall v \in \set*{ \Delta } \cup \bracket*{ s^*, a } \\
        &{\zmarg{ \AliceMechanism }^{+}\paren*{ v }
        \in \bracket*{1 - \frac{ R_{\mathsf{LA}}^* }{ v - \Delta },
        \frac{ \utilityMechanism{ \AliceMechanism_{\mathsf{LA}}^\ast }{ a } }{ b }}}
        && \forall v \in \oc*{ \Delta, s^* },
    \end{align*}
    where $\AliceMechanism_{\mathsf{LA}}^\ast$ is the optimal
    Large-Alice-Equalizer of \cref{def:canonical-optimal-equalizers},
    $R_{\mathsf{LA}}^\ast$ is the revenue of Bob when best-responding to it,
    and $s^\ast = \Delta + \frac{b \cdot R_{\mathsf{LA}}^\ast}{b - \utilityMechanism{ \AliceMechanism_{\mathsf{LA}}^\ast }{ a }}$.

    \item When $C \cdot b \leq a$, $\AliceMechanism$ is a posted price of $a - b$.

    \end{enumerate}
\end{proposition}

\begin{proof}[{Proof of \cref{prop:main-optimal-mechanisms}}]
    These characterizations are proved in \cref{sec:stackelberg:alice:lb} for $a<b$
    and \cref{sec:stackelberg:alice:la} for $a\geq b$.
    For $a\geq Cb$, the allocation and utility conditions in the second case,
    with $R^\ast=0$ and $s^\ast=s(0)=\Delta+b/C$ as in \cref{eq:s:la},
    reduce to $\AliceMechanism=\FP{a-b}$.
\end{proof}

{This characterization shows that the initial straight segment in
\cref{fig:introduction-alice-pricing-menu} can be replaced only by
discounted options, illustrated in \textcolor[HTML]{08745A}{green}.}
{Indeed, $\zmarg{ \AliceMechanism^\ast_{\mathsf{LB}} }^+\paren*{ v }$
jumps directly from $0$ to the upper bound
$\frac{\utilityMechanism{ \AliceMechanism_{\mathsf{LB}}^\ast }{ a }}{b}$.}
{Thus, for $a<b$, every Alice-optimal mechanism $\AliceMechanism$ satisfies
$\utilityMechanism{\AliceMechanism}{v} \geq \utilityMechanism{\AliceMechanism^\ast_{ \mathsf{LB}}}{v}$
for all $v$, justifying our choice of $\AliceMechanism^\ast_{\mathsf{LB}}$
as the canonical mechanism.}
{The analogous statement holds for $b \leq a < C \cdot b$
with $\AliceMechanism_{ \mathsf{LA}}^\ast$.}

\section{The Aligned Model}
\label{sec:alice-opt-stalkeberg-fixed-order}
\label{sec:alice-opt-stalkeberg-AB}

The goal of this section is to characterize all Alice-optimal
mechanisms and derive her optimal revenue when the buyer must visit
Alice first. We call this the \emph{aligned model}. 
Recall that $a>0$ and $b>0$ denote the buyer's values for
Alice's and Bob's goods, respectively,
and that $\Delta=|a-b|$ denotes
their difference.
The residual values after receiving the other good are
$\residualValue{a}{b}=(a-b)_+$ and $\residualValue{b}{a}=(b-a)_+$.
{$\FP{p}$ denotes a posted price $p$.}
For $i\in\{A,B\}$, let
$\revDuoOrder{\AB}{i}{\AliceMechanism}{\BobMechanism}$ denote
seller $i$'s revenue under the buyer's prescribed tie-breaking rules 
and when the buyer is forced to visit Alice first.
Recall that $\optOrder{a}{b}{\AB}$ denotes
the supremum of Alice's expected revenue over Stackelberg equilibria
when a buyer with values $a,b > 0$ is forced to visit Alice first.

\introductionaligned*

\begin{figure}[H]
  \centering
  \resizebox{0.65\linewidth}{!}{%
    \begingroup%
\definecolor{AlignedMenuBlue}{HTML}{246298}%
\definecolor{AlignedMenuOrange}{HTML}{B95B17}%
\definecolor{AlignedMenuGuide}{HTML}{C9D1D9}%
\definecolor{AlignedMenuGrid}{HTML}{D5D5D5}%
\def\AlignedMenuWidth{10.8cm}%
\def\AlignedMenuHeight{5.2cm}%
\def\AlignedMenuXAxisTitleY{-0.23}%
\def\AlignedMenuYAxisTitleX{-0.10}%
\def\AlignedMenuSegmentLabelY{-0.08}%
\def\AlignedMenuSegmentFontSize{9}%
\def\AlignedMenuSegmentLineSpacing{10}%
\def\AlignedMenuSegmentLabelXShift{0pt}%
\def\AlignedMenuCurvedLabelXShift{0pt}%
\def\AlignedMenuCurvedLabelYShift{0pt}%
\def\AlignedMenuExtensionLabelXShift{0pt}%
\def\AlignedMenuExtensionLabelYShift{0pt}%
\def\AlignedMenuShowXOne{1}%
\def\AlignedMenuSharedX{0.30}%
\def\AlignedMenuSharedY{0.48}%
\def\AlignedMenuExtensionX{0.835}%
\def\AlignedMenuExtensionY{0.39}%
\def\AlignedMenuLegendX{0.035}%
\def\AlignedMenuLegendY{0.945}%
\def\AlignedMenuLegendStep{0.077}%
\def\AlignedMenuParametersX{0.035}%
\def\AlignedMenuParametersY{0.79}%
\def\AlignedMenuParametersGap{0.06em}%
\def\AlignedMenuELabelXShift{-6pt}%
\def\AlignedMenuELabelYShift{0.5pt}%
\def\AlignedMenuSharedTargetX{0.40}%
\def\AlignedMenuArrowTipGap{3pt}%
\def\AlignedMenuExtensionBraceWidth{0.9}%
\def\AlignedMenuExtensionBraceGap{5pt}%
\def\AlignedMenuExtensionBraceAmplitude{5pt}%
\def\AlignedMenuExtensionBraceLineWidth{0.85pt}%
\def\AlignedMenuTextHaloScale{2.5}%
\def\AlignedMenuTextHaloStrength{3}%
\def\AlignedMenuTextHaloRadius{0.65pt}%
\def\AlignedMenuTextHaloMiddle{0.40pt}%
\def\AlignedMenuTextHaloCore{0.18pt}%
\pgfmathsetlengthmacro{\AlignedMenuHaloOuterWidth}{max(0,\AlignedMenuTextHaloScale)*\AlignedMenuTextHaloRadius}%
\pgfmathsetlengthmacro{\AlignedMenuHaloMiddleWidth}{max(0,\AlignedMenuTextHaloScale)*\AlignedMenuTextHaloMiddle}%
\pgfmathsetlengthmacro{\AlignedMenuHaloInnerWidth}{max(0,\AlignedMenuTextHaloScale)*\AlignedMenuTextHaloCore}%
\pgfmathsetmacro{\AlignedMenuHaloOuterOpacity}{min(1,max(0,0.15*\AlignedMenuTextHaloStrength))}%
\pgfmathsetmacro{\AlignedMenuHaloMiddleOpacity}{min(1,max(0,0.30*\AlignedMenuTextHaloStrength))}%
\pgfmathsetmacro{\AlignedMenuHaloInnerOpacity}{min(1,max(0,0.70*\AlignedMenuTextHaloStrength))}%
\newcommand*\AlignedMenuTextHalo[1]{%
  \begingroup
  \ifdim\AlignedMenuHaloOuterWidth>0pt\relax
  \rlap{\begingroup\transparent{\AlignedMenuHaloOuterOpacity}\color{white}%
    \contourlength{\AlignedMenuHaloOuterWidth}\contour{white}{#1}\endgroup}%
  \rlap{\begingroup\transparent{\AlignedMenuHaloMiddleOpacity}\color{white}%
    \contourlength{\AlignedMenuHaloMiddleWidth}\contour{white}{#1}\endgroup}%
  \rlap{\begingroup\transparent{\AlignedMenuHaloInnerOpacity}\color{white}%
    \contourlength{\AlignedMenuHaloInnerWidth}\contour{white}{#1}\endgroup}%
  \fi
  #1%
  \endgroup
}%
\def\AlignedMenuBobRevenue{0.3678794411714423}%
\def\AlignedMenuEquilibrium{0.6321205588285577}%
\def\AlignedMenuEquilibriumPayment{0.3678794411714423}%
\def\AlignedMenuUtility{0.2642411176571153}%
\def\AlignedMenuTerminalPayment{0.7357588823428847}%
\begin{tikzpicture}[
  baseline=(AlignedMenuMain.south west),
  aligned menu text/.style={inner sep=1.5pt,outer sep=0pt},
  declare function={
    alignedmenuprice(\x)=-\AlignedMenuBobRevenue*ln(1-\x);
    alignedmenutangent(\x)=\x-\AlignedMenuUtility;
  },
]
\begin{axis}[
  name=AlignedMenuMain,
  width=\AlignedMenuWidth, height=\AlignedMenuHeight, scale only axis,
  xmin=0, xmax=1.025, ymin=0, ymax=0.88,
  axis lines=left, axis line style={line width=0.8pt,-},
  tick align=outside, tick style={black,line width=0.8pt},
  major tick length=3.5pt,
  xtick={0,0.25,0.5,0.75,1},
  xticklabels={$0$,{},$1/2$,{},{$\ifnum\AlignedMenuShowXOne=1\relax 1\fi$}},
  ytick={0,0.25,0.5,0.75},
  yticklabels={$0$,$a/4$,$a/2$,$3\cdot a/4$},
  tick label style={font=\fontsize{8}{9}\selectfont},
  xticklabel style={yshift=-1pt},
  yticklabel style={xshift=-1pt},
  xlabel={Allocation probability, $x$},
  xlabel style={at={(axis description cs:0.5,\AlignedMenuXAxisTitleY)},
    anchor=north,inner sep=0pt,font=\fontsize{9}{10}\selectfont},
  ylabel={Expected payment, $\tau$},
  ylabel style={at={(axis description cs:\AlignedMenuYAxisTitleX,0.5)},
    anchor=south,font=\fontsize{9}{10}\selectfont},
  grid=major, grid style={AlignedMenuGrid,line width=0.45pt,opacity=0.65},
  axis on top=false, clip=false,
  every axis plot/.append style={forget plot},
]
\draw[AlignedMenuGuide,line width=0.65pt,dash pattern=on 2pt off 4pt]
  (axis cs:\AlignedMenuEquilibrium,0) --
  (axis cs:\AlignedMenuEquilibrium,\AlignedMenuEquilibriumPayment);
\draw[AlignedMenuGuide,line width=0.65pt,dash pattern=on 2pt off 4pt]
  (axis cs:1,0) -- (axis cs:1,\AlignedMenuTerminalPayment);
\addplot[AlignedMenuBlue,line width=1.6pt,
  domain=0:\AlignedMenuEquilibrium,samples=121]
  {alignedmenuprice(x)};
\addplot[AlignedMenuOrange,line width=1.4pt,dash pattern=on 5pt off 3pt,
  domain=\AlignedMenuEquilibrium:1,samples=2]
  {alignedmenutangent(x)};
\addplot[only marks,mark=*,mark size=1.5pt,AlignedMenuBlue]
  coordinates {(0,0) (\AlignedMenuEquilibrium,\AlignedMenuEquilibriumPayment)};
\addplot[only marks,mark=o,mark size=1.7pt,AlignedMenuOrange,
  mark options={solid,fill=white,line width=0.8pt}]
  coordinates {(1,\AlignedMenuTerminalPayment)};
\node[aligned menu text,font=\fontsize{9}{10}\selectfont,text=AlignedMenuBlue,
  anchor=south,xshift=\AlignedMenuELabelXShift,yshift=\AlignedMenuELabelYShift]
  at (axis cs:\AlignedMenuEquilibrium,\AlignedMenuEquilibriumPayment)
  {\AlignedMenuTextHalo{$E'$}};

\node[aligned menu text,align=center,font=\fontsize{8}{9}\selectfont\bfseries,
  text=AlignedMenuBlue] (AlignedMenuShared)
  at (axis description cs:\AlignedMenuSharedX,\AlignedMenuSharedY)
  {\AlignedMenuTextHalo{Shared by all}\\\AlignedMenuTextHalo{Optimal Mechanisms}};
\draw[AlignedMenuBlue!85!black,line width=0.95pt,
  -{Stealth[length=5pt,width=3.5pt]},shorten >=\AlignedMenuArrowTipGap]
  (AlignedMenuShared.south) to[out=-90,in=140]
  (axis cs:\AlignedMenuSharedTargetX,{alignedmenuprice(\AlignedMenuSharedTargetX)});
\node[aligned menu text,align=center,font=\fontsize{8}{9}\selectfont,
  text=AlignedMenuOrange] (AlignedMenuExtension)
  at (axis description cs:\AlignedMenuExtensionX,\AlignedMenuExtensionY)
  {\AlignedMenuTextHalo{Choose where to stop}\\\AlignedMenuTextHalo{on the tangent}};
\pgfmathsetmacro{\AlignedMenuExtensionBraceStartX}{%
  (1+\AlignedMenuEquilibrium-\AlignedMenuExtensionBraceWidth*(1-\AlignedMenuEquilibrium))/2}%
\pgfmathsetmacro{\AlignedMenuExtensionBraceEndX}{%
  (1+\AlignedMenuEquilibrium+\AlignedMenuExtensionBraceWidth*(1-\AlignedMenuEquilibrium))/2}%
\pgfmathsetlengthmacro{\AlignedMenuExtensionBraceDepth}{%
  \AlignedMenuExtensionBraceGap+\AlignedMenuExtensionBraceAmplitude}%
\coordinate (AlignedMenuExtensionBraceStart)
  at (axis cs:\AlignedMenuExtensionBraceStartX,{alignedmenutangent(\AlignedMenuExtensionBraceStartX)});
\coordinate (AlignedMenuExtensionBraceEnd)
  at (axis cs:\AlignedMenuExtensionBraceEndX,{alignedmenutangent(\AlignedMenuExtensionBraceEndX)});
\draw[AlignedMenuOrange!85!black,
  line width=\AlignedMenuExtensionBraceLineWidth,
  decorate,decoration={brace,mirror,
    amplitude=\AlignedMenuExtensionBraceAmplitude,
    raise=\AlignedMenuExtensionBraceGap}]
  (AlignedMenuExtensionBraceStart) -- (AlignedMenuExtensionBraceEnd)
  node[midway,sloped,below=\AlignedMenuExtensionBraceDepth,
    inner sep=0pt,outer sep=0pt] (AlignedMenuExtensionBraceCenter) {};
\draw[AlignedMenuOrange!85!black,
  line width=\AlignedMenuExtensionBraceLineWidth]
  (AlignedMenuExtension.north)
  \pgfextra{%
    \pgfmathanglebetweenpoints
      {\pgfpointanchor{AlignedMenuExtensionBraceStart}{center}}%
      {\pgfpointanchor{AlignedMenuExtensionBraceEnd}{center}}%
    \pgfmathsetmacro{\AlignedMenuExtensionBraceApproachAngle}{\pgfmathresult-90}%
  }
  to[out=90,in=\AlignedMenuExtensionBraceApproachAngle]
  (AlignedMenuExtensionBraceCenter.center);

\draw[AlignedMenuBlue,line width=2.1pt]
  (axis cs:0,0) -- (axis cs:\AlignedMenuEquilibrium,0);
\draw[AlignedMenuOrange,line width=2.1pt]
  (axis cs:\AlignedMenuEquilibrium,0) -- (axis cs:1,0);
\node[anchor=north,align=center,inner sep=0pt,
  font=\fontsize{\AlignedMenuSegmentFontSize}{\AlignedMenuSegmentLineSpacing}\selectfont,
  text=AlignedMenuBlue,
  xshift={\AlignedMenuSegmentLabelXShift+\AlignedMenuCurvedLabelXShift},
  yshift=\AlignedMenuCurvedLabelYShift]
  at (axis cs:{\AlignedMenuEquilibrium/2},\AlignedMenuSegmentLabelY)
  {Curved segment};
\node[anchor=north,align=center,inner sep=0pt,
  font=\fontsize{\AlignedMenuSegmentFontSize}{\AlignedMenuSegmentLineSpacing}\selectfont,
  text=AlignedMenuOrange,
  xshift={\AlignedMenuSegmentLabelXShift+\AlignedMenuExtensionLabelXShift},
  yshift=\AlignedMenuExtensionLabelYShift]
  at (axis cs:{(\AlignedMenuEquilibrium+1)/2},\AlignedMenuSegmentLabelY)
  {Optional\\extension};

\draw[AlignedMenuBlue,line width=1.6pt]
  (axis description cs:\AlignedMenuLegendX,\AlignedMenuLegendY) --
  (axis description cs:{\AlignedMenuLegendX+0.08},\AlignedMenuLegendY);
\node[aligned menu text,anchor=west,font=\fontsize{8}{9}\selectfont,xshift=3pt]
  at (axis description cs:{\AlignedMenuLegendX+0.08},\AlignedMenuLegendY)
  {\AlignedMenuTextHalo{Canonical menu}};
\draw[AlignedMenuOrange,line width=1.4pt,dash pattern=on 5pt off 3pt]
  (axis description cs:\AlignedMenuLegendX,{\AlignedMenuLegendY-\AlignedMenuLegendStep}) --
  (axis description cs:{\AlignedMenuLegendX+0.08},{\AlignedMenuLegendY-\AlignedMenuLegendStep});
\node[aligned menu text,anchor=west,font=\fontsize{8}{9}\selectfont,xshift=3pt]
  at (axis description cs:{\AlignedMenuLegendX+0.08},{\AlignedMenuLegendY-\AlignedMenuLegendStep})
  {\AlignedMenuTextHalo{Optional tangent extension}};
\node[aligned menu text,anchor=north west,font=\fontsize{8}{9}\selectfont]
  at (axis description cs:\AlignedMenuParametersX,\AlignedMenuParametersY)
  {\AlignedMenuTextHalo{$a=1,\hspace{\AlignedMenuParametersGap} b=1$}};
\end{axis}
\end{tikzpicture}%
\endgroup%
%
  }
  \caption{{Alice's optimal pricing
  {menu} in the aligned model,
  illustrated at $a=b=1$.
  {Every optimal menu shares the \textcolor{blue}{blue curved portion}
  through the equilibrium lottery $E'$; after irrelevant options are removed,
  optional extensions follow the \textcolor{orange}{orange tangent}
  and may stop anywhere up to allocation one.}}}
  \label{fig:aligned-optimal-pricing-menus}
\end{figure}
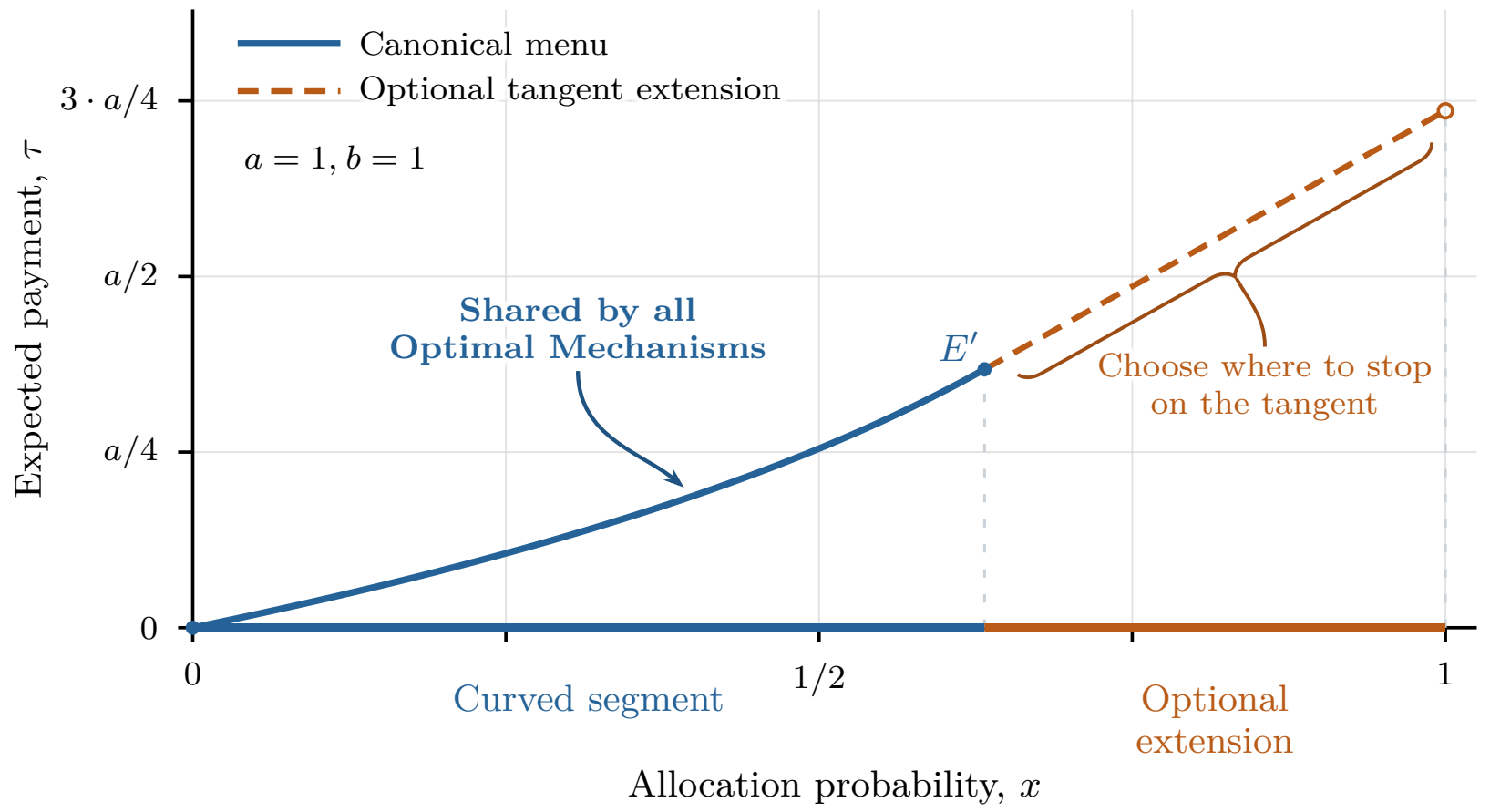

We informally describe the characterization of all Alice-optimal
mechanisms in the introduction.
There, we also compare Alice's optimal revenue in the aligned 
and main models; see \cref{fig:introduction-aligned-variable-revenue-difference}.

{The rest of this section is devoted to proving
\cref{thm:introduction-aligned-rephrased}.}
The proof has a similar structure to the proof of the main model, so we omit a proof outline here. A brief description of the differences {of this model from the main model} appears in
\cref{sec:introduction-aligned-proof-overview}.
As most proofs in this section are short, they all appear below.

\subsection{\texorpdfstring{{Proof of \cref{thm:introduction-aligned-rephrased}}}{Proof of Theorem \getrefnumber{thm:introduction-aligned-rephrased}}}
\label{sec:aligned:proof}

{We begin the proof by characterizing the buyer's behavior. 
Recall that the augmentation-induced utility gap measures the increase in
utility when the buyer's value rises from $(w-v)_+$ to $w$: for
$v,w\geq0$, $\aug_{\mathcal M}(v,w)=\utilityMechanism{\mathcal M}{w}
-\utilityMechanism{\mathcal M}{(w-v)_+}$.}
The following corollary is a direct implication of \cref{lemma:buyer} with the order Alice \textrightarrow\ Bob imposed.

\begin{corollary}
  \label{cor:aligned-buyer}
  \label{obs:alice-opt-stalkeberg-AB-a-at-least-b-ALICE-ALLOCATION}
  \label{obs:alice-opt-stalkeberg-AB-a-at-most-b-ALICE-ALLOCATION}
  {Let Alice and Bob use mechanisms $\AliceMechanism$ and
  $\BobMechanism$ in the aligned model, and define
  $\effectiveValue{a}{\BobMechanism}=a-\aug_{\BobMechanism}\paren*{a,b}$.
  \begin{enumerate}
    \item The buyer's maximum utility is
    \[
      \utilityAB{\AliceMechanism}{\BobMechanism}
        =\utilityMechanism{\AliceMechanism}{\effectiveValue{a}{\BobMechanism}}
         +\utilityMechanism{\BobMechanism}{b}.
    \]
    \item The buyer first picks the option $x_A$ from Alice, where
    \[
      x_A=\begin{cases}
        \zmarg{\AliceMechanism}^-\paren*{\effectiveValue{a}{\BobMechanism}},
          &\text{if }\zmarg{\BobMechanism}^{+}\paren*{b}
                       >\zmarg{\BobMechanism}^{+}\paren*{\residualValue{b}{a}},\\
        \zmarg{\AliceMechanism}^{+}\paren*{\effectiveValue{a}{\BobMechanism}},
          &\text{if }\zmarg{\BobMechanism}^{+}\paren*{b}
                       =\zmarg{\BobMechanism}^{+}\paren*{\residualValue{b}{a}}.
      \end{cases}
    \]
    He then picks $\zmarg{\BobMechanism}^{+}\paren*{\residualValue{b}{a}}$ from Bob if
    he receives Alice's good, and $\zmarg{\BobMechanism}^{+}\paren*{b}$
    otherwise.
    \item The sellers' revenues are
    \begin{align*}
      \revDuoOrder{\AB}{A}{\AliceMechanism}{\BobMechanism}
        &=\AliceMechanism\paren*{x_A},\\
      \revDuoOrder{\AB}{B}{\AliceMechanism}{\BobMechanism}
        &=(1-x_A)\cdot\BobMechanism\paren*{\zmarg{\BobMechanism}^{+}\paren*{b}}
          +x_A\cdot\BobMechanism\paren*{\zmarg{\BobMechanism}^{+}\paren*{\residualValue{b}{a}}}.
    \end{align*}
  \end{enumerate}}
\end{corollary}

As the buyer must select the order Alice \textrightarrow\ Bob, it is enough to consider the family of double-lottery mechanisms presented in the main model.

\doublelotterymechanism*

As the order Alice \textrightarrow\ Bob is imposed, Bob need not worry about accidentally inducing the buyer to switch to Bob \textrightarrow\ Alice when designing $\DL_{x,q}$. This means that {Bob} can increase $x$ beyond the threshold $\x_{ \AliceMechanism }(q)$ to its largest permissible value of $\frac{ b - q }{ b - \residualValue{b}{a} }$ without worrying about it causing the buyer to interact with Bob first. Thus, the revenue computation in \cref{lemma:dl} applies throughout the feasible range of $x$.

\begin{corollary}
  \label{cor:aligned-dl-revenue}
  {Let $\AliceMechanism$ be a mechanism for Alice and
  $q\in\bracket*{\residualValue{b}{a},b}$. The following hold in the aligned model:
  \begin{enumerate}
    \item If $a\geq b$, for all
    $x\in\bracket*{0,\frac{b-q}{b-\residualValue{b}{a}}}$, we have
    $\revDuoOrder{\AB}{B}{\AliceMechanism}{\DL_{x,q}}
      =q\cdot\paren*{1-\zmarg{\AliceMechanism}^-\paren*{\Delta+q}}$.
    \item If $a\leq b$, we have
    $\revDuoOrder{\AB}{B}{\AliceMechanism}{\DL_{\frac{b-q}{b-\residualValue{b}{a}},q}}
      =q-\zmarg{\AliceMechanism}^-\paren*{q-\Delta}
           \cdot\paren*{q-\Delta\cdot\frac{b-q}{b-\residualValue{b}{a}}}$.
  \end{enumerate}}
\end{corollary}

\begin{proof}
  {The same revenue calculation as in the proof of
  \cref{lemma:dl}, with order $\AB$ imposed, gives both formulas
  for $q>\residualValue{b}{a}$. At $q=\residualValue{b}{a}$, $\DL_{x,\residualValue{b}{a}}=\FP{\residualValue{b}{a}}$ gives Bob revenue
  $\residualValue{b}{a}$, as claimed.}
\end{proof}

This {corollary} exemplifies our finding presented in \cref{fig:introduction-aligned-variable-revenue-difference}, which shows that in some cases Alice can gain more revenue in the main model. When the order Alice \textrightarrow\ Bob is prescribed, Bob can increase the value of $x$ above $\x_{\AliceMechanism}(q)$ and {up to} $\frac{b - q}{b - \residualValue{b}{a}}$, without worry, thus providing Bob more freedom when designing mechanisms for the order Alice \textrightarrow\ {Bob}. In turn, making the corresponding constraint on Alice's {mechanism} more severe.

Returning to the proof, we stick to the structure of the main model's proof by considering a pair of mechanisms $\AliceMechanism$ and $\BobMechanism$ that constitute a Stackelberg equilibrium. Let $R$ be Bob's revenue in this equilibrium. We bound Alice's revenue in terms of $R$ and $\utilityMechanism{\AliceMechanism}{a}$. Then, using the fact that every Bob mechanism can gain {at most} $R$, including the double-lottery mechanisms presented above {(\cref{cor:aligned-dl-revenue})}, we get a lower bound for $\utilityMechanism{\AliceMechanism}{a}$ in terms of $R$.

\begin{corollary}
  \label{cor:aligned-alice-revenue-bound}
  \label{lem:aligned-lb-alice-revenue-bound}
  {Let $\AliceMechanism$ and $\BobMechanism$ be mechanisms
  for Alice and Bob in the aligned model, and set
  $R=\revDuoOrder{\AB}{B}{\AliceMechanism}{\BobMechanism}$ and
  $\effectiveValue{a}{\BobMechanism}=a-\aug_{\BobMechanism}\paren*{a,b}$.
  The following hold:
  \begin{enumerate}
    \item If $a\geq b$, we have
    $\revDuoOrder{\AB}{A}{\AliceMechanism}{\BobMechanism}
       \leq a\cdot\paren*{1-\frac{R}{b}}
             -\utilityMechanism{\AliceMechanism}{a}$.
    \item If $a\leq b$, we have $\effectiveValue{a}{\BobMechanism}\geq R-\Delta$ and
    $\revDuoOrder{\AB}{A}{\AliceMechanism}{\BobMechanism}
       \leq\frac{a}{b}\cdot\paren*{\effectiveValue{a}{\BobMechanism}-R+\Delta}
             -\utilityMechanism{\AliceMechanism}{\effectiveValue{a}{\BobMechanism}}$.
  \end{enumerate}}
\end{corollary}

\begin{proof}
  {For $a\geq b$, 
  if $\zmarg{\BobMechanism}^{+}(b)>\zmarg{\BobMechanism}^{+}(0)$,
  then $x_A = \zmarg{\AliceMechanism}^-(\effectiveValue{a}{\BobMechanism})$ 
  and the $\AB$-order computation 
  from the proof of \cref{lemma:alice-structure-2:la}
  follows throughout.
  Otherwise, if $\zmarg{\BobMechanism}^{+}(b)=\zmarg{\BobMechanism}^{+}(0)$,
  both continuation payments to Bob are zero, so $R=0$.
  The payment identity
  and the $1$-Lipschitz property give
  $
    \revDuoOrder{\AB}{A}{\AliceMechanism}{\BobMechanism}
      = \effectiveValue{a}{\BobMechanism} \cdot x_A-\utilityMechanism{\AliceMechanism}{\effectiveValue{a}{\BobMechanism}}
      \leq \effectiveValue{a}{\BobMechanism}-\utilityMechanism{\AliceMechanism}{\effectiveValue{a}{\BobMechanism}}
      \leq a-\utilityMechanism{\AliceMechanism}{a},
  $ 
  as required.
  For $a\leq b$, monotonicity and Myerson's payment identity bound
  Bob's payments at $\Delta$ and $b$ by
  $\Delta\cdot(1-\effectiveValue{a}{\BobMechanism}/a)$ and $\Delta+\effectiveValue{a}{\BobMechanism}$, respectively.
  Hence
  \begin{equation}
    \label{eq:aligned-lb-bob-revenue-bound}
    R\leq(1-x_A)\cdot(\Delta+\effectiveValue{a}{\BobMechanism})
           +x_A\cdot\Delta\cdot\paren*{1-\frac{\effectiveValue{a}{\BobMechanism}}{a}}
      =\Delta+\effectiveValue{a}{\BobMechanism}\cdot\paren*{1-\frac{b}{a}\cdot x_A}.
  \end{equation}
  This gives $\effectiveValue{a}{\BobMechanism}\geq R-\Delta$; rearranging and using
  $\revDuoOrder{\AB}{A}{\AliceMechanism}{\BobMechanism}
    =\effectiveValue{a}{\BobMechanism}\cdot x_A-\utilityMechanism{\AliceMechanism}{\effectiveValue{a}{\BobMechanism}}$
  proves the second claim.}
\end{proof}

\begin{lemma}
  \label{lem:aligned-alice-utility-bound}
  \label{lem:aligned-lb-alice-utility-bound}
  {Let $R\in\oc*{\residualValue{b}{a},b}$ and $\AliceMechanism$ be a
  mechanism for Alice such that, for all mechanisms $\BobMechanism$
  for Bob in the aligned model, we have
  $\revDuoOrder{\AB}{B}{\AliceMechanism}{\BobMechanism}\leq R$.
  The following hold:
  \begin{enumerate}
    \item If $a\geq b$, we have
    $\utilityMechanism{\AliceMechanism}{a}
       \geq\int_{R+\Delta}^{a}
          \paren*{1-\frac{R}{v-\Delta}}\,\del v$.
    \item If $a\leq b$, for all
    $v\in\bracket*{R-\Delta,a}$, we have
    $\utilityMechanism{\AliceMechanism}{v}
       \geq\frac{a}{b}\cdot\int_{R-\Delta}^{v}
          \paren*{1-\frac{R-\Delta}{t}}\,\del t$.
  \end{enumerate}}
\end{lemma}

\begin{proof}
  {Apply \cref{cor:aligned-dl-revenue} to
  $\DL_{0,v-\Delta}$ in the first case and
  $\DL_{1-v/a,\Delta+v}$ in the second. Since each deviation
  earns at most $R$, rearranging gives the corresponding pointwise
  lower bound on $\zmarg{\AliceMechanism}^-(v)$.
  Integrating gives both claims.}
\end{proof}

We are now ready to prove \cref{thm:introduction-aligned-rephrased}. We divide the proof into two cases, $a \geq b$ and $a < b$, for convenience. For each case, combining the two lemmas above provides us with an upper bound on the revenue Alice can gain in terms of $R$. Optimizing over $R$ provides an upper bound on Alice's optimal revenue. For each case, we then design a mechanism which matches this upper bound. Finally, we extend our result to a characterization of all Alice-optimal mechanisms.

\paragraph{\texorpdfstring{{The case $a\geq b$.}}{The case a >= b.}}

\phantomsection
\label{proof:stackelberg-aligned-at-least-b}
\begin{proof}[Proof of
  \cref{thm:introduction-aligned-rephrased}
  when $a\geq b$]
  {Let $\AliceMechanism$ and $\BobMechanism$ be an arbitrary
  pair of on-path mechanisms in a Stackelberg equilibrium of the
  aligned model. Let $R=\revDuoOrder{\AB}{B}{\AliceMechanism}{\BobMechanism}
  \in\bracket*{0,b}$ be Bob's revenue. We first deal with the case
  $R=0$. Every fixed price $q\in\oc*{0,b}$ then gives Bob zero
  revenue, so \cref{cor:aligned-dl-revenue}
  {with $x=0$ (i.e., $\DL_{0,q}=\FP{q}$)} implies
  $\zmarg{\AliceMechanism}^-(\Delta+q)=1$. Integrating gives
  $\utilityMechanism{\AliceMechanism}{a}\geq b$, and
  \cref{cor:aligned-alice-revenue-bound} bounds Alice's revenue
  by $\Delta$.

  For the rest of the upper-bound argument, assume $R>0$.
  Since $\BobMechanism$ is a best response, no mechanism for Bob
  earns more than $R$. Using
  \cref{cor:aligned-alice-revenue-bound,lem:aligned-alice-utility-bound},
  we get
  \begin{align*}
    \revDuoOrder{\AB}{A}{\AliceMechanism}{\BobMechanism}
      &\leq a\cdot\paren*{1-\frac{R}{b}}
             -\int_{R+\Delta}^a
                \paren*{1-\frac{R}{v-\Delta}}\,\del v\\
      &=a\cdot\paren*{1-\frac{R}{b}}-b+R+R\cdot\ln\frac{b}{R}\\
      &=\Delta+R\cdot\paren*{1-\frac{a}{b}+\ln\frac{b}{R}}
        =:F(R).
  \end{align*}
  Define $F(0)=\Delta$ by continuity. For $R>0$,
  $F'(R)=\ln(b/R)-a/b$ and $F''(R)=-1/R<0$.
  Thus the unique maximum on $[0,b]$ is attained at
  $R^*=b\cdot e^{-a/b}$, and we obtain
  \begin{equation}
    \label{eq:aligned-la-revenue-upper-bound}
    \optOrder{a}{b}{\AB}\leq F(R^*)=\Delta+b\cdot e^{-a/b}.
  \end{equation}
  The same argument bounds Alice's revenue by $F(R)$ for every
  mechanism pair in which Bob best-responds.}

{To finish the proof, we show that the bound on
$\optOrder{a}{b}{\AB}$ above is tight. For this, we define
Aligned-Large-Alice-Equalizer mechanisms.}

\begin{restatable}[{{Aligned-Large-Alice-Equalizer}}]{definition}{aliceoptmechanismaatleastb}
  \label{def:alice-opt-stalkeberg-AB-a-at-least-b-ALICE-OPTIMAL-MECHANISM-CLASS}
  {Let $a\geq b$. Given a parameter $R\in\bracket*{0,b}$,
  define $\AliceMechanism_R^{\mathsf{ALA}}$ as follows.
  If $R>0$, its allocation rule is
  \[
    \zmarg{\AliceMechanism_R^{\mathsf{ALA}}}^{+}\paren*{v}
      =\begin{cases}
        0,&0\leq v<R+\Delta,\\[5pt]
        1-\dfrac{R}{v-\Delta},&R+\Delta\leq v\leq a.
      \end{cases}
  \]
  After $a$, the allocation retains its value at $a$.
  At $R=0$, let $\AliceMechanism_0^{\mathsf{ALA}}=\FP{\Delta}$.}
\end{restatable}

Here, $\mathsf{ALA}$ stands for Aligned-Large-Alice.
{In words, for $0<R<b$, the allocation is zero up to
$R+\Delta$, then rises continuously as $v$ increases to $a$, and
remains constant thereafter. The allocation rule is monotone, so
\cref{lem:prelim-MYERSON} gives a proper pricing rule implementing
it. Also, the definition implies
\begin{equation}
  \label{eq:aligned-la-equalizer-util}
  \zmarg{\AliceMechanism_R^{\mathsf{ALA}}}^-(a)=1-\frac{R}{b},
  \qquad
  \utilityMechanism{\AliceMechanism_R^{\mathsf{ALA}}}{a}
    =b-R-R\cdot\ln\frac{b}{R},
\end{equation}
where $R\ln(b/R)=0$ at $R=0$. We next show that Bob can never earn
more than $R$, and that $\FP{b}$ attains this revenue.}

\begin{lemma}
  \label{lem:stackelberg-aligned-at-least-b-summary}
  {Let $a\geq b$ and $R\in\bracket*{0,b}$. For all
  mechanisms $\BobMechanism$ for Bob in the aligned model,
  \[
    \revDuoOrder{\AB}{B}{\AliceMechanism_R^{\mathsf{ALA}}}{\BobMechanism}
      \leq\revDuoOrder{\AB}{B}{\AliceMechanism_R^{\mathsf{ALA}}}{\FP{b}}=R.
  \]
  Moreover, when $\BobMechanism$ is an Alice-favorable best response,
  we have
  \[
    \revDuoOrder{\AB}{A}{\AliceMechanism_R^{\mathsf{ALA}}}{\BobMechanism}
      =a\cdot\paren*{1-\frac{R}{b}}
        -\utilityMechanism{\AliceMechanism_R^{\mathsf{ALA}}}{a}.
  \]}
\end{lemma}

\begin{proof}
  {Fix any mechanism $\BobMechanism$ for Bob and set
  $q=b-\utilityMechanism{\BobMechanism}{b}\in\bracket*{0,b}$.
  If $q=0$, Bob's revenue is zero. Otherwise, Myerson's payment
  identity bounds his payment at value $b$ by $q$, and his payment
  at $\residualValue{b}{a}=0$ is zero. Since $\effectiveValue{a}{\BobMechanism}=\Delta+q$,
  \cref{cor:aligned-buyer} therefore gives
  \[
    \revDuoOrder{\AB}{B}{\AliceMechanism_R^{\mathsf{ALA}}}{\BobMechanism}
      \leq q\cdot\paren*{1-
          \zmarg{\AliceMechanism_R^{\mathsf{ALA}}}^-(\Delta+q)}
      =\min\paren*{q,R}\leq R.
  \]
  Next, \cref{cor:aligned-buyer,eq:aligned-la-equalizer-util}
  imply that $\FP{b}$ gives Bob revenue $R$ and Alice revenue
  $a\cdot(1-R/b)-\utilityMechanism{\AliceMechanism_R^{\mathsf{ALA}}}{a}$.
  This is the highest payment in Alice's menu, since her allocation
  rule is constant after $a$. As Bob selects an Alice-favorable
  best response, Alice receives this payment.}
\end{proof}

{By \cref{eq:aligned-la-equalizer-util,lem:stackelberg-aligned-at-least-b-summary},
$\AliceMechanism_R^{\mathsf{ALA}}$ gives Alice revenue $F(R)$ against
$\FP{b}$. This is the same expression optimized above, so choosing
$R^*=b\cdot e^{-a/b}$ attains the upper bound.
To obtain a Stackelberg equilibrium, use
\cref{obs:buyer-tie-breaks-for-alice} to select Alice-favorable
responses after every Alice mechanism, choosing $\FP{b}$ after
$\AliceMechanism_{R^*}^{\mathsf{ALA}}$. The upper-bound argument
rules out any profitable deviation by Alice. Hence
$\optOrder{a}{b}{\AB}=\Delta+b\cdot e^{-a/b}>\Delta$,
{establishing the optimal revenue.}}

Let $(\AliceMechanism,\BobMechanism)$ be an arbitrary
Alice-optimal Stackelberg equilibrium in the aligned model, and let
$R=\revDuoOrder{\AB}{B}{\AliceMechanism}{\BobMechanism}$ be Bob's revenue.
Let $\AliceMechanism^*=\AliceMechanism_{R^*}^{\mathsf{ALA}}$
and put $x_{eq}=1-R^*/b$.
Strict concavity of $F(R)$ forces $R=R^*$ in every optimal
pair. Equality in the upper-bound chain
forces
$\utilityMechanism{\AliceMechanism}{a}
 =\utilityMechanism{\AliceMechanism^*}{a}$.
The allocation lower bounds therefore have zero integrated slack.
Right-continuity then gives the necessary condition
\begin{equation}
  \label{eq:aligned-optimal-allocation-characterization}
  \zmarg{\AliceMechanism}^{+}(v)=\zmarg{\AliceMechanism^*}^{+}(v)
  \quad(0\leq v<a),\qquad
  \zmarg{\AliceMechanism}^-(a)=x_{eq}.
\end{equation}
Conversely, let $\AliceMechanism$ be any mechanism satisfying
\cref{eq:aligned-optimal-allocation-characterization}.
The argument bounding Bob's revenue in the proof of
\cref{lem:stackelberg-aligned-at-least-b-summary} applies to every
such mechanism. Thus Bob's revenue is at most $R^*$, while
$\FP{b}$ attains $R^*$ and is therefore a best response.
Against $\FP{b}$, Alice earns
$a \cdot x_{eq}-\utilityMechanism{\AliceMechanism}{a}=F(R^*)$,
exactly the revenue she obtains by offering $\AliceMechanism^*$.
An Alice-favorable best response gives her at least this revenue,
and the upper bound proved above rules out a higher revenue.
Thus she earns exactly $F(R^*)$ against every Alice-favorable best
response, proving that \cref{eq:aligned-optimal-allocation-characterization}
characterizes all Alice-optimal mechanisms.
In proper-pricing form, it fixes the equalizer's menu through
allocation $x_{eq}$, permitting any admissible proper convex extension beyond it.
Such extensions lie on or above the tangent of slope $a$;
the largest maximizer at $a$ can be any value in $[x_{eq},1]$.

Finally, define the buyer's effective value for Alice's good by
$\effectiveValue{a}{\BobMechanism}=a-\aug_{\BobMechanism}\paren*{a,b}$.
The inequality in \cref{cor:aligned-alice-revenue-bound} is strict if
$\effectiveValue{a}{\BobMechanism}<a$. Optimality therefore forces $\effectiveValue{a}{\BobMechanism}=a$,
which implies that $\utilityMechanism{\BobMechanism}{b}
=\utilityMechanism{\BobMechanism}{\residualValue{b}{a}}=0$
and $\pmarg{\BobMechanism}^{+}(\residualValue{b}{a}) = 0$.
As the buyer tie-breaks in favor of Bob, 
he selects the lowest utility-maximizing allocation 
from Alice, $x_{eq}$,
and so
$\pmarg{\BobMechanism}^{+}(b) \cdot (1 - x_{eq}) = R^* =  b \cdot (1-x_{eq})$ forces
Bob's payment and allocation at $b$ to be $b$ and one.
These conditions
can hold only if he offers the posted price $b$, and earns $R^*=b\cdot e^{-a/b}$.
All optimal equilibria therefore give Alice her optimum, Bob $R^*$,
and the buyer $\utilityMechanism{\AliceMechanism^*}{a}$, as claimed.
\end{proof}

\paragraph{\texorpdfstring{{The case $a<b$.}}{The case a < b.}}

\phantomsection
\label{proof:stackelberg-aligned-at-most-b}
\begin{proof}[Proof of
  \cref{thm:introduction-aligned-rephrased}
  when $a<b$]
  {Let $\AliceMechanism$ and $\BobMechanism$ be an arbitrary
  pair of on-path mechanisms in a Stackelberg equilibrium of the
  aligned model. Let $R=\revDuoOrder{\AB}{B}{\AliceMechanism}{\BobMechanism}$
  be Bob's revenue and $\effectiveValue{a}{\BobMechanism}=a-\aug_{\BobMechanism}\paren*{a,b}$.
  Bob can guarantee $\Delta$ by using $\FP{\Delta}$, so
  $R\in\bracket*{\Delta,b}$. We first deal with the case
  $R=\Delta$. Since no deviation gives Bob more than $\Delta$,
  \cref{cor:aligned-dl-revenue}
  {with $q=\Delta+t\in\oc*{\Delta,b}$} implies
  $\zmarg{\AliceMechanism}^-(t)\geq a/b$ for every
  $t\in\oc*{0,a}$. Integrating gives
  $\utilityMechanism{\AliceMechanism}{\effectiveValue{a}{\BobMechanism}}\geq(a/b)\cdot\effectiveValue{a}{\BobMechanism}$,
  and \cref{cor:aligned-alice-revenue-bound} bounds Alice's
  revenue by zero.

  For the rest of the upper-bound argument, assume $R>\Delta$.
  Since $\BobMechanism$ is a best response, no mechanism for Bob
  earns more than $R$. By \cref{cor:aligned-alice-revenue-bound},
  $R-\Delta\leq\effectiveValue{a}{\BobMechanism}\leq a$. Combining this corollary with
  \cref{lem:aligned-alice-utility-bound} at $v=\effectiveValue{a}{\BobMechanism}$, we get
  \begin{align*}
    \revDuoOrder{\AB}{A}{\AliceMechanism}{\BobMechanism}
      &\leq\frac{a}{b}\cdot\paren*{\effectiveValue{a}{\BobMechanism}-R+\Delta}
          -\frac{a}{b}\cdot\int_{R-\Delta}^{\effectiveValue{a}{\BobMechanism}}
             \paren*{1-\frac{R-\Delta}{t}}\,\del t\\
      &=\frac{a}{b}\cdot(R-\Delta)\cdot
          \ln\frac{\effectiveValue{a}{\BobMechanism}}{R-\Delta}\\
      &\leq\frac{a}{b}\cdot(R-\Delta)\cdot
          \ln\frac{a}{R-\Delta}=:F(R).
  \end{align*}
  Define $F(\Delta)=0$ by continuity. For $R>\Delta$,
  $F'(R)=(a/b)\cdot(\ln(a/(R-\Delta))-1)$ and
  $F''(R)=-a/(b\cdot(R-\Delta))<0$.
  Thus the unique maximum on $[\Delta,b]$ is attained at
  $R^*=\Delta+a/e$, and we obtain
  \begin{equation}
    \label{eq:aligned-lb-revenue-upper-bound}
    \optOrder{a}{b}{\AB}\leq F(R^*)=\frac{a^2}{e\cdot b}.
  \end{equation}
  The same argument bounds Alice's revenue by $F(R)$ for every
  mechanism pair in which Bob best-responds.}

{To finish the proof, we show that the bound on
$\optOrder{a}{b}{\AB}$ above is tight. For this, we define
Aligned-Large-Bob-Equalizer mechanisms.}

\begin{restatable}[{{Aligned-Large-Bob-Equalizer}}]{definition}{aliceoptmechanismaatmostb}
  \label{def:alice-opt-stalkeberg-AB-a-at-most-b-ALICE-OPTIMAL-MECHANISM-CLASS}
  {Let $a<b$. Given a parameter $R\in\bracket*{\Delta,b}$,
  define $\AliceMechanism_R^{\mathsf{ALB}}$ as follows.
  If $R>\Delta$, its allocation rule is
  \[
    \zmarg{\AliceMechanism_R^{\mathsf{ALB}}}^{+}\paren*{v}
      =\begin{cases}
        0,&0\leq v<R-\Delta,\\[5pt]
        \dfrac{a}{b}\cdot\paren*{1-\dfrac{R-\Delta}{v}},
          &R-\Delta\leq v\leq a.
      \end{cases}
  \]
  After $a$, the allocation retains its value at $a$.
  At $R=\Delta$, let
  $\AliceMechanism_\Delta^{\mathsf{ALB}}=\SL_{a/b,0}$,
  which offers allocation $a/b$ for free.}
\end{restatable}

Here, $\mathsf{ALB}$ stands for Aligned-Large-Bob.
{In words, for $\Delta<R<b$, the allocation is zero up to
$R-\Delta$, then rises continuously as $v$ increases to $a$, and
remains constant thereafter. The allocation rule is monotone, so
\cref{lem:prelim-MYERSON} gives a proper pricing rule implementing
it. Also, the definition implies
\begin{equation}
  \label{eq:aligned-lb-equalizer-util}
  \zmarg{\AliceMechanism_R^{\mathsf{ALB}}}^-(a)=1-\frac{R}{b},
  \qquad
  \utilityMechanism{\AliceMechanism_R^{\mathsf{ALB}}}{a}
    =\frac{a}{b}\cdot\paren*{b-R-(R-\Delta)\cdot
        \ln\frac{a}{R-\Delta}},
\end{equation}
where $(R-\Delta)\ln(a/(R-\Delta))=0$ at $R=\Delta$.
We next show that Bob can never earn more than $R$, and that
$\FP{b}$ attains this revenue.}

\begin{lemma}
  \label{lem:stackelberg-aligned-at-most-b-summary}
  {Let $a<b$ and $R\in\bracket*{\Delta,b}$. For all
  mechanisms $\BobMechanism$ for Bob in the aligned model,
  \[
    \revDuoOrder{\AB}{B}{\AliceMechanism_R^{\mathsf{ALB}}}{\BobMechanism}
      \leq\revDuoOrder{\AB}{B}{\AliceMechanism_R^{\mathsf{ALB}}}{\FP{b}}=R.
  \]
  Moreover, when $\BobMechanism$ is an Alice-favorable best response,
  we have
  \[
    \revDuoOrder{\AB}{A}{\AliceMechanism_R^{\mathsf{ALB}}}{\BobMechanism}
      =a\cdot\paren*{1-\frac{R}{b}}
        -\utilityMechanism{\AliceMechanism_R^{\mathsf{ALB}}}{a}.
  \]}
\end{lemma}

\begin{proof}
  {Fix any mechanism $\BobMechanism$ for Bob and set
  $\effectiveValue{a}{\BobMechanism}=a-\aug_{\BobMechanism}\paren*{a,b}\in\bracket*{0,a}$.
  If $\effectiveValue{a}{\BobMechanism}=0$, \cref{eq:aligned-lb-bob-revenue-bound} bounds
  Bob's revenue by $\Delta\leq R$. Otherwise, the same inequality
  and $x_A\geq\zmarg{\AliceMechanism_R^{\mathsf{ALB}}}^-(\effectiveValue{a}{\BobMechanism})$
  give
  \[
    \revDuoOrder{\AB}{B}{\AliceMechanism_R^{\mathsf{ALB}}}{\BobMechanism}
      \leq\Delta+\effectiveValue{a}{\BobMechanism}\cdot\paren*{1-\frac{b}{a}\cdot
             \zmarg{\AliceMechanism_R^{\mathsf{ALB}}}^-(\effectiveValue{a}{\BobMechanism})}
      =\Delta+\min\paren*{\effectiveValue{a}{\BobMechanism},R-\Delta}\leq R.
  \]
  Next, \cref{cor:aligned-buyer,eq:aligned-lb-equalizer-util}
  imply that $\FP{b}$ gives Bob revenue $R$ and Alice revenue
  $a\cdot(1-R/b)-\utilityMechanism{\AliceMechanism_R^{\mathsf{ALB}}}{a}$.
  This is the highest payment in Alice's menu, since her allocation
  rule is constant after $a$. As Bob selects an Alice-favorable
  best response, Alice receives this payment.}
\end{proof}

{By \cref{eq:aligned-lb-equalizer-util,lem:stackelberg-aligned-at-most-b-summary},
$\AliceMechanism_R^{\mathsf{ALB}}$ gives Alice revenue $F(R)$ against
$\FP{b}$. This is the same expression optimized above, so choosing
$R^*=\Delta+a/e$ attains the upper bound.
To obtain a Stackelberg equilibrium, use
\cref{obs:buyer-tie-breaks-for-alice} to select Alice-favorable
responses after every Alice mechanism, choosing $\FP{b}$ after
$\AliceMechanism_{R^*}^{\mathsf{ALB}}$. The upper-bound argument
rules out any profitable deviation by Alice. Hence
$\optOrder{a}{b}{\AB}=a^2/(e\cdot b)>0$, strictly exceeding
her Bertrand revenue and %
{establishing the optimal revenue.}}

{Let $(\AliceMechanism,\BobMechanism)$ be an arbitrary
Alice-optimal Stackelberg equilibrium in the aligned model, and let
$R=\revDuoOrder{\AB}{B}{\AliceMechanism}{\BobMechanism}$ be Bob's revenue.
Let $\AliceMechanism^*=\AliceMechanism_{R^*}^{\mathsf{ALB}}$
and $x_{eq}=1-R^*/b$.
Strict concavity of $F(R)$ forces $R=R^*$ in every optimal
pair. Equality in the upper-bound chain forces
$\utilityMechanism{\AliceMechanism}{a}
=\utilityMechanism{\AliceMechanism^*}{a}$.
Thus the allocation lower bounds have zero integrated slack;
right-continuity gives
$\zmarg{\AliceMechanism}^{+}(v)=\zmarg{\AliceMechanism^*}^{+}(v)$
for $0\leq v<a$, and $\zmarg{\AliceMechanism}^-(a)=x_{eq}$.
Conversely, let $\AliceMechanism$ be any mechanism satisfying these
equalities. The argument bounding Bob's revenue in the proof of
\cref{lem:stackelberg-aligned-at-most-b-summary} applies to every
such mechanism. Thus Bob's revenue is at most $R^*$, while
$\FP{b}$ attains $R^*$ and is therefore a best response.
Against $\FP{b}$, Alice earns
$a \cdot x_{eq}-\utilityMechanism{\AliceMechanism}{a}=F(R^*)$,
exactly the revenue she obtains by offering $\AliceMechanism^*$.
An Alice-favorable best response gives her at least this revenue,
and the upper bound proved above rules out a higher revenue.
Thus she earns exactly $F(R^*)$ against every Alice-favorable best
response, proving that these equalities characterize all
Alice-optimal mechanisms:
the proper menu coincides with the equalizer through allocation $x_{eq}$,
with any admissible proper convex extension beyond it.
Such extensions lie on or above the tangent of slope $a$,
and $\zmarg{\AliceMechanism}^{+}(a)$ can be any value in $[x_{eq},1]$.}

{%
Finally, define the buyer's effective value for Alice's good by
$\effectiveValue{a}{\BobMechanism}=a-\aug_{\BobMechanism}\paren*{a,b}$.
The logarithmic upper bound above, obtained using
\cref{cor:aligned-alice-revenue-bound,lem:aligned-alice-utility-bound},
is strict if $\effectiveValue{a}{\BobMechanism}<a$, since $R-\Delta>0$.
Optimality therefore forces $\effectiveValue{a}{\BobMechanism}=a$, which implies
{$\utilityMechanism{\BobMechanism}{b}
=\utilityMechanism{\BobMechanism}{\residualValue{b}{a}}=0$
by convexity and normalization, and $\pmarg{\BobMechanism}^{+}(\residualValue{b}{a})=0$.}
{As the buyer tie-breaks in favor of Bob,
he selects the lowest utility-maximizing allocation from Alice,
$x_{eq}$, and so}
{$\pmarg{\BobMechanism}^{+}(b)\cdot(1-x_{eq})=R^*=b\cdot(1-x_{eq})$} forces Bob's
payment and allocation at $b$ to be $b$ and one.
{These conditions
can hold only if he offers} the posted price $b$,
and earns $R^*=\Delta+a/e$.
All optimal equilibria therefore give Alice her optimum, Bob $R^*$,
and the buyer $\utilityMechanism{\AliceMechanism^*}{a}$, as claimed.}
\end{proof}

\subsection{\texorpdfstring{{Characterization of Optimal Mechanisms}}{Characterization of Optimal Mechanisms}}
\label{sec:aligned:alice:summary}

{We summarize the two characterizations established above.}

\begin{proposition}
  \label{prop:aligned-optimal-mechanisms}
  {Consider an instance of the duopoly game in the aligned model
  with fixed buyer values $a,b>0$, and write $\Delta=|a-b|$.
  A mechanism $\AliceMechanism$ is optimal for Alice if and only if
  {the following condition holds:}
  \begin{enumerate}
    \item When $a<b$, its allocation rule satisfies
    $\zmarg{\AliceMechanism}^{+}(v)=\zmarg{\AliceMechanism_{\mathsf{ALB}}^\ast}^{+}(v)$
    for all $v\in\co*{0,a}$, and
    $\zmarg{\AliceMechanism}^-(a)=\frac{a}{b}\cdot\paren*{1-\frac{1}{e}}${,
    where $R_{\mathsf{ALB}}^\ast=\Delta+a/e$ and
    $\AliceMechanism_{\mathsf{ALB}}^\ast=\AliceMechanism_{R_{\mathsf{ALB}}^\ast}^{\mathsf{ALB}}$
    as in \cref{def:alice-opt-stalkeberg-AB-a-at-most-b-ALICE-OPTIMAL-MECHANISM-CLASS}}. 
    \item When $a\geq b$, its allocation rule satisfies
    $\zmarg{\AliceMechanism}^{+}(v)=\zmarg{\AliceMechanism_{\mathsf{ALA}}^\ast}^{+}(v)$
    for all $v\in\co*{0,a}$, and
    $\zmarg{\AliceMechanism}^-(a)=1-e^{-a/b}$,
    where $R_{\mathsf{ALA}}^\ast=b\cdot e^{-a/b}$ and
    $\AliceMechanism_{\mathsf{ALA}}^\ast=\AliceMechanism_{R_{\mathsf{ALA}}^\ast}^{\mathsf{ALA}}$
    as in \cref{def:alice-opt-stalkeberg-AB-a-at-least-b-ALICE-OPTIMAL-MECHANISM-CLASS}.
  \end{enumerate}
  In either case, the largest maximizer $\zmarg{\AliceMechanism}^{+}(a)$
  may be any value in $\bracket*{\zmarg{\AliceMechanism}^-(a),1}$.}
\end{proposition}

\begin{proof}
  {The claim follows from the characterizations in
  \cref{sec:aligned:proof}, after substituting the optimal parameters
  {given above}.}
\end{proof}

Optimal mechanisms need not be essentially equivalent to
the canonical mechanisms defined above. For example, when Bob
offers only the null option, the buyer selects
$\zmarg{\AliceMechanism}^{+}(a)$, which may differ across optimal
mechanisms.
\section{The Misaligned Model}
\label{sec:alice-opt-stalkeberg-BA}

The goal of this section is to find Alice's optimal 
revenue when the buyer must visit Bob first.
We call this the \emph{misaligned model}.
Recall that $a > 0$ and $b > 0$
denote
the buyer's values for Alice's and Bob's goods
and that $\residualValue{a}{b}=(a-b)_+$ and $\residualValue{b}{a}=(b-a)_+$ are the
residual values after receiving the other good.
{$\FP{p}$ denotes a posted price $p$.}
{For a pair of mechanisms $(\AliceMechanism,\BobMechanism)$,
we write $\revDuoOrder{\BA}{i}{\AliceMechanism}{\BobMechanism}$ for
seller $i$'s expected revenue under the prescribed order $\BA$,
where $i\in\{A,B\}$.}
Let $\optOrder{a}{b}{\BA}$ denote
the supremum of Alice's expected revenue over Stackelberg equilibria
when a buyer with values $a,b > 0$ is forced to visit Bob first.

\introductionmisaligned*

{The rest of this section is devoted to proving
\cref{thm:introduction-misaligned-rephrased}.
Since most proofs in this section are short, we present all of them here rather than in a separate appendix.
We begin with an outline of the proof.}

\paragraph{Alice's Optimal Revenue.}
{When the buyer visits Bob, he has an effective type $\effectiveValue{b}{\AliceMechanism}$
which encapsulates his willingness to buy from Bob
given his outside option from Alice.
This is the most he is willing to pay for Bob's good, so Bob can
extract it by posting that price.
If the effective type is positive, this is Bob's
unique best response.
The buyer then receives Bob's good and reaches Alice with residual
value $\residualValue{a}{b}$ for her good.
The buyer's effective type depends on the gap in utility Alice 
provides to a buyer with value $a$ versus a buyer with value $\residualValue{a}{b}$.
In order to reset 
the buyer's effective type, this gap must be equal to $b$,
implying that Alice must offer at least this utility.
Thus, in either case, Alice
cannot earn more than $\residualValue{a}{b} = (a-b)_+$.}
Alice can always attain this bound by posting price $\residualValue{a}{b}$, 
so our result implies that every
Stackelberg equilibrium gives her 
the same expected revenue.

\paragraph{Equilibrium when $a \leq b$.}
When $a\le b$, she earns zero against every best response, even when offering randomized mechanisms, and is
therefore indifferent among all admissible commitments.
Her choice nevertheless affects Bob's expected revenue and the buyer's
expected utility, allowing both to vary across equilibria. This
contrasts with the fixed payoffs in all other Alice-optimal equilibria considered throughout the paper.

{For concreteness, let $a=4$ and $b=10$. If Alice posts price $0$,
the buyer can obtain utility $4$ from her alone, so Bob's unique
optimal posted price is $6$.
If Alice instead posts price $4$, Bob's optimal posted price is $10$.
Alice earns zero in both cases, while Bob's revenue rises from his
Bertrand revenue of $6$ to $10$.
More generally, for $a\le b$ and {$u\in[0,a]$}, the on-path pair
{$(\FP{a-u},\FP{b-u})$} attains the payoff triple
{$(0,b-u,u)$}, realizing the entire interval in the theorem.}

We can therefore characterize all Stackelberg equilibria when $a<b$.
Alice may choose any admissible mechanism, while Bob must respond to
every possible commitment with his unique best response: a posted price
that leaves the buyer the same utility as Alice's mechanism alone.
The same characterization extends to $a=b$ as well, 
except when Alice offers her good for free.
Against such a commitment, Bob earns
zero under every admissible mechanism, so his response is unrestricted.
Interestingly, the commitment power of Alice is crucial in allowing Bob
to beat his Bertrand revenue. 
Alice's commitment to price $4$ sustains Bob's higher revenue
in the example above.
This could not occur in the simultaneous model. 

\paragraph{Equilibrium when $a > b$.}
{When $a>b$, Alice's unique optimal mechanism is a posted price
of $a - b$.
Bob may choose any admissible mechanism on path and earns zero revenue.
Regardless of Bob's choice, the buyer purchases Alice's good with
certainty, giving her revenue $a-b$ and providing the buyer with
a utility of $b$.}

\subsection{\texorpdfstring{{Proof of \cref{thm:introduction-misaligned-rephrased}}}{Proof of Theorem \getrefnumber{thm:introduction-misaligned-rephrased}}}
\label{sec:misaligned:proof}

{We now proceed with the formal proof, beginning with the
buyer's behavior. Recall that the augmentation-induced utility gap
measures the increase in
utility from $\mathcal M$ when the buyer's value rises from $(w-v)_+$
to $w$: for $v,w\geq0$,
$\aug_{\mathcal M}(v,w)=\utilityMechanism{\mathcal M}{w}
-\utilityMechanism{\mathcal M}{(w-v)_+}$.}
From \cref{lemma:buyer} with the order Bob \textrightarrow\ Alice imposed, we immediately get the following.

\begin{corollary}
  \label{cor:alice-opt-stalkeberg-BA-ALLOCATION-AND-REVENUE}
{Let Alice and Bob use mechanisms $\AliceMechanism$ and
$\BobMechanism$ respectively. Define the quantity
$\effectiveValue{b}{\AliceMechanism}=b-\aug_{\AliceMechanism}\paren*{b,a}$.
\begin{enumerate}
\item \label{item:misaligned:buyer:util} The maximum utility the buyer can
obtain when interacting with Bob first is:
\[
\utilityBA{\AliceMechanism}{\BobMechanism}
  =\utilityMechanism{\BobMechanism}{\effectiveValue{b}{\AliceMechanism}}
   +\utilityMechanism{\AliceMechanism}{a}.
\]
\item \label{item:misaligned:buyer:allocation} When the buyer interacts
with Bob first, the utility in \cref{item:misaligned:buyer:util} is
attained by first picking the option $x_B$ from Bob, where
\[
x_B=\allocDuo{B}{\AliceMechanism}{\BobMechanism}
   =\zmarg{\BobMechanism}^{+}\paren*{\effectiveValue{b}{\AliceMechanism}},
\]
and then picking the option $\zmarg{\AliceMechanism}^{+}\paren*{\residualValue{a}{b}}$ from
Alice if he receives Bob's good, and the option
$\zmarg{\AliceMechanism}^{+}\paren*{a}$ from Alice otherwise.
\item \label{item:misaligned:buyer:payment} When the buyer interacts
with Bob first, his payments to Alice and Bob are (here, $x_B$ is as
defined above):
\begin{align*}
\revDuoOrder{\BA}{A}{\AliceMechanism}{\BobMechanism}
  &=\paren*{1-x_B}\cdot\AliceMechanism\paren*{\zmarg{\AliceMechanism}^{+}\paren*{a}}
    +x_B\cdot\AliceMechanism\paren*{\zmarg{\AliceMechanism}^{+}\paren*{\residualValue{a}{b}}},\\
\revDuoOrder{\BA}{B}{\AliceMechanism}{\BobMechanism}
  &=\BobMechanism\paren*{x_B}.
\end{align*}
\end{enumerate}}
\end{corollary}

As discussed above, Bob's response is aimed at targeting the effective value of the buyer. For each commitment $\AliceMechanism$ of Alice, the effective value of the buyer when interacting with Bob is given by
$
  \effectiveValue{b}{\AliceMechanism}=b-\aug_{\AliceMechanism}(b,a).
$

\begin{lemma}
  \label{lem:alice-opt-stalkeberg-BA-BOB-FP-BR}
  Fix $\AliceMechanism$.
  If ${\effectiveValue{b}{\AliceMechanism}}>0$, Bob's unique best response is the posted price
  $\FP{{\effectiveValue{b}{\AliceMechanism}}}$.
  If ${\effectiveValue{b}{\AliceMechanism}}=0$, every admissible mechanism is a best response.
  In both cases, every best response gives Bob revenue ${\effectiveValue{b}{\AliceMechanism}}$ and
  the buyer utility $\utilityMechanism{\AliceMechanism}{a}$.
\end{lemma}

\begin{proof}
  The effective value
  ${\effectiveValue{b}{\AliceMechanism}}=b-{\aug_{\AliceMechanism}(b,a)}$
  is nonnegative. By individual rationality, the buyer's payment
  to Bob is at most ${\effectiveValue{b}{\AliceMechanism}}$ times his allocation probability, and
  hence at most ${\effectiveValue{b}{\AliceMechanism}}$. The mechanism $\FP{{\effectiveValue{b}{\AliceMechanism}}}$ attains this bound:
  if ${\effectiveValue{b}{\AliceMechanism}}>0$, Bob's revenue tie-breaking selects allocation one;
  if ${\effectiveValue{b}{\AliceMechanism}}=0$, \cref{cor:alice-opt-stalkeberg-BA-ALLOCATION-AND-REVENUE}
  gives the same allocation.

  If ${\effectiveValue{b}{\AliceMechanism}}>0$, every best response attaining revenue ${\effectiveValue{b}{\AliceMechanism}}$ must allocate
  Bob's good with probability one.
  {It must also leave zero utility at effective value ${\effectiveValue{b}{\AliceMechanism}}$,
  so ${\BobMechanism}(1)={\effectiveValue{b}{\AliceMechanism}}$ and
  ${\BobMechanism}(x)\geq {\effectiveValue{b}{\AliceMechanism}}\cdot x$ for all $x\in[0,1]$.
  Convexity and ${\BobMechanism}(0)=0$ give the reverse
  inequality. Hence ${\BobMechanism}(x)={\effectiveValue{b}{\AliceMechanism}}\cdot x$,
    so the unique best-response is $\FP{{\effectiveValue{b}{\AliceMechanism}}}$.}

  {If ${\effectiveValue{b}{\AliceMechanism}}=0$, nonnegative payments and individual rationality
  force Bob's revenue to be zero under every mechanism, so every
  mechanism is a best response. In either case, the buyer's utility is
  $\utilityMechanism{\AliceMechanism}{a}
   +\utilityMechanism{\BobMechanism}{{\effectiveValue{b}{\AliceMechanism}}}
   =\utilityMechanism{\AliceMechanism}{a}$.}
\end{proof}

The buyer receiving Bob's item with certainty whenever $\effectiveValue{b}{\AliceMechanism} > 0$
almost immediately implies that Alice's revenue is bounded by $\residualValue{a}{b}$. 
All that remains is to show that if $\effectiveValue{b}{\AliceMechanism} = 0$,
Alice must offer a large utility, bounding her revenue for this case as well.

\begin{lemma}
  \label{lem:alice-opt-stalkeberg-BA-ALICE-REVENUE}
  For every mechanism $\AliceMechanism$ and every best
  response $\BobMechanism$ to it,
  \[
    \revDuoOrder{\BA}{A}{\AliceMechanism}{\BobMechanism}
      =\pmarg{\AliceMechanism}^{+}({\residualValue{a}{b}})
      \leq{\residualValue{a}{b}}=(a-b)_+.
  \]

\end{lemma}

\begin{proof}
  {If $\effectiveValue{b}{\AliceMechanism}>0$, every best response allocates
  Bob's good with certainty, leaving residual value ${\residualValue{a}{b}}$ for
  Alice's good. If $\effectiveValue{b}{\AliceMechanism}=0$,
  {the equality $\utilityMechanism{\AliceMechanism}{a}
  -\utilityMechanism{\AliceMechanism}{\residualValue{a}{b}}=b$ saturates the
  $1$-Lipschitz bound, making allocation one optimal at both values.
  {By \cref{item:misaligned:buyer:allocation} of
  \cref{cor:alice-opt-stalkeberg-BA-ALLOCATION-AND-REVENUE},
  the buyer selects the largest utility-maximizing allocation
  when visiting Alice second, hence allocation one at both values.}
  The payment identity then}
  gives $\pmarg{\AliceMechanism}^{+}(a)
  =\pmarg{\AliceMechanism}^{+}({\residualValue{a}{b}})$, so Alice's revenue is the
  same regardless of Bob's allocation. Finally, individual rationality
  gives $\pmarg{\AliceMechanism}^{+}({\residualValue{a}{b}})\leq{\residualValue{a}{b}}$.}

\end{proof}

After bounding Alice's revenue, we can find exactly which mechanisms achieve this revenue, allowing us to characterize all Stackelberg equilibria.

\begin{lemma}
  \label{lem:misaligned-equilibrium-characterization}
  {A strategy profile
  $(\AliceMechanism,\mathbb{M})$ is a Stackelberg equilibrium if and
  only if both of the following conditions hold:}
  \begin{align*}
    \AliceMechanism&=\FP{a-b}
      &&\text{if }a>b,\\
    \mathbb{M}(\AliceMechanism')&=\FP{\effectiveValue{b}{\AliceMechanism'}}
      &&\text{for every }\AliceMechanism'\in\AllMechanisms
        \text{ with }\effectiveValue{b}{\AliceMechanism'}>0.
  \end{align*}
  {Thus, when $a>b$, Alice posts price $a-b$ and Bob may
  choose any admissible mechanism on path.
  When $a\leq b$, Alice may choose any admissible mechanism;
  Bob's on-path response must be $\FP{\effectiveValue{b}{\AliceMechanism}}$ if
  $\effectiveValue{b}{\AliceMechanism}>0$, and may be any admissible mechanism if
  $\effectiveValue{b}{\AliceMechanism}=0$.}
  {In the latter regime, $\effectiveValue{b}{\AliceMechanism}=0$ occurs exactly
  when $a=b$ and Alice's mechanism is 
  $\FP{0}$, so she offers her good for free.}
\end{lemma}

\begin{proof}
  {The second condition is exactly Bob's best-response
  requirement after every possible commitment, by
  \cref{cor:alice-opt-stalkeberg-BA-ALLOCATION-AND-REVENUE,lem:alice-opt-stalkeberg-BA-BOB-FP-BR}.
  Against any such strategy,
  \cref{lem:alice-opt-stalkeberg-BA-ALICE-REVENUE} gives Alice revenue
  $\pmarg{\AliceMechanism}^{+}({\residualValue{a}{b}})\leq{\residualValue{a}{b}}$.
  Since $\FP{{\residualValue{a}{b}}}$ attains this bound, her optimality is
  equivalent to $\pmarg{\AliceMechanism}^{+}({\residualValue{a}{b}})={\residualValue{a}{b}}$.
  If $a\leq b$, then ${\residualValue{a}{b}}=0$ and this equality holds for every
  mechanism. If $a>b$, the payment identity
  \[
    {\residualValue{a}{b}}={\residualValue{a}{b}}\cdot\zmarg{\AliceMechanism}^{+}({\residualValue{a}{b}})
      -\utilityMechanism{\AliceMechanism}{{\residualValue{a}{b}}}
  \]
  forces $\zmarg{\AliceMechanism}^{+}({\residualValue{a}{b}})=1$ and
  $\utilityMechanism{\AliceMechanism}{{\residualValue{a}{b}}}=0$.
  Thus ${\AliceMechanism}(1)={\residualValue{a}{b}}$ and
  ${\AliceMechanism}(x)\geq{\residualValue{a}{b}}\cdot x$ for all $x\in[0,1]$.
  Convexity and ${\AliceMechanism}(0)=0$ give the reverse
  inequality, so ${\AliceMechanism}(x)={\residualValue{a}{b}}\cdot x$,
  meaning that Alice's mechanism is $\FP{a-b}$, establishing the first condition.
  For this mechanism, $\effectiveValue{b}{\AliceMechanism}=b-(a-{\residualValue{a}{b}})=0$,
  so Bob may choose any admissible mechanism on path.}

  {Finally, when $a\le b$,
  $\effectiveValue{b}{\AliceMechanism}=b-\utilityMechanism{\AliceMechanism}{a}\ge b-a$.
  Hence $\effectiveValue{b}{\AliceMechanism}=0$ requires $a=b$ and
  $\utilityMechanism{\AliceMechanism}{a}=a$.
  Since $a>0$, this forces full allocation at zero payment.
  Thus ${\AliceMechanism}(1)=0$, and nonnegative, nondecreasing
  prices give ${\AliceMechanism}(x)=0$ throughout $[0,1]$.
  Alice's mechanism is therefore $\FP{0}$;
  conversely, this mechanism gives $\effectiveValue{b}{\AliceMechanism}=0$ when $a=b$.}
\end{proof}

We are now ready to prove \cref{thm:introduction-misaligned-rephrased}.

\phantomsection
\label{proof:stackelberg-misaligned}
\begin{proof}[Proof of \cref{thm:introduction-misaligned-rephrased}]
  {Let $\mathbb{M}(\AliceMechanism')=\FP{\effectiveValue{b}{\AliceMechanism'}}$
  for every $\AliceMechanism'\in\AllMechanisms$.
  By \cref{lem:misaligned-equilibrium-characterization},
  $(\FP{{\residualValue{a}{b}}},\mathbb{M})$ is an equilibrium, and Alice earns
  ${\residualValue{a}{b}}$ in every equilibrium.
  It remains to characterize Bob's revenue and the buyer's utility.}

  {If $a>b$, Alice's mechanism is 
  $\FP{a-b}$ by \cref{lem:misaligned-equilibrium-characterization}.
  Hence $\utilityMechanism{\AliceMechanism}{{\residualValue{a}{b}}}=0$,
  $\utilityMechanism{\AliceMechanism}{a}=b$ and
  $\effectiveValue{b}{\AliceMechanism}=0$.
  {By \cref{lem:alice-opt-stalkeberg-BA-BOB-FP-BR}, every
  admissible mechanism is a best response for Bob, gives him zero
  revenue, and leaves the buyer utility $b$.
  The payoff triple is therefore $(a-b,0,b)$.}}

  {If $a\leq b$, then ${\residualValue{a}{b}}=0$ and
  $\pmarg{\AliceMechanism}^{+}(0)=\utilityMechanism{\AliceMechanism}{0}=0$
  for every mechanism. Thus every Alice mechanism is optimal, with
  payoff triple {$(0,b-u,u)$, where $u=\utilityMechanism{\AliceMechanism}{a}$}.
  Since {$0\leq u\leq a$},
  Bob's revenue {$\effectiveValue{b}{\AliceMechanism}=b-u$}
  lies in $[b-a,b]$.
  For each {$u\in[0,a]$, $(\FP{a-u},\mathbb{M})$} is an equilibrium by
  \cref{lem:misaligned-equilibrium-characterization}, with on-path response
  {$\mathbb{M}(\FP{a-u})=\FP{b-u}$} and payoff triple
  {$(0,b-u,u)$}. Thus every triple in the corresponding set is attained.}
\end{proof}

\section{The Simultaneous Model}
\label{sec:nash}

The goal of this section is to show that when Alice 
and Bob choose their mechanisms simultaneously,
every pure Nash equilibrium is essentially a Bertrand competition, 
even when the mechanisms themselves may still be randomized.
We call this model the \emph{simultaneous model}.
Recall that $a,b>0$ are the buyer's values for Alice's
and Bob's goods, and $\residualValue{a}{b}=(a-b)_+$ and $\residualValue{b}{a}=(b-a)_+$ are the residual
values after receiving the other good.
A pair
$(\AliceMechanism,\BobMechanism)$ is a pure Nash equilibrium if neither seller
can increase expected revenue by changing mechanisms.

\introductionsimultaneous*

For completeness, we
first establish the existence of a pure Nash equilibrium with {the Bertrand revenues}.

\begin{restatable}{observation}{nashbertrandequilibrium}
  \label{lem:nash-a-at-least-b-BERTRAND-EQUILIBRIUM}
  There exists a {pure} Nash equilibrium $(\AliceMechanism,\BobMechanism)$
  with revenues ${\residualValue{a}{b}} = (a - b)_+$ 
  and {$\residualValue{b}{a} = (b - a)_+$}, for Alice and Bob respectively.
\end{restatable}

The rest of this section is devoted to proving
\cref{thm:introduction-simultaneous}. 
All omitted proofs appear in
\cref{sec:nash-app}. We begin with an outline of the proof.

Henceforth, let $a\geq b$, so ${\residualValue{a}{b}}=a-b$ and $\residualValue{b}{a} = 0$.
The revenue-bound
arguments below use only the buyer's utility maximization, and hence
also hold after interchanging the sellers. We may therefore prove the
revenue characterization under this assumption without loss of
generality.

Our guiding idea is to show that if an equilibrium 
does not provide the sellers' Bertrand revenues,
then the seller visited second can strictly increase his revenue. 
The main difficulty is that when the seller visited second deviates, 
the allocation probability chosen from the seller visited first might increase,
making the deviation not profitable. 
We show that the allocation probability chosen from the first seller 
in equilibrium is essentially maximal, ruling out this possibility.

\paragraph{Simultaneous Revenue Increase.}
We do this in several steps. 
First, we reduce the problem to equilibria in which both
sellers earn more than their Bertrand revenues.
It is enough to show that Alice receives a revenue of ${\residualValue{a}{b}}$ 
if and only if Bob receives a revenue of zero.
One direction is simple. 
If Bob earns zero in equilibrium, 
deviating to any positive posted price, however small, 
must also yield zero revenue.
This implies that Alice must offer a utility of at least $b$
and achieves a revenue of at most ${\residualValue{a}{b}} = a-b$. 
It is not difficult to verify that Alice's revenue is at least ${\residualValue{a}{b}}$, 
completing this direction.  

The second direction is more complex. To see this, notice that in equilibrium, 
Bob might gain positive revenue while offering some allocation for free. 
Therefore, if Alice deviates, the buyer might instead first
buy Bob's free lottery, dramatically cutting Alice's revenue.
This motivates a new idea: using Bob's largest free
allocation to control the buyer's response.
We define $y_0$
to be the largest free allocation offered by Bob's equilibrium mechanism.
As Bob's effective type tends to zero, every optimal
allocation from his menu converges to {$y_0$}.
Additionally, when $a>b$, the buyer may still pay Alice after winning Bob's
item, unlike in Lucier and Saxena's equal-value setting; this motivates
a double-lottery deviation.
Alice offers two options distinguishing
between buyers who won Bob's good
and those who did not.
This double lottery is designed to make the  
buyer's effective type when facing Bob first arbitrarily small,
so the allocation chosen by the buyer approaches $y_0$. The payments are
chosen so that, if Bob remains first, Alice's expected gain after
his failure outweighs her expected discount after his success.
They also make skipping Alice or taking her low option first
worse than a Bob-first plan. Thus any globally utility-maximizing
Alice-first plan buys her high option, whose payment exceeds
{$\residualValue{a}{b}$}. Alice therefore earns more than
{$\residualValue{a}{b}$} in either
order, contradicting equilibrium.
Consequently, it remains to exclude the possibility of an equilibrium which
provides Alice with strictly more than ${\residualValue{a}{b}}$ revenue and Bob with a positive revenue.
We suppose such an equilibrium exists and seek a profitable deviation.

\paragraph{Standalone Optimality.}
We then show that, in such an equilibrium, if Alice (resp. Bob) is approached first, 
then the allocation chosen from her (resp. him) is
optimal for a buyer with an effective value of $a$ (resp. $b$). 
Thus, roughly speaking, the allocation is the same 
one the buyer would choose when
facing Alice (resp. Bob) as a monopolist. 
Thus, we call this allocation \emph{standalone optimal}.
As the set of optimal allocations for a buyer is monotone
in the buyer's effective type, a standalone optimal allocation 
is essentially maximal.

{
{We explain why equilibrium forces this property.
In equilibrium the buyer must be indifferent between
the two orders, as if there is slack one of 
the sellers may use it to strictly increase their revenue. 
Moreover, it turns out that if $x$ is the
allocation chosen from Alice in equilibrium, then there exists 
an action plan which selects the order $\BA$ and chooses an 
allocation of $x$ from Alice, when losing Bob's lottery. 
Therefore, $x$ must be standalone optimal; otherwise, 
we can construct an action plan which strictly beats the action plan
chosen by the buyer, a contradiction to the utility-maximization of the buyer.
An identical argument for $\BA$ holds.}
}

\paragraph{The Contradiction.}
Standalone optimality now allows us to construct the required
deviations. For order $\AB$, this is simple. 
Suppose Bob posts a price above his original
expected payment after Alice's failure. 
If Alice remains first, her effective value
is at most $a$, so her allocation cannot increase. Bob therefore
earns more after each failure of Alice's lottery, and these failures
occur at least as often as before. If the buyer instead visits Bob
first, he buys Bob's good surely, which also increases Bob's revenue.

{
{Order $\BA$ requires a different deviation, since Alice can
offer a menu which attempts to screen between buyers who won and lost
Bob's lottery.}
We construct Alice's deviation by scaling down the allocation
and payment of the option bought after Bob succeeds, and replacing
the option bought after he fails by a sure allocation at a higher
payment. The two offers are chosen so that a plan visiting Bob first
and then Alice gives more utility than either skipping Alice or
buying her cheaper option first. The higher payment increases
Alice's revenue when Bob's lottery fails.
If Bob remains first, his new effective value is at most $b$,
so standalone optimality prevents his allocation from increasing.
Alice's more expensive option is therefore reached at least as often
as before, and the parameters are chosen so that its higher payment
more than compensates for the discount on the other option.
The construction also ensures that, if the buyer switches to Alice
first, he selects her more expensive option and again increases her
revenue. All in all, neither order can occur in the hypothetical
equilibrium, completing the argument.
}

The rest of this section develops the proof 
and is organized as follows.
In \cref{sec:nash-buyer}, we describe the buyer's optimal choices and
utility in each order using the buyer-behavior results from the
main model.
In \cref{sec:nash-thresholds}, we establish the revenue thresholds
that reduce the problem to excluding simultaneous gains over the
Bertrand
revenues.
Finally, \cref{sec:nash-no-joint} rules out such gains, completing the
proof of \cref{thm:introduction-simultaneous}.

\subsection{Buyer Behavior}
\label{sec:nash-buyer}

{The buyer-behavior analysis is identical to that in
\cref{sec:stackelberg:buyer}: once the mechanisms are fixed, the
buyer's behavior does not depend on the order in which the sellers
chose them. We connect the buyer's duopolistic choices to utility
maximization in the monopolist setting using the effective values
$\effectiveValue{a}{\BobMechanism}=a-\aug_{\BobMechanism}(a,b)$ and
$\effectiveValue{b}{\AliceMechanism}=b-\aug_{\AliceMechanism}(b,a)$ when visiting Alice or Bob
first, respectively.
Recall that, for $v,w\geq0$,
$\aug_{\mathcal M}(v,w)=\utilityMechanism{\mathcal M}{w}
-\utilityMechanism{\mathcal M}{(w-v)_+}$ is the augmentation-induced
utility gap. Thus, \cref{lemma:buyer} applies unchanged, giving the
buyer's utility, allocations, and payments in each order.}

\buyerbehavior*

{We write
$\utilityDuo{\AliceMechanism}{\BobMechanism}
  =\max\{\utilityAB{\AliceMechanism}{\BobMechanism},
          \utilityBA{\AliceMechanism}{\BobMechanism}\}$
for the buyer's maximum utility from the pair, using the order
utilities from \cref{sec:stackelberg:buyer}.
A lottery option
$\langle x,\tau\rangle$ allocates with probability $x$ and charges total
expected payment $\tau$.
}

\subsection{{Revenue Thresholds}}
\label{sec:nash-thresholds}

To restrict the equilibrium we must rule out, we begin by showing that sellers can only beat their Bertrand revenues together.

\begin{restatable}[Revenue thresholds]{proposition}{nashrevenuethresholds}
{
\label{lem:nash-a-at-least-b-REVENUE-THRESHOLD-EQUIVALENCE}
  Every pure Nash equilibrium $(\AliceMechanism,\BobMechanism)$
  satisfies $\revDuo{A}{\AliceMechanism}{\BobMechanism}\geq{\residualValue{a}{b}}$ and
  \[
    \revDuo{A}{\AliceMechanism}{\BobMechanism}={\residualValue{a}{b}}
    \quad\Longleftrightarrow\quad
    \revDuo{B}{\AliceMechanism}{\BobMechanism}=0.
  \]
}
\end{restatable}

{
  {\emph{Alice earns at least ${\residualValue{a}{b}}$.}
  {If ${\residualValue{a}{b}}>0$, Alice can guarantee any positive price below
  ${\residualValue{a}{b}}$, even after the buyer obtains Bob's good.}
  Taking prices up to ${\residualValue{a}{b}}$ gives the equilibrium
  lower bound; nonnegative revenues cover ${\residualValue{a}{b}}=0$.}

  {\emph{Alice earns exactly ${\residualValue{a}{b}}$ when Bob earns zero.}
  If Bob earns zero, posting any sufficiently small positive price
  $\varepsilon$ must still yield zero revenue. The buyer must then
  obtain Alice's good surely, since otherwise Bob provides a valuable
  fallback after her lottery fails. Buying only from Bob gives utility
  $b-\varepsilon$, so Alice's unchanged full-allocation option must
  give at least $b-\varepsilon$ for every such $\varepsilon$.
  Letting $\varepsilon\downarrow0$, this option guarantees utility
  at least $b$ in the original market. Since Bob earns zero, Alice's
  original revenue equals expected consumption value minus buyer
  utility. The former is at most $a$ and the latter at least $b$, so
  her revenue is at most $a-b={\residualValue{a}{b}}$. Together with the lower bound,
  this gives equality.}

  \emph{Bob earns zero when Alice earns ${\residualValue{a}{b}}$.}
  {We say that an allocation $z$ is \emph{optimal} for a mechanism
  $\mathcal M$ at value $v$ if it maximizes
  $v \cdot z-{\mathcal M}(z)$ over feasible allocations.}
  For the converse, suppose Alice earns ${\residualValue{a}{b}}$ while Bob earns a
  positive amount. Bob's standalone utility must be positive, since
  otherwise Alice can profitably post a price above ${\residualValue{a}{b}}$.
  Let $y_0$ be Bob's largest free allocation. As his
  effective value tends to zero, every optimal allocation tends to
  $y_0$. This limits how useful Bob is when visited first after Alice
  offers a nearly certain lottery. If
  $\utilityMechanism{\BobMechanism}{b}>a \cdot y_0$, Alice can use such a
  lottery to charge above ${\residualValue{a}{b}}$ while making $\AB$ strictly
  preferable: Bob is more valuable as a fallback after her failure.
  Hence $\utilityMechanism{\BobMechanism}{b}\leq a \cdot y_0$.
  {If the original order is $\AB$, Alice's chosen allocation
  must be optimal at value $a$. To see why, let
  $x=\allocDuo{A}{\AliceMechanism}{\BobMechanism}$.
  Since Alice's payment is ${\residualValue{a}{b}}$, the original plan gives utility
  $a \cdot x-{\residualValue{a}{b}}+(1-x) \cdot \utilityMechanism{\BobMechanism}{b}$.
  Instead, take Bob's free option first and buy the same Alice option
  only after failure. This gives utility $y_0 \cdot b+(1-y_0) \cdot (a \cdot x-{\residualValue{a}{b}})$:
  success gives Bob's good for free, and failure leaves Alice's
  original option at value $a$. 
  The alternative
  utility minus the original utility is
  \begin{align*}
    &y_0 \cdot b+(1-y_0) \cdot (a \cdot x-{\residualValue{a}{b}})
      -\bigl[a \cdot x-{\residualValue{a}{b}}+(1-x) \cdot \utilityMechanism{\BobMechanism}{b}\bigr]\\
    &\qquad=y_0 \cdot (a-a \cdot x)-(1-x) \cdot \utilityMechanism{\BobMechanism}{b}
      =(1-x) \cdot \bigl[a \cdot y_0-\utilityMechanism{\BobMechanism}{b}\bigr]\geq0.
  \end{align*}
  Thus this alternative gives at least the original utility; since
  the original plan is optimal, the two utilities must be equal.
  Moreover, $y_0<1$, since a free full allocation from Bob would
  preclude his positive revenue. Alice is therefore reached with
  positive probability in the alternative plan. If her option were
  not optimal at value $a$, choosing a better option after Bob's
  failure would strictly improve this already optimal plan, a
  contradiction.}
  Bob can profitably switch to a posted price: Alice's
  allocation cannot increase if she remains first, while switching
  orders forces a sure purchase from Bob.

  If the original order is $\BA$, the utility and revenue bounds
  instead leave room for Alice to offer a discounted, nearly certain
  lottery and a sure allocation charging above ${\residualValue{a}{b}}$. This drives
  Bob's effective value towards zero, so his allocation approaches
  $y_0$, and the two payments can be chosen to increase Alice's expected
  revenue in $\BA$. A switch to $\AB$ can only select her high option,
  which also pays above ${\residualValue{a}{b}}$. Either order contradicts equilibrium.
}

\subsection{{Excluding Simultaneous Revenue Improvement}}

\label{sec:nash-no-joint}

{
By \cref{lem:nash-a-at-least-b-REVENUE-THRESHOLD-EQUIVALENCE}, it remains
to exclude an equilibrium in which Alice earns more than ${\residualValue{a}{b}}$
and Bob earns a positive amount. In any such hypothetical equilibrium,
the buyer strictly benefits from access to both sellers:
$\utilityDuo{\AliceMechanism}{\BobMechanism}
  >\max\{\utilityMechanism{\AliceMechanism}{a},
          \utilityMechanism{\BobMechanism}{b}\}$.
Otherwise a seller could improve revenue with a fixed price that
slightly improves the buyer's utility; this is proved in
\cref{lem:nash-strict-buyer-utility}.

We next hope to show that the allocation chosen from the first seller is standalone optimal. 
An allocation is \emph{standalone optimal} if it is optimal
at the seller's standalone value: $a$ for Alice and $b$ for Bob.}

\begin{restatable}[Standalone optimality of the first allocation]{lemma}{nashstandaloneoptimality}
{
\label{lem:nash-a-at-least-b-UTILITY-MAXIMIZING-ALLOCATIONS}
  Let $(\AliceMechanism,\BobMechanism)$ be a pure Nash equilibrium
  in which Alice's revenue exceeds ${\residualValue{a}{b}}$ and Bob's revenue is
  positive. The allocation purchased from the first seller is also optimal at
  that seller's standalone value. Specifically,
  \begin{align*}
    &\allocDuo{A}{\AliceMechanism}{\BobMechanism}
      \in\arg\max_{x\in[0,1]}
           {\{a \cdot x-{\AliceMechanism}(x)\}}
      \quad\text{if the order is }\AB, \\
    &\allocDuo{B}{\AliceMechanism}{\BobMechanism}
      \in\arg\max_{y\in[0,1]}
           {\{b \cdot y-{\BobMechanism}(y)\}}
      \quad\text{if the order is }\BA.
  \end{align*}
}
\end{restatable}

{
  For simplicity, assume the original order is $\AB$; the
  $\BA$ case is analogous.
  Let $x=\allocDuo{A}{\AliceMechanism}{\BobMechanism}<1$.
  We show that, under the original mechanisms, there is also an
  optimal $\BA$ plan that buys Alice's original option $x$ after
  Bob's lottery fails. Additionally, for a mechanism $\mathcal{M}$ and a feasible allocation $z$, let
  $\allocationUtilityMechanism{\mathcal M}{v}{z}=v\cdot z-\mathcal M(z)$
  denote the buyer's utility at value $v$ from selecting $z$.

  {Let Alice replace her menu by the single lottery
  $\langle x+\delta,{\AliceMechanism}(x)+a \cdot \delta\rangle$
  for small $\delta>0$. 
  Denote this mechanism $\AliceMechanism'$. 
  Alice's deviation
  $\AliceMechanism'$ is designed such that
  the extra allocation $\delta$ costs
  $a \cdot \delta$, exactly its value to a buyer with value $a$.
  Thus,
  $ \utilityMechanism{\AliceMechanism'}{a}
  = a \cdot (x+\delta)-[{\AliceMechanism}(x)+a \cdot \delta]
  = a \cdot x-{\AliceMechanism}(x)
  = \utilityMechanism{\AliceMechanism}{a,x}
  $.
  Notice that buying this lottery first and retaining Bob's original continuation
  loses only $\delta \cdot \utilityMechanism{\BobMechanism}{b}$ in utility.

Consider the action-plan chosen by the buyer
  when facing $\AliceMechanism'$. 
  If $\AB$ remains selected, the buyer 
  buys Alice's full lottery and
  increases her revenue. 
  Thus, equilibrium forces $\BA$.}
  {In this order, after Bob's lottery fails, the buyer 
  selects Alice's lottery option.
  If Bob's lottery succeeds, it is not 
  difficult to verify that the 
  buyer chooses not to interact with Alice.
  {Thus, the buyer visits Bob first and selects an allocation
  $y^\ast$ that maximizes his total expected utility. He then buys
  Alice's lottery with allocation $x+\delta$ only if Bob's lottery fails.}
  As $\utilityMechanism{\AliceMechanism'}{a}
  = \utilityMechanism{\AliceMechanism}{a,x}$, 
  when choosing the order $\BA$, 
  replacing Alice's perturbed lottery
  by her original option $x$
  (when facing Alice's original mechanism)
  preserves the buyer's utility.
  Moreover, this utility is at least
  $\utilityDuo{\AliceMechanism}{\BobMechanism}
    -\delta \cdot \utilityMechanism{\BobMechanism}{b}$.
  {Letting $\delta\downarrow0$ shows that, under the original
  mechanisms, the buyer can attain his maximum utility by visiting
  Bob first and buying Alice's original option $x$ only if Bob's
  lottery fails, as claimed.}
  The $\BA$ plan must reach Alice 
  with positive probability;
  otherwise its utility could not exceed Bob's standalone utility.
  Finally, if $x$ were not optimal at value $a$, choosing a better Alice
  option after Bob's failure would strictly improve this optimal
  plan, a contradiction.}
}

We are now ready to rule out all equilibria in which both agents beat their Bertrand revenues.

\begin{restatable}[No simultaneous revenue improvement]{lemma}{nashnojointimprovement}
{
\label{lem:nash-a-at-least-b-NO-JOINT-REVENUE-IMPROVEMENT}
  No pure Nash equilibrium $(\AliceMechanism,\BobMechanism)$ satisfies
  both $\revDuo{A}{\AliceMechanism}{\BobMechanism}>{\residualValue{a}{b}}$ and
  $\revDuo{B}{\AliceMechanism}{\BobMechanism}>0$.
}
\end{restatable}

{
  Suppose such an equilibrium exists. We construct a profitable
  deviation in either selected order, using the standalone optimality
  established above.

  {\emph{Order $\AB$.}} Bob can post a price slightly above
  $b-\utilityMechanism{\BobMechanism}{b}$, still below $b$.
  This exceeds his original expected payment after Alice's failure. If Alice
  remains first, her effective value stays below $a$, so standalone
  optimality prevents her allocation from increasing. Bob then earns
  more after each failure, and failures occur at least as often.
  If Bob becomes first,
  the buyer buys his good surely. Thus the deviation is profitable
  in either order, as formalized in
  \cref{lem:nash-bob-posted-price-deviation}.

  {\emph{Order $\BA$.}} Alice replaces her menu by two options. She reduces
  and discounts the low option bought after Bob succeeds and replaces
  the high option bought after his
  failure by a sure allocation at a higher payment. The parameters
  balance two requirements: the higher payment must offset the loss
  on the low option, and an available $\BA$ plan must beat both
  skipping Alice and taking her low option first.
  If $\BA$ remains selected, Bob's effective value is below $b$;
  his standalone optimality prevents his allocation from increasing,
  making Alice's continuation with the higher payment at least as likely as
  before. If $\AB$ is selected, the strict comparisons force purchase
  of Alice's high option. Her revenue therefore increases in either
  order.
}

The proof of \cref{thm:introduction-simultaneous} follows almost immediately. All that is left is to justify our assumption that $a \geq b$.

\phantomsection
\label{proof:nash-equilibrium}

\begin{proof}[Proof of \cref{thm:introduction-simultaneous}]
  Let $(\AliceMechanism,\BobMechanism)$ be a pure Nash equilibrium.
  For $a\geq b$, \cref{lem:nash-a-at-least-b-REVENUE-THRESHOLD-EQUIVALENCE,lem:nash-a-at-least-b-NO-JOINT-REVENUE-IMPROVEMENT}
  imply that
  \[
    \revDuo{A}{\AliceMechanism}{\BobMechanism}={\residualValue{a}{b}},
    \qquad \revDuo{B}{\AliceMechanism}{\BobMechanism}=0.
  \]
  The revenue-bound proofs use only utility maximization, so
  interchanging the sellers proves the characterization when $a<b$.
\end{proof}

\section*{AI Methodology}
\addcontentsline{toc}{section}{AI Methodology}
\markright{AI Methodology}

{The authors developed {the research questions, the model, and the main proof strategy.}
\aiedit{They also developed}
{most of the} proof ideas for the main results
(\cref{thm:introduction-variable-order-rephrased,thm:introduction-simultaneous,thm:introduction-aligned-rephrased,thm:introduction-misaligned-rephrased}), while two technical claims, \cref{lemma:cap} and \cref{lem:nash-alice-threshold-revenue}, were proven by OpenAI's Codex (GPT-6 Astra). 
AI assisted with refining proof arguments, reviewing the paper, improving the presentation, and creating figures.
Codex also contributed in proving specified claims using routine arguments,
{\cref{obs:buyer-tie-breaks-for-alice,obs:tie-breaking-revenue-comparison,obs:uniqe-transfer-and-alloc-rules}}, and solving an ODE in \cref{lemma:solve-ode,lemma:zeta:lb,lemma:kappa:lb}.
The authors verified the correctness %
of all content, %
and take full responsibility for the content of the paper.}

\bibliographystyle{alpha}
\bibliography{refs}

\begin{thebibliography}{LOBR15}

\bibitem[AU05]{arribas2005nash}
Iv{\'a}n Arribas and Amparo Urbano.
\newblock {Nash} equilibria in a model of multiproduct price competition: An
  assignment problem.
\newblock {\em Journal of Mathematical Economics}, 41(3):351--385, 2005.

\bibitem[BGN22]{BabaioffGN17}
Moshe Babaioff, Yannai~A. Gonczarowski, and Noam Nisan.
\newblock The menu-size complexity of revenue approximation.
\newblock {\em Games and Economic Behavior}, 134:281--307, 2022.

\bibitem[BLN13]{babaioff2013bertrand}
Moshe Babaioff, Brendan Lucier, and Noam Nisan.
\newblock {Bertrand} networks.
\newblock In {\em Proceedings of the 14th {ACM} Conference on Electronic
  Commerce ({EC})}, pages 33--34. Association for Computing Machinery, 2013.
\newblock Full version:
  \href{https://arxiv.org/abs/1304.6806}{arXiv:1304.6806}.

\bibitem[BNPL14]{babaioff2014price}
Moshe Babaioff, Noam Nisan, and Renato Paes~Leme.
\newblock Price competition in online combinatorial markets.
\newblock In {\em Proceedings of the 23rd International Conference on World
  Wide Web ({WWW})}, pages 711--722. Association for Computing Machinery, 2014.

\bibitem[BS99]{burguet1999imperfect}
Roberto Burguet and J{\'o}zsef S{\'a}kovics.
\newblock Imperfect competition in auction designs.
\newblock {\em International Economic Review}, 40(1):231--247, 1999.

\bibitem[CBL24]{cai2024bundling}
Linda Cai, Moshe Babaioff, and Brendan Lucier.
\newblock Bundling in oligopoly: Revenue maximization with single-item
  competitors.
\newblock In {\em Proceedings of the 25th {ACM} Conference on Economics and
  Computation ({EC})}, page 465. Association for Computing Machinery, 2024.
\newblock Full version:
  \href{https://arxiv.org/abs/2406.13835}{arXiv:2406.13835}.

\bibitem[CN09]{chawla2009price}
Shuchi Chawla and Feng Niu.
\newblock The price of anarchy in {Bertrand} games.
\newblock In {\em Proceedings of the 10th {ACM} Conference on Electronic
  Commerce ({EC})}, pages 305--314. Association for Computing Machinery, 2009.

\bibitem[CR89]{champsaur1989multiproduct}
Paul Champsaur and Jean-Charles Rochet.
\newblock Multiproduct duopolists.
\newblock {\em Econometrica}, 57(3):533--557, 1989.

\bibitem[CR08]{chawla2008bertrand}
Shuchi Chawla and Tim Roughgarden.
\newblock {Bertrand} competition in networks.
\newblock In Burkhard Monien and Ulf-Peter Schroeder, editors, {\em Algorithmic
  Game Theory: First International Symposium ({SAGT} 2008)}, volume 4997 of
  {\em Lecture Notes in Computer Science}, pages 70--82. Springer, 2008.

\bibitem[CZ17]{CaiZ17}
Yang Cai and Mingfei Zhao.
\newblock Simple mechanisms for subadditive buyers via duality.
\newblock In {\em Proceedings of the 49th Annual {ACM} {SIGACT} Symposium on
  Theory of Computing ({STOC})}, pages 170--183. Association for Computing
  Machinery, 2017.

\bibitem[Dam12]{damianov2012seller}
Damian~S. Damianov.
\newblock Seller competition by mechanism design.
\newblock {\em Economic Theory}, 51(1):105--137, 2012.

\bibitem[DDT17]{DaskalakisDT17}
Constantinos Daskalakis, Alan Deckelbaum, and Christos Tzamos.
\newblock Strong duality for a multiple-good monopolist.
\newblock {\em Econometrica}, 85(3):735--767, 2017.

\bibitem[GILL24]{gonczarowski2024revenue}
Yannai~A. Gonczarowski, Nicole Immorlica, Yingkai Li, and Brendan Lucier.
\newblock Revenue maximization for buyers with costly participation.
\newblock In {\em Proceedings of the 2024 Annual {ACM-SIAM} Symposium on
  Discrete Algorithms ({SODA})}, pages 41--73. Society for Industrial and
  Applied Mathematics, 2024.

\bibitem[GT79]{gabszewicz1979price}
Jean~Jaskold Gabszewicz and Jacques-Fran{\c{c}}ois Thisse.
\newblock Price competition, quality and income disparities.
\newblock {\em Journal of Economic Theory}, 20(3):340--359, 1979.

\bibitem[HN19]{HartN13}
Sergiu Hart and Noam Nisan.
\newblock Selling multiple correlated goods: Revenue maximization and menu-size
  complexity.
\newblock {\em Journal of Economic Theory}, 183:991--1029, 2019.

\bibitem[JR01]{jehle2001advanced}
Geoffrey~A. Jehle and Philip~J. Reny.
\newblock {\em Advanced Microeconomic Theory}.
\newblock Addison-Wesley, Boston, MA, second edition, 2001.

\bibitem[LOBR15]{lev2015pricing}
Omer Lev, Joel Oren, Craig Boutilier, and Jeffrey~S. Rosenschein.
\newblock The pricing war continues: On competitive multi-item pricing.
\newblock In {\em Proceedings of the 29th {AAAI} Conference on Artificial
  Intelligence}, pages 972--978. AAAI Press, 2015.

\bibitem[LS25]{lucier2025approximately}
Brendan Lucier and Raghuvansh~Raj Saxena.
\newblock Approximately optimal mechanism design for competing sellers.
\newblock In {\em Proceedings of the 26th {ACM} Conference on Economics and
  Computation ({EC})}, page 477. Association for Computing Machinery, 2025.
\newblock Full version:
  \href{https://arxiv.org/abs/2505.19453}{arXiv:2505.19453}.

\bibitem[McA93]{mcafee1993mechanism}
R.~Preston McAfee.
\newblock Mechanism design by competing sellers.
\newblock {\em Econometrica}, 61(6):1281--1312, 1993.

\bibitem[MO01]{moscarini2001price}
Giuseppe Moscarini and Marco Ottaviani.
\newblock Price competition for an informed buyer.
\newblock {\em Journal of Economic Theory}, 101(2):457--493, 2001.

\bibitem[MR78]{mussa1978monopoly}
Michael Mussa and Sherwin Rosen.
\newblock Monopoly and product quality.
\newblock {\em Journal of Economic Theory}, 18(2):301--317, 1978.

\bibitem[MS03]{martimort2003contractual}
David Martimort and Lars~A. Stole.
\newblock Contractual externalities and common agency equilibria.
\newblock {\em Advances in Theoretical Economics}, 3(1), 2003.
\newblock Article 4.

\bibitem[MV07]{ManelliV07}
Alejandro~M. Manelli and Daniel~R. Vincent.
\newblock Multidimensional mechanism design: Revenue maximization and the
  multiple-good monopoly.
\newblock {\em Journal of Economic Theory}, 137(1):153--185, 2007.

\bibitem[Mye81]{Mye81}
Roger~B. Myerson.
\newblock Optimal auction design.
\newblock {\em Mathematics of Operations Research}, 6(1):58--73, 1981.

\bibitem[PS97]{peters1997competition}
Michael Peters and Sergei Severinov.
\newblock Competition among sellers who offer auctions instead of prices.
\newblock {\em Journal of Economic Theory}, 75(1):141--179, 1997.

\bibitem[RS02]{rochet2002nonlinear}
Jean-Charles Rochet and Lars~A. Stole.
\newblock Nonlinear pricing with random participation.
\newblock {\em The Review of Economic Studies}, 69(1):277--311, 2002.

\bibitem[SS82]{shaked1982relaxing}
Avner Shaked and John Sutton.
\newblock Relaxing price competition through product differentiation.
\newblock {\em The Review of Economic Studies}, 49(1):3--13, 1982.

\bibitem[Sto95]{stole1995nonlinear}
Lars~A. Stole.
\newblock Nonlinear pricing and oligopoly.
\newblock {\em Journal of Economics \& Management Strategy}, 4(4):529--562,
  1995.

\bibitem[Tha04]{Thanassoulis04}
John Thanassoulis.
\newblock Haggling over substitutes.
\newblock {\em Journal of Economic Theory}, 117(2):217--245, 2004.

\bibitem[TUW97]{tauman1997multiproduct}
Yair Tauman, Amparo Urbano, and Junichi Watanabe.
\newblock A model of multiproduct price competition.
\newblock {\em Journal of Economic Theory}, 77(2):377--401, 1997.

\bibitem[Vir10]{virag2010competing}
G{\'a}bor Vir{\'a}g.
\newblock Competing auctions: Finite markets and convergence.
\newblock {\em Theoretical Economics}, 5(2):241--274, 2010.

\bibitem[Viv99]{vives1999oligopoly}
Xavier Vives.
\newblock {\em Oligopoly Pricing: Old Ideas and New Tools}.
\newblock MIT Press, Cambridge, MA, 1999.

\bibitem[vS34]{stackelberg1934marktform}
Heinrich von Stackelberg.
\newblock {\em Marktform und Gleichgewicht}.
\newblock Julius Springer, Wien und Berlin, 1934.

\end{thebibliography}

\appendix
\section{Missing Proofs: Model and Preliminaries}
\label{sec:prelim-app}

The following observation establishes that the buyer's optimal continuation
messages can be chosen to depend only on whether he received the first
seller's good, independently of the transfer to that seller.

\begin{observation}[Transfer-Independent Continuation]
    \label{obs:independent-of-transfer}
    Suppose Alice and Bob choose the mechanisms 
    $\mathcal{M}_\Alice$ and $\mathcal{M}_\Bob$, respectively.
    For convenience, let $v_\Alice=a$ and $v_\Bob=b$, and fix the seller
    $i\in\{\Alice,\Bob\}$ approached first.
    For an outcome $(q_i,t_i)$ of the first interaction, let
    $C_{-i}(q_i,t_i)$ denote the set of messages to the remaining
    seller that maximize the buyer's expected utility after that outcome,
    \begin{align*}
        C_{-i}(q_i,t_i)
        = \operatorname*{arg\,max}_{\sigma\in\Sigma}
        \mathbb{E}_{(q_{-i},t_{-i})\sim\mathcal{M}_{-i}(\sigma)}
        \left[\max\{q_i \cdot v_{i}, q_{-i} \cdot v_{-i}\}\,-t_i - t_{-i}\right].
    \end{align*}
    Then, for every $q_i\in\{0,1\}$ and
    $t_i,t_i'\geq 0$,
    $C_{-i}(q_i,t_i)=C_{-i}(q_i,t_i')$.
\end{observation}

\begin{proof}
    Fix $q_i\in\{0,1\}$. For every message $\sigma\in\Sigma$,
    the buyer's expected utility after outcome $(q_i,t_i)$ is
    \begin{align*}
        &\mathbb{E}_{(q_{-i},t_{-i})\sim\mathcal{M}_{-i}(\sigma)}
        \left[\max\{q_i \cdot v_i,q_{-i} \cdot v_{-i}\}-t_i-t_{-i}\right]
        = -t_i
        +\mathbb{E}_{(q_{-i},t_{-i})\sim\mathcal{M}_{-i}(\sigma)}
        \left[\max\{q_i \cdot v_i,q_{-i} \cdot v_{-i}\}-t_{-i}\right].
    \end{align*}
    By the definition of a mechanism, the distribution
    $\mathcal{M}_{-i}(\sigma)$ depends only on $\sigma$, so the
    expectation on the right-hand side does not depend on $t_i$.
    Moreover, the term $-t_i$ is constant across all messages.
    Adding this constant does not change the set of maximizers,
    and hence $C_{-i}(q_i,t_i)=C_{-i}(q_i,t'_i)$.
\end{proof}

{For the fixed buyer values $a,b$, we make the dependence of
revenues and buyer utility on the action plan explicit. Given mechanisms
$\AliceMechanism,\BobMechanism$ and an action plan $\pi$, write
\begin{align*}
    \revTri{i}{\AliceMechanism}{\BobMechanism}{\pi}
    &:=\mathbb{E}_{\AliceMechanism,\BobMechanism,\pi}[t_i],
    \qquad i\in\{\Alice,\Bob\},\\
    \utilityTri{\AliceMechanism}{\BobMechanism}{\pi}
    &:=\mathbb{E}_{\AliceMechanism,\BobMechanism,\pi}
        [\max\{a \cdot q_\Alice,b \cdot q_\Bob\}-t_\Alice-t_\Bob]
\end{align*}
for seller $i$'s expected revenue and the buyer's expected utility,
respectively. The expectations are over the outcomes induced by the
mechanisms and the buyer's plan. Sending the
null message gives a transfer of zero. Under a strategy profile
$(\AliceMechanism,\mathbb{M},\hat{\mathbb{M}})$, Bob's mechanism is
$\BobMechanism=\mathbb{M}(\AliceMechanism)$ and the buyer's plan is
$\pi=\hat{\mathbb{M}}(\AliceMechanism,\BobMechanism)$.
The two-argument revenue notation in \cref{sec:prelim} suppresses
$\pi$ when the buyer's strategy is fixed.}

\begin{observation}[Existence and Tie-Breaking for Bob and the Buyer]
    \label{obs:buyer-tie-breaks-for-alice}
    {Suppose the buyer has values $a$ and $b$.}
    \begin{enumerate}[label=(\roman*)]
        \item For every mechanism pair, there exists a buyer
        action plan successively maximizing his expected utility,
        Bob's expected revenue, Alice's expected revenue, the allocation
        probability from Bob, and the allocation probability from Alice,
        finally preferring $\AB$ to $\BA$.

        \item {Fix any buyer strategy selecting plans with the
        priorities in (i) after every mechanism pair.} For every Alice
        mechanism, Bob has a best response maximizing his expected
        revenue, and Alice's expected revenue attains a maximum
        among these best responses.

    \end{enumerate}
    {Both conclusions hold in the aligned and misaligned
    models as well, with buyer plans restricted to the prescribed order.}\footnote{{The final tie-break in favor of visiting Alice first is irrelevant when the order is fixed.}}
\end{observation}

\begin{proof}
    {Let $\Pi^\ast(\AliceMechanism,\BobMechanism)$ denote the buyer plans
    maximizing $\utilityTri{\AliceMechanism}{\BobMechanism}{\pi}$.
    {Here $\Pi^\ast$ consists of ex-ante utility maximizers,
    with unrestricted choices on zero-probability branches;
    optimal completion is performed afterwards.}
    Write $X_i(\AliceMechanism,\BobMechanism,\pi)$
    for the overall probability of receiving seller $i$'s good.}

    We prove (i) and (ii) by establishing attainment of the
    successive maximizations.
    As in \cref{obs:independent-of-transfer}, a sunk transfer does not
    affect the buyer's continuation comparisons. It also contributes
    only a constant to either seller's revenue. Thus continuation
    messages can be selected independently of that transfer when
    successively maximizing
    \[
    (\utilityTri{\AliceMechanism}{\BobMechanism}{\pi},\revTri{\Bob}{\AliceMechanism}{\BobMechanism}{\pi},
      \revTri{\Alice}{\AliceMechanism}{\BobMechanism}{\pi},X_\Bob(\AliceMechanism,\BobMechanism,\pi),X_\Alice(\AliceMechanism,\BobMechanism,\pi)).
    \]
    This selection preserves the interaction order, so the final
    preference for $\AB$ can be applied afterwards.
    Put $v_\Alice=a$ and $v_\Bob=b$. Owning seller $i$'s good can
    increase the buyer's optimal continuation utility by at most
    $v_i$. Consequently, a message with expected payment greater
    than $v_i$ is strictly worse than the null message, whether seller
    $i$ is approached first or second. For a mechanism $\mathcal M_i$,
    consider the auxiliary menu
    \[
        K_i=\left\{(x,p)\in[0,1]\times[0,v_i]:
        \exists\mu\in\operatorname{im}(\mathcal M_i),\quad
        \mathbb E_\mu[q]=x,\quad \mathbb E_\mu[t]\leq p\right\}.
    \]
    Here compactness of outcome distributions is understood in the
    topology of weak convergence. Expected allocation is continuous
    and expected nonnegative payment is lower semicontinuous in this
    topology.
    {Indeed, the epigraph
    \[
        E_i=\left\{(\mu,p)\in\operatorname{im}(\mathcal M_i)\times[0,v_i]:
        \mathbb E_\mu[t]\leq p\right\}
    \]
    is closed in the compact product by lower semicontinuity.
    Its image under the continuous map
    $(\mu,p)\mapsto(\mathbb E_\mu[q],p)$ is exactly $K_i$,
    so $K_i$ is compact.}
    An auxiliary option with payment strictly above its witness's
    expected payment is strictly dominated by that witness, since
    they have the same allocation probability. Such an option cannot
    be used with positive probability in a utility-maximizing plan.
    Thus these menus preserve optimal buyer utility and the seller
    revenues and allocations selected by the stated priorities.
    The replacements also preserve each plan's interaction order,
    and hence preserve the final order comparison.
    {After selecting an optimal plan, we complete its
    zero-probability branches optimally within these compact menus,
    applying the stated tie-breaking. Such choices exist by compactness
    and leave expected payoffs and overall allocation probabilities
    unchanged.}

    For a fixed mechanism pair, transfer-independent plans range over
    a finite disjoint union of compact products of $K_\Alice$ and $K_\Bob$,
    with the interaction order as a discrete coordinate.
    Utilities, revenues, and allocation probabilities are continuous
    functions of the selected allocation probabilities and expected
    payments. The indicator of order $\AB$ is also continuous on
    this space. Successively maximizing these continuous functions
    leaves a nonempty compact set at each step, including the final
    order comparison. This proves (i). The final comparison does not
    change the utility, revenues, or allocation probabilities selected
    at the preceding steps.
    {Fix any strategy $\hat{\mathbb{M}}^\ast$
    selecting plans with these priorities at every mechanism pair.}

    Fix now an arbitrary Alice mechanism $\AliceMechanism$. To optimize Bob's
    revenue and then Alice's over Bob mechanisms and utility-maximizing
    buyer plans, it suffices to consider Bob menus with at most two
    non-null options. Indeed, a transfer-independent plan uses at most
    two Bob messages. {Messages used with positive probability
    have expected payments at most $b$,
    {and the same bound holds for any unused Bob continuation
    after optimal completion}.}
    Retaining these messages and the null message
    preserves the plan's feasibility and utility optimality, since
    it only removes alternatives. Replace the retained messages by
    lotteries with the same allocation probabilities and deterministic
    payments equal to their expectations. This preserves the optimal
    utility and the payoffs of the chosen plan. For any original plan
    with transfer-dependent continuations, choosing optimal
    continuations independently of the transfer first weakly improves
    the pair $(\revTri{\Bob}{\AliceMechanism}{\BobMechanism}{\pi},\revTri{\Alice}{\AliceMechanism}{\BobMechanism}{\pi})$
    in lexicographic order.

    These two Bob options range over the compact parameter space
    $([0,1]\times[0,b])^2$, with the null option included separately.
    Buyer plans range over a fixed finite union of products of
    $K_\Alice$ and the index set $\{0,1,2\}$ for Bob's null and two
    non-null options. Their utilities and revenues are
    continuous in the menu parameters and the plan. The buyer's
    maximum utility is therefore continuous in the menu parameters,
    so the graph of utility-maximizing plans is compact. Maximizing
    Bob's revenue and then Alice's revenue on this graph proves that
    both of the following maxima are attained over the full mechanism
    class:
    \begin{align*}
        W(\AliceMechanism)&=\max_{\substack{\BobMechanism\in\AllMechanisms\\
                            \pi\in\Pi^\ast(\AliceMechanism,\BobMechanism)}} \revTri{\Bob}{\AliceMechanism}{\BobMechanism}{\pi},\\
        T(\AliceMechanism)&=\max_{\substack{\BobMechanism\in\AllMechanisms,\ \pi\in\Pi^\ast(\AliceMechanism,\BobMechanism)\\
                            \revTri{\Bob}{\AliceMechanism}{\BobMechanism}{\pi}=W(\AliceMechanism)}} \revTri{\Alice}{\AliceMechanism}{\BobMechanism}{\pi}.
    \end{align*}
    At a mechanism attaining $(W(\AliceMechanism),T(\AliceMechanism))$, the buyer strategy
    $\hat{\mathbb{M}}^\ast$ must also give these revenues: its
    priorities cannot increase either coordinate beyond the respective
    maximum. Choose such a mechanism as $\mathbb{M}^\ast(\AliceMechanism)$.
    Thus $\mathbb{M}^\ast$ best-responds and breaks ties in Alice's
    favor after every $\AliceMechanism$ under $\hat{\mathbb{M}}^\ast$: Bob's maximum
    revenue is $W(\AliceMechanism)$, and Alice's maximum revenue over his best
    responses is $T(\AliceMechanism)$. Since the
    buyer strategy was arbitrary subject to the stated priorities,
    this proves (ii).

    {For either fixed-order model, restricting buyer plans to
    the prescribed order retains the corresponding compact component
    of each plan space. The final preference between orders is then
    vacuous, and the same menu reduction and attainment arguments
    prove both (i) and (ii).}
\end{proof}

\begin{observation}[Revenue Comparison under Favorable Tie-Breaking]
    \label{obs:tie-breaking-revenue-comparison}
    {Let $\hat{\mathbb{M}}$ be any buyer strategy
    maximizing his expected utility after every mechanism pair,
    and let $\mathbb{M}$ be any Bob strategy best-responding after
    every Alice mechanism given $\hat{\mathbb{M}}$ (no tie-breaking
    restrictions are imposed). Then there exist a buyer strategy
    $\hat{\mathbb{M}}^\ast$ 
    and a Bob strategy $\mathbb{M}^\ast$
    satisfying the priorities in
    \cref{obs:buyer-tie-breaks-for-alice} such that for every $\AliceMechanism$,
    \[
        \revTri{A}{\AliceMechanism}{\BobMechanism^\ast}{\pi^\ast}
        \geq\revTri{A}{\AliceMechanism}{\BobMechanism}{\pi},
    \]
    where $\BobMechanism=\mathbb{M}(\AliceMechanism)$,
    $\pi=\hat{\mathbb{M}}(\AliceMechanism,\BobMechanism)$,
    $\BobMechanism^\ast=\mathbb{M}^\ast(\AliceMechanism)$, and
    $\pi^\ast=\hat{\mathbb{M}}^\ast(\AliceMechanism,\BobMechanism^\ast)$.}
    The same conclusion holds in the aligned and misaligned
    models, with buyer plans and strategies restricted to the
    prescribed order.
\end{observation}

\begin{proof}
    {Use the notation $\Pi^\ast,W$, and $T$ from the
    proof of \cref{obs:buyer-tie-breaks-for-alice}. In particular,
    $W(\AliceMechanism)$ is Bob's maximum revenue over all mechanisms and
    utility-maximizing buyer plans, and $T(\AliceMechanism)$ is Alice's maximum
    revenue among these outcomes giving Bob $W(\AliceMechanism)$.}

    {Fix buyer and Bob strategies
    $\hat{\mathbb{M}}$ and $\mathbb{M}$ satisfying the stated
    best-response conditions, and fix any Alice mechanism $\AliceMechanism$.
    We first show that Bob's revenue after $\AliceMechanism$ is $W(\AliceMechanism)$, which will
    bound Alice's revenue by $T(\AliceMechanism)$.} Fix any $\BobMechanism$ and
    $\pi\in\Pi^\ast(\AliceMechanism,\BobMechanism)$. For $\varepsilon\in(0,1)$, let
    $\BobMechanism_\varepsilon$ multiply every realized payment in $\BobMechanism$ by
    $1-\varepsilon$, leaving allocations unchanged. This is an
    admissible mechanism, since this continuous transformation
    preserves compactness of its image and the null option.
    Write $\pi_\varepsilon=\hat{\mathbb{M}}(\AliceMechanism,\BobMechanism_\varepsilon)$.
    If a plan depends on observed transfers
    {(as may occur in general subgame-perfect equilibria before
    applying \cref{obs:independent-of-transfer})}, identify its histories
    in the two mechanisms via the invertible rescaling
    $t_\Bob\mapsto(1-\varepsilon) \cdot t_\Bob$.
    For a corresponding action plan $\rho$ with finite expected
    transfers,
    \[
        \utilityTri{\AliceMechanism}{\BobMechanism_\varepsilon}{\rho}
        =\utilityTri{\AliceMechanism}{\BobMechanism}{\rho}+\varepsilon \cdot \revTri{\Bob}{\AliceMechanism}{\BobMechanism}{\rho}.
    \]
    Both optimizing plans being compared have finite expected
    transfers, since the buyer can decline both offers and
    $1-\varepsilon>0$.
    Optimality of $\pi_\varepsilon$ for the discounted mechanism and
    of $\pi$ for the original mechanism gives
    \begin{align*}
        \utilityTri{\AliceMechanism}{\BobMechanism}{\pi_\varepsilon}
            +\varepsilon \cdot \revTri{\Bob}{\AliceMechanism}{\BobMechanism}{\pi_\varepsilon}
        &\geq\utilityTri{\AliceMechanism}{\BobMechanism}{\pi}+\varepsilon \cdot \revTri{\Bob}{\AliceMechanism}{\BobMechanism}{\pi},\\
        \utilityTri{\AliceMechanism}{\BobMechanism}{\pi_\varepsilon}&\leq\utilityTri{\AliceMechanism}{\BobMechanism}{\pi}.
    \end{align*}
    Hence Bob's revenue from this deviation is at least
    $(1-\varepsilon) \cdot \revTri{\Bob}{\AliceMechanism}{\BobMechanism}{\pi}$. Since the original strategy
    $\mathbb{M}$ best-responds after every $\AliceMechanism$, letting
    $\varepsilon\downarrow0$ and maximizing over $(\BobMechanism,\pi)$ shows that
    its revenue is at least $W(\AliceMechanism)$. The reverse inequality follows
    from the definition of $W(\AliceMechanism)$, proving equality.

    {For each Alice mechanism $\AliceMechanism$, the original strategies
    therefore give Bob $W(\AliceMechanism)$, so Alice's revenue is at most $T(\AliceMechanism)$
    by its definition. The strategies
    $\hat{\mathbb{M}}^\ast$ and $\mathbb{M}^\ast$ constructed in the
    proof of \cref{obs:buyer-tie-breaks-for-alice} give her exactly
    $T(\AliceMechanism)$ for every $\AliceMechanism$, proving the
    comparison simultaneously for all Alice mechanisms.}

    {For either fixed-order model, define $\Pi^\ast,W$, and $T$
    using only buyer plans with the prescribed order. Discounting
    Bob's payments preserves this order and the correspondence
    between feasible plans, so the same inequalities prove the
    comparison. The fixed-order extension of
    \cref{obs:buyer-tie-breaks-for-alice} supplies the required
    buyer and Bob strategies.}
\end{proof}

Recall from \cref{sec:prelim} that a mechanism
$\AliceMechanism$ is \emph{optimal} for Alice if there exists a Bob
strategy $\mathbb{M}$ such that $(\AliceMechanism,\mathbb{M})$ is a
Stackelberg equilibrium maximizing her expected revenue over all
Stackelberg equilibria of the induced game, under the prescribed
buyer behavior and Bob tie-breaking rule.
We call a mechanism $\AliceMechanism$
\emph{optimal over all subgame-perfect equilibria} if there exist a Bob
strategy $\mathbb{M}$ and a buyer strategy $\hat{\mathbb{M}}$ such that
$(\AliceMechanism,\mathbb{M},\hat{\mathbb{M}})$ is a subgame-perfect
equilibrium maximizing Alice's expected revenue among all
subgame-perfect equilibria, without imposing tie-breaking restrictions.

\begin{corollary}[Optimal Mechanisms and Tie-Breaking]
    \label{cor:optimal-mechanisms-tie-breaking}
    Fix buyer values $a,b>0$.
    A mechanism is optimal under
    the prescribed buyer and Bob tie-breaking rules if and only if
    it is optimal over all subgame-perfect equilibria.
    Alice obtains the same expected revenue in both cases.
    The same conclusion holds in the aligned and
    misaligned models.
\end{corollary}

\begin{proof}
    By \cref{obs:buyer-tie-breaks-for-alice},
    the tie-breaking rules can always be imposed 
    and by \cref{obs:tie-breaking-revenue-comparison}
    they cannot decrease Alice's revenue,
    proving the forward implication.
    Conversely, if imposing the tie-breaking rules
    after some mechanism
    improved upon an Alice-optimal subgame-perfect equilibrium,
    using them only there and letting Alice choose that mechanism
    would yield a more profitable equilibrium.
    Hence Alice's original mechanism is optimal under the prescribed
    rules, with the same revenue. Both observations also cover the
    fixed-order models.
\end{proof}

{We now make the allocation notation depend explicitly on the
buyer's plan. Given mechanisms $\AliceMechanism,\BobMechanism$ and
$\pi=(i,\sigma_i,\sigma^W_{-i},\sigma^L_{-i})$, write
\begin{align*}
    \allocTri{i}{\AliceMechanism}{\BobMechanism}{\pi}
        &:=\mathbb E_{(q,t)\sim\mathcal M_i(\sigma_i)}[q],\\
    \allocTriCond{S}{-i}{\AliceMechanism}{\BobMechanism}{\pi}
        &:=\mathbb E_{(q,t)\sim\mathcal M_{-i}(\sigma^S_{-i})}[q],
        \qquad S\in\{W,L\}.
\end{align*}
These are the first seller's allocation probability and the second
seller's allocation probabilities after winning or losing the first
good. They refer to the messages specified by the plan, so they remain
defined even on an unused branch.
The following observation establishes uniqueness of the selected
revenues, interaction order, and allocations on every reached branch.}

\begin{observation}[Unique Revenues, Allocations, and Interaction Order]
    \label{obs:uniqe-transfer-and-alloc-rules}
    {Fix mechanisms $\AliceMechanism,\BobMechanism$, and let
    $\hat{\mathbb{M}},\hat{\mathbb{M}}'$ be two buyer strategies
    satisfying all priorities in \cref{sec:prelim}, including the final
    preference for $\AB$. Their selected plans
    $\pi=\hat{\mathbb{M}}(\AliceMechanism,\BobMechanism)$ and
    $\pi'=\hat{\mathbb{M}}'(\AliceMechanism,\BobMechanism)$
    have the same expected buyer utility, expected revenue to each
    seller, interaction order, and allocation probability from the first
    seller. The second seller's conditional allocation probabilities also
    agree on every branch reached with positive probability.
    If $i$ is the common first seller, write
    $x := \allocTri{i}{\AliceMechanism}{\BobMechanism}{\pi}
      =\allocTri{i}{\AliceMechanism}{\BobMechanism}{\pi'}$.
    Then, for $j\in\{\Alice,\Bob\}$,
    \[
        \revTri{j}{\AliceMechanism}{\BobMechanism}{\pi}
            =\revTri{j}{\AliceMechanism}{\BobMechanism}{\pi'},
        \qquad
        \begin{aligned}
        \allocTriCond{W}{-i}{\AliceMechanism}{\BobMechanism}{\pi}
            &=\allocTriCond{W}{-i}{\AliceMechanism}{\BobMechanism}{\pi'}
            &&\text{if }x>0,\\
        \allocTriCond{L}{-i}{\AliceMechanism}{\BobMechanism}{\pi}
            &=\allocTriCond{L}{-i}{\AliceMechanism}{\BobMechanism}{\pi'}
            &&\text{if }x<1.
        \end{aligned}
    \]}
\end{observation}

\begin{proof}
    {For any feasible plan $\rho$, write $X_j(\rho)$ for seller $j$'s
    overall allocation probability, obtained by averaging
    over the two branches when $j$ is second.
    The buyer's priorities mean that $\pi$ and $\pi'$ both
    lexicographically maximize
    \[
        \bigl(\utilityTri{\AliceMechanism}{\BobMechanism}{\rho},
              \revTri{\Bob}{\AliceMechanism}{\BobMechanism}{\rho},
              \revTri{\Alice}{\AliceMechanism}{\BobMechanism}{\rho},
              X_\Bob(\rho),X_\Alice(\rho),
              \mathbf{1}_{\{\rho\text{ has order }\AB\}}\bigr).
    \]
    Their vectors must agree: otherwise, the first differing coordinate
    would give one plan priority over the other. This proves equality
    of utility, revenues, overall allocations, and order, and hence
    equality of the first seller's allocation $x$.

    To identify the second seller's allocations on reached branches,
    put $v_\Alice=a$, $v_\Bob=b$, and let $Y$ be that seller's common
    overall allocation probability. For either $\rho\in\{\pi,\pi'\}$,
    the buyer's expected gross value gives
    \[
        \utilityTri{\AliceMechanism}{\BobMechanism}{\rho}
        +\revTri{\Alice}{\AliceMechanism}{\BobMechanism}{\rho}
        +\revTri{\Bob}{\AliceMechanism}{\BobMechanism}{\rho}
        =v_i \cdot x+v_{-i} \cdot Y
          -\min\{a,b\} \cdot x \cdot \allocTriCond{W}{-i}{\AliceMechanism}{\BobMechanism}{\rho}.
    \]
    Since $a,b>0$ and all other quantities agree, the two plans have
    the same product
    $x \cdot \allocTriCond{W}{-i}{\AliceMechanism}{\BobMechanism}{\rho}$.
    Subtracting this product from $Y$ also gives equality of
    $(1-x) \cdot \allocTriCond{L}{-i}{\AliceMechanism}{\BobMechanism}{\rho}$.
    Dividing by the corresponding branch probability whenever it is
    positive proves the remaining claims.}
\end{proof}

The final order preference only selects among plans agreeing in the
first five coordinates. Adding it therefore leaves the selected utility,
revenues, and marginal allocations unchanged at every mechanism pair.
In particular, it leaves Bob's best responses and the sellers' equilibrium
revenues unchanged.
{It also uniquely determines the buyer's interaction order at every
mechanism pair.}
{For example, if Alice offers only the null option, moving her
interaction before Bob's preserves the first five coordinates, so the
buyer always selects $\AB$.}
{Recall that two mechanisms offered by the same seller are
\emph{essentially equivalent} if, against every mechanism of the
other seller, they induce the same buyer utility, each seller's expected
revenue, interaction order, first-seller allocation probability, and
second-seller conditional allocation probability on each branch reached
with positive probability under the stated buyer tie-breaking rules
(\cref{def:strategic-equivalence}).}

\taxationprinciple*

\begin{proof}
{Fix seller $i$, put $v_\Alice=a$ and $v_\Bob=b$, and let
\[
    K_i=\left\{(x,p)\in[0,1]\times[0,v_i]:
    \exists\mu\in\operatorname{im}(\mathcal M_i),\quad
    \mathbb E_\mu[q]=x,\quad \mathbb E_\mu[t]\leq p\right\}.
\]
As shown in the proof of \cref{obs:buyer-tie-breaks-for-alice},
$K_i$ is compact, and replacing $\mathcal M_i$ by this menu preserves
the successively maximized utility, revenues, and overall allocation
probabilities separately within each interaction order.}

{Let $C_i=\operatorname{conv}K_i$ and
$\bar x_i=\max\{x:(x,p)\in C_i\}$. Define
\[
    \Pmarg{\mathcal M_i}(x)=
    \begin{cases}
        \min\{p:(x,p)\in C_i\},&0\leq x\leq\bar x_i,\\
        \infty,&x>\bar x_i.
    \end{cases}
\]
The compactness of $C_i$ makes this minimum attained and the function
lower semicontinuous. It is convex, nonnegative, and zero at zero,
hence non-decreasing and continuous on $[0,\bar x_i]$, including
its endpoints. Thus it is proper, with a compact finite-price graph.
Implementing each option by a Bernoulli allocation and its deterministic
payment gives a mechanism with compact image and the null option.\footnote{
{Since $\Sigma$ is an uncountable standard Borel space, there is a
measurable surjection from $\Sigma$ onto any compact metrizable menu
containing the null outcome, mapping $\emptyset$ to that outcome.}}}

{Fix any opposing mechanism and either interaction order,
representing its options by its compact menu as above. By
\cref{obs:independent-of-transfer,obs:buyer-tie-breaks-for-alice},
we may use transfer-independent continuations. If the first seller is
$f$ and the second is $s$, write $(x,p)$ for the first option and
$(y^W,p^W),(y^L,p^L)$ for the continuation options. The buyer's utility is
\[
    v_f x-p+x\bigl((v_s-v_f)_+y^W-p^W\bigr)
    +(1-x)\bigl(v_s y^L-p^L\bigr).
\]
The sellers' revenues are $p$ and $xp^W+(1-x)p^L$, and their overall
allocation probabilities are $x$ and $xy^W+(1-x)y^L$.
Consequently, the vector of these five quantities, in the buyer's
priority order, is affine separately in each selected option.
Expanding convex combinations of options
{separately at each decision node (including the winning and losing
continuation nodes when $i$ is second), with the opposing seller's options
fixed,} therefore expresses every
vector attainable using $C_i$ as a convex combination of vectors
attainable using $K_i$, with the same interaction order. Since
$K_i\subseteq C_i$, their lexicographic maxima coincide.}

{An option strictly above the graph of
$\Pmarg{\mathcal M_i}$ cannot be selected on a reached branch:
the graph offers the same allocation more cheaply, strictly increasing
utility.
{After selecting an optimal plan, we complete its
zero-probability branches at seller $i$ optimally on this compact graph, applying
the stated tie-breaking. This leaves expected payoffs and overall
allocation probabilities unchanged.}
Restricting $C_i$
to this graph therefore preserves the same lexicographic maximum.
Applying the final preference for $\AB$ preserves the selected order
as well. The gross-value identity in the proof of
\cref{obs:uniqe-transfer-and-alloc-rules} then gives equality of the
second seller's conditional allocations on every reached branch.
The construction depends only on $\mathcal M_i$ and $v_i$, so it
establishes essential equivalence against every opposing mechanism.
Applying this replacement to either seller proves the observation.}
\end{proof}

\begin{corollary}[Taxation Principle for Fixed Orders]
    \label{cor:taxation-fixed-order}
    {For the fixed buyer values $a,b$, the proper pricing
    representation $\Pmarg{\mathcal M}$ constructed in the proof of
    \cref{obs:taxation} is also {essentially equivalent} to
    $\mathcal M$ in each fixed-order variant. Consequently,
    restricting both sellers to such pricing mechanisms preserves
    their best-response and
    revenue-optimization problems in either variant.}
\end{corollary}

\begin{proof}
    {The proof of \cref{obs:taxation} preserves, separately
    within each order, the successive maxima of buyer utility, Bob's
    revenue, Alice's revenue, Bob's allocation probability, and Alice's
    allocation probability. When the order is fixed, the replacement
    therefore preserves these quantities against every opposing
    mechanism. {By the same gross-value identity used in the proof
    of \cref{obs:uniqe-transfer-and-alloc-rules}, it also preserves the
    second seller's conditional allocation on every branch reached
    with positive probability, establishing essential equivalence in
    that variant.}
    Applying the replacement to either seller proves the final claim.}
\end{proof}

\myersonlemma*

{We omit the proof, which follows from Myerson's
characterization~\cite{Mye81}.}

\section{Missing Proofs from \cref{sec:stackelberg}}
\label{app:stackelberg}

\subsection{Missing Proofs from \cref{sec:stackelberg:buyer}}
\label{app:stackelberg:buyer}

\begin{proof}[Proof of \cref{lemma:buyer}]
For \cref{item:buyer:util}, we only show the first equation as the proof of the second equation is analogous. Suppose the buyer interacts with Alice first and picks the option $x_A$, and then interacts with Bob by picking the option $x_B^W$ if he receives Alice's item and picking the option $x_B^L$ if he does not. As $\residualValue{b}{a}$ is the extra value the buyer gets from receiving item $b$ given he already has item $a$, we get that his utility from the above interaction is: 
\[
a x_A - \AliceMechanism\paren*{ x_A } + x_A \cdot \paren*{ \residualValue{b}{a} x_B^W - \BobMechanism\paren*{ x_B^W } } + \paren*{ 1 - x_A } \cdot \paren*{ b x_B^L - \BobMechanism\paren*{ x_B^L } } .
\]
Observe that $x_B^W$ and $x_B^L$ only appear in the second and third terms respectively with a non-negative multiplier. Using this and the definition of $\utilityMechanism{ \mathcal{M} }{ v }$, we get:
\begin{align*}
\utilityAB{ \AliceMechanism }{ \BobMechanism } &= \max_{ x_A } \paren*{ a x_A - \AliceMechanism\paren*{ x_A } + x_A \cdot \max_{ x_B^W } \paren*{ \residualValue{b}{a} x_B^W - \BobMechanism\paren*{ x_B^W } } + \paren*{ 1 - x_A } \cdot \max_{ x_B^L } \paren*{ b x_B^L - \BobMechanism\paren*{ x_B^L } } } \\
&= \max_{ x_A } \paren*{ a x_A - \AliceMechanism\paren*{ x_A } + x_A \cdot \utilityMechanism{ \BobMechanism }{ \residualValue{b}{a} } + \paren*{ 1 - x_A } \cdot \utilityMechanism{ \BobMechanism }{ b } } .
\end{align*}
Rearranging and using $\effectiveValue{a}{\BobMechanism} = a - \aug_{ \BobMechanism }\paren*{ a, b }$ together with \cref{eq:aug}, we get:
\[
\utilityAB{ \AliceMechanism }{ \BobMechanism } = \max_{ x_A } \paren*{ \effectiveValue{a}{\BobMechanism} x_A - \AliceMechanism\paren*{ x_A } } + \utilityMechanism{ \BobMechanism }{ b } .
\]
Again using the definition of $\utilityMechanism{ \mathcal{M} }{ v }$, we get $\utilityAB{ \AliceMechanism }{ \BobMechanism } = \utilityMechanism{ \AliceMechanism }{ \effectiveValue{a}{\BobMechanism} } + \utilityMechanism{ \BobMechanism }{ b }$, as desired. For \cref{item:buyer:allocation}, we recall that the buyer in our duopolist setting tie-breaks in favor of maximizing Bob's and Alice's revenues, and then in favor of their allocation probabilities, in that order. As the tie-breaking in $\zmarg{ \mathcal{M} }^{+}\paren*{ v }$ is also in favor of larger values, this means that the values $x_B^W$ and $x_B^L$ chosen by the buyer will be $\zmarg{ \BobMechanism }^{+}\paren*{ \residualValue{b}{a} }$ and $\zmarg{ \BobMechanism }^{+}\paren*{ b }$ respectively. Then, the buyer will choose $x_A$ to maximize $\effectiveValue{a}{\BobMechanism} x_A - \AliceMechanism\paren*{ x_A }$. This value is $\zmarg{ \AliceMechanism }^{+}\paren*{ \effectiveValue{a}{\BobMechanism} }$ but the buyer would prefer picking the smallest maximizer $\zmarg{ \AliceMechanism }^-\paren*{ \effectiveValue{a}{\BobMechanism} }$ in case it gives Bob higher revenue (equivalently, larger allocation). Thus, \cref{item:buyer:allocation} follows. A similar proof works for the case when the buyer interacts with Bob first. Finally, \cref{item:buyer:payment} follows from \cref{item:buyer:allocation} and the observation that, when $\zmarg{ \BobMechanism }^{+}\paren*{ b } = \zmarg{ \BobMechanism }^{+}\paren*{ \residualValue{b}{a} }$, the payment to Bob does not depend on the probability of getting the item from Alice.
\end{proof}

\subsection{Missing Proofs from \cref{sec:stackelberg:bob}}
\label{app:stackelberg:bob}

\subsubsection{Missing Proofs from \cref{sec:stackelberg:bob:bob}}
\label{app:stackelberg:bob:bob}

\begin{lemma}
\label{lemma:sl-params}
We have $\utilityMechanism{ \AliceMechanism }{ \vthreshold_{ \AliceMechanism }(x) } = \utilityMechanism{ \AliceMechanism }{ a } - x \cdot \aug_{ \AliceMechanism }\paren*{ b, a }$ and $\max\paren*{ a - ax, a - bx } \leq \vthreshold_{ \AliceMechanism }(x) \leq a - x \cdot \aug_{ \AliceMechanism }\paren*{ b, a }$.
\end{lemma}
\begin{proof}
The first part is a consequence of the fact that utilities are continuous. For the second part, use the first part and note that
\begin{align*}
\utilityMechanism{ \AliceMechanism }{ \vthreshold_{ \AliceMechanism }(x) } &= \utilityMechanism{ \AliceMechanism }{ a } - x \cdot \aug_{ \AliceMechanism }\paren*{ b, a } \\
&\geq \paren*{ 1 - x } \cdot \utilityMechanism{ \AliceMechanism }{ a } + x \cdot \utilityMechanism{ \AliceMechanism }{ \max\paren*{ 0, a - b } }\tag{\cref{eq:aug}} \\
&\geq \utilityMechanism{ \AliceMechanism }{ \max\paren*{ a - ax, a - bx } } \tag{Convexity of utilities} .
\end{align*}
We are done by the fact that utilities are monotone. To finish, note that the upper bound follows from the fact that utilities are $1$-Lipschitz. Indeed, $v$ cannot be larger as for all $v > a - x \cdot \aug_{ \AliceMechanism }\paren*{ b, a }$, we have:
\[
\utilityMechanism{ \AliceMechanism }{ a } - \utilityMechanism{ \AliceMechanism }{ v } \leq a - v < x \cdot \aug_{ \AliceMechanism }\paren*{ b, a } .
\]
\end{proof}

\begin{proof}[Proof of \cref{lemma:sl}]
Define the values $\effectiveValue{a}{\SL_{x,p}} = a - \aug_{ \SL_{ x, p } }\paren*{ a, b }$ and $\effectiveValue{b}{\AliceMechanism} = b - \aug_{ \AliceMechanism }\paren*{ b, a }$. From \cref{item:buyer:util} of \cref{lemma:buyer}, we have:
\[
\utilityAB{ \AliceMechanism }{ \SL_{ x, p } } - \utilityBA{ \AliceMechanism }{ \SL_{ x, p } } = \utilityMechanism{ \SL_{ x, p } }{ b } - \utilityMechanism{ \SL_{ x, p } }{ \effectiveValue{b}{\AliceMechanism} } - \utilityMechanism{ \AliceMechanism }{ a } + \utilityMechanism{ \AliceMechanism }{ \effectiveValue{a}{\SL_{x,p}} } .
\]
Note that if $\utilityMechanism{ \AliceMechanism }{ a } = 0$, then the monotonicity of utility implies that the last two terms above vanish and that $\effectiveValue{b}{\AliceMechanism} = b$. From \cref{def:sl-params}, this means that $\p_{ \AliceMechanism }(x) = b$ and we are done as both sides of the equivalence are true. Similarly, if $x = 0$, the first two terms on the right hand side vanish and we have $\effectiveValue{a}{\SL_{x,p}} = a$. Using \cref{def:sl-params}, we have $\p_{ \AliceMechanism }(x) = b - \aug_{ \AliceMechanism }\paren*{ b, a }$ and both sides of the equivalence are again true.

Henceforth, we assume that $\utilityMechanism{ \AliceMechanism }{ a }, x > 0$. As $p \leq \effectiveValue{b}{\AliceMechanism} \leq b$, the first two terms simplify to $x \cdot \aug_{ \AliceMechanism }\paren*{ b, a }$. Similarly, as $p \geq \residualValue{b}{a}$, we have that $\utilityMechanism{ \SL_{ x, p } }{ \residualValue{b}{a} } = 0$. Putting this into \cref{eq:aug}, we get that $\effectiveValue{a}{\SL_{x,p}} = a - \utilityMechanism{ \SL_{ x, p } }{ b } = a - x \cdot \paren*{ b - p }$. Plugging these in, we get:
\begin{align*}
\utilityAB{ \AliceMechanism }{ \SL_{ x, p } } - \utilityBA{ \AliceMechanism }{ \SL_{ x, p } } &= x \cdot \aug_{ \AliceMechanism }\paren*{ b, a } - \utilityMechanism{ \AliceMechanism }{ a } + \utilityMechanism{ \AliceMechanism }{ a - x \cdot \paren*{ b - p } } \\
&= \utilityMechanism{ \AliceMechanism }{ a - x \cdot \paren*{ b - p } } - \utilityMechanism{ \AliceMechanism }{ \vthreshold_{ \AliceMechanism }(x) } \tag{\cref{lemma:sl-params}} .
\end{align*}
Now, as we assume that $\utilityMechanism{ \AliceMechanism }{ a } > 0$, our definition of $\vthreshold_{ \AliceMechanism }(x)$ in \cref{def:sl-params} implies that $v \leq \vthreshold_{ \AliceMechanism }(x)$ if and only if $\utilityMechanism{ \AliceMechanism }{ v } \leq \utilityMechanism{ \AliceMechanism }{ \vthreshold_{ \AliceMechanism }(x) }$. This means that:
\begin{align*}
\utilityAB{ \AliceMechanism }{ \SL_{ x, p } } \leq \utilityBA{ \AliceMechanism }{ \SL_{ x, p } } &\iff \utilityMechanism{ \AliceMechanism }{ a - x \cdot \paren*{ b - p } } \leq \utilityMechanism{ \AliceMechanism }{ \vthreshold_{ \AliceMechanism }(x) } \\
&\iff a - x \cdot \paren*{ b - p } \leq \vthreshold_{ \AliceMechanism }(x) \\
&\iff p \leq \p_{ \AliceMechanism }(x) \tag{As $x > 0$} .
\end{align*}
For the ``moreover'' part, note that if $x < 1$, then $\utilityMechanism{ \AliceMechanism }{ \vthreshold_{ \AliceMechanism }(x) } > 0$. This means that $v < \vthreshold_{ \AliceMechanism }(x)$ if and only if $\utilityMechanism{ \AliceMechanism }{ v } < \utilityMechanism{ \AliceMechanism }{ \vthreshold_{ \AliceMechanism }(x) }$. This allows us to repeat the derivation above but with strict inequalities, finishing the proof.
\end{proof}

\begin{proof}[Proof of \cref{lemma:bob-first}]
We know that the buyer interacts with Bob first. From \cref{lemma:buyer}, we get that the buyer picks the option $x = \zmarg{ \BobMechanism }^{+}\paren*{ \effectiveValue{b}{\AliceMechanism} }$ from Bob, where $\effectiveValue{b}{\AliceMechanism} = b - \aug_{ \AliceMechanism }\paren*{ b, a }$. Let $p = \frac{ \BobMechanism\paren*{ x } }{ x }$ so that the buyer pays Bob $xp$. Here, we assume $x > 0$ as otherwise, the lemma is straightforward. Thus, the pair $\paren*{ x, xp }$ is the option the buyer picks from Bob's mechanism. Now, note that we can ``remove'' all other options from Bob's mechanism while preserving the buyer behavior. Indeed, removing options only decreases the space of possible actions the buyer can take, and if the original action was optimal in the larger space, it continues to be optimal in the smaller space. Thus, the buyer behavior will be the same when faced with the mechanisms $\AliceMechanism$ and $\SL_{ x, p }$ and therefore, Bob's revenue will also be the same. Moreover, Myerson's payment identity implies that $p \leq \effectiveValue{b}{\AliceMechanism}$. We must also have $p \leq \p_{ \AliceMechanism }(x)$ as otherwise, \cref{lemma:sl} would contradict the fact that the buyer goes to Bob first. It follows that $\revDuo{ B }{ \AliceMechanism }{ \SL_{ x, p } } \leq x \cdot \p_{ \AliceMechanism }(x)$.
\end{proof}

\subsubsection{Missing Proofs from \cref{sec:stackelberg:bob:alice}}
\label{app:stackelberg:bob:alice}

\begin{lemma}
\label{lemma:dl-util}
Let $x \in \bracket*{ 0, 1 }$ and $q \in \bracket*{ \residualValue{b}{a}, b - x \cdot \paren*{ b - \residualValue{b}{a} } }$. For all $v \geq 0$, we have that:
\[
\utilityMechanism{ \DL_{ x, q } }{ v } = \begin{cases}
0 , &\text{~if~} v < \residualValue{b}{a} \\
x \cdot \paren*{ v - \residualValue{b}{a} } , &\text{~else if~} x = 1 \text{~or~} v < \frac{ q - x\residualValue{b}{a} }{ 1 - x } \\
v - q , &\text{~otherwise~}
\end{cases} .
\]
\end{lemma}
\begin{proof}
By definition of $\utilityMechanism{ \DL_{ x, q } }{ v }$, we have that
\[
\utilityMechanism{ \DL_{ x, q } }{ v } = \max_{ z \in \bracket*{ 0, 1 } } v z - \DL_{ x, q }\paren*{ z } = \max_{ z \in \bracket*{ 0, 1 } } \begin{cases}
z \cdot \paren*{ v - \residualValue{b}{a} }, &\text{~if~} 0 \leq z \leq x \\
z \cdot \paren*{ v - \frac{ q - x\residualValue{b}{a} }{ 1 - x } } + x \cdot \frac{ q - \residualValue{b}{a} }{ 1 - x }, &\text{~if~} x < z \leq 1
\end{cases} .
\]
Note from $\residualValue{b}{a} \leq q$ that $\residualValue{b}{a} \leq \frac{ q - x\residualValue{b}{a} }{ 1 - x }$ for all $x < 1$. Now, if $v < \residualValue{b}{a}$, both the cases above are (strictly) decreasing and therefore the maximum over all $z \in \bracket*{ 0, 1 }$ is just $0$. 
Otherwise, if $x = 1$, then only the first case applies and the maximum is attained at $1$. On the other hand, if $x < 1$ and $\residualValue{b}{a} \leq v < \frac{ q - x\residualValue{b}{a} }{ 1 - x }$, the first case is non-decreasing while the second case is strictly decreasing. Thus, the maximum over all $z \in \bracket*{ 0, 1 }$ is attained at $z = x$. Finally, if $x < 1$ and $\frac{ q - x\residualValue{b}{a} }{ 1 - x } \leq v$, both cases are non-decreasing and the maximum is attained at $1$. The lemma follows.
\end{proof}

\begin{lemma}
\label{lemma:dl-params}
Let $\AliceMechanism$ be a mechanism for Alice and $\effectiveValue{b}{\AliceMechanism} = b - \aug_{ \AliceMechanism }\paren*{ b, a }$. For all $q \in \bracket*{ \residualValue{b}{a}, b }$, we have:
\begin{enumerate}
\item \label{item:dl-params:1} It holds that $\x_{ \AliceMechanism }(q) \leq \frac{ b - q }{ b - \residualValue{b}{a} }$.
\item \label{item:dl-params:2} If $\aug_{ \AliceMechanism }\paren*{ b, a } < \min\paren*{ a, b }$, we also have $\frac{ \effectiveValue{b}{\AliceMechanism} - q }{ \effectiveValue{b}{\AliceMechanism} - \residualValue{b}{a} } \leq \x_{ \AliceMechanism }(q)$.
\end{enumerate}
\end{lemma}
\begin{proof}
We prove both parts separately.
\begin{enumerate}
\item This is trivial if $\aug_{ \AliceMechanism }\paren*{ b, a } = \min\paren*{ a, b }$ so we assume otherwise. From \cref{def:dl-params,eq:aug}, we have that:
\[
\x_{ \AliceMechanism }(q) = \frac{ b - q - \utilityMechanism{ \AliceMechanism }{ a } + \utilityMechanism{ \AliceMechanism }{ a - b + q } }{ \min\paren*{ a, b } - \aug_{ \AliceMechanism }\paren*{ \min\paren*{ a, b }, a } } .
\]
Now, note from $\residualValue{b}{a} \leq q$ that $b - q \leq b - \residualValue{b}{a} = \min\paren*{ a, b }$ by \cref{eq:primes}. This means that $\frac{ b - q }{ b - \residualValue{b}{a} } \in \bracket*{ 0, 1 }$. Defining this ratio to be $\lambda$, we get:
\begin{align*}
\x_{ \AliceMechanism }(q) &= \frac{ b - q - \utilityMechanism{ \AliceMechanism }{ a } + \utilityMechanism{ \AliceMechanism }{ a - \lambda \cdot \paren*{ b - \residualValue{b}{a} } } }{ \min\paren*{ a, b } - \aug_{ \AliceMechanism }\paren*{ \min\paren*{ a, b }, a } } \\
&\leq \frac{ b - q - \utilityMechanism{ \AliceMechanism }{ a } + \paren*{ 1 - \lambda } \cdot \utilityMechanism{ \AliceMechanism }{ a } + \lambda \cdot \utilityMechanism{ \AliceMechanism }{ a - b + \residualValue{b}{a} } }{ \min\paren*{ a, b } - \aug_{ \AliceMechanism }\paren*{ \min\paren*{ a, b }, a } } \tag{As utility functions are convex} \\
&\leq \frac{ \lambda \cdot \min\paren*{ a, b } - \lambda \cdot \paren*{ \utilityMechanism{ \AliceMechanism }{ a } - \utilityMechanism{ \AliceMechanism }{ \max\paren*{ 0, a - b } } } }{ \min\paren*{ a, b } - \aug_{ \AliceMechanism }\paren*{ \min\paren*{ a, b }, a } } \tag{\cref{eq:primes}} \\
&\leq \lambda . \tag{\cref{eq:aug}} 
\end{align*}
\item Note that the assumption $\aug_{ \AliceMechanism }\paren*{ b, a } < \min\paren*{ a, b }$ is necessary for the denominator to be non-zero. As above, we get:
\begin{align*}
\x_{ \AliceMechanism }(q) &= \frac{ b - q - \utilityMechanism{ \AliceMechanism }{ a } + \utilityMechanism{ \AliceMechanism }{ a - b + q } }{ \min\paren*{ a, b } - \aug_{ \AliceMechanism }\paren*{ \min\paren*{ a, b }, a } } \\
&= \frac{ \effectiveValue{b}{\AliceMechanism} - q - \utilityMechanism{ \AliceMechanism }{ \max\paren*{ 0, a - b } } + \utilityMechanism{ \AliceMechanism }{ a - b + q } }{ \min\paren*{ a, b } - \aug_{ \AliceMechanism }\paren*{ b, a } } \tag{\cref{eq:aug}} \\
&\geq \frac{ \effectiveValue{b}{\AliceMechanism} - q }{ \min\paren*{ a, b } - \aug_{ \AliceMechanism }\paren*{ b, a } } \tag{As utility functions are monotone} .
\end{align*}
\cref{eq:primes} allows us to write the denominator as $\effectiveValue{b}{\AliceMechanism} - \residualValue{b}{a}$ finishing the proof.
\end{enumerate}
\end{proof}

\begin{proof}[Proof of \cref{lemma:dl}]
Define the values $\effectiveValue{a}{\DL_{x,q}} = a - \aug_{ \DL_{ x, q } }\paren*{ a, b }$ and $\effectiveValue{b}{\AliceMechanism} = b - \aug_{ \AliceMechanism }\paren*{ b, a }$. From \cref{item:buyer:util} of \cref{lemma:buyer}, we have:
\[
\utilityBA{ \AliceMechanism }{ \DL_{ x, q } } - \utilityAB{ \AliceMechanism }{ \DL_{ x, q } } = \utilityMechanism{ \DL_{ x, q } }{ \effectiveValue{b}{\AliceMechanism} } + \utilityMechanism{ \AliceMechanism }{ a } - \utilityMechanism{ \AliceMechanism }{ \effectiveValue{a}{\DL_{x,q}} } - \utilityMechanism{ \DL_{ x, q } }{ b } .
\]
We use utilities computed in \cref{lemma:dl-util} to simplify the right hand side in the above equation. It gives $\utilityMechanism{ \DL_{ x, q } }{ b } = b - q$ and $\utilityMechanism{ \DL_{ x, q } }{ \residualValue{b}{a} } = 0$ and $\utilityMechanism{ \DL_{ x, q } }{ \effectiveValue{b}{\AliceMechanism} } = \max\paren*{ x \cdot \paren*{ \effectiveValue{b}{\AliceMechanism} - \residualValue{b}{a} }, \effectiveValue{b}{\AliceMechanism} - q }$. Combining with \cref{eq:aug}, we get that $\effectiveValue{a}{\DL_{x,q}} = a - b + q$. Plugging in, we get:
\begin{equation}
\label{eq:alice-first}
\utilityBA{ \AliceMechanism }{ \DL_{ x, q } } - \utilityAB{ \AliceMechanism }{ \DL_{ x, q } } = \max\paren*{ x \cdot \paren*{ \effectiveValue{b}{\AliceMechanism} - \residualValue{b}{a} }, \effectiveValue{b}{\AliceMechanism} - q } + \utilityMechanism{ \AliceMechanism }{ a } - \utilityMechanism{ \AliceMechanism }{ a - b + q } - b + q .
\end{equation}
We now consider the following cases:
\begin{itemize}
\item \textbf{When $\aug_{ \AliceMechanism }\paren*{ b, a } = \min\paren*{ a, b }$:} This means $\effectiveValue{b}{\AliceMechanism} = \residualValue{b}{a} \leq q$ and, as the utility function is convex and $1$-Lipschitz, also that the last four terms in \cref{eq:alice-first} vanish. Thus, the right hand side in \cref{eq:alice-first} is $0$ so both sides of the desired equivalence are true. For the moreover part, we consider both parts separately.
\begin{enumerate}
\item Let $a \geq b$ and $x \in \bracket*{ 0, \x_{ \AliceMechanism }(q) }$. From \cref{eq:primes}, this means that $\residualValue{b}{a} = 0$. Note that the result is trivial if $q = 0$ so we assume that $q > 0$. Then, since $\aug_{ \AliceMechanism }\paren*{ b, a } = \min\paren*{ a, b }$ and $q > 0$, we have that $\zmarg{ \AliceMechanism }^-\paren*{ a - b + q } = 1$. Thus, if the buyer goes to Alice first, he never buys from Bob, and if the buyer goes to Bob first, $q > \effectiveValue{b}{\AliceMechanism} = \residualValue{b}{a}$ implies that he never buys from Bob. The result follows.

\item Let $a \leq b$ and $x = \x_{ \AliceMechanism }(q)$. From \cref{eq:primes}, this means that $\residualValue{b}{a} = \Delta \geq 0$. If $q = \Delta$, the result is straightforward, so we assume $q > \Delta$. Recall that $\effectiveValue{b}{\AliceMechanism} = \residualValue{b}{a} \leq q$ and note that the fact that $\aug_{ \AliceMechanism }\paren*{ b, a } = \min\paren*{ a, b }$ implies that $\zmarg{ \AliceMechanism }^-\paren*{ a - b + q } = 1$. When the buyer goes to Alice first, the fact that $\zmarg{ \AliceMechanism }^-\paren*{ a - b + q } = 1$ implies that the buyer always gets the item. This means that, when he goes to Bob, he gets the item with probability $x$ at a price of $x\residualValue{b}{a}$. On the other hand, when the buyer goes to Bob first, he first gets the item with probability $x$ at a price of $x\residualValue{b}{a}$ and then, if he is not allocated the item, he goes to Alice and picks the option $\zmarg{ \AliceMechanism }^{+}\paren*{ a } = 1$. Thus, in either case, Bob's revenue is $x\residualValue{b}{a} = \Delta \cdot \x_{ \AliceMechanism }(q)$, as desired.
\end{enumerate}
\item \textbf{When $\aug_{ \AliceMechanism }\paren*{ b, a } < \min\paren*{ a, b }$ and $x \cdot \paren*{ \effectiveValue{b}{\AliceMechanism} - \residualValue{b}{a} } \leq \effectiveValue{b}{\AliceMechanism} - q$:} In this case, we have from \cref{lemma:dl-params} that $x \leq \x_{ \AliceMechanism }(q)$. Thus, it suffices to show that the right hand side of \cref{eq:alice-first} is non-positive. This is because $\effectiveValue{b}{\AliceMechanism} = b - \aug_{ \AliceMechanism }\paren*{ b, a }$ implies:
\[
\utilityBA{ \AliceMechanism }{ \DL_{ x, q } } - \utilityAB{ \AliceMechanism }{ \DL_{ x, q } } = \utilityMechanism{ \AliceMechanism }{ \max\paren*{ 0, a - b } } - \utilityMechanism{ \AliceMechanism }{ a - b + q } ,
\]
For the moreover part, note that if the right hand side above is negative, the buyer buys from Alice first. In this case, the revenue bound for both parts follows from \cref{lemma:buyer}. So we assume that the right hand side above is $0$. If $q > \residualValue{b}{a}$, this is only possible if $\utilityMechanism{ \AliceMechanism }{ a - b + q } = 0$ and therefore, we must have $\zmarg{ \AliceMechanism }^-\paren*{ a - b + q } = 0$. Then, if the buyer buys from Alice first, he does not get the item and therefore buys from Bob with probability $1$ at a price of $q$. If he goes to Bob first, $x \geq 0$ implies $\effectiveValue{b}{\AliceMechanism} \geq q$ and therefore the buyer again buys from Bob with probability $1$ at a price of $q$. Thus, Bob's revenue is $q$ in either case, and both parts follow.

Henceforth, we assume that the right hand side above is $0$ and $q = \residualValue{b}{a}$ and consider both parts separately. For the first part, note that $a \geq b$ implies $q = 0$ and therefore the result is trivial. For the second part, if $q = \residualValue{b}{a}$, we have $\x_{ \AliceMechanism }(q) = 1$ implying that we have to show $\revDuo{ B }{ \AliceMechanism }{ \DL_{ \x_{ \AliceMechanism }(q), q } } = q$. Now, if the buyer goes to Bob first, $x \geq 0$ implies $\effectiveValue{b}{\AliceMechanism} \geq q$ and therefore the buyer buys from Bob with probability $1$ at a price of $q$. If the buyer buys from Alice first, he may or may not get Alice's item, but in either case, he will pay $q$ to Bob. Thus, $\revDuo{ B }{ \AliceMechanism }{ \DL_{ \x_{ \AliceMechanism }(q), q } } = q$, as desired.

\item \textbf{When $\aug_{ \AliceMechanism }\paren*{ b, a } < \min\paren*{ a, b }$ and $x \cdot \paren*{ \effectiveValue{b}{\AliceMechanism} - \residualValue{b}{a} } > \effectiveValue{b}{\AliceMechanism} - q$:} In this case, we can continue \cref{eq:alice-first} as:
\begin{align*}
\utilityBA{ \AliceMechanism }{ \DL_{ x, q } } - \utilityAB{ \AliceMechanism }{ \DL_{ x, q } } &= x \cdot \paren*{ \effectiveValue{b}{\AliceMechanism} - \residualValue{b}{a} } + \utilityMechanism{ \AliceMechanism }{ a } - \utilityMechanism{ \AliceMechanism }{ a - b + q } - b + q \\
&= x \cdot \paren*{ \min\paren*{ a, b } - \aug_{ \AliceMechanism }\paren*{ b, a } } + \utilityMechanism{ \AliceMechanism }{ a } - \utilityMechanism{ \AliceMechanism }{ a - b + q } - b + q \\
&= x \cdot \paren*{ \min\paren*{ a, b } - \aug_{ \AliceMechanism }\paren*{ \min\paren*{ a, b }, a } } - \paren*{ b - q - \aug_{ \AliceMechanism }\paren*{ b - q, a } } \tag{\cref{eq:aug}} .
\end{align*}
This is non-positive if and only if $x \leq \x_{ \AliceMechanism }(q)$, as desired. For the moreover part, we consider both parts separately.
\begin{enumerate}
\item Let $a \geq b$ and $x \in \bracket*{ 0, \x_{ \AliceMechanism }(q) }$. From \cref{eq:primes}, this means that $\residualValue{b}{a} = 0$. If $q = 0$, the result is trivial, so assume that $q > 0$. If the right hand side above is negative, the buyer interacts with Alice first and the revenue bound follows. As $x \in \bracket*{ 0, \x_{ \AliceMechanism }(q) }$, the only other case is that the right hand side is $0$ and the buyer gets the same utility regardless of which seller he interacts with first. When he interacts with Alice first, we have by \cref{item:buyer:payment} of \cref{lemma:buyer} that Bob's revenue is $q \cdot \paren*{ 1 - \zmarg{ \AliceMechanism }^-\paren*{ \Delta + q } }$. On the other hand, when he interacts with Bob first, $x \cdot \paren*{ \effectiveValue{b}{\AliceMechanism} - \residualValue{b}{a} } > \effectiveValue{b}{\AliceMechanism} - q$ implies that he pays him nothing. As the buyer tie-breaks in favor of higher revenue for Bob, we get that Bob's revenue is the larger of these two, which is $q \cdot \paren*{ 1 - \zmarg{ \AliceMechanism }^-\paren*{ \Delta + q } }$, as desired.

\item Let $a \leq b$ and $x = \x_{ \AliceMechanism }(q)$. From \cref{eq:primes}, this means that $\residualValue{b}{a} = \Delta \geq 0$ and from \cref{def:dl-params}, this means that the buyer gets the same utility regardless of which seller he interacts with first. If $q = 0$, we must have $\residualValue{b}{a} = 0$ and the result is trivial. Assume that $q > 0$. When he interacts with Alice first, we have by \cref{item:buyer:payment} of \cref{lemma:buyer} that Bob's revenue is $q - \zmarg{ \AliceMechanism }^-\paren*{ a - b + q } \cdot \paren*{ q - \Delta \cdot \x_{ \AliceMechanism }(q) }$. On the other hand, when he interacts with Bob first, $x \cdot \paren*{ \effectiveValue{b}{\AliceMechanism} - \residualValue{b}{a} } > \effectiveValue{b}{\AliceMechanism} - q$ implies that he pays him $x\residualValue{b}{a} \leq q$. As the buyer tie-breaks in favor of higher revenue for Bob, we get that Bob's revenue is the larger of these two, which is $q - \zmarg{ \AliceMechanism }^-\paren*{ a - b + q } \cdot \paren*{ q - \Delta \cdot \x_{ \AliceMechanism }(q) }$.
\end{enumerate}
\end{itemize}
\end{proof}

\begin{proof}[Proof of \cref{lemma:alice-first}]
We know that the buyer interacts with Alice first. From \cref{lemma:buyer}, we get that the buyer picks the option $x^w = \zmarg{ \BobMechanism }^{+}\paren*{ \residualValue{b}{a} }$ from Bob when he gets Alice's item and the option $x^l = \zmarg{ \BobMechanism }^{+}\paren*{ b }$ from Bob when he does not. Define $q^l = \BobMechanism\paren*{ x^l }$ and $p^w = \frac{ \BobMechanism\paren*{ x^w } }{ x^w }$ if $x^w > 0$ and $p^w = \residualValue{b}{a}$ otherwise. This ensures that $\BobMechanism\paren*{ x^w } = x^w p^w$ in both cases. It follows that $\paren*{ x^w, x^w p^w }$ and $\paren*{ x^l, q^l }$ are options in Bob's menu. Using this, we get from \cref{lemma:buyer} and the fact that the buyer goes to Alice first when faced with $\AliceMechanism$ and $\BobMechanism$ that (letting $\effectiveValue{a}{\BobMechanism} = a - \aug_{ \BobMechanism }\paren*{ a, b }$ and $\effectiveValue{b}{\AliceMechanism} = b - \aug_{ \AliceMechanism }\paren*{ b, a }$):
\[
x^w \cdot \paren*{ \effectiveValue{b}{\AliceMechanism} - p^w } + \utilityMechanism{ \AliceMechanism }{ a } \leq \utilityMechanism{ \BobMechanism }{ \effectiveValue{b}{\AliceMechanism} } + \utilityMechanism{ \AliceMechanism }{ a } \leq \utilityMechanism{ \AliceMechanism }{ \effectiveValue{a}{\BobMechanism} } + \utilityMechanism{ \BobMechanism }{ b } .
\]
From Myerson's payment identity and incentive compatibility, we get that $\utilityMechanism{ \BobMechanism }{ \residualValue{b}{a} } = x^w \cdot \paren*{ \residualValue{b}{a} - p^w } \geq 0$ and $\utilityMechanism{ \BobMechanism }{ b } = b x^l - q^l \geq x^w \cdot \paren*{ b - p^w } \geq 0$. This gives:
\[
x^w \cdot \paren*{ \effectiveValue{b}{\AliceMechanism} - \residualValue{b}{a} } + \utilityMechanism{ \AliceMechanism }{ a } - \utilityMechanism{ \AliceMechanism }{ \effectiveValue{a}{\BobMechanism} } \leq b x^l - q^l - x^w \cdot \paren*{ \residualValue{b}{a} - p^w } = \utilityMechanism{ \BobMechanism }{ b } - \utilityMechanism{ \BobMechanism }{ \residualValue{b}{a} } = \aug_{ \BobMechanism }\paren*{ a, b } = a - \effectiveValue{a}{\BobMechanism} ,
\]
where we use \cref{eq:aug}. We define $q$ so that the right hand side above is $b - q$ and note from \cref{eq:aug} that $q \in \bracket*{ \residualValue{b}{a}, b }$. We claim that $x^w \leq \x_{ \AliceMechanism }(q)$. For this note that, if the second case of \cref{def:dl-params} applies, this is true because $b - q + x^w \cdot \paren*{ \residualValue{b}{a} - p^w } = b x^l - q^l \geq x^w \cdot \paren*{ b - p^w }$ by the incentive compatibility inequality mentioned above. This rearranges to $x^w \leq \frac{ b - q }{ b - \residualValue{b}{a} }$, as desired for \cref{def:dl-params}. If the first case applies, note from the above equation and \cref{eq:aug} that:
\[
\frac{ b - q - \aug_{ \AliceMechanism }\paren*{ b - q, a } }{ \min\paren*{ a, b } - \aug_{ \AliceMechanism }\paren*{ \min\paren*{ a, b }, a } } = \frac{ b - q - \utilityMechanism{ \AliceMechanism }{ a } + \utilityMechanism{ \AliceMechanism }{ a - b + q } }{ \effectiveValue{b}{\AliceMechanism} - \residualValue{b}{a} } \geq x^w
\]
by the equation above. We are now ready to show the required bound. From \cref{item:buyer:payment} of \cref{lemma:buyer}, we have that Bob's revenue is:
\[
\revDuo{ B }{ \AliceMechanism }{ \BobMechanism } = \paren*{ 1 - \zmarg{ \AliceMechanism }^-\paren*{ \effectiveValue{a}{\BobMechanism} } } \cdot q^l + \zmarg{ \AliceMechanism }^-\paren*{ \effectiveValue{a}{\BobMechanism} } \cdot x^w p^w .
\]
If $a \geq b$, then $p^w = \residualValue{b}{a} = 0$ and the second term above vanishes. The bound therefore follows from \cref{item:dl:la} of \cref{lemma:dl}. If $a \leq b$, we use $x^w \leq \x_{ \AliceMechanism }(q)$ and $p^w \leq \residualValue{b}{a} = \Delta$ to conclude the bound from \cref{item:dl:lb} of \cref{lemma:dl}.
\end{proof}

\subsection{Missing Proofs from \cref{sec:stackelberg:alice}}
\label{app:stackelberg:alice}
\subsubsection{Missing Proofs from \cref{sec:stackelberg:alice:la}}
\label{app:stackelberg:alice:la}

\begin{proof}[Proof of \cref{lemma:alice-structure-1:la}]
As the result is trivial for $R = b$, we assume $R < b$ throughout. We also assume $x < 1$ as the result for all other values of $x$ implies the result for $x = 1$ by continuity. We define $s = R + \Delta + b \cdot \paren*{ 1 - x }$ for convenience. Note that $x \geq \frac{ R }{ b }$ implies that $s \leq a$. We first claim that $\utilityMechanism{ \AliceMechanism }{ s } \geq \utilityMechanism{ \AliceMechanism }{ a } \cdot \paren*{ 1 - x }$. Indeed, suppose for contradiction that we have $\utilityMechanism{ \AliceMechanism }{ s } < \utilityMechanism{ \AliceMechanism }{ a } \cdot \paren*{ 1 - x } \leq \utilityMechanism{ \AliceMechanism }{ a } - x \cdot \aug_{ \AliceMechanism }\paren*{ b, a }$ by \cref{eq:aug}. Then, from the monotonicity of utility and \cref{lemma:sl-params}, it follows that $s < \vthreshold_{ \AliceMechanism }(x)$. As $a - bx \leq s$, we have from \cref{def:sl-params} that $\p_{ \AliceMechanism }(x) > b - \frac{ a - s }{ x } \geq 0$. From \cref{lemma:sl}, this means that for all $p < \p_{ \AliceMechanism }(x)$, when Bob uses the mechanism $\SL_{ x, p }$, the buyer goes to Bob first and pays him $xp$. By the assumption in the lemma, we get that $x \cdot \p_{ \AliceMechanism }(x) \leq R$. But this means that $R > bx - a + s = R$, a contradiction. From this claim, we get:
\[
\utilityMechanism{ \AliceMechanism }{ a } = \utilityMechanism{ \AliceMechanism }{ s } + \int_s^a \zmarg{ \AliceMechanism }^{+}\paren*{ v } \del v \geq \utilityMechanism{ \AliceMechanism }{ a } \cdot \paren*{ 1 - x } + \int_s^a \zmarg{ \AliceMechanism }^{+}\paren*{ v } \del v .
\]
We now lower bound $\zmarg{ \AliceMechanism }^{+}\paren*{ v }$. For this, note from \cref{item:dl:la} of \cref{lemma:dl} that $\revDuo{ B }{ \AliceMechanism }{ \DL_{ 0, q } } = q \cdot \paren*{ 1 - \zmarg{ \AliceMechanism }^-\paren*{ \Delta + q } }$ for all $q \in \oc*{ 0, b }$. The assumption in the lemma then implies that this is at most $R$ and we get $\zmarg{ \AliceMechanism }^{+}\paren*{ \Delta + q } \geq 1 - \frac{ R }{ q }$. Plugging in,
\[
x \cdot \utilityMechanism{ \AliceMechanism }{ a } \geq \int_s^a \paren*{ 1 - \frac{ R }{ v - \Delta } } \del v .
\]
We are done as $x > 0$. 
\end{proof}

\begin{proof}[Proof of \cref{lemma:alice-structure-2:la}]
If $R = 0$, the lemma is trivial, so we assume $R > 0$. Throughout this proof, we make use of Myerson's payment identity, namely that for all mechanisms $\mathcal{M}$ and all $v \geq 0$, it holds that:
\[
\mathcal{M}\paren*{ \zmarg{ \mathcal{M} }^{+}\paren*{ v } } = v \cdot \zmarg{ \mathcal{M} }^{+}\paren*{ v } - \utilityMechanism{ \mathcal{M} }{ v } .
\]
Let $\AliceMechanism$, $\BobMechanism$, and $R$ be as in the lemma statement. We divide the proof into two cases, based on which seller the buyer approaches first. First, we consider the case when the buyer approaches Bob first. In this case, let $\effectiveValue{b}{\AliceMechanism} = b - \aug_{ \AliceMechanism }\paren*{ b, a }$. We have from \cref{item:buyer:payment} of \cref{lemma:buyer} that:
\begin{align*}
\revDuo{ A }{ \AliceMechanism }{ \BobMechanism } &= \zmarg{ \BobMechanism }^{+}\paren*{ \effectiveValue{b}{\AliceMechanism} } \cdot \AliceMechanism\paren*{ \zmarg{ \AliceMechanism }^{+}\paren*{ \residualValue{a}{b} } } \\
&\qquad {}+ \paren*{ 1 - \zmarg{ \BobMechanism }^{+}\paren*{ \effectiveValue{b}{\AliceMechanism} } } \cdot \AliceMechanism\paren*{ \zmarg{ \AliceMechanism }^{+}\paren*{ a } } \\
&= \zmarg{ \BobMechanism }^{+}\paren*{ \effectiveValue{b}{\AliceMechanism} } \cdot \paren*{ \residualValue{a}{b} \cdot \zmarg{ \AliceMechanism }^{+}\paren*{ \residualValue{a}{b} } - \utilityMechanism{ \AliceMechanism }{ \residualValue{a}{b} } } \\
&\qquad {}+ \paren*{ 1 - \zmarg{ \BobMechanism }^{+}\paren*{ \effectiveValue{b}{\AliceMechanism} } } \cdot \paren*{ a \cdot \zmarg{ \AliceMechanism }^{+}\paren*{ a } - \utilityMechanism{ \AliceMechanism }{ a } } \displaybreak[1] \\
&= \zmarg{ \BobMechanism }^{+}\paren*{ \effectiveValue{b}{\AliceMechanism} } \cdot \paren*{ a \cdot \zmarg{ \AliceMechanism }^{+}\paren*{ \residualValue{a}{b} } - b \cdot \zmarg{ \AliceMechanism }^{+}\paren*{ \residualValue{a}{b} } - \utilityMechanism{ \AliceMechanism }{ a } + \aug_{ \AliceMechanism }\paren*{ b, a } } \\
&\qquad {}+ \paren*{ 1 - \zmarg{ \BobMechanism }^{+}\paren*{ \effectiveValue{b}{\AliceMechanism} } } \cdot \paren*{ a \cdot \zmarg{ \AliceMechanism }^{+}\paren*{ a } - \utilityMechanism{ \AliceMechanism }{ a } } \tag{\cref{eq:primes,eq:aug}} \\
&= \zmarg{ \BobMechanism }^{+}\paren*{ \effectiveValue{b}{\AliceMechanism} } \cdot \paren*{ a \cdot \zmarg{ \AliceMechanism }^{+}\paren*{ \residualValue{a}{b} } - b \cdot \zmarg{ \AliceMechanism }^{+}\paren*{ \residualValue{a}{b} } + \aug_{ \AliceMechanism }\paren*{ b, a } } \\
&\qquad {}+ \paren*{ 1 - \zmarg{ \BobMechanism }^{+}\paren*{ \effectiveValue{b}{\AliceMechanism} } } \cdot a \cdot \zmarg{ \AliceMechanism }^{+}\paren*{ a } - \utilityMechanism{ \AliceMechanism }{ a } \\
&\leq \zmarg{ \BobMechanism }^{+}\paren*{ \effectiveValue{b}{\AliceMechanism} } \cdot \frac{ a }{ b } \cdot \aug_{ \AliceMechanism }\paren*{ b, a } + \paren*{ 1 - \zmarg{ \BobMechanism }^{+}\paren*{ \effectiveValue{b}{\AliceMechanism} } } \cdot a - \utilityMechanism{ \AliceMechanism }{ a } \tag{As $a \geq b$ and $\aug_{ \AliceMechanism }\paren*{ b, a } \geq b \cdot \zmarg{ \AliceMechanism }^{+}\paren*{ \residualValue{a}{b} }$} \\
&\leq a \cdot \paren*{ 1 - \frac{ \zmarg{ \BobMechanism }^{+}\paren*{ \effectiveValue{b}{\AliceMechanism} } }{ b } \cdot \effectiveValue{b}{\AliceMechanism} } - \utilityMechanism{ \AliceMechanism }{ a } \tag{As $\effectiveValue{b}{\AliceMechanism} = b - \aug_{ \AliceMechanism }\paren*{ b, a }$} \\
&\leq a \cdot \paren*{ 1 - \frac{ R }{ b } } - \utilityMechanism{ \AliceMechanism }{ a } \tag{\cref{lemma:buyer}, \cref{item:buyer:payment}} .
\end{align*}
Now, we consider the case when the buyer approaches Alice first. In this case, let $\effectiveValue{a}{\BobMechanism} = a - \aug_{ \BobMechanism }\paren*{ a, b }$ as in \cref{lemma:buyer} and note from \cref{eq:aug} that $\effectiveValue{a}{\BobMechanism} = a - \utilityMechanism{ \BobMechanism }{ b }$. As $R > 0$, we have from \cref{item:buyer:payment} of \cref{lemma:buyer} that:
\begin{align*}
\revDuo{ A }{ \AliceMechanism }{ \BobMechanism } &= \AliceMechanism\paren*{ \zmarg{ \AliceMechanism }^-\paren*{ \effectiveValue{a}{\BobMechanism} } } \\
&= \effectiveValue{a}{\BobMechanism} \cdot \zmarg{ \AliceMechanism }^-\paren*{ \effectiveValue{a}{\BobMechanism} } - \utilityMechanism{ \AliceMechanism }{ \effectiveValue{a}{\BobMechanism} } \\
&\leq \effectiveValue{a}{\BobMechanism} \cdot \zmarg{ \AliceMechanism }^-\paren*{ \effectiveValue{a}{\BobMechanism} } - \utilityMechanism{ \AliceMechanism }{ a } + a - \effectiveValue{a}{\BobMechanism} \tag{As utilities are $1$-Lipschitz} \\
&\leq a - \effectiveValue{a}{\BobMechanism} \cdot \paren*{ 1 - \zmarg{ \AliceMechanism }^-\paren*{ \effectiveValue{a}{\BobMechanism} } } - \utilityMechanism{ \AliceMechanism }{ a } \\
&\leq a - \frac{ a }{ b } \cdot \paren*{ b - \utilityMechanism{ \BobMechanism }{ b } } \cdot \paren*{ 1 - \zmarg{ \AliceMechanism }^-\paren*{ \effectiveValue{a}{\BobMechanism} } } - \utilityMechanism{ \AliceMechanism }{ a } \tag{As $\effectiveValue{a}{\BobMechanism} = a - \utilityMechanism{ \BobMechanism }{ b }$ and $a \geq b$} \\
&\leq a \cdot \paren*{ 1 - \frac{ R }{ b } } - \utilityMechanism{ \AliceMechanism }{ a } \tag{\cref{lemma:buyer}, \cref{item:buyer:payment}} .
\end{align*}
\end{proof}

\begin{proof}[Proof of \cref{lemma:la-equalizer}]
We first show that for all mechanisms $\BobMechanism$, it holds that $\revDuo{ B }{ \AliceMechanism_R^{ \mathsf{LA} } }{ \BobMechanism } \leq R$. Fix an arbitrary mechanism $\BobMechanism$ and consider first the case when the buyer interacts with Bob first when faced with the mechanisms $\AliceMechanism_R^{ \mathsf{LA} }$ and $\BobMechanism$. From \cref{lemma:bob-first}, we have that there exists $x \in \bracket*{ 0, 1 }$ such that $\revDuo{ B }{ \AliceMechanism_R^{ \mathsf{LA} } }{ \BobMechanism } \leq x \cdot \p_{ \AliceMechanism_R^{ \mathsf{LA} } }(x)$. As the claim is trivial otherwise, we assume that $x > 0$. From \cref{def:sl-params}, we have that $\revDuo{ B }{ \AliceMechanism_R^{ \mathsf{LA} } }{ \BobMechanism } \leq bx - a + \vthreshold_{ \AliceMechanism_R^{ \mathsf{LA} } }(x)$. Thus, it suffices to show that $\vthreshold_{ \AliceMechanism_R^{ \mathsf{LA} } }(x) \leq R + a - bx$. For this, recall \cref{def:sl-params} and note that, for all $v \in \oc*{ R + a - bx, a }$, we have $R + \Delta < v \leq a$ implying by \cref{def:la-equalizer} that $\zmarg{ \AliceMechanism_R^{ \mathsf{LA} } }^-\paren*{ v } > 0$
\[
\utilityMechanism{ \AliceMechanism_R^{ \mathsf{LA} } }{ v } > \utilityMechanism{ \AliceMechanism_R^{ \mathsf{LA} } }{ R + a - bx } \geq b \cdot \paren*{ 1 - x } \cdot \frac{ \utilityMechanism{ \AliceMechanism_R^{ \mathsf{LA} } }{ a } }{ b } = \paren*{ 1 - x } \cdot \utilityMechanism{ \AliceMechanism_R^{ \mathsf{LA} } }{ a } ,
\] 
where we use \cref{eq:la-equalizer-util} in the second step. This means that $v > \vthreshold_{ \AliceMechanism_R^{ \mathsf{LA} } }(x)$, as desired. Now, consider the case when the buyer interacts with Alice first. Here, by \cref{lemma:alice-first,lemma:dl}, we have that there exists $q \in \bracket*{ \residualValue{b}{a}, b }$ such that $\revDuo{ B }{ \AliceMechanism_R^{ \mathsf{LA} } }{ \BobMechanism } \leq \revDuo{ B }{ \AliceMechanism_R^{ \mathsf{LA} } }{ \DL_{ \x_{ \AliceMechanism_R^{ \mathsf{LA} } }(q), q } } = q \cdot \paren*{ 1 - \zmarg{ \AliceMechanism_R^{ \mathsf{LA} } }^-\paren*{ a - b + q } }$. As the claim is trivial otherwise, assume $q > 0$. By \cref{def:la-equalizer}, we can continue as $\revDuo{ B }{ \AliceMechanism_R^{ \mathsf{LA} } }{ \BobMechanism } \leq q \cdot \frac{ R }{ q } = R$.

Next, note that either $\aug_{ \AliceMechanism_R^{ \mathsf{LA} } }\paren*{ b, a } > 0$ or $R = b$. In either case, \cref{lemma:buyer} implies $\revDuo{ B }{ \AliceMechanism_R^{ \mathsf{LA} } }{ \FP{b} } = R$ and $\revDuo{ A }{ \AliceMechanism_R^{ \mathsf{LA} } }{ \FP{b} } = a \cdot \paren*{ 1 - \frac{ R }{ b } } - \utilityMechanism{ \AliceMechanism_R^{ \mathsf{LA} } }{ a }$. As the latter is the highest payment any buyer can make to Alice and as Bob tie-breaks in favor of Alice's revenue, even if his best response is not $\FP{b}$, it is impossible for Alice to get a higher payment and the proof is complete.
\end{proof}

We are now ready to prove that Alice's optimal revenue when $a \geq b$ can be achieved by a Large-Alice equalizer.
\paragraph{Proving the optimality of Large-Alice equalizer mechanisms when $a \geq b$.}
We begin by bounding Alice's optimal revenue. Let $\AliceMechanism$ and $\BobMechanism$ be an arbitrary pair of mechanisms for Alice and Bob respectively \ygedit{with $\BobMechanism$ a best response to $\AliceMechanism$}. Let $R = \revDuo{ B }{ \AliceMechanism }{ \BobMechanism } \in \bracket*{ 0, b }$ be Bob's \ygedit{best-response revenue}. We first deal with the easy cases where either $\utilityMechanism{ \AliceMechanism }{ a } = 0$ or $R = 0$. Suppose that $\utilityMechanism{ \AliceMechanism }{ a } = 0$. Then, if Bob uses the mechanism $\FP{b}$, he can get a revenue of $b$ and Alice will get nothing. As this is the only mechanism that yields Bob a revenue of $b$, this is indeed what Bob will do and Alice will get zero revenue. Now, suppose that $R = 0$. This means that no matter what mechanism Bob uses, he does not get any revenue. In particular, for any $\delta > 0$ using $\FP{\delta}$ does not give him any revenue. This means that Alice's payment is at most $\Delta$.

For the rest of the proof, we assume that $\utilityMechanism{ \AliceMechanism }{ a } > 0$ and $R > 0$. We first show that, for all $R \in \bracket*{ 0, b }$, there is a unique value $s = s(R) \in \bracket*{ R + \Delta, a }$ that satisfies:
\begin{equation}
\label{eq:s:la}
\frac{ b + R }{ s - \Delta } - \ln \frac{ b }{ s - \Delta } = 2 .
\end{equation}
This is easy to see when $R = 0$, so we assume $R > 0$. For this, let $\zeta(s)$ denote the left hand side above and note that $\frac{ \del \zeta(s) }{ \del s } = \frac{ 1 }{ s - \Delta } - \frac{ b + R }{ \paren*{ s - \Delta }^2 } = \frac{ s - a - R }{ \paren*{ s - \Delta }^2 } < 0$. This means $\zeta(s)$ is strictly decreasing so if such an $s$ exists, it must be unique. The reason $s$ exists is because $\zeta\paren*{ R + \Delta } = 1 + \frac{ b }{ R } - \ln \frac{ b }{ R } \geq 2$ and $\zeta(a) = 1 + \frac{ R }{ b } \leq 2$ implying by continuity that there must exist $s \in \bracket*{ R + \Delta, a }$ such that $\zeta(s) = 2$. Now, we use \cref{lemma:alice-structure-1:la} with $x = \frac{ a + R - s }{ b }$ and \cref{lemma:alice-structure-2:la} to get:
\begin{align*}
\revDuo{ A }{ \AliceMechanism }{ \BobMechanism } &\leq a \cdot \paren*{ 1 - \frac{ R }{ b } } - \frac{ b }{ a + R - s } \cdot \int_s^a \paren*{ 1 - \frac{ R }{ v - \Delta } } \del v \\
&\leq a \cdot \paren*{ 1 - \frac{ R }{ b } } - \frac{ b }{ a + R - s } \cdot \paren*{ a - s - R \cdot \ln \frac{ b }{ s - \Delta } } \\
&\leq a \cdot \paren*{ 1 - \frac{ R }{ b } } - \frac{ b }{ a + R - s } \cdot \paren*{ a - s - R \cdot \paren*{ \frac{ b + R }{ s - \Delta } - 2 } } \tag{\cref{eq:s:la}} \\
&\leq a \cdot \paren*{ 1 - \frac{ R }{ b } } - b + R \cdot \frac{ b }{ s - \Delta } \\
&\leq a \cdot \paren*{ 1 - \frac{ R }{ b } } - b \cdot \paren*{ 1 - \frac{ R }{ s - \Delta } } .
\end{align*}
To continue, note that $s$ is determined by $R$ and therefore the right hand side above is a function of $R$. Let $c \in \paren*{ 3.146, 3.147 }$ satisfy $c - \ln c = 2$. In \cref{lemma:optimize:la} (stated and proved below), we show that when $\frac{ a }{ b } \geq c$, the right hand side is always at most $\Delta$ and is equal to $\Delta$ when $R = 0$. On the other hand, when $\frac{ a }{ b } < c$, define $\varsigma^* \in \co*{ 1, c }$ to be the unique value satisfying $\varsigma^2 - \varsigma = \frac{ a }{ b } \cdot \paren*{ 1 + \ln \varsigma }$. Then, the right hand side is at most $\Delta + \frac{ b^2 }{ a \varsigma^* } \cdot \paren*{ \varsigma^* - 1 } \cdot \paren*{ \varsigma^* - \frac{ a }{ b } }^2$ with the maximum being attained at $R = b \cdot \paren*{ \frac{ b }{ a } - \frac{ 1 }{ \varsigma^* } } \cdot \paren*{ \varsigma^* - 1 }$. This gives the following bound \ygedit{for every such pair, and hence for $\opt{a}{b}$}.
\begin{equation}
\label{eq:opt:la}
\opt{ a }{ b } \leq \begin{cases}
\Delta, &\text{~if~} \frac{ a }{ b } \geq c \\
\Delta + \frac{ b^2 }{ a \varsigma^* } \cdot \paren*{ \varsigma^* - 1 } \cdot \paren*{ \varsigma^* - \frac{ a }{ b } }^2, &\text{~if~} \frac{ a }{ b } < c 
\end{cases} .
\end{equation}
To finish the proof, note from \cref{lemma:la-equalizer} that the
{revenue bound} above is attained by Large-Alice-equalizer mechanisms.
\ygedit{As Alice {cannot} beat this bound,
{using \cref{obs:buyer-tie-breaks-for-alice} to design an Alice-favorable strategy for Bob, we have found an Stackelberg equilibrium which acheives this bound.}}
Therefore, the bound is tight and the proof is complete.

\begin{lemma}
\label{lemma:optimize:la}
For all $R \in \bracket*{ 0, b }$, let $s$ be as in \cref{eq:s:la}. Let $c \in \paren*{ 3.146, 3.147 }$ satisfy $c - \ln c = 2$. We have:
\[
a \cdot \paren*{ 1 - \frac{ R }{ b } } - b + R \cdot \frac{ b }{ s - \Delta } \leq \begin{cases}
\Delta, &\text{~if~} \frac{ a }{ b } \geq c \\
\Delta + \frac{ b^2 }{ a \varsigma^* } \cdot \paren*{ \varsigma^* - 1 } \cdot \paren*{ \varsigma^* - \frac{ a }{ b } }^2, &\text{~if~} \frac{ a }{ b } < c 
\end{cases} .
\]
Here, $\varsigma^* \in \co*{ 1, c }$ is the unique value satisfying $\varsigma^2 - \varsigma = \frac{ a }{ b } \cdot \paren*{ 1 + \ln \varsigma }$. Moreover, the maxima in the two cases are attained at $R = 0$ and $R = b \cdot \paren*{ \frac{ b }{ a } - \frac{ 1 }{ \varsigma^* } } \cdot \paren*{ \varsigma^* - 1 }$ respectively.
\end{lemma}
\begin{proof}
Let $\varsigma = \frac{ b }{ s - \Delta }$ and $\xi = R \cdot \paren*{ \varsigma - \frac{ a }{ b } }$ so that the left hand side above is $\Delta + \xi$. Note from \cref{eq:s:la} that as $R$ increases in the interval $\bracket*{ 0, b }$, we have that $\varsigma$ decreases in the interval $\bracket*{ 1, c }$. Thus, $\varsigma$ determines $R$ and we can find the optimum with respect to $\varsigma$. We have from \cref{eq:s:la} that:
\[
\xi = R \cdot \paren*{ \varsigma - \frac{ a }{ b } } = \paren*{ b - \frac{ a }{ \varsigma } } \cdot \paren*{ 2 - \varsigma + \ln \varsigma } .
\]
This means that
\begin{align*}
\frac{ \del \xi }{ \del \varsigma } &= \paren*{ b - \frac{ a }{ \varsigma } } \cdot \paren*{ \frac{ 1 }{ \varsigma } - 1 } + \frac{ a }{ \varsigma^2 } \cdot \paren*{ 2 - \varsigma + \ln \varsigma } \\
&= \frac{ 1 }{ \varsigma^2 } \cdot \paren*{ \paren*{ b \varsigma - a } \cdot \paren*{ 1 - \varsigma } + a \cdot \paren*{ 2 - \varsigma + \ln \varsigma } } \\
&= \frac{ b }{ \varsigma^2 } \cdot \paren*{ \varsigma - \varsigma^2 + \frac{ a }{ b } \cdot \paren*{ 1 + \ln \varsigma } } \\
&= b \cdot \frac{ 1 + \ln \varsigma }{ \varsigma^2 } \cdot \paren*{ \frac{ \varsigma - \varsigma^2 }{ 1 + \ln \varsigma } + \frac{ a }{ b } } .
\end{align*}
Now, note from our definition of $c$ that $c' - \ln c' < 2$ for all $c' \in \co*{ 1, c }$. Thus, for all $\varsigma \in \co*{ 1, c }$, we have that $\frac{ \del \xi }{ \del \varsigma } \geq \frac{ b }{ \varsigma^2 } \cdot \paren*{ \frac{ a }{ b } - \varsigma } \cdot \paren*{ \varsigma - 1 }$. In particular, when $\frac{ a }{ b } \geq c$, we have that $\xi$ is increasing with $\varsigma$ and maximized at the right endpoint. At this point $R = 0$ and $s = \Delta + \frac{ b }{ c }$. It remains to consider the case $\frac{ a }{ b } < c$. In this case, setting $\varsigma = 1$ and $\varsigma = c$ in the equation above gives positive and negative values respectively. Moreover, the last factor in the above equation is decreasing and therefore, there is a unique value $\varsigma^*$ for which $\frac{ \del \xi }{ \del \varsigma } = 0$. This value maximizes $\xi$ and satisfies $\varsigma^2 - \varsigma = \frac{ a }{ b } \cdot \paren*{ 1 + \ln \varsigma }$ and implies $R = b \cdot \frac{ 2 - \varsigma^* + \ln \varsigma^* }{ \varsigma^* } = b \cdot \paren*{ \frac{ b }{ a } - \frac{ 1 }{ \varsigma^* } } \cdot \paren*{ \varsigma^* - 1 }$ and $s = \Delta + \frac{ b }{ \varsigma^* }$, finishing the proof.
\end{proof}

\subsubsection{Missing Proofs from \cref{sec:stackelberg:alice:lb}}
\label{app:stackelberg:alice:lb}

\begin{proof}[Proof of \cref{lemma:alice-structure-1:lb}]
We prove both parts separately.
\begin{enumerate}
\item We assume $v > R - \Delta$ as the result is trivial otherwise. Define $x = \frac{ R + a - v }{ b }$ and note that $x \in \co*{ \frac{ R }{ b }, 1 }$. In this notation, we have to show that $\utilityMechanism{ \AliceMechanism }{ v } \geq \paren*{ 1 - x } \cdot \utilityMechanism{ \AliceMechanism }{ a }$. Suppose not. Then, we have $\utilityMechanism{ \AliceMechanism }{ v } < \paren*{ 1 - x } \cdot \utilityMechanism{ \AliceMechanism }{ a } = \utilityMechanism{ \AliceMechanism }{ \vthreshold_{ \AliceMechanism }(x) }$ by \cref{eq:aug,lemma:sl-params}. By the monotonicity of utility, this implies that $v < \vthreshold_{ \AliceMechanism }(x)$. Next, note from $R \geq \Delta$ that $v \geq b - bx \geq a - ax$. Thus, \cref{def:sl-params} implies that $\p_{ \AliceMechanism }(x) > b - \frac{ a - v }{ x } \geq \residualValue{b}{a}$. From \cref{lemma:sl}, this means that for all $p \in \paren*{ \residualValue{b}{a}, \p_{ \AliceMechanism }(x) }$, when Bob uses the mechanism $\SL_{ x, p }$, the buyer goes to Bob first and pays him $xp$. By the assumption in the lemma, we get that $x \cdot \p_{ \AliceMechanism }(x) \leq R$. However, this means that $R > bx - a + v = R$, a contradiction.
\item For this part, we use \cref{item:dl:lb} of \cref{lemma:dl}. From the assumption in the lemma, we get $v + \Delta - \zmarg{ \AliceMechanism }^-\paren*{ v } \cdot \paren*{ v + \Delta - \Delta \cdot \x_{ \AliceMechanism }\paren*{ v + \Delta } } \leq R$. We now use \cref{def:dl-params} to get $v + \Delta - \zmarg{ \AliceMechanism }^-\paren*{ v } \cdot \paren*{ v + \Delta \cdot \frac{ v - \utilityMechanism{ \AliceMechanism }{ v } }{ a - \utilityMechanism{ \AliceMechanism }{ a } } } \leq R$. Rearranging gives the result.
\end{enumerate}
\end{proof}

\begin{proof}[Proof of \cref{lemma:alice-structure-2:lb}]
Recall that $\DL_{ 0, b } = \FP{b}$. Also, note from \cref{def:dl-params,eq:aug} that $\x_{ \AliceMechanism }(b) = 0$. With \cref{item:dl:lb} of \cref{lemma:dl}, this gives $R = \revDuo{ B }{ \AliceMechanism }{ \DL_{ 0, b } } = b \cdot \paren*{ 1 - \zmarg{ \AliceMechanism }^-\paren*{ a } }$. Now, assuming the buyer is going to Alice first, we have from \cref{item:buyer:payment} of \cref{lemma:buyer} that:
\[
\revDuo{ A }{ \AliceMechanism }{ \DL_{ 0, b } } = \AliceMechanism\paren*{ \zmarg{ \AliceMechanism }^-\paren*{ a } } = a \cdot \zmarg{ \AliceMechanism }^-\paren*{ a } - \utilityMechanism{ \AliceMechanism }{ a } = a \cdot \paren*{ 1 - \frac{ R }{ b } } - \utilityMechanism{ \AliceMechanism }{ a } .
\]
Assume now that the buyer goes to Bob first. As we prefer to tie-break in favor of going to Alice first, the buyer goes to Bob first only if he wants to buy Bob's item for a price of $b$ and get zero utility. As the buyer is utility maximizing, this means that $R = b$ and $\utilityMechanism{ \AliceMechanism }{ a } = 0$. As Alice gets no revenue in this case, the proof is complete.
\end{proof}

\begin{proof}[Proof of \cref{lemma:cap}]
Fix an arbitrary mechanism $\tilde{ \BobMechanism }$ for Bob. First, we consider the case when the buyer goes to Bob first when faced with the mechanisms $\tilde{ \AliceMechanism }$ and $\tilde{ \BobMechanism }$. In this case, we have by \cref{lemma:bob-first} that there exists $\tilde{x} \in \bracket*{ 0, 1 }$ such that $\revDuo{ B }{ \tilde{ \AliceMechanism } }{ \tilde{ \BobMechanism } } \leq \tilde{x} \cdot \p_{ \tilde{ \AliceMechanism } }\paren*{ \tilde{x} }$. We can assume $\tilde{x} > 0$ as otherwise there is nothing to show. By \cref{def:sl-params,lemma:sl-params}, we have that $\revDuo{ B }{ \tilde{ \AliceMechanism } }{ \tilde{ \BobMechanism } } \leq \tilde{x} \cdot b - a + \vthreshold_{ \tilde{ \AliceMechanism } }\paren*{ \tilde{x} }$, where $\vthreshold_{ \tilde{ \AliceMechanism } }\paren*{ \tilde{x} }$ is the largest value satisfying $\utilityMechanism{ \tilde{ \AliceMechanism } }{ \vthreshold_{ \tilde{ \AliceMechanism } }\paren*{ \tilde{x} } } = \utilityMechanism{ \tilde{ \AliceMechanism } }{ a } \cdot \paren*{ 1 - \tilde{x} }$. Next, we claim that for all $v \in \bracket*{ 0, a }$, the function below is non-negative.
\[
\frac{ \utilityMechanism{ \tilde{ \AliceMechanism } }{ v } }{ \utilityMechanism{ \tilde{ \AliceMechanism } }{ a } } - \frac{ \utilityMechanism{ \AliceMechanism }{ v } }{ \utilityMechanism{ \AliceMechanism }{ a } } .
\]
Indeed, for $v \geq 0$ such that $\zmarg{ \AliceMechanism }^{+}\paren*{ v } \leq x$, this is because \cref{eq:cap} implies that the numerators match while the first denominator is no larger than the second. For $v \geq 0$, such that $\zmarg{ \AliceMechanism }^{+}\paren*{ v } > x$, the fact that $\utilityMechanism{ \tilde{ \AliceMechanism } }{ v }$ is affine in this range implies that overall, the function is concave. Thus, we have a concave function that starts at a non-negative value in a given range and goes to $0$ at $v = a$. This is possible only if the function is non-negative throughout, as desired.

Setting $v = \vthreshold_{ \tilde{ \AliceMechanism } }\paren*{ \tilde{x} }$ and using the non-negativity of the function, we have that $\vthreshold_{ \tilde{ \AliceMechanism } }\paren*{ \tilde{x} } \leq \vthreshold_{ \AliceMechanism }\paren*{ \tilde{x} }$ from \cref{def:sl-params}. In turn, we have that $\tilde{x} \cdot \p_{ \tilde{ \AliceMechanism } }\paren*{ \tilde{x} } \leq \tilde{x} \cdot \p_{ \AliceMechanism }\paren*{ \tilde{x} }$. However, this is upper bounded by $R \leq b \cdot \paren*{ 1 - x }$. Indeed, if $\p_{ \AliceMechanism }\paren*{ \tilde{x} } \leq \residualValue{b}{a} = \Delta$, we are done because $R > \Delta$. If not, we are done by the fact that $R$ is the best revenue Bob can get when Alice uses $\AliceMechanism$ and \cref{lemma:sl}.

Now, consider the case when the buyer goes to Alice first. In this case, we have by \cref{lemma:dl,lemma:alice-first} that there exists $\tilde{q} \in \bracket*{ \residualValue{b}{a}, b }$ such that $\revDuo{ B }{ \tilde{ \AliceMechanism } }{ \tilde{ \BobMechanism } } \leq \revDuo{ B }{ \tilde{ \AliceMechanism } }{ \DL_{ \x_{ \tilde{ \AliceMechanism } }\paren*{ \tilde{q} }, \tilde{q} } } = \tilde{q} - \zmarg{ \tilde{ \AliceMechanism } }^-\paren*{ \tilde{q} - \Delta } \cdot \paren*{ \tilde{q} - \Delta \cdot \x_{ \tilde{ \AliceMechanism } }\paren*{ \tilde{q} } }$. By \cref{def:dl-params}, this can be rewritten as:
\[
\revDuo{ B }{ \tilde{ \AliceMechanism } }{ \tilde{ \BobMechanism } } \leq \tilde{q} - \zmarg{ \tilde{ \AliceMechanism } }^-\paren*{ \tilde{q} - \Delta } \cdot \paren*{ \tilde{q} - \Delta + \Delta \cdot \frac{ \tilde{q} - \Delta - \utilityMechanism{ \tilde{ \AliceMechanism } }{ \tilde{q} - \Delta } }{ a - \utilityMechanism{ \tilde{ \AliceMechanism } }{ a } } } .
\]
Setting $v = \tilde{q} - \Delta$ for convenience and using Myerson's payment identity, we get the following equivalent forms.
\begin{align*}
\revDuo{ B }{ \tilde{ \AliceMechanism } }{ \tilde{ \BobMechanism } } &\leq \Delta + v \cdot \paren*{ 1 - \zmarg{ \tilde{ \AliceMechanism } }^-\paren*{ v } } - \zmarg{ \tilde{ \AliceMechanism } }^-\paren*{ v } \cdot \Delta \cdot \frac{ v - \utilityMechanism{ \tilde{ \AliceMechanism } }{ v } }{ a - \utilityMechanism{ \tilde{ \AliceMechanism } }{ a } } \\
&= \Delta + v \cdot \paren*{ 1 - \zmarg{ \tilde{ \AliceMechanism } }^-\paren*{ v } } - \zmarg{ \tilde{ \AliceMechanism } }^-\paren*{ v } \cdot \Delta \cdot \frac{ v \cdot \paren*{ 1 - \zmarg{ \tilde{ \AliceMechanism } }^-\paren*{ v } } + \tilde{ \AliceMechanism }\paren*{ \zmarg{ \tilde{ \AliceMechanism } }^-\paren*{ v } } }{ a - \utilityMechanism{ \tilde{ \AliceMechanism } }{ a } } \\
&= \Delta + v \cdot \paren*{ 1 - \zmarg{ \tilde{ \AliceMechanism } }^-\paren*{ v } } \cdot \paren*{ 1 - \frac{ \zmarg{ \tilde{ \AliceMechanism } }^-\paren*{ v } \cdot \Delta }{ a - \utilityMechanism{ \tilde{ \AliceMechanism } }{ a } } } - \zmarg{ \tilde{ \AliceMechanism } }^-\paren*{ v } \cdot \Delta \cdot \frac{ \tilde{ \AliceMechanism }\paren*{ \zmarg{ \tilde{ \AliceMechanism } }^-\paren*{ v } } }{ a - \utilityMechanism{ \tilde{ \AliceMechanism } }{ a } } .
\end{align*}
Now, from the first form, note that for $v \geq 0$ such that $\zmarg{ \AliceMechanism }^-\paren*{ v } \leq x$, which means that $\zmarg{ \AliceMechanism }^-\paren*{ v } = \zmarg{ \tilde{ \AliceMechanism } }^-\paren*{ v }$, we have that {the expression above is weakly larger than the corresponding expression for $\AliceMechanism$, and the difference is nondecreasing in $v$.}
{Indeed, passing from $\AliceMechanism$ to $\tilde{\AliceMechanism}$ replaces the denominator $a-\utilityMechanism{ \AliceMechanism }{ a }$ by the weakly larger denominator $a-\utilityMechanism{ \tilde{\AliceMechanism} }{ a }$, so the subtracted fraction weakly decreases.} {Consequently, this change weakly increases the whole expression. Moreover, the amount of this increase is nondecreasing in $v$, because both $\zmarg{ \AliceMechanism }^-\paren*{v}$ and $v-\utilityMechanism{ \AliceMechanism }{ v }$ are nonnegative and nondecreasing.} By \cref{lemma:alice-first} and the definition of $x$, the corresponding expression for $\AliceMechanism$ was maximized at $v$ such that $\zmarg{ \AliceMechanism }^-\paren*{ v } = x$, {so the new expression at this $v$ is at least as large as its value at every smaller argument.}

From the third form, note that for $v \geq 0$ such that $\zmarg{ \AliceMechanism }^-\paren*{ v } \geq x$, which means that $\zmarg{ \tilde{ \AliceMechanism } }^-\paren*{ v } = x$, the expression increases with $v$. Indeed, the first and third terms do not change with $v$ while the second one has multipliers that are easily shown to be positive. Indeed, if they were negative, then the fact that setting $\tilde{ \BobMechanism } = \BobMechanism$ gives Bob revenue $R > \Delta$ would be a contradiction. Combining, we get that the above expression is maximized at $v = a$, which means (as $\zmarg{ \tilde{ \AliceMechanism } }^-\paren*{ a } = x$ by monotonicity and \cref{eq:cap}):
\[
\revDuo{ B }{ \tilde{ \AliceMechanism } }{ \tilde{ \BobMechanism } } \leq \Delta + a \cdot \paren*{ 1 - x } - x \cdot \Delta = b \cdot \paren*{ 1 - x } .
\]
This proves the first part of the lemma. For the ``moreover'' part, we simply invoke \cref{lemma:alice-structure-2:lb}.
\end{proof}

\begin{proof}[Proof of \cref{lemma:lb-equalizer}]
We assume $x > 0$ as the result is trivial otherwise. We first show that for all mechanisms $\BobMechanism$, it holds that $\revDuo{ B }{ \AliceMechanism_R^{ \mathsf{LB} } }{ \BobMechanism } \leq R$. Fix an arbitrary mechanism $\BobMechanism$ and consider first the case when the buyer interacts with Bob first when faced with the mechanisms $\AliceMechanism_R^{ \mathsf{LB} }$ and $\BobMechanism$. From \cref{lemma:bob-first}, we have that there exists $x' \in \bracket*{ 0, 1 }$ such that $\revDuo{ B }{ \AliceMechanism_R^{ \mathsf{LB} } }{ \BobMechanism } \leq x' \cdot \p_{ \AliceMechanism_R^{ \mathsf{LB} } }\paren*{ x' }$. As the claim is trivial otherwise, we assume that $x' > 0$. From \cref{def:sl-params}, we have that $\revDuo{ B }{ \AliceMechanism_R^{ \mathsf{LB} } }{ \BobMechanism } \leq bx' - a + \vthreshold_{ \AliceMechanism_R^{ \mathsf{LB} } }\paren*{ x' }$. Thus, it suffices to show that $\vthreshold_{ \AliceMechanism_R^{ \mathsf{LB} } }\paren*{ x' } \leq R + a - bx'$. For this, recall \cref{def:sl-params} and note that, for all $v \in \oc*{ R + a - bx', a }$, we have $R - \Delta < v \leq a$ implying by \cref{def:lb-equalizer} that $\zmarg{ \AliceMechanism_R^{ \mathsf{LB} } }^-\paren*{ v } \geq u$
\[
\utilityMechanism{ \AliceMechanism_R^{ \mathsf{LB} } }{ v } > \utilityMechanism{ \AliceMechanism_R^{ \mathsf{LB} } }{ R + a - bx' } \geq b \cdot \paren*{ 1 - x' } \cdot u = \paren*{ 1 - x' } \cdot \utilityMechanism{ \AliceMechanism_R^{ \mathsf{LB} } }{ a } ,
\] 
This means that $v > \vthreshold_{ \AliceMechanism_R^{ \mathsf{LB} } }\paren*{ x' }$, as desired. {For use in the proof of
\cref{thm:introduction-variable-order-rephrased},
we note that
the same Bob-first argument applies to any mechanism
$\AliceMechanism$ whose utility agrees with that of
$\AliceMechanism_R^{\mathsf{LB}}$ on $\bracket*{s_u,a}$.
Indeed, utility matching gives
$\utilityMechanism{ \AliceMechanism }{ a }=bu$ and
$\zmarg{ \AliceMechanism }^{+}\paren*{s_u}=u$,
so convexity yields
\[
\utilityMechanism{ \AliceMechanism }{ v }
\geq \utilityMechanism{ \AliceMechanism }{ s_u }
     +u(v-s_u)
=(v-R+\Delta)u
\qquad\forall v\in\bracket*{0,a}.
\]
Since $u>0$, this yields the same strict utility inequality
whenever $v>R+a-bx'$.} We return to the proof. Now, consider the case when the buyer interacts with Alice first. Here, by \cref{lemma:alice-first,lemma:dl}, we have that there exists $q \in \bracket*{ \residualValue{b}{a}, b }$ such that $\revDuo{ B }{ \AliceMechanism_R^{ \mathsf{LB} } }{ \BobMechanism } \leq \revDuo{ B }{ \AliceMechanism_R^{ \mathsf{LB} } }{ \DL_{ \x_{ \AliceMechanism_R^{ \mathsf{LB} } }(q), q } } = q - \zmarg{ \AliceMechanism_R^{ \mathsf{LB} } }^-\paren*{ q - \Delta } \cdot \paren*{ q - \Delta \cdot \x_{ \AliceMechanism_R^{ \mathsf{LB} } }(q) }$. By \cref{def:dl-params} and setting $v = q - \Delta$, this gives $\revDuo{ B }{ \AliceMechanism_R^{ \mathsf{LB} } }{ \BobMechanism } \leq \Delta + v - \zmarg{ \AliceMechanism_R^{ \mathsf{LB} } }^-\paren*{ v } \cdot \paren*{ v + \Delta \cdot \frac{ v - \utilityMechanism{ \AliceMechanism_R^{ \mathsf{LB} } }{ v } }{ a - bu } }$. By \cref{def:lb-equalizer}, this gives $\revDuo{ B }{ \AliceMechanism_R^{ \mathsf{LB} } }{ \BobMechanism } \leq \Delta + v - v + R - \Delta = R$.

Next, note that either $\aug_{ \AliceMechanism_R^{ \mathsf{LB} } }\paren*{ b, a } > 0$ or $R = b$. In either case, \cref{lemma:buyer,lemma:alice-structure-2:lb} imply
\begin{align*}
\revDuo{ B }{ \AliceMechanism_R^{ \mathsf{LB} } }{ \FP{b} } &= R, \\
\revDuo{ A }{ \AliceMechanism_R^{ \mathsf{LB} } }{ \FP{b} } &= a \cdot \paren*{ 1 - \frac{ R }{ b } } - \utilityMechanism{ \AliceMechanism_R^{ \mathsf{LB} } }{ a } = ax - bu.
\end{align*}
As the latter is the highest payment any buyer can make to Alice and as Bob tie-breaks in favor of Alice's revenue, even if his best response is not $\FP{b}$, it is impossible for Alice to get a higher payment and the proof is complete.
\end{proof}

We are now ready to prove that Alice's optimal revenue when $a < b$ can be achieved by a Large-Bob equalizer.
\paragraph{Proving the optimality of Large-Bob equalizer mechanisms when $a < b$.}
We begin by bounding Alice's optimal revenue. Let $\AliceMechanism$ and $\BobMechanism$ be an arbitrary pair of mechanisms for Alice and Bob respectively \ygedit{with $\BobMechanism$ a best response to $\AliceMechanism$}. 
\ygedit{By \cref{lemma:lb-equalizer} and the construction at the end of this proof, Alice can earn at least $\frac{a^2}{4b}$. It therefore suffices to bound pairs giving her at least this revenue.}
Let $R = \revDuo{ B }{ \AliceMechanism }{ \BobMechanism }$ be Bob's \ygedit{best-response revenue}. Note that $R \in \bracket*{ \Delta, b }$.
We first deal with the easy cases where either $\utilityMechanism{ \AliceMechanism }{ a } \in \set*{ 0, a }$. For this, observe first that if $\utilityMechanism{ \AliceMechanism }{ a } = a$, then Alice is giving her item for free and gets no revenue\ygedit{, a contradiction to our revenue assumption (that it is
{at least} $\frac{a^2}{4b}$).} Next, suppose that $\utilityMechanism{ \AliceMechanism }{ a } = 0$. In this case, Bob can get a revenue of $b$ by using the mechanism $\FP{b}$ and Alice will get nothing. As this is the only mechanism that yields Bob a revenue of $b$, this is indeed what Bob will do and Alice will get zero revenue, \ygedit{leading to a similar contradiction.} %

For the rest of the proof, we assume that $0 < \utilityMechanism{ \AliceMechanism }{ a } < a$. We first claim that the buyer must go to Alice first. Indeed, suppose for contradiction that the buyer goes to Bob first. By \cref{lemma:bob-first}, we have that there exists $x \in \bracket*{ 0, 1 }$ such that $R \leq x \cdot \p_{ \AliceMechanism }(x)$. We claim that $0 < x < 1$. This is because, if $x = 0$, Bob will get zero revenue contradicting the fact that $R \geq \Delta > 0$ and if $x = 1$, Bob will always sell the item and as $b > a$, this implies that Alice gets no revenue, \ygedit{contradicting our revenue assumption}. This means that we can apply \cref{lemma:sl} and get $R = x \cdot \p_{ \AliceMechanism }(x)$. From \cref{def:sl-params}, this means that $R = bx - a + \vthreshold_{ \AliceMechanism }(x)$. Using \cref{lemma:sl-params}, we have that $\utilityMechanism{ \AliceMechanism }{ \vthreshold_{ \AliceMechanism }(x) } = \frac{ \vthreshold_{ \AliceMechanism }(x) + \Delta - R }{ b } \cdot \utilityMechanism{ \AliceMechanism }{ a }$. It follows that $\vthreshold_{ \AliceMechanism }(x) < a$. Moreover, the inequality in \cref{item:alice-structure-1:lb:1} of \cref{lemma:alice-structure-1:lb} is tight when $v = \vthreshold_{ \AliceMechanism }(x)$. As the inequality holds in a neighborhood around $v = \vthreshold_{ \AliceMechanism }(x)$, we get that $\zmarg{ \AliceMechanism }^-\paren*{ \vthreshold_{ \AliceMechanism }(x) } \leq \frac{ \utilityMechanism{ \AliceMechanism }{ a } }{ b }$. We now combine this with \cref{item:alice-structure-1:lb:2} of \cref{lemma:alice-structure-1:lb} to get $\frac{ v + \Delta - R }{ v + \Delta \cdot \frac{ v - \utilityMechanism{ \AliceMechanism }{ v } }{ a - \utilityMechanism{ \AliceMechanism }{ a } } } \leq \frac{ \utilityMechanism{ \AliceMechanism }{ a } }{ b }$ for $v = \vthreshold_{ \AliceMechanism }(x)$. As the denominator is (strictly) increasing in $v$ and $\vthreshold_{ \AliceMechanism }(x) < a$, we get $1 - x < \frac{ \utilityMechanism{ \AliceMechanism }{ a } }{ b }$. Now, as $a < b$, observe that the probability of the buyer going to Alice afterwards is just the probability that he does not get the item from Bob, which is $1 - x$. When he does go to Alice, the payment he makes to her is at most $\AliceMechanism\paren*{ \zmarg{ \AliceMechanism }^{+}\paren*{ a } } \leq a - \utilityMechanism{ \AliceMechanism }{ a }$. Thus, Alice's expected revenue is:
\[
\revDuo{ A }{ \AliceMechanism }{ \BobMechanism } \leq \paren*{ 1 - x } \cdot \paren*{ a - \utilityMechanism{ \AliceMechanism }{ a } } < \frac{ \utilityMechanism{ \AliceMechanism }{ a } }{ b } \cdot \paren*{ a - \utilityMechanism{ \AliceMechanism }{ a } } \leq \frac{ a^2 }{ 4b } .
\]
Thus, \ygedit{our revenue assumption is contradicted}. This means that the buyer goes to Alice first.

In this case, we first show that $R = \Delta$ cannot happen. Suppose it does. Then, we have from \cref{item:alice-structure-1:lb:2} of \cref{lemma:alice-structure-1:lb} that for all $v \in \oc*{ 0, a }$, we have $v \leq \zmarg{ \AliceMechanism }^-\paren*{ v } \cdot \paren*{ v + \frac{ \Delta }{ a - \utilityMechanism{ \AliceMechanism }{ a } } \cdot \paren*{ v - \utilityMechanism{ \AliceMechanism }{ v } } }$. Let $\theta = \inf_{ v \in \oc*{ 0, a } } \zmarg{ \AliceMechanism }^-\paren*{ v }$ and observe that for all $v \in \oc*{ 0, a }$, we have $\utilityMechanism{ \AliceMechanism }{ v } \geq \theta v$. Plugging this in, and dividing by $v$, we get that for all $v \in \oc*{ 0, a }$, we have $1 \leq \zmarg{ \AliceMechanism }^-\paren*{ v } \cdot \paren*{ 1 + \frac{ \Delta }{ a - \utilityMechanism{ \AliceMechanism }{ a } } \cdot \paren*{ 1 - \theta } }$. As $v \in \oc*{ 0, a }$ is arbitrary, it must be the case that $1 \leq \theta \cdot \paren*{ 1 + \frac{ \Delta }{ a - \utilityMechanism{ \AliceMechanism }{ a } } \cdot \paren*{ 1 - \theta } }$ which implies that either $\theta = 1$ or we have $1 \leq \frac{ \theta \Delta }{ a - \utilityMechanism{ \AliceMechanism }{ a } }$. The former implies that Alice gets zero revenue, so we assume the latter. Next, note from the fact that \cref{lemma:alice-first} is tight and \cref{lemma:dl} that we have $\effectiveValue{a}{\BobMechanism} = \zmarg{ \AliceMechanism }^-\paren*{ \effectiveValue{a}{\BobMechanism} } \cdot \paren*{ \effectiveValue{a}{\BobMechanism} + \frac{ \Delta }{ a - \utilityMechanism{ \AliceMechanism }{ a } } \cdot \paren*{ \effectiveValue{a}{\BobMechanism} - \utilityMechanism{ \AliceMechanism }{ \effectiveValue{a}{\BobMechanism} } } }$. By Myerson's payment identity, this gives $\effectiveValue{a}{\BobMechanism} \cdot \paren*{ 1 - \zmarg{ \AliceMechanism }^-\paren*{ \effectiveValue{a}{\BobMechanism} } } \cdot \paren*{ 1 - \frac{ \zmarg{ \AliceMechanism }^-\paren*{ \effectiveValue{a}{\BobMechanism} } \cdot \Delta }{ a - \utilityMechanism{ \AliceMechanism }{ a } } } = \frac{ \zmarg{ \AliceMechanism }^-\paren*{ \effectiveValue{a}{\BobMechanism} } \cdot \Delta }{ a - \utilityMechanism{ \AliceMechanism }{ a } } \cdot \AliceMechanism\paren*{ \zmarg{ \AliceMechanism }^-\paren*{ \effectiveValue{a}{\BobMechanism} } }$. \ygedit{Alice gets positive revenue by assumption}, so, by \cref{item:buyer:payment} of \cref{lemma:buyer}, that revenue must be $\AliceMechanism\paren*{ \zmarg{ \AliceMechanism }^-\paren*{ \effectiveValue{a}{\BobMechanism} } }$ and the right hand side must be positive and $\effectiveValue{a}{\BobMechanism} > 0$. It follows that the left hand side must be positive as well, contradicting our bound on $\theta$.

Now that we know $R > \Delta$, let $x \in \bracket*{ 0, 1 }$ be the option the buyer chooses from Alice. As Alice gets non-zero revenue \ygedit{(by assumption)} and $R > \Delta$, we have $0 < x < 1$.
Individual rationality at $\Delta$ and incentive compatibility
between $\Delta$ and $b$ bound Bob's respective payments by
$\Delta\zmarg{\BobMechanism}^{+}(\Delta)$ and
$b\zmarg{\BobMechanism}^{+}(b)-a\zmarg{\BobMechanism}^{+}(\Delta)$.
Hence
$
R\leq b(1-x)\zmarg{\BobMechanism}^{+}(b)+(bx-a)\zmarg{\BobMechanism}^{+}(\Delta)
\leq b(1-x)+\max\{bx-a,0\}$. 
In particular, this means that $x < \frac{ a }{ b }$ as otherwise we have $R \leq \Delta < R$, a contradiction. This means that $R \leq b \cdot \paren*{ 1 - x }$. As stated in \cref{sec:stackelberg:alice:lb}, the next step in our proof is to construct a capped mechanism $\tilde{ \AliceMechanism }$ using \cref{eq:cap}. From \cref{lemma:cap}, we have that this capped mechanism has the property that \ygedit{$\DL_{0,b}$ is a best response to $\tilde{\AliceMechanism}$}. (Recall that $\DL_{ 0, b } = \FP{b}$ is just the fixed price mechanism with price $b$.) Furthermore, \ygedit{for this pair}, Bob's revenue is $\revDuo{ B }{ \tilde{ \AliceMechanism } }{ \DL_{ 0, b } } = b \cdot \paren*{ 1 - x }$ and Alice's revenue is $\revDuo{ A }{ \tilde{ \AliceMechanism } }{ \DL_{ 0, b } } = ax - \utilityMechanism{ \tilde{ \AliceMechanism } }{ a }$.

\ygedit{Since $x$ was chosen at effective value at most $a$, it remains optimal at $a$ among retained options, giving $\utilityMechanism{\tilde{\AliceMechanism}}{a}=ax-\AliceMechanism(x)$. Hence} we have $\revDuo{ A }{ \AliceMechanism }{ \BobMechanism } = \revDuo{ A }{ \tilde{ \AliceMechanism } }{ \DL_{ 0, b } }$. We now upper bound $\revDuo{ A }{ \tilde{ \AliceMechanism } }{ \DL_{ 0, b } }$ by giving a lower bound on $\utilityMechanism{ \tilde{ \AliceMechanism } }{ a }$. Specifically, we will show that $\utilityMechanism{ \tilde{ \AliceMechanism } }{ a } \geq bu$, where $u = u(x) \in \paren*{ 0, \frac{ ax }{ b } }$ is the unique value that satisfies:
\begin{equation}
\label{eq:u:lb}
\ln \frac{ a - \paren*{ 2b - a } u }{ a - bx } = \int_u^x \frac{ 1 + \alpha_u - \alpha_u y }{ \paren*{ 1 - y } \paren*{ 1 - \alpha_u y } } \del y , \qquad \text{where} \qquad \alpha_u = \frac{ b - a }{ a - bu } .
\end{equation}
For all $u \in \oc*{ 0, \frac{ ax }{ b } }$, define:
\begin{equation}
\label{eq:s:lb}
s_u = \paren*{ a - bx } \cdot \frac{ a - \paren*{ b - a } u }{ a - \paren*{ 2b - a } u } .
\end{equation}
Note that for all $u \in \bracket*{ 0, \frac{ ax }{ b } }$, we have $\alpha_u \in \paren*{ 0, 1/x }$. Indeed, observe that $\alpha_u$ is increasing in $u$ and setting $u = 0$ gives $\alpha_u > 0$ and setting $u = \frac{ ax }{ b } < \frac{ a^2 }{ b^2 }$ gives $\alpha_u < \frac{ b^2 - ab }{ ab - a^2 } = \frac{ b }{ a } < \frac{ 1 }{ x }$. For these values of $u$, we also have $a - \paren*{ 2b - a } u > \frac{ a }{ b^2 } \cdot \paren*{ b - a }^2 > 0$. This means that all the factors above are positive for all $u \in \bracket*{ 0, \frac{ ax }{ b } }$. This also means that $s_u > a - bx$ for all $u \in \oc*{ 0, \frac{ ax }{ b } }$. To upper bound $s_u$, note that $s_u + bx - a = \paren*{ a - bx } \cdot \frac{ bu }{ a - \paren*{ 2b - a } u } < \paren*{ a - bx } \cdot \frac{ bx }{ b - \paren*{ 2b - a } x } < bx$ by cross-multiplying and the fact that $u < \frac{ ax }{ b }$. Thus, we have $s_u \in \oc*{ a - bx, a }$ for $u \in \oc*{ 0, \frac{ ax }{ b } }$.

To see why $u$ exists and is unique, let $\zeta_x(u)$ denote the left hand side of \cref{eq:u:lb} above minus the right hand side. We show in \cref{lemma:zeta:lb} that the derivative of $\zeta_x(u)$ is negative whenever $a < b$. Thus, if such a $u$ exists it must be unique. To see why such a $u$ exists, we also show in \cref{lemma:zeta:lb} that $\zeta_x(u)$ is (strictly) decreasing and that $\zeta_x(0) > 0$ and $\zeta_x\paren*{ \frac{ ax }{ b } } < 0$. Combining, we get that such a $u$ exists and is unique. 
\begin{lemma}
\label{lemma:zeta:lb}
For all $u \in \paren*{ 0, \frac{ ax }{ b } }$, we have:
\[
\frac{ \partial \zeta_x(u) }{ \partial u } = - \frac{ b - a }{ \paren*{ 1 - u } \paren*{ a - \paren*{ 2b - a } u } } - \frac{ b \paren*{ b - a } }{ \paren*{ a - bu }^2 } \cdot \int_u^x \frac{ 1 }{ \paren*{ 1 - y } \paren*{ 1 - \alpha_u y }^2 } \del y < 0 .
\]
Moreover, we have $\zeta_x(0) > 0$ and $\zeta_x\paren*{ \frac{ ax }{ b } } < 0$.
\end{lemma}

For the rest of the proof, we define $\tilde{u} = \frac{ \utilityMechanism{ \tilde{ \AliceMechanism } }{ a } }{ b }$. Note that $\tilde{u} \in \oc*{ 0, \frac{ ax }{ b } }$ due to \cref{eq:cap}. From \cref{item:alice-structure-1:lb:1} of \cref{lemma:alice-structure-1:lb}, we have that $\utilityMechanism{ \tilde{ \AliceMechanism } }{ s_{ \tilde{u} } } \geq \paren*{ s_{ \tilde{u} } + bx - a } \cdot \tilde{u}$. From \cref{item:alice-structure-1:lb:2} of \cref{lemma:alice-structure-1:lb}, we have for all $v \in \oc*{ 0, a }$ that:
\[
\zmarg{ \tilde{ \AliceMechanism } }^{+}\paren*{ v } \geq \frac{ v + bx - a }{ v + \alpha_{ \tilde{u} } \cdot \paren*{ v - \utilityMechanism{ \tilde{ \AliceMechanism } }{ v } } } ,
\]
as $\alpha_{ \tilde{u} } = \frac{ b - a }{ a - b \tilde{u} }$. In \cref{lemma:solve-ode}, we show that there is a smooth function $U(v)$ that satisfies $U\paren*{ s_{ \tilde{u} } } = \paren*{ s_{ \tilde{u} } + bx - a } \cdot \tilde{u}$ and makes the above inequality tight. Formally, for all $v \in \bracket*{ s_{ \tilde{u} }, a }$, letting $z(v) = \frac{ \del U(v) }{ \del v }$, we have:
\[
z(v) = \frac{ v + bx - a }{ v + \alpha_{ \tilde{u} } \cdot \paren*{ v - U(v) } } .
\]
\begin{lemma}
\label{lemma:solve-ode}
There exists a smooth function $U : \bracket*{ s_{ \tilde{u} }, a } \to \mathbb{R}$ such that $U\paren*{ s_{ \tilde{u} } } = \paren*{ s_{ \tilde{u} } + bx - a } \cdot \tilde{u}$ and, denoting $z(v) = \frac{ \del U(v) }{ \del v }$, we have for all $v \in \bracket*{ s_{ \tilde{u} }, a }$ that:
\begin{equation}
\label{eq:solve-ode}
z(v) = \frac{ v + bx - a }{ v + \alpha_{ \tilde{u} } \cdot \paren*{ v - U(v) } } .
\end{equation}
Moreover, we have $u \leq \tilde{u} \iff U(a) \leq b \tilde{u}$ and the left inequality is tight if and only if the right one is.
\end{lemma}

\ygedit{The capped utility is Lipschitz, hence absolutely continuous,
and satisfies the differential inequality almost everywhere.
$U(v)\leq v$ and $\utilityMechanism{\tilde{\AliceMechanism}}{v}\leq v$
for $v\in\bracket*{s_{\tilde{u}},a}$, so the denominators
are bounded below by $s_{\tilde{u}}>0$. The right-hand side
of \cref{eq:solve-ode} is locally Lipschitz and increasing in utility.
Thus, we can apply the standard comparison theorem.
Moreover, monotonicity in utility makes
$\utilityMechanism{\tilde{\AliceMechanism}}{v}-U(v)$
nondecreasing, so equality at $a$ forces equality throughout.}
In particular,
we get that $U(a) \leq \utilityMechanism{ \tilde{ \AliceMechanism } }{ a } = b \tilde{u}$. From the ``moreover'' part of \cref{lemma:solve-ode}, this means that $u \leq \tilde{u}$. It follows that 
\[
\revDuo{ A }{ \tilde{ \AliceMechanism } }{ \DL_{ 0, b } } \leq ax - bu .
\]
Recall that $u = u(x)$ is defined in terms of $x$ in \cref{eq:u:lb}. As $x$ approaches $0$, so does $u$ and the right hand side above also approaches $0$. As $x$ approaches $\frac{ a }{ b }$, the left hand side of \cref{eq:u:lb} shoots to infinity. Thus, so should the right hand side, which is only possible if $u$ approaches $\frac{ a^2 }{ b^2 }$ so that $\alpha_u$ approaches $\frac{ 1 }{ x }$. This means that Alice's revenue again approaches $0$. By continuity, we get that there must be a value $x^* \in \paren*{ 0, \frac{ a }{ b } }$ such that $\left. \frac{ \del u }{ \del x } \right|_{ x = x^* } = \frac{ a }{ b }$. We will show that at any such point, we have $\left. \frac{ \del^2 u }{ \del x^2 } \right|_{ x = x^* } > 0$. This implies that any such point is a maximizer and also implies that the maximizer is unique. Indeed, if not, then there would be a minimizer sitting between two maximizers for which the double derivative will have the wrong sign.

To this end, first apply the multivariable chain rule to \cref{eq:u:lb} and get $0 = \frac{ \del \zeta_x(u) }{ \del x } = \frac{ \partial \zeta_x(u) }{ \partial x } + \frac{ \partial \zeta_x(u) }{ \partial u } \cdot \frac{ \del u }{ \del x }$. This means that
\begin{align*}
- \frac{ \partial \zeta_x(u) }{ \partial u } \cdot \frac{ \del u }{ \del x } &= \frac{ \partial \zeta_x(u) }{ \partial x } \\
&= \frac{ b }{ a - bx } - \frac{ 1 + \alpha_u - \alpha_u x }{ \paren*{ 1 - x } \paren*{ 1 - \alpha_u x } } \tag{\cref{eq:u:lb}} \\
&= \frac{ b - a - a \alpha_u \cdot \paren*{ 1 - x } }{ \paren*{ a - bx } \paren*{ 1 - x } \paren*{ 1 - \alpha_u x } } \\
&= \frac{ \alpha_u \paren*{ ax - bu } }{ \paren*{ a - bx } \paren*{ 1 - x } \paren*{ 1 - \alpha_u x } } \tag{As $\alpha_u = \frac{ b - a }{ a - bu }$} .
\end{align*}
Now, define:
\begin{equation}
\label{eq:kappa:lb}
\kappa_x(u) = \frac{ \paren*{ a - bx } \paren*{ 1 - x } \paren*{ 1 - \alpha_u x } }{ \alpha_u } \cdot \frac{ \partial \zeta_x(u) }{ \partial u } .
\end{equation} 
This allows us to rearrange the above equation and get $\frac{ \del u }{ \del x } = - \frac{ ax - bu }{ \kappa_x(u) }$. Differentiating, we get:
\begin{align*}
\frac{ \del^2 u }{ \del x^2 } &= - \frac{ a - b \frac{ \del u }{ \del x } }{ \kappa_x(u) } + \frac{ ax - bu }{ \kappa_x(u)^2 } \frac{ \del \kappa_x(u) }{ \del x } \\
&= - \frac{ a - b \frac{ \del u }{ \del x } }{ \kappa_x(u) } + \frac{ ax - bu }{ \kappa_x(u)^2 } \cdot \paren*{ \frac{ \partial \kappa_x(u) }{ \partial x } + \frac{ \partial \kappa_x(u) }{ \partial u } \frac{ \del u }{ \del x } } \tag{Multivariable chain rule} .
\end{align*}
For any point $x^*$ satisfying $\left. \frac{ \del u }{ \del x } \right|_{ x = x^* } = \frac{ a }{ b }$, we get:
\begin{align*}
\left. \frac{ \del^2 u }{ \del x^2 } \right|_{ x = x^* } &= \left. \frac{ ax - bu }{ \kappa_x(u)^2 } \cdot \paren*{ \frac{ \partial \kappa_x(u) }{ \partial x } + \frac{ \partial \kappa_x(u) }{ \partial u } \frac{ \del u }{ \del x } } \right|_{ x = x^* } \\
&= \left. \frac{ ax - bu }{ \kappa_x(u)^2 } \cdot \paren*{ \frac{ \partial \kappa_x(u) }{ \partial x } + \frac{ \partial \kappa_x(u) }{ \partial u } \frac{ a }{ b } } \right|_{ x = x^* } .
\end{align*}
Let $u^* = u\paren*{ x^* }$ and $\xi = ax^* - b u^* > 0$. Note that, by the multivariable chain rule, we have $\frac{ \del \kappa_x\paren*{ \frac{ ax }{ b } - \frac{ \xi }{ b } } }{ \del x } = \left. \frac{ \partial \kappa_x(u) }{ \partial x } + \frac{ \partial \kappa_x(u) }{ \partial u } \frac{ a }{ b } \right|_{ u = \frac{ ax }{ b } - \frac{ \xi }{ b } }$. Setting $x = x^*$ in this, we get: 
\[
\left. \frac{ \del^2 u }{ \del x^2 } \right|_{ x = x^* } = \frac{ \xi }{ \kappa_{ x^* }\paren*{ u^* }^2 } \cdot \left. \frac{ \del \kappa_x\paren*{ \frac{ ax }{ b } - \frac{ \xi }{ b } } }{ \del x } \right|_{ x = x^* } .
\]
Thus, it suffices to show that for all $\frac{ \del \kappa_x\paren*{ \frac{ ax }{ b } - \frac{ \xi }{ b } } }{ \del x }$ is always positive. This is done in \cref{lemma:kappa:lb} below.

\begin{lemma}
\label{lemma:kappa:lb}
For all $\xi > 0$ and $\frac{ \xi }{ a } < x < \frac{ a }{ b }$, we have $\frac{ \del \kappa_x\paren*{ \frac{ ax }{ b } - \frac{ \xi }{ b } } }{ \del x } > 0$.
\end{lemma}

Overall, we get that, if $x^*$ denotes the (unique) maximizer of $ax - bu(x)$, then we have 
\begin{equation}
\label{eq:opt:lb}
\opt{ a }{ b } \leq a x^* - b u \paren*{ x^* }.
\end{equation}

To finish the proof, we show that the bound on $\opt{ a }{ b }$ above is tight. For this, we recall Large-Bob-equalizer mechanisms defined in \cref{def:lb-equalizer}. Recall that \cref{lemma:lb-equalizer} shows that, for all $x \in \co*{ 0, \frac{ a }{ b } }$, \ygedit{Alice can earn $ax-bu$ against Bob's selected best response}. This is the same quantity we optimized to get \cref{eq:opt:lb}, and thus the bound there is tight. \ygedit{As in the Large-Alice case, the bound for every best response makes the maximizing equalizer a Stackelberg equilibrium.} Finally, we \ygedit{verify the revenue guarantee used above} by showing that there is a mechanism that gives her revenue at least $\frac{ a^2 }{ 4b } > 0$. For this, consider the Large-Bob-equalizer mechanism with $R = \Delta + \frac{ a }{ 2 }$ so that $x = 1 - \frac{ \Delta + a/2 }{ b } = \frac{ a }{ 2b }$. For this value of $x$, \cref{def:lb-equalizer} implies that $u \leq \frac{ ax }{ 2b }$ as Alice's utility at $a$ is at most the largest value of her allocation multiplied by $b - R$, which is the length of the interval where her allocation is non-zero. This means that Alice's revenue is at least $\frac{ ax }{ 2 } = \frac{ a^2 }{ 4b }$.

\begin{proof}[Proof of \cref{lemma:zeta:lb}]
Differentiating the expression for $\zeta_x(u)$ gives (keeping $x$ fixed):
\begin{align*}
\frac{ \partial \zeta_x(u) }{ \partial u } &= - \frac{ 2b - a }{ a - \paren*{ 2b - a } u } + \frac{ 1 + \alpha_u - \alpha_u u }{ \paren*{ 1 - u } \paren*{ 1 - u \alpha_u } } - \frac{ \del \alpha_u }{ \del u } \cdot \int_u^x \frac{ 1 }{ \paren*{ 1 - y } \paren*{ 1 - \alpha_u y }^2 } \del y \\
&= - \frac{ 2b - a }{ a - \paren*{ 2b - a } u } + \frac{ 1 + \alpha_u - \alpha_u u }{ \paren*{ 1 - u } \paren*{ 1 - u \alpha_u } } - \frac{ b \paren*{ b - a } }{ \paren*{ a - bu }^2 } \cdot \int_u^x \frac{ 1 }{ \paren*{ 1 - y } \paren*{ 1 - \alpha_u y }^2 } \del y \tag{As $\alpha_u = \frac{ b - a }{ a - bu }$} \\
&= - \frac{ 2b - a }{ a - \paren*{ 2b - a } u } + \frac{ b - \paren*{ 2b - a } u }{ \paren*{ 1 - u } \paren*{ a - \paren*{ 2b - a } u } } - \frac{ b \paren*{ b - a } }{ \paren*{ a - bu }^2 } \cdot \int_u^x \frac{ 1 }{ \paren*{ 1 - y } \paren*{ 1 - \alpha_u y }^2 } \del y \tag{As $\alpha_u = \frac{ b - a }{ a - bu }$} \\
&= - \frac{ b - a }{ \paren*{ 1 - u } \paren*{ a - \paren*{ 2b - a } u } } - \frac{ b \paren*{ b - a } }{ \paren*{ a - bu }^2 } \cdot \int_u^x \frac{ 1 }{ \paren*{ 1 - y } \paren*{ 1 - \alpha_u y }^2 } \del y \\
&< 0 ,
\end{align*}
as all the factors here are positive, so the last term is negative. For the ``moreover'' part, we have:
\begin{align*}
\zeta_x(0) &= \ln \frac{ a }{ a - bx } - \int_0^x \frac{ 1 + \alpha_0 - \alpha_0 y }{ \paren*{ 1 - y } \paren*{ 1 - \alpha_0 y } } \del y \\
&= \int_0^x \frac{ b }{ a - by } \del y - \int_0^x \frac{ 1 + \alpha_0 - \alpha_0 y }{ \paren*{ 1 - y } \paren*{ 1 - \alpha_0 y } } \del y \\
&= \int_0^x \frac{ b - a - \alpha_0 a + \alpha_0 ay }{ \paren*{ 1 - y } \paren*{ a - by } \paren*{ 1 - \alpha_0 y } } \del y \\
&= \int_0^x \frac{ \alpha_0 ay }{ \paren*{ 1 - y } \paren*{ a - by } \paren*{ 1 - \alpha_0 y } } \del y \tag{As $\alpha_0 = \frac{ b - a }{ a }$} \\
&> 0 ,
\end{align*}
as all the factors here are positive. We also have:
\begin{align*}
\zeta_x\paren*{ \frac{ ax }{ b } } &= \ln \frac{ a - \paren*{ 2b - a } \frac{ ax }{ b } }{ a - bx } - \int_{ \frac{ ax }{ b } }^x \frac{ 1 + \alpha - \alpha y }{ \paren*{ 1 - y } \paren*{ 1 - \alpha y } } \del y , \qquad \text{where} \qquad \alpha = \frac{ b - a }{ a - ax } \\
&= \alpha \cdot \int_{ \frac{ ax }{ b } }^x \frac{ 1 }{ 1 - \alpha y } \del y - \int_{ \frac{ ax }{ b } }^x \frac{ 1 + \alpha - \alpha y }{ \paren*{ 1 - y } \paren*{ 1 - \alpha y } } \del y \tag{As $\alpha = \frac{ b - a }{ a - ax }$} \\
&= - \int_{ \frac{ ax }{ b } }^x \frac{ 1 }{ \paren*{ 1 - y } \paren*{ 1 - \alpha y } } \del y \\
&< 0 ,
\end{align*}
as all the factors here are positive.
\end{proof}

\begin{proof}[Proof of \cref{lemma:solve-ode}]
We first show that $U\paren*{ s_{ \tilde{u} } } = \paren*{ s_{ \tilde{u} } + bx - a } \cdot \tilde{u}$ implies that $z\paren*{ s_{ \tilde{u} } } = \tilde{u}$. Indeed, we have:
\begin{align*}
z\paren*{ s_{ \tilde{u} } } &= \frac{ s_{ \tilde{u} } + bx - a }{ s_{ \tilde{u} } + \alpha_{ \tilde{u} } \cdot \paren*{ s_{ \tilde{u} } - \paren*{ s_{ \tilde{u} } + bx - a } \cdot \tilde{u} } } \\
&= \frac{ b \tilde{u} }{ \paren*{ a - \paren*{ b - a } \tilde{u} } + \alpha_{ \tilde{u} } \cdot \paren*{ \paren*{ a - \paren*{ b - a } \tilde{u} } - b \tilde{u}^2 } } \tag{\cref{eq:s:lb}} \\
&= \frac{ b \tilde{u} \cdot \paren*{ a - b \tilde{u} } }{ \paren*{ a - \paren*{ b - a } \tilde{u} } \paren*{ b - b \tilde{u} } - \paren*{ b - a } \cdot b \tilde{u}^2 } \tag{As $\alpha_{ \tilde{u} } = \frac{ b - a }{ a - b \tilde{u} }$} \\
&= \tilde{u} .
\end{align*}
Rearranging the equation in the lemma statement to isolate $U(v)$, we get:
\[
U(v) = v \cdot \frac{ 1 + \alpha_{ \tilde{u} } }{ \alpha_{ \tilde{u} } } - \frac{ v + bx - a }{ \alpha_{ \tilde{u} } \cdot z(v) } .
\]
Differentiating this, we get:
\[
z(v) = \frac{ 1 + \alpha_{ \tilde{u} } }{ \alpha_{ \tilde{u} } } - \frac{ 1 }{ \alpha_{ \tilde{u} } \cdot z(v) } + \frac{ v + bx - a }{ \alpha_{ \tilde{u} } \cdot z(v)^2 } \cdot \frac{ \del z(v) }{ \del v } .
\]
Multiplying throughout by $\alpha_{ \tilde{u} } \cdot z(v)^2$ and rearranging, we get:
\begin{align*}
\frac{ \del v }{ v + bx - a } &= \frac{ \del z(v) }{ \alpha_{ \tilde{u} } z(v)^3 - \paren*{ 1 + \alpha_{ \tilde{u} } } z(v)^2 + z(v) } \\
&= \frac{ \del z(v) }{ z(v) \paren*{ 1 - z(v) } \paren*{ 1 - \alpha_{ \tilde{u} } z(v) } } \\
&= \frac{ \del z(v) }{ z(v) } + \frac{ 1 + \alpha_{ \tilde{u} } - \alpha_{ \tilde{u} } z(v) }{ \paren*{ 1 - z(v) } \paren*{ 1 - \alpha_{ \tilde{u} } z(v) } } \del z(v) .
\end{align*}
Integrating both sides and setting $z\paren*{ s_{ \tilde{u} } } = \tilde{u}$, we get that, for all $v \in \bracket*{ s_{ \tilde{u} }, a }$:
\[
\ln \frac{ v + bx - a }{ s_{ \tilde{u} } + bx - a } = \ln \frac{ z(v) }{ \tilde{u} } + \int_{ \tilde{u} }^{ z(v) } \frac{ 1 + \alpha_{ \tilde{u} } - \alpha_{ \tilde{u} } y }{ \paren*{ 1 - y } \paren*{ 1 - \alpha_{ \tilde{u} } y } } \del y .
\]
Note that the right hand side, when viewed as a function of $z(v)$ is a smoothly increasing function for the range $z(v) \in \co*{ \tilde{u}, \min\paren*{ 1, 1/\alpha_{ \tilde{u} } } }$, where it increases from $0$ to $\infty$. Thus, the right hand side is a bijection from $z(v) \in \co*{ \tilde{u}, \min\paren*{ 1, 1/\alpha_{ \tilde{u} } } }$ to $\co*{ 0, \infty }$. It follows that for all $v \in \bracket*{ s_{ \tilde{u} }, a }$, a value $z(v)$ in this range is smoothly defined. From the definition of $U(v)$ in terms of $z(v)$, it can be verified that $U(v)$ satisfies the desired properties. For the ``moreover'' part, we first use \cref{eq:s:lb} to get that, for all $v \in \bracket*{ s_{ \tilde{u} }, a }$:
\[
\ln \frac{ a - \paren*{ 2b - a } \tilde{u} }{ a - bx } = \ln \frac{ b z(v) }{ v + bx - a } + \int_{ \tilde{u} }^{ z(v) } \frac{ 1 + \alpha_{ \tilde{u} } - \alpha_{ \tilde{u} } y }{ \paren*{ 1 - y } \paren*{ 1 - \alpha_{ \tilde{u} } y } } \del y .
\]
Recalling the function $\zeta_x(u)$ defined after \cref{eq:u:lb}, this gives:
\[
\zeta_x\paren*{ \tilde{u} } = \ln \frac{ b z(v) }{ v + bx - a } + \int_x^{ z(v) } \frac{ 1 + \alpha_{ \tilde{u} } - \alpha_{ \tilde{u} } y }{ \paren*{ 1 - y } \paren*{ 1 - \alpha_{ \tilde{u} } y } } \del y .
\]
From \cref{lemma:zeta:lb}, we have that $\zeta_x(u)$ is (strictly) decreasing and is $0$ at $u$. Thus, we have 
\begin{align*}
u \leq \tilde{u} &\iff \zeta_x\paren*{ \tilde{u} } \leq 0 \\
&\iff \ln \frac{ z(a) }{ x } + \int_x^{ z(a) } \frac{ 1 + \alpha_{ \tilde{u} } - \alpha_{ \tilde{u} } y }{ \paren*{ 1 - y } \paren*{ 1 - \alpha_{ \tilde{u} } y } } \del y \leq 0 \tag{Setting $v = a$ in the above equation} \\
&\iff z(a) \leq x \tag{As all factors are positive} \\
&\iff U(a) \leq b \tilde{u} \tag{As $z(a) = \frac{ bx }{ a + \alpha_{ \tilde{u} } \cdot \paren*{ a - U(a) } }$} .
\end{align*}
Observe that a similar derivation is possible with all the $\leq$ replaced by $=$. This finishes the proof of the lemma.
\end{proof}

\begin{proof}[Proof of \cref{lemma:kappa:lb}]
Note from \cref{lemma:zeta:lb} that the partial derivative in \cref{eq:kappa:lb} consists of two negative terms. We differentiate both these terms separately and then combine the results. Call the terms $\kappa^{(1)}_x\paren*{ \frac{ ax }{ b } - \frac{ \xi }{ b } }$ and $\kappa^{(2)}_x\paren*{ \frac{ ax }{ b } - \frac{ \xi }{ b } }$ respectively. For the first term, taking the logarithm and differentiating, we have:
\[
\frac{ 1 }{ \kappa^{(1)}_x\paren*{ \frac{ ax }{ b } - \frac{ \xi }{ b } } } \cdot \frac{ \del \kappa^{(1)}_x\paren*{ \frac{ ax }{ b } - \frac{ \xi }{ b } } }{ \del x } = - \frac{ b }{ a - bx } - \frac{ 1 }{ 1 - x } + \frac{ a }{ b - ax + \xi } - \frac{ b }{ a - bx + \xi } + \frac{ \paren*{ 2b - a } a }{ ab - \paren*{ 2b - a } \paren*{ ax - \xi } } .
\]
For the second term, we first simplify to get:
\begin{align*}
\kappa^{(2)}_x\paren*{ \frac{ ax }{ b } - \frac{ \xi }{ b } } &= - \paren*{ a - bx } \paren*{ 1 - x } \cdot \frac{ a - bx + \xi }{ b - a } \cdot \frac{ b \paren*{ b - a } }{ \paren*{ a - ax + \xi }^2 } \cdot \int_{ \frac{ ax }{ b } - \frac{ \xi }{ b } }^x \frac{ 1 }{ \paren*{ 1 - y } \paren*{ 1 - \alpha_{ \frac{ ax }{ b } - \frac{ \xi }{ b } } y }^2 } \del y \\
&= - \paren*{ a - bx } \paren*{ 1 - x } \cdot \frac{ a - bx + \xi }{ b - a } \cdot \frac{ b \paren*{ b - a } }{ \paren*{ a - ax + \xi }^2 } \cdot \int_{ \frac{ ax }{ b } - \frac{ \xi }{ b } }^x \frac{ 1 }{ \paren*{ 1 - y } \paren*{ 1 - \frac{ b - a }{ a - ax + \xi } y }^2 } \del y \tag{As $\alpha_u = \frac{ b - a }{ a - bu }$} \\
&= - b \paren*{ a - bx } \paren*{ 1 - x } \cdot \paren*{ a - bx + \xi } \cdot \int_{ \frac{ ax }{ b } - \frac{ \xi }{ b } }^x \frac{ 1 }{ \paren*{ 1 - y } \paren*{ a - ax + \xi - \paren*{ b - a } y }^2 } \del y \\
&= - b \paren*{ a - bx } \paren*{ 1 - x } \cdot \paren*{ a - bx + \xi } \cdot \int_0^{ x - \frac{ ax }{ b } + \frac{ \xi }{ b } } \frac{ 1 }{ \paren*{ 1 - x + y } \paren*{ a - bx + \xi + \paren*{ b - a } y }^2 } \del y \tag{Reparametrizing $y \to x - y$} \\
&= - b \cdot \int_0^{ x - \frac{ ax }{ b } + \frac{ \xi }{ b } } \frac{ \paren*{ a - bx } \paren*{ 1 - x }  \paren*{ a - bx + \xi } }{ \paren*{ 1 - x + y } \paren*{ a - bx + \xi + \paren*{ b - a } y }^2 } \del y .
\end{align*}
Differentiating this, we get:
\begin{multline*}
\frac{ \del \kappa^{(2)}_x\paren*{ \frac{ ax }{ b } - \frac{ \xi }{ b } } }{ \del x } = - \paren*{ b - a } \cdot \frac{ \paren*{ a - bx } \paren*{ 1 - x }  \paren*{ a - bx + \xi } }{ \paren*{ 1 - \frac{ ax }{ b } + \frac{ \xi }{ b } } \paren*{ a - bx + \xi + \paren*{ b - a } \paren*{ x - \frac{ ax }{ b } + \frac{ \xi }{ b } } }^2 } \\
- b \cdot \int_0^{ x - \frac{ ax }{ b } + \frac{ \xi }{ b } } \frac{ \partial }{ \partial x } \paren*{ \frac{ \paren*{ a - bx } \paren*{ 1 - x }  \paren*{ a - bx + \xi } }{ \paren*{ 1 - x + y } \paren*{ a - bx + \xi + \paren*{ b - a } y }^2 } } \del y .
\end{multline*}
This is the same as:
\[
\frac{ \del \kappa^{(2)}_x\paren*{ \frac{ ax }{ b } - \frac{ \xi }{ b } } }{ \del x } = \paren*{ b - a } \cdot \frac{ \kappa^{(1)}_x\paren*{ \frac{ ax }{ b } - \frac{ \xi }{ b } } }{ a - \paren*{ 2b - a } \paren*{ \frac{ ax }{ b } - \frac{ \xi }{ b } } } - b \cdot \int_0^{ x - \frac{ ax }{ b } + \frac{ \xi }{ b } } \frac{ \partial }{ \partial x } \paren*{ \frac{ \paren*{ a - bx } \paren*{ 1 - x }  \paren*{ a - bx + \xi } }{ \paren*{ 1 - x + y } \paren*{ a - bx + \xi + \paren*{ b - a } y }^2 } } \del y .
\]
We will later show that the integral above is negative, say $- \nabla$ for $\nabla > 0$. Assuming it for now, and combining the two terms, we get:
\begin{multline*}
\frac{ \del \kappa_x\paren*{ \frac{ ax }{ b } - \frac{ \xi }{ b } } }{ \del x } = b \nabla + \kappa^{(1)}_x\paren*{ \frac{ ax }{ b } - \frac{ \xi }{ b } } \cdot \paren*{ \frac{ b \paren*{ b - a } }{ ab - \paren*{ 2b - a } \paren*{ ax - \xi } } - \frac{ b }{ a - bx } + \frac{ a }{ b - ax + \xi } - \frac{ 1 }{ 1 - x } } \\
+ \kappa^{(1)}_x\paren*{ \frac{ ax }{ b } - \frac{ \xi }{ b } } \cdot \paren*{ \frac{ \paren*{ 2b - a } a }{ ab - \paren*{ 2b - a } \paren*{ ax - \xi } } - \frac{ b }{ a - bx + \xi } } .
\end{multline*}
We now consider the terms in pairs in the order they are written. For the first pair, note that:
\begin{multline*}
\frac{ b \paren*{ b - a } }{ ab - \paren*{ 2b - a } \paren*{ ax - \xi } } - \frac{ b }{ a - bx } \leq \frac{ b \paren*{ b - a } }{ ab - \paren*{ 2b - a } ax } - \frac{ b }{ a - bx } \\
= \frac{ b }{ a \paren*{ a - bx } \paren*{ b - \paren*{ 2b - a } x } } \cdot \paren*{ - a \paren*{ a - bx } - x \paren*{ a - b }^2 } < 0 .
\end{multline*}
For the second pair, note that:
\[
\frac{ a }{ b - ax + \xi } - \frac{ 1 }{ 1 - x } \leq \frac{ a }{ b - ax } - \frac{ 1 }{ 1 - x } = \frac{ a - b }{ \paren*{ b - ax } \paren*{ 1 - x } } < 0 .
\]
For the third pair, note that:
\begin{multline*}
\frac{ \paren*{ 2b - a } a }{ ab - \paren*{ 2b - a } \paren*{ ax - \xi } } - \frac{ b }{ a - bx + \xi } = \frac{ a \paren*{ 2b - a } \paren*{ a - bx } + a \paren*{ 2b - a } \xi - ab^2 + b \paren*{ 2b - a } ax - b \paren*{ 2b - a } \xi }{ \paren*{ ab - \paren*{ 2b - a } \paren*{ ax - \xi } } \paren*{ a - bx + \xi } } \\
= \frac{ -a \paren*{ b - a }^2 - \paren*{ b - a } \paren*{ 2b - a } \xi }{ \paren*{ ab - \paren*{ 2b - a } \paren*{ ax - \xi } } \paren*{ a - bx + \xi } } < 0 .
\end{multline*}
As all three terms are negative, and so is $\kappa^{(1)}_x\paren*{ \frac{ ax }{ b } - \frac{ \xi }{ b } }$, we get that the product is positive. We now finish by showing that $\nabla > 0$. For this, it suffices to show that the partial derivative inside the integral is negative. Taking the logarithm, it suffices to show that
\[
 - \frac{ 1 }{ 1 - x } - \frac{ b }{ a - bx } - \frac{ b }{ a - bx + \xi } + \frac{ 1 }{ 1 - x + y } + \frac{ 2b }{ a - bx + \xi + \paren*{ b - a } y } < 0
\]
This is easily seen by matching the positive and negative terms in the order they are written (the last term has a multiplier of $2$ and we treat it like two terms). 
\end{proof}

\section{\texorpdfstring{{Missing Proofs from \cref{sec:nash}}}{Missing Proofs from Section \getrefnumber{sec:nash}}}
\label{sec:nash-app}

{This section contains the omitted proofs from \cref{sec:nash}.}
We denote a feasible action plan in order $\AB$ by
${(\AB,x,y_W,y_L)}$: $x$ is Alice's allocation, and ${y_W},{y_L}$ are Bob's
allocations after winning and losing Alice's lottery, respectively.
Similarly, ${(\BA,y,x_W,x_L)}$ denotes Bob's allocation $y$, followed
by Alice's allocation ${x_W}$ after winning Bob's lottery and ${x_L}$
after losing it. For a mechanism $\mathcal{M}$ and a feasible allocation $z$, let
$\allocationUtilityMechanism{\mathcal M}{v}{z}=v\cdot z-\mathcal M(z)$
denote the buyer's utility at value $v$ from selecting $z$. The corresponding utilities are
{\begin{align*}
  {\utility_{\AliceMechanism,\BobMechanism}^{\AB}}\paren*{x,y_W,y_L}
    &=\allocationUtilityMechanism{\AliceMechanism}{a}{x}
      +x \cdot \allocationUtilityMechanism{\BobMechanism}{{\residualValue{b}{a}}}{y_W}
      +(1-x) \cdot \allocationUtilityMechanism{\BobMechanism}{b}{y_L},\\
  {\utility_{\AliceMechanism,\BobMechanism}^{\BA}}\paren*{y,x_W,x_L}
    &=\allocationUtilityMechanism{\BobMechanism}{b}{y}
      +y \cdot \allocationUtilityMechanism{\AliceMechanism}{{\residualValue{a}{b}}}{x_W}
      +(1-y) \cdot \allocationUtilityMechanism{\AliceMechanism}{a}{x_L}.
\end{align*}}
All allocations in these expressions have finite prices. For the selected plan
under $(\AliceMechanism,\BobMechanism)$, the first allocation is
$x=\allocDuo{A}{\AliceMechanism}{\BobMechanism}$ in order $\AB$ or
$y=\allocDuo{B}{\AliceMechanism}{\BobMechanism}$ in order $\BA$.
We write the selected continuation allocations as
${y_W}(\AliceMechanism,\BobMechanism),{y_L}(\AliceMechanism,\BobMechanism)$
or ${x_W}(\AliceMechanism,\BobMechanism),{x_L}(\AliceMechanism,\BobMechanism)$,
respectively, omitting the arguments only after fixing the mechanism
pair. All maxima over plan allocations below range over feasible
allocations in $[0,1]$.

All menus below include the null option and mixtures
of their listed options; zero payments are allowed.

We also use two elementary facts that hold for every optimal
allocation, including ties. If $z'$ and $z$ are optimal for
$\mathcal M$ at values $v'<v$, adding their optimality inequalities
gives $(v-v') \cdot (z-z')\geq0$, hence $z'\leq z$. If
$\utilityMechanism{\mathcal M}{v}=0$, every feasible $z>0$ instead
gives $\allocationUtilityMechanism{\mathcal M}{v'}{z}
\leq(v'-v) \cdot z<0$ at $0\leq v'<v$. Thus zero is the unique optimal
allocation at every smaller nonnegative value.

\phantomsection
\label{proof:nash-bertrand-equilibrium}

{
\begin{proof}[{Proof of \cref{lem:nash-a-at-least-b-BERTRAND-EQUILIBRIUM}}]
  {First suppose $a\geq b$.}
  {Let $\AliceMechanism=\FP{\residualValue{a}{b}}$ and
  $\BobMechanism=\FP{0}$.}
  Buying from Alice gives utility $b$ and revenue ${\residualValue{a}{b}}$ for Alice.
  No plan gives higher utility, and {since Bob's revenue is always zero,
  the buyer's secondary tie-breaking priority for Alice's revenue}
  selects these revenues; when ${\residualValue{a}{b}}=0$, all payments are zero.
  Against Bob's free good, the buyer obtains utility at least $b$, while
  welfare is at most $a$. Thus any deviation $\AliceMechanism'$ satisfies
  $\revDuo{A}{\AliceMechanism'}{\BobMechanism}\leq a-b={\residualValue{a}{b}}$.

  To bound a deviation $\BobMechanism'$ by Bob, note that Alice's
  fixed-price menu charges ${\residualValue{a}{b}}$ times her allocation probability.
  The expected value of any action plan is at most $b$ plus ${\residualValue{a}{b}}$
  times that probability: the buyer's value is at most $b$ when he
  does not obtain Alice's good and exactly $a=b+{\residualValue{a}{b}}$ otherwise.
  Consequently,
  \[
    \utilityDuo{\AliceMechanism}{\BobMechanism'}
    \leq b+\revDuo{A}{\AliceMechanism}{\BobMechanism'}
             -\revDuo{A}{\AliceMechanism}{\BobMechanism'}
             -\revDuo{B}{\AliceMechanism}{\BobMechanism'}
    =b-\revDuo{B}{\AliceMechanism}{\BobMechanism'}.
  \]
  Buying only from Alice still gives utility $b$, so Bob's deviation
  cannot earn positive revenue.

  {Now suppose $a<b$.
  Let $\AliceMechanism=\FP{0}$ and
  $\BobMechanism=\FP{\residualValue{b}{a}}$.
  Buying from Bob gives utility $a$ and revenue $\residualValue{b}{a}$ for Bob.
  No plan gives higher utility, and the buyer's primary tie-breaking
  priority for Bob's revenue selects these revenues.
  The same deviation bounds, with the sellers
  interchanged, show that neither seller can
  profitably deviate.}
\end{proof}
}

{For the remaining proofs, we retain the assumption $a\geq b$.}

\subsection{\texorpdfstring{{Missing Proofs from \cref{sec:nash-thresholds}}}{Missing Proofs from Section \getrefnumber{sec:nash-thresholds}}}

\label{sec:nash-app-thresholds}

{
We prove \cref{lem:nash-a-at-least-b-REVENUE-THRESHOLD-EQUIVALENCE}
through two supporting lemmas
{(\cref{lem:nash-zero-bob-revenue,lem:nash-alice-threshold-revenue})}.
We first establish a posted-price
deviation for Bob, used in the threshold argument and again in
\cref{sec:nash-app-no-joint}.
}

{
\begin{lemma}[A profitable deviation for Bob]
  \label{lem:nash-bob-posted-price-deviation}
  Suppose $(\AliceMechanism,\BobMechanism)$ selects $\AB$, and let
  $x=\allocDuo{A}{\AliceMechanism}{\BobMechanism}<1$.
  {If $x$ is optimal for $\AliceMechanism$ at value $a$ and
  $\utilityMechanism{\BobMechanism}{b}>0$, then Bob has a deviation
  that strictly increases his revenue.}
\end{lemma}
\begin{proof}
  Choose
  $0<\varepsilon<\utilityMechanism{\BobMechanism}{b}$ and let Bob deviate to
  $
    \BobMechanism'=\FP{p}
  $
  where
  $
    p=b-\utilityMechanism{\BobMechanism}{b}+\varepsilon.
  $
  This price exceeds Bob's original payment conditional on Alice's
  failure, while still giving positive utility from Bob alone.
  Indeed, $0<p<b$, and that original payment is at most
  $b-\utilityMechanism{\BobMechanism}{b}$. Hence
  \[
    \revDuo{B}{\AliceMechanism}{\BobMechanism}
      \leq(1-x) \cdot \bigl(b-\utilityMechanism{\BobMechanism}{b}\bigr)
      <(1-x) \cdot p.
  \]
  We show that the deviation earns at least $(1-x) \cdot p$ in either order.

  If $(\AliceMechanism,\BobMechanism')$ selects $\AB$, Alice faces
  value $a-\utilityMechanism{\BobMechanism}{b}+\varepsilon<a$.
  Since $x$ is optimal at $a$ for her unchanged mechanism, monotonicity
  gives $x'=\allocDuo{A}{\AliceMechanism}{\BobMechanism'}\leq x$.
  Bob sells after Alice's lottery fails, earning
  $(1-x') \cdot p\geq(1-x) \cdot p$.

  If the deviation selects $\BA$, we show that Bob sells surely.
  Skipping Bob gives
  $\utilityMechanism{\AliceMechanism}{a}$, while the feasible plan
  $(\AB,x,0,1)$ under $(\AliceMechanism,\BobMechanism')$ gives
  \[
    {\utility_{\AliceMechanism,\BobMechanism'}^{\AB}}\paren*{x,0,1}
      =\utilityMechanism{\AliceMechanism}{a}
        +(1-x) \cdot \bigl(\utilityMechanism{\BobMechanism}{b}-\varepsilon\bigr)
      >\utilityMechanism{\AliceMechanism}{a}.
  \]
  Thus any selected $\BA$ plan must give more utility than skipping
  Bob. His fixed-price offer makes the buyer's utility affine in
  Bob's allocation, so its slope is positive and Bob sells surely.
  His revenue is then $p\geq(1-x) \cdot p$, again exceeding his original
  revenue.
\end{proof}
}

{
\begin{lemma}
  \label{lem:nash-zero-bob-revenue}
  {Every pure Nash equilibrium $(\AliceMechanism,\BobMechanism)$
  satisfies $\revDuo{A}{\AliceMechanism}{\BobMechanism}\geq{\residualValue{a}{b}}$.
  If $\revDuo{B}{\AliceMechanism}{\BobMechanism}=0$, then
  $\revDuo{A}{\AliceMechanism}{\BobMechanism}={\residualValue{a}{b}}$.}
\end{lemma}
\begin{proof}
  Fix a pure Nash equilibrium $(\AliceMechanism,\BobMechanism)$.
  Alice can guarantee any positive price below ${\residualValue{a}{b}}$, since her
  good is strictly worth buying even after obtaining Bob's good.
  Taking prices up to ${\residualValue{a}{b}}$ gives
  $\revDuo{A}{\AliceMechanism}{\BobMechanism}\geq{\residualValue{a}{b}}$; for
  ${\residualValue{a}{b}}=0$, this follows from nonnegative payments.

  {Suppose now that $\revDuo{B}{\AliceMechanism}{\BobMechanism}=0$.}
  For every $\varepsilon\in(0,b)$, Bob's deviation to
  $\BobMechanism'=\FP{\varepsilon}$ must still yield zero revenue.
  {Bob is skipped on every reached branch. Let $z$ be
  the allocation purchased from Alice: $z=x$ in order $\AB$
  and $z=x_L$ in order $\BA$.
  For either case, if $z<1$, the feasible plan $(\AB,z,0,1)$ increases the
  buyer's utility by $(1-z)(b-\varepsilon)>0$, a contradiction.
  Thus $z=1$ is purchased at a finite price, so
  $\AliceMechanism(1)<\infty$.
  Comparing this choice with buying only from Bob,}
  \[
    {\utility_{\AliceMechanism,\BobMechanism'}^{\AB}}\paren*{1,0,0}
      =a-{\AliceMechanism}(1)
      \geq{\utility_{\AliceMechanism,\BobMechanism'}^{\AB}}\paren*{0,0,1}
      =b-\varepsilon.
  \]
  Letting $\varepsilon\downarrow0$ yields
  ${\AliceMechanism}(1)\leq{\residualValue{a}{b}}$. Since the pricing function
  is nondecreasing, each payment to Alice in the original pair is at
  most this amount. Thus
  $\revDuo{A}{\AliceMechanism}{\BobMechanism}\leq{\residualValue{a}{b}}$.
  {Together with the lower bound, this gives equality.}
\end{proof}
}

{
\begin{lemma}
  \label{lem:nash-alice-threshold-revenue}
  {If a pure Nash equilibrium $(\AliceMechanism,\BobMechanism)$
  satisfies $\revDuo{A}{\AliceMechanism}{\BobMechanism}={\residualValue{a}{b}}$, then
  $\revDuo{B}{\AliceMechanism}{\BobMechanism}=0$.}
\end{lemma}
\begin{proof}
{ Suppose, for a contradiction, that a pure Nash equilibrium
  $(\AliceMechanism,\BobMechanism)$ satisfies
  $\revDuo{A}{\AliceMechanism}{\BobMechanism}={\residualValue{a}{b}}$ and
  $\revDuo{B}{\AliceMechanism}{\BobMechanism}>0$.
  The proof structure is as follows:
  We begin by deriving a restriction on Bob's menu that holds regardless
  of the selected order. We then use it to find a profitable Bob
  deviation in order $\AB$, and construct a profitable Alice
  deviation in order $\BA$.

  We first establish $0<\utilityMechanism{\BobMechanism}{b}<b$.
  If $\utilityMechanism{\BobMechanism}{b}=0$, Alice can post a price
  ${\residualValue{a}{b}}<p<a$. In order $\AB$, her good is bought surely. In order
  $\BA$, Bob faces value $p-{\residualValue{a}{b}}<b$, so his unique optimal
  allocation is zero; Alice again
  sells surely. Either order raises her revenue above ${\residualValue{a}{b}}$.
  The upper bound follows because
  $\utilityMechanism{\BobMechanism}{b}=b$ would require a free full
  allocation from Bob, precluding positive Bob revenue.

  Both Alice deviations below make Bob's effective value tend to
  zero. To control his response, let
  $y_0=\max\{z:{\BobMechanism}(z)=0\}$ be his largest free
  allocation, which exists by compactness and satisfies $y_0<1$.
  For $t>0$, let
  $y(t)\in\arg\max_y\allocationUtilityMechanism{\BobMechanism}{t}{y}$.
  We will use the limits
  \begin{equation}
    \label{eq:nash-free-allocation-limit}
    y(t)\longrightarrow y_0,\qquad
    \frac{\utilityMechanism{\BobMechanism}{t}}{t}\longrightarrow y_0
    \quad(t\downarrow0),
  \end{equation}
  valid for every choice of maximizing allocation. Indeed,
  optimality relative to the free option gives $y(t)\geq y_0$ and
  ${\BobMechanism}(y(t))\leq t$. Compactness makes every
  accumulation point free, so $y(t)\to y_0$. The utility limit follows
  from nonnegative payments and
  $y_0\leq\utilityMechanism{\BobMechanism}{t}/t
    =y(t)-{\BobMechanism}(y(t))/t\leq y(t)$.

  We next show that $\utilityMechanism{\BobMechanism}{b}\leq a \cdot y_0$.
  Otherwise, Alice can charge more than ${\residualValue{a}{b}}$ for an almost
  certain allocation while making $\AB$ strictly preferable.
  Choose $k>0$ with
  $y_0 \cdot (a+k)<\utilityMechanism{\BobMechanism}{b}$, and let
  $\AliceMechanism'$ offer
  $\langle1-\eta,{\residualValue{a}{b}}+k \cdot \eta\rangle$ for small $\eta>0$.
  We verify the desired order by comparing both utilities with the
  utility from Alice alone. {Her continuation utilities are
  $\utilityMechanism{\AliceMechanism'}{{\residualValue{a}{b}}}=0$ and
  $\utilityMechanism{\AliceMechanism'}{a}=b-\eta \cdot (a+k)>0$.}
  \ygcomment{} In order $\AB$, Bob is available after Alice's lottery fails,
  adding $\eta \cdot \utilityMechanism{\BobMechanism}{b}$ to that baseline.
  This plan's utility tends to
  $b>\utilityMechanism{\BobMechanism}{b}$, so the full lottery is
  uniquely optimal in this order. In order $\BA$, Bob instead faces
  effective value $\eta \cdot (a+k)$. Thus
  \begin{align*}
    \utilityAB{\AliceMechanism'}{\BobMechanism}
      &=b-\eta \cdot (a+k)+\eta \cdot \utilityMechanism{\BobMechanism}{b},\\
    \utilityBA{\AliceMechanism'}{\BobMechanism}
      &=b-\eta \cdot (a+k)+\utilityMechanism{\BobMechanism}{\eta \cdot (a+k)}.
  \end{align*}
  Subtracting and using \eqref{eq:nash-free-allocation-limit} gives
  \[
    \utilityAB{\AliceMechanism'}{\BobMechanism}
      -\utilityBA{\AliceMechanism'}{\BobMechanism}
      =\eta \cdot \bigl[\utilityMechanism{\BobMechanism}{b}
                       -y_0 \cdot (a+k)+o(1)\bigr]>0.
  \]
  Hence Alice earns ${\residualValue{a}{b}}+k \cdot \eta>{\residualValue{a}{b}}$, a contradiction.
  We have proved $\utilityMechanism{\BobMechanism}{b}\leq a \cdot y_0$;
  since Bob's utility is positive, also $y_0>0$.
  We now distinguish the two original equilibrium orders.

  \emph{Case 1: Equilibrium Induces $\AB$.}
  Let $x=\allocDuo{A}{\AliceMechanism}{\BobMechanism}$, whose payment
  is ${\residualValue{a}{b}}$. Positive Bob revenue implies $x<1$.
  To apply \cref{lem:nash-bob-posted-price-deviation}, it remains to
  show that $x$ is optimal for Alice's menu at value $a$.
  The buyer can take Bob's free allocation first and, after failure,
  take an optimal standalone option from Alice. Therefore,
  \[
    a \cdot x-{\residualValue{a}{b}}+(1-x) \cdot \utilityMechanism{\BobMechanism}{b}
      =\utilityDuo{\AliceMechanism}{\BobMechanism}
      \geq y_0 \cdot b+(1-y_0) \cdot \utilityMechanism{\AliceMechanism}{a}.
  \]
  Rearranging and using the bound just proved yields
  \[
    0\leq(1-y_0) \cdot \bigl[\utilityMechanism{\AliceMechanism}{a}
                         -(a \cdot x-{\residualValue{a}{b}})\bigr]
      \leq(1-x) \cdot \bigl[\utilityMechanism{\BobMechanism}{b}-a \cdot y_0\bigr]
      \leq0.
  \]
  Since $y_0<1$, this forces
  $\utilityMechanism{\AliceMechanism}{a}=a \cdot x-{\residualValue{a}{b}}$.
  \Cref{lem:nash-bob-posted-price-deviation} now gives a profitable
  deviation for Bob, contradicting equilibrium.

  \emph{Case 2: Equilibrium Induces $\BA$.}
  We first establish
  $b-\utilityMechanism{\BobMechanism}{b}>{\residualValue{a}{b}} \cdot y_0$.
  This will provide room for Alice to raise her revenue with two
  options, one chosen after Bob succeeds and the other after he fails.
  Let $y=\allocDuo{B}{\AliceMechanism}{\BobMechanism}$ and
  $x_W=x_W(\AliceMechanism,\BobMechanism)$.
  Positive Bob payment implies $y>y_0$. With probability $y \cdot (1-x_W)$,
  the buyer obtains Bob's good but not Alice's, giving value
  $b$; all other outcomes give value at most $a$. Subtracting the
  sellers' revenues from this expected value gives
  \[
    \utilityDuo{\AliceMechanism}{\BobMechanism}
      \leq y \cdot (1 - x_W) \cdot b + (1 - y \cdot (1 - x_W)) \cdot a - {\residualValue{a}{b}}
            -\revDuo{B}{\AliceMechanism}{\BobMechanism}
      \leq b-y \cdot {\residualValue{a}{b}} \cdot (1-x_W)
            -\revDuo{B}{\AliceMechanism}{\BobMechanism}.
  \]
  For a lower bound, the buyer can take Alice's option $x_W$ first
  and visit Bob after failure. Since this option is optimal at
  value ${\residualValue{a}{b}}$, its payment is at most ${\residualValue{a}{b}} \cdot x_W$. Hence
  \begin{align*}
    \utilityDuo{\AliceMechanism}{\BobMechanism}
      \geq a \cdot x_W-{\AliceMechanism}(x_W)
                   +(1-x_W) \cdot \utilityMechanism{\BobMechanism}{b}
      \geq b-\bigl(b-\utilityMechanism{\BobMechanism}{b}\bigr) \cdot (1-x_W).
  \end{align*}
  Combining the two bounds gives
  \[
    \bigl(b-\utilityMechanism{\BobMechanism}{b}-y \cdot {\residualValue{a}{b}}\bigr) \cdot (1-x_W)
      \geq\revDuo{B}{\AliceMechanism}{\BobMechanism}>0,
  \]
  proving $b-\utilityMechanism{\BobMechanism}{b}>y \cdot {\residualValue{a}{b}}\geq y_0 \cdot {\residualValue{a}{b}}$.

  We now design Alice's deviation to meet two requirements: she must
  earn more than ${\residualValue{a}{b}}$ if the buyer selects $\BA$, and any globally
  optimal $\AB$ plan must buy her high option. {Choose $c$ so that
  ${\residualValue{a}{b}} \cdot y_0<(1-y_0) \cdot c<b-\utilityMechanism{\BobMechanism}{b}$.}
  The lower bound will make the gain from her high payment outweigh
  the discount on her low payment in order $\BA$. The upper bound
  will make $\BA$ preferable to taking her low option first.
  Since $\utilityMechanism{\BobMechanism}{b}\geq b \cdot y_0$, we also have
  $0<c<b$. For sufficiently small $\eta>0$, let $\AliceMechanism'$
  offer the low and high options
  \[
    \langle1-\eta,{\residualValue{a}{b}} \cdot (1-\eta)-{\residualValue{a}{b}} \cdot \eta^2\rangle,
    \qquad
    \langle1,{\residualValue{a}{b}}+c \cdot \eta-{\residualValue{a}{b}} \cdot \eta^2\rangle.
  \]

  These options give the required continuation choices. Upgrading
  from low to high adds allocation $\eta$ at payment
  $({\residualValue{a}{b}}+c) \cdot \eta$, with ${\residualValue{a}{b}}<{\residualValue{a}{b}}+c<a$. Thus the low option is
  optimal at value ${\residualValue{a}{b}}$, and the high option is uniquely optimal
  at value $a$ for small $\eta$.
  {Thus,
  $\utilityMechanism{\AliceMechanism'}{{\residualValue{a}{b}}}={\residualValue{a}{b}} \cdot \eta^2$ and
  $\utilityMechanism{\AliceMechanism'}{a}=b-c \cdot \eta+{\residualValue{a}{b}} \cdot \eta^2$.}
  Bob's effective value in order $\BA$ is consequently $c \cdot \eta$,
  so his allocation tends to $y_0$.

  To verify the first requirement, suppose the deviation selects
  $\BA$, and write $y(c \cdot \eta)$ for Bob's selected allocation.
  Alice's revenue gain is
  \begin{align*}
    \revDuo{A}{\AliceMechanism'}{\BobMechanism}-{\residualValue{a}{b}}
      &=y(c \cdot \eta) \cdot (-{\residualValue{a}{b}} \cdot \eta-{\residualValue{a}{b}} \cdot \eta^2)
           +(1-y(c \cdot \eta)) \cdot (c \cdot \eta-{\residualValue{a}{b}} \cdot \eta^2)\\
      &=\eta \cdot \bigl[c-(c+{\residualValue{a}{b}}) \cdot y(c \cdot \eta)\bigr]-{\residualValue{a}{b}} \cdot \eta^2>0,
  \end{align*}
  by \eqref{eq:nash-free-allocation-limit} and $(1-y_0) \cdot c>{\residualValue{a}{b}} \cdot y_0$.
  Thus selecting $\BA$ contradicts Alice's best-response condition.

  To verify the second requirement, we rule out both skipping Alice
  and taking her low option first. Skipping Alice gives
  $\utilityMechanism{\BobMechanism}{b}$, whereas
  $\utilityBA{\AliceMechanism'}{\BobMechanism}\to
    b>\utilityMechanism{\BobMechanism}{b}$.
  For the low option, choose
  $y_b\in\arg\max_y\allocationUtilityMechanism{\BobMechanism}{b}{y}$.
  The best $\AB$ plan starting with that option uses $y_b$ after
  failure. Its utility and the optimal $\BA$ utility are
  \begin{align*}
    {\utility_{\AliceMechanism',\BobMechanism}^{\AB}}\paren*{1-\eta,0,y_b}
      &=b+\eta \cdot \bigl(\utilityMechanism{\BobMechanism}{b}-b\bigr)
             +{\residualValue{a}{b}} \cdot \eta^2,\\
    \utilityBA{\AliceMechanism'}{\BobMechanism}
      &=b-c \cdot \eta+{\residualValue{a}{b}} \cdot \eta^2
             +\utilityMechanism{\BobMechanism}{c \cdot \eta}.
  \end{align*}
  Subtracting gives
  \begin{align*}
    &\utilityBA{\AliceMechanism'}{\BobMechanism}
       -{\utility_{\AliceMechanism',\BobMechanism}^{\AB}}\paren*{1-\eta,0,y_b}\\
    &\quad=\eta \cdot \bigl(b-\utilityMechanism{\BobMechanism}{b}-c\bigr)
            +\utilityMechanism{\BobMechanism}{c \cdot \eta}\\
    &\quad=\eta \cdot \bigl[b-\utilityMechanism{\BobMechanism}{b}
                         -(1-y_0) \cdot c+o(1)\bigr]>0,
  \end{align*}
  by the upper bound on $c$ and \eqref{eq:nash-free-allocation-limit}.
  Utility is affine over mixtures of Alice's options, so every
  globally optimal $\AB$ plan must take her high option.
  Its payment ${\residualValue{a}{b}}+c \cdot \eta-{\residualValue{a}{b}} \cdot \eta^2$ exceeds ${\residualValue{a}{b}}$ for small
  $\eta$, again contradicting equilibrium.
}
\end{proof}
}

\phantomsection

\label{proof:nash-revenue-thresholds}

{
\begin{proof}[{Proof of \cref{lem:nash-a-at-least-b-REVENUE-THRESHOLD-EQUIVALENCE}}]
  This follows directly from
  \cref{lem:nash-zero-bob-revenue,lem:nash-alice-threshold-revenue}.
\end{proof}
}

\subsection{\texorpdfstring{{Missing Proofs from \cref{sec:nash-no-joint}}}{Missing Proofs from Section \getrefnumber{sec:nash-no-joint}}}
\label{sec:nash-app-improvement}

\label{sec:nash-app-structure}

{
\begin{lemma}[Strict gains from visiting both sellers]
  \label{lem:nash-strict-buyer-utility}
  Let $(\AliceMechanism,\BobMechanism)$ be a pure Nash equilibrium
  with $\revDuo{A}{\AliceMechanism}{\BobMechanism}>{\residualValue{a}{b}}$ and
  $\revDuo{B}{\AliceMechanism}{\BobMechanism}>0$. Then
  \[
    \utilityDuo{\AliceMechanism}{\BobMechanism}
      >\max\bigl\{\utilityMechanism{\AliceMechanism}{a},
                    \utilityMechanism{\BobMechanism}{b}\bigr\}.
  \]
\end{lemma}
\begin{proof}
  Fix such an equilibrium $(\AliceMechanism,\BobMechanism)$ and
  suppose, for a contradiction, that the asserted inequality fails.
  The buyer can visit either seller alone, so his utility is at least
  each standalone utility. It must therefore equal one of them.

  First, let
  $\utilityDuo{\AliceMechanism}{\BobMechanism}
      =\utilityMechanism{\AliceMechanism}{a}$.
  Expected value is at most $a$, and Bob's revenue is positive. Thus
  {Alice can choose
  $0<\varepsilon<\revDuo{B}{\AliceMechanism}{\BobMechanism}$, giving}
  {\[
    p=a-\utilityDuo{\AliceMechanism}{\BobMechanism}-\varepsilon
       >a-\utilityDuo{\AliceMechanism}{\BobMechanism}
          -\revDuo{B}{\AliceMechanism}{\BobMechanism}
       \geq\revDuo{A}{\AliceMechanism}{\BobMechanism}>{\residualValue{a}{b}}.
  \]}
  Let $\AliceMechanism'=\FP{p}$. Originally,
  {\cref{lemma:buyer}} implies
  $\utilityMechanism{\BobMechanism}{b-
       \utilityMechanism{\AliceMechanism}{a}
       +\utilityMechanism{\AliceMechanism}{{\residualValue{a}{b}}}}=0$.
  After the deviation, Bob's effective value in order $\BA$ is
  $b-\utilityDuo{\AliceMechanism}{\BobMechanism}-\varepsilon$,
  which is nonnegative because $p>{\residualValue{a}{b}}$ and strictly smaller than
  that original value. His unique optimal allocation is therefore
  zero, and Alice sells surely. In order $\AB$, buying Alice surely
  gives utility $\utilityDuo{\AliceMechanism}{\BobMechanism}+\varepsilon$,
  strictly above $\utilityMechanism{\BobMechanism}{b}$. Again her
  full good is bought, contradicting equilibrium.

  Second, let
  $\utilityDuo{\AliceMechanism}{\BobMechanism}
      =\utilityMechanism{\BobMechanism}{b}$.
  The welfare bound and Alice's revenue being strictly above ${\residualValue{a}{b}}$
  imply
  \[
    \revDuo{B}{\AliceMechanism}{\BobMechanism}
      \leq a-\revDuo{A}{\AliceMechanism}{\BobMechanism}
              -\utilityDuo{\AliceMechanism}{\BobMechanism}
      <b-\utilityDuo{\AliceMechanism}{\BobMechanism}.
  \]
  Choose
  $0<\varepsilon<b-\utilityDuo{\AliceMechanism}{\BobMechanism}
  -\revDuo{B}{\AliceMechanism}{\BobMechanism}$
  and let Bob post
  $
  p=b-\utilityDuo{\AliceMechanism}{\BobMechanism}-\varepsilon
  >\revDuo{B}{\AliceMechanism}{\BobMechanism}.
  $
  By {\cref{lemma:buyer}}, originally
  $\utilityMechanism{\AliceMechanism}{a-
        \utilityDuo{\AliceMechanism}{\BobMechanism}}=0$.
  In order $\AB$, Alice's effective value falls by $\varepsilon$,
  so she is skipped and Bob sells surely. In order $\BA$, skipping
  Bob gives $\utilityMechanism{\AliceMechanism}{a}$, whereas buying
  him alone gives the strictly larger utility
  $\utilityDuo{\AliceMechanism}{\BobMechanism}+\varepsilon$.
  Since the fixed-price objective is affine in Bob's allocation, any
  optimal plan in this order also buys him surely. His revenue
  increases in either order, giving the required contradiction.
\end{proof}
}

\phantomsection

\label{proof:nash-standalone-optimality}

{
\begin{proof}[{Proof of \cref{lem:nash-a-at-least-b-UTILITY-MAXIMIZING-ALLOCATIONS}}]

  Fix such an equilibrium $(\AliceMechanism,\BobMechanism)$.
  Suppose first that its selected order is $\AB$, and let
  $x=\allocDuo{A}{\AliceMechanism}{\BobMechanism}$ and
  ${y_L}={y_L}(\AliceMechanism,\BobMechanism)$.
  Positive Bob revenue implies $x<1$, and
  {$\utilityDuo{\AliceMechanism}{\BobMechanism}
      =\allocationUtilityMechanism{\AliceMechanism}{a}{x}
          +(1-x) \cdot \utilityMechanism{\BobMechanism}{b}$.}
  For sufficiently small $0<\delta<1-x$, let $\AliceMechanism'$ offer
  $\langle x+\delta,{\AliceMechanism}(x)+a \cdot \delta\rangle$.
  This lottery is designed to preserve exactly the utility of
  Alice's original option $x$ at value $a${:
  $\utilityMechanism{\AliceMechanism'}{a}
      =\allocationUtilityMechanism{\AliceMechanism'}{a}{x+\delta}
      =\allocationUtilityMechanism{\AliceMechanism}{a}{x}$.}
  At value ${\residualValue{a}{b}}$, it is never bought, since
  ${\AliceMechanism}(x)=\revDuo{A}{\AliceMechanism}{\BobMechanism}
       >{\residualValue{a}{b}}$.
  The feasible plan ${(\AB,x+\delta,0,y_L)}$ under
  $(\AliceMechanism',\BobMechanism)$ gives
  ${\utility_{\AliceMechanism',\BobMechanism}^{\AB}}\paren*{x+\delta,0,y_L}
      =\utilityDuo{\AliceMechanism}{\BobMechanism}
         -\delta \cdot \utilityMechanism{\BobMechanism}{b}
      >\utilityMechanism{\BobMechanism}{b}$
  for small $\delta$, by \cref{lem:nash-strict-buyer-utility}.
  Thus selecting $\AB$ under $(\AliceMechanism',\BobMechanism)$
  entails buying Alice's full lottery and increasing her revenue.
  Equilibrium therefore forces $\BA$. Utility preservation gives
  \[
    \utilityBA{\AliceMechanism'}{\BobMechanism}
      =\max_y{\utility_{\AliceMechanism',\BobMechanism}^{\BA}}\paren*{y,0,x+\delta}
      =\max_y{\utility_{\AliceMechanism,\BobMechanism}^{\BA}}\paren*{y,0,x}
      \leq\utilityDuo{\AliceMechanism}{\BobMechanism}.
  \]
  The last maximum is independent of $\delta$ and at least the
  preceding $\AB$ utility. Letting $\delta\downarrow0$ therefore gives
  \[
    \max_y{\utility_{\AliceMechanism,\BobMechanism}^{\BA}}\paren*{y,0,x}
      =\utilityDuo{\AliceMechanism}{\BobMechanism}.
  \]
  Choose a maximizing $y^\ast$. It is below one: otherwise the plan
  gives at most $\utilityMechanism{\BobMechanism}{b}$, strictly less
  than the right-hand side by \cref{lem:nash-strict-buyer-utility}.
  If $x$ is not standalone optimal, choose an optimal allocation
  $\widehat x$ from $\AliceMechanism$ at value $a$. Replacing Alice's
  continuation in ${(\BA,y^\ast,0,x)}$ under the original mechanisms gives
  \begin{align*}
    &{\utility_{\AliceMechanism,\BobMechanism}^{\BA}}\paren*{y^\ast,0,\widehat x}\\
    &\quad=\utilityDuo{\AliceMechanism}{\BobMechanism}
      +(1-y^\ast) \cdot \bigl[
        \utilityMechanism{\AliceMechanism}{a}
        -\allocationUtilityMechanism{\AliceMechanism}{a}{x}\bigr]
      >\utilityDuo{\AliceMechanism}{\BobMechanism}.
  \end{align*}
  The gain is strict because the improved continuation is reached
  with probability $1-y^\ast>0$. This contradicts optimality of the
  original buyer utility.

  {The case of order $\BA$ is analogous. Let
  $y=\allocDuo{B}{\AliceMechanism}{\BobMechanism}$; then $y<1$,
  since otherwise Alice earns at most ${\residualValue{a}{b}}$.
  For small $0<\delta<1-y$, let $\BobMechanism'$ offer
  $\langle y+\delta,{\BobMechanism}(y)+b \cdot \delta\rangle$,
  preserving the utility of Bob's chosen lottery at value $b$.
  The plan $(\BA,y+\delta,x_W,x_L)$ under
  $(\AliceMechanism,\BobMechanism')$, using Alice's original
  continuations, loses
  $\delta \cdot {\aug_{\AliceMechanism}(b,a)}$ in utility.
  For small $\delta$, it still beats visiting Alice alone, so
  selecting $\BA$ would increase Bob's revenue. Equilibrium therefore
  forces $\AB$, and the same squeeze argument gives
  \[
    \max_x{\utility_{\AliceMechanism,\BobMechanism}^{\AB}}\paren*{x,0,y}
      =\utilityDuo{\AliceMechanism}{\BobMechanism}.
  \]
  Any maximizing $x^\ast$ is below one, since $x^\ast=1$ gives at most
  $\utilityMechanism{\AliceMechanism}{a}
    <\utilityDuo{\AliceMechanism}{\BobMechanism}$.
  If $y$ were not standalone optimal, improving Bob's continuation
  would increase utility by
  $(1-x^\ast) \cdot [\utilityMechanism{\BobMechanism}{b}
    -\allocationUtilityMechanism{\BobMechanism}{b}{y}]>0$,
  again contradicting optimality under the original mechanisms.}
\end{proof}
}

\label{sec:nash-app-no-joint}

\phantomsection

\label{proof:nash-no-joint-improvement}

{
\begin{proof}[{Proof of \cref{lem:nash-a-at-least-b-NO-JOINT-REVENUE-IMPROVEMENT}}]
  Suppose such an equilibrium $(\AliceMechanism,\BobMechanism)$
  exists. By \cref{lem:nash-strict-buyer-utility,lem:nash-a-at-least-b-UTILITY-MAXIMIZING-ALLOCATIONS},
  the buyer's utility strictly exceeds both standalone utilities, and
  the first allocation is standalone optimal.

  {We show that one seller can profitably deviate in each of the
  two possible selected orders.}

  {
  \emph{Order $\AB$.}
  We apply \cref{lem:nash-bob-posted-price-deviation}.
  Let $x=\allocDuo{A}{\AliceMechanism}{\BobMechanism}$.
  Its standalone optimality is already known, so it remains to check
  $x<1$ and $\utilityMechanism{\BobMechanism}{b}>0$.
  Both follow from
  $\utilityDuo{\AliceMechanism}{\BobMechanism}
    =\utilityMechanism{\AliceMechanism}{a}
       +(1-x) \cdot \utilityMechanism{\BobMechanism}{b}
    >\utilityMechanism{\AliceMechanism}{a}$.
  The lemma therefore gives a profitable deviation for Bob,
  contradicting equilibrium.
  }

  {
  \emph{Order $\BA$.}
  We construct a profitable deviation for Alice that works whether the
  buyer keeps order $\BA$ or switches to $\AB$.
  Let $y=\allocDuo{B}{\AliceMechanism}{\BobMechanism}$,
  $x_W=x_W(\AliceMechanism,\BobMechanism)$, and
  $x_L=x_L(\AliceMechanism,\BobMechanism)$. Write Alice's selected
  continuation options as $\langle x_W,\ell\rangle$
  after Bob succeeds and $\langle x_L,h\rangle$ after he fails.
  Positive Bob revenue gives $y>0$, while Alice's revenue exceeding
  ${\residualValue{a}{b}}$ gives $y<1$. Thus
  \[
    \ell\leq{\residualValue{a}{b}}<\revDuo{A}{\AliceMechanism}{\BobMechanism}<h,
    \qquad
    \revDuo{A}{\AliceMechanism}{\BobMechanism}=y \cdot \ell+(1-y) \cdot h.
  \]
  The continuation choices give
  $\utilityMechanism{\AliceMechanism}{{\residualValue{a}{b}}}={\residualValue{a}{b}} \cdot x_W-\ell$
  and $\utilityMechanism{\AliceMechanism}{a}=a \cdot x_L-h$; Bob's
  standalone optimality gives
  $\utilityMechanism{\BobMechanism}{b}=b \cdot y-{\BobMechanism}(y)$.

  To control a possible switch to $\AB$, compare the original plan
  with taking Alice's low option first. Since $y$ is optimal for Bob
  at $b$, the plan $(\AB,x_W,0,y)$ maximizes utility conditional on
  that first option. {The two utilities are
  {
  \begin{align*}
    &\utilityDuo{\AliceMechanism}{\BobMechanism}
       ={\utility_{\AliceMechanism,\BobMechanism}^{\BA}}\paren*{y,x_W,x_L}
       =\utilityMechanism{\BobMechanism}{b}
        +y \cdot \utilityMechanism{\AliceMechanism}{{\residualValue{a}{b}}}
        +(1-y) \cdot \utilityMechanism{\AliceMechanism}{a},\\
    &{\utility_{\AliceMechanism,\BobMechanism}^{\AB}}\paren*{x_W,0,y}
       =a \cdot x_W-\ell+(1-x_W) \cdot \utilityMechanism{\BobMechanism}{b}
       =\utilityMechanism{\AliceMechanism}{{\residualValue{a}{b}}}+b \cdot x_W
        +(1-x_W) \cdot \utilityMechanism{\BobMechanism}{b}.
  \end{align*}
  }
  Subtracting these expressions gives the nonnegative utility gap}
  {
  \begin{equation}
    \label{eq:nash-ba-utility-gap}
      G:=\utilityDuo{\AliceMechanism}{\BobMechanism}
          -{\utility_{\AliceMechanism,\BobMechanism}^{\AB}}\paren*{x_W,0,y}
        =(1-y) \cdot \bigl(\utilityMechanism{\AliceMechanism}{a}
                      -\utilityMechanism{\AliceMechanism}{{\residualValue{a}{b}}}\bigr)
          -x_W \cdot \bigl(b-\utilityMechanism{\BobMechanism}{b}\bigr)\geq0.
  \end{equation}
  }

  {Our deviation will shrink the low allocation and payment by roughly
  a factor $1-\eta$ and replace the high option by a sure allocation
  at payment $h+c \cdot \eta$. To offset the lost low payment while keeping
  the plan taking the low option first suboptimal, we seek
  $\frac{y \cdot \ell}{1-y}<c<\utilityMechanism{\AliceMechanism}{a}$.
  The lower bound will ensure a revenue gain.
  The upper bound is designed to limit the increase in the high option's
  payment so that the proposed $\BA$ plan is strictly better than
  taking Alice's low option first.
  To show that this interval is nonempty,
  we prove $(1-y) \cdot \utilityMechanism{\AliceMechanism}{a}>y \cdot \ell$.
  First, use the definition of $G$ and then
  $\ell={\residualValue{a}{b}} \cdot x_W-\utilityMechanism{\AliceMechanism}{{\residualValue{a}{b}}}$ to obtain
  \begin{align*}
    (1-y) \cdot \utilityMechanism{\AliceMechanism}{a}-y \cdot \ell
      &=G+(1-y) \cdot \utilityMechanism{\AliceMechanism}{{\residualValue{a}{b}}}
         +x_W \cdot \bigl(b-\utilityMechanism{\BobMechanism}{b}\bigr)-y \cdot \ell\\
      &=G+\utilityMechanism{\AliceMechanism}{{\residualValue{a}{b}}}
         +x_W \cdot \bigl(b-\utilityMechanism{\BobMechanism}{b}-y \cdot {\residualValue{a}{b}}\bigr).
  \end{align*}
  Next, subtract $x_W$ times the left-hand side from both sides.
  Using $\utilityMechanism{\BobMechanism}{b}=b \cdot y-{\BobMechanism}(y)$,
  $a=b+{\residualValue{a}{b}}$, and
  $a-\utilityMechanism{\AliceMechanism}{a}=h+a \cdot (1-x_L)$ gives
  \begin{align*}
    &(1-x_W) \cdot \bigl[(1-y) \cdot \utilityMechanism{\AliceMechanism}{a}-y \cdot \ell\bigr]\\
    &\quad=G+\utilityMechanism{\AliceMechanism}{{\residualValue{a}{b}}}
       +x_W \cdot \bigl[b-\utilityMechanism{\BobMechanism}{b}-y \cdot {\residualValue{a}{b}}
                   -(1-y) \cdot \utilityMechanism{\AliceMechanism}{a}+y \cdot \ell\bigr]\\
    &\quad=G+\utilityMechanism{\AliceMechanism}{{\residualValue{a}{b}}}
       +x_W \cdot \bigl[(1-y) \cdot (a-\utilityMechanism{\AliceMechanism}{a})
                   +y \cdot \ell-{\residualValue{a}{b}}+{\BobMechanism}(y)\bigr]\\
    &\quad=G+\utilityMechanism{\AliceMechanism}{{\residualValue{a}{b}}}
       +x_W \cdot \bigl[(1-y) \cdot h+y \cdot \ell-{\residualValue{a}{b}}+{\BobMechanism}(y)
                   +(1-y) \cdot a \cdot (1-x_L)\bigr]\\
    &\quad=G+\utilityMechanism{\AliceMechanism}{{\residualValue{a}{b}}}
       +x_W \cdot \bigl[\revDuo{A}{\AliceMechanism}{\BobMechanism}
                   -{\residualValue{a}{b}}+{\BobMechanism}(y)
                   +(1-y) \cdot a \cdot (1-x_L)\bigr]>0.
  \end{align*}
  {We now show that the last expression is strictly positive.
  Note that every term is nonnegative.} Thus, if $x_W>0$,
  the bracketed term is positive because Alice's original revenue
  exceeds ${\residualValue{a}{b}}$; if $x_W=0$, then
  $\ell=\utilityMechanism{\AliceMechanism}{{\residualValue{a}{b}}}=0$ and
  $G=\utilityDuo{\AliceMechanism}{\BobMechanism}
       -\utilityMechanism{\BobMechanism}{b}>0$.
  Hence $x_W<1$ and
  $(1-y) \cdot \utilityMechanism{\AliceMechanism}{a}>y \cdot \ell$, as required.}

  Fix such a $c$, and let $\AliceMechanism'$ offer
  $\langle x_W',\ell'\rangle$ and $\langle1,h'\rangle$, where
  \[
    x_W'=(1-\eta) \cdot x_W,\qquad
    \ell'=(1-\eta-\eta^2) \cdot \ell,\qquad h'=h+c \cdot \eta
  \]
  for sufficiently small $0<\eta<1/2$.
  {{At value ${\residualValue{a}{b}}$, the low option
  $\langle x_W',\ell'\rangle$ is optimal: it gives nonnegative utility,
  whereas the high option gives negative utility since
  $h'>h>{\residualValue{a}{b}}$. Every tied continuation pays $\ell'$.
  At value $a$, the high option $\langle1,h'\rangle$ is uniquely optimal
  for sufficiently small $\eta$.} Consequently,
  $\utilityMechanism{\AliceMechanism'}{{\residualValue{a}{b}}}
    =(1-\eta) \cdot \utilityMechanism{\AliceMechanism}{{\residualValue{a}{b}}}+\eta^2 \cdot \ell$
  and $\utilityMechanism{\AliceMechanism'}{a}
    =\utilityMechanism{\AliceMechanism}{a}+a \cdot (1-x_L)-c \cdot \eta$;
  their difference is strictly positive for small $\eta$.}

  Next, we rule out selecting an $\AB$ plan that skips Alice or takes
  her low option first. It suffices to show that the feasible plan
  $(\BA,y,x_W',1)$ under $(\AliceMechanism',\BobMechanism)$ beats
  both alternatives. {Using the utility expressions
  as above, then substituting the new continuation utilities and
  $x_W'=(1-\eta) \cdot x_W$, gives
  \begin{align*}
    &{\utility_{\AliceMechanism',\BobMechanism}^{\BA}}\paren*{y,x_W',1}
       -{\utility_{\AliceMechanism',\BobMechanism}^{\AB}}\paren*{x_W',0,y}\\
    &\quad=(1-y) \cdot \bigl(\utilityMechanism{\AliceMechanism'}{a}
                       -\utilityMechanism{\AliceMechanism'}{{\residualValue{a}{b}}}\bigr)
         -x_W' \cdot \bigl(b-\utilityMechanism{\BobMechanism}{b}\bigr)\\
    &\quad=(1-y) \cdot \bigl[\utilityMechanism{\AliceMechanism}{a}
          +a \cdot (1-x_L)-c \cdot \eta
          -(1-\eta) \cdot \utilityMechanism{\AliceMechanism}{{\residualValue{a}{b}}}-\eta^2 \cdot \ell\bigr]
          -(1-\eta) \cdot x_W \cdot \bigl(b-\utilityMechanism{\BobMechanism}{b}\bigr)\\
    &\quad=(1-\eta) \cdot \Bigl[(1-y) \cdot \bigl(\utilityMechanism{\AliceMechanism}{a}
                       -\utilityMechanism{\AliceMechanism}{{\residualValue{a}{b}}}\bigr)
          -x_W \cdot \bigl(b-\utilityMechanism{\BobMechanism}{b}\bigr)\Bigr]\\
    &\qquad\quad +(1-y) \cdot \bigl[a \cdot (1-x_L)
                    +\eta \cdot \utilityMechanism{\AliceMechanism}{a}
                    -c \cdot \eta-\eta^2 \cdot \ell\bigr]\\
    &\quad=(1-\eta) \cdot G+(1-y) \cdot a \cdot (1-x_L)
       +(1-y) \cdot \eta \cdot \bigl[\utilityMechanism{\AliceMechanism}{a}
                         -c-\eta \cdot \ell\bigr]>0.
  \end{align*}
  Here the last equality uses the definition of $G$. The first two
  terms are nonnegative, and the last is positive for small $\eta$
  because $c<\utilityMechanism{\AliceMechanism}{a}$.} Bob's standalone
  optimality again makes the subtracted term the best utility from
  taking the low option first. The candidate $\BA$ plan also beats
  Bob alone: as $\eta\downarrow0$, its utility tends to
  $\utilityDuo{\AliceMechanism}{\BobMechanism}+(1-y) \cdot a \cdot (1-x_L)
     >\utilityMechanism{\BobMechanism}{b}$.

  Finally, Alice earns more in either selected order under
  $(\AliceMechanism',\BobMechanism)$. If the buyer selects $\BA$,
  Bob faces value
  $b-{\aug_{\AliceMechanism'}(b,a)}<b$.
  Since $y$ is optimal for Bob at $b$, monotonicity gives
  $y'=\allocDuo{B}{\AliceMechanism'}{\BobMechanism}\leq y$.
  As $h'>\ell'$, the lower bound on $c$ now yields
  \begin{align*}
    \revDuo{A}{\AliceMechanism'}{\BobMechanism}
      &=y' \cdot \ell'+(1-y') \cdot h'\geq y \cdot \ell'+(1-y) \cdot h'\\
      &=\revDuo{A}{\AliceMechanism}{\BobMechanism}
        +\eta \cdot [(1-y) \cdot c-y \cdot \ell]-y \cdot \eta^2 \cdot \ell
        >\revDuo{A}{\AliceMechanism}{\BobMechanism}.
  \end{align*}
  If the buyer selects $\AB$, the strict comparisons above exclude
  skipping Alice and taking her low option. Since utility is affine
  over mixtures of her options, any selected plan in this order
  must take the high option, paying
  $h'>h>\revDuo{A}{\AliceMechanism}{\BobMechanism}$.
  Thus Alice has a profitable deviation regardless of the buyer's
  optimal response, contradicting equilibrium.
  }
\end{proof}
}

\end{document}